\documentclass[article]{JHEP3}
\pdfoutput=1

\usepackage{amsmath,amssymb,amsthm,amscd}
\usepackage{epsfig}
\usepackage{psfrag,comment}
\usepackage{graphicx}

\newcommand{\CB}{{\cal B}}
\newcommand{\CC}{{\cal C}}

\newcommand{\CF}{{\cal F}}

\newcommand{\CL}{{\cal L}}

\newcommand{\CO}{{\cal O}}
\newcommand{\CP}{{\cal P}}

\newcommand{\CR}{{\cal R}}
\newcommand{\CS}{{\cal S}}

\newcommand{\CX}{{\cal X}}

\def\BN{{\mathbb N}}
\def\BZ{{\mathbb Z}}
\def\BR{{\mathbb R}}
\def\BC{{\mathbb C}}
\def\BP{{\mathbb P}}

\def\BS{{\mathbb S}}

\newcommand{\be}{\begin{equation}}
\newcommand{\ee}{\end{equation}}
\newcommand{\ba}{\begin{aligned}}
\newcommand{\ea}{\end{aligned}}
\newcommand{\bea}{\begin{eqnarray}}
\newcommand{\eea}{\end{eqnarray}}
\newcommand{\bean}{\begin{eqnarray*}}
\newcommand{\eean}{\end{eqnarray*}}

\def\r{\right\rangle}
\def\a{\alpha'}
\def\1{\mathbf{1}}
\def\0{|\1\r}
\def\im{{\mathbb{I}}{\mathrm{m}}}
\def\re{{\mathbb{R}}{\mathrm{e}}}

\def\disc{{\mathrm{Disc\,}}}
\newcommand{\tr}{{\rm Tr}}
\newcommand{\rme}{{\rm e}}
\newcommand{\rmi}{{\rm i}}
\newcommand{\rmd}{{\rm d}}

\renewcommand{\mod}{\mbox{~mod~}}

\def\XXint#1#2#3{{\setbox0=\hbox{$#1{#2#3}{\int}$}
     \vcenter{\hbox{$#2#3$}}\kern-.5\wd0}}

\newdimen\tableauside\tableauside=1.0ex
\newdimen\tableaurule\tableaurule=0.4pt
\newdimen\tableaustep
\def\phantomhrule#1{\hbox{\vbox to0pt{\hrule height\tableaurule width#1\vss}}}
\def\phantomvrule#1{\vbox{\hbox to0pt{\vrule width\tableaurule height#1\hss}}}
\def\sqr{\vbox{%
  \phantomhrule\tableaustep
  \hbox{\phantomvrule\tableaustep\kern\tableaustep\phantomvrule\tableaustep}%
  \hbox{\vbox{\phantomhrule\tableauside}\kern-\tableaurule}}}
\def\squares#1{\hbox{\count0=#1\noindent\loop\sqr
  \advance\count0 by-1 \ifnum\count0>0\repeat}}
\def\tableau#1{\vcenter{\offinterlineskip
  \tableaustep=\tableauside\advance\tableaustep by-\tableaurule
  \kern\normallineskip\hbox
    {\kern\normallineskip\vbox
      {\gettableau#1 0 }%
     \kern\normallineskip\kern\tableaurule}%
  \kern\normallineskip\kern\tableaurule}}
\def\gettableau#1{\ifnum#1=0\let\next=\null\else
\squares{#1}\let\next=\gettableau\fi\next}
\preprint{
{\small{\texttt{arXiv:1106.5922[hep-th]}}}}

\title{The Resurgence of Instantons in String Theory}

\author{In\^es Aniceto, Ricardo Schiappa and Marcel Vonk
\\
CAMGSD, Departamento de Matem\'atica, Instituto Superior T\'ecnico,\\ 
Av. Rovisco Pais 1, 1049--001 Lisboa, Portugal\\
\\
\email{ianiceto@math.ist.utl.pt}, \quad
\email{schiappa@math.ist.utl.pt}, \quad
\email{mvonk@math.ist.utl.pt}

}

\abstract{
Nonperturbative effects in string theory are usually associated to D--branes. In many cases it can be explicitly shown that D--brane instantons control the large--order behavior of string perturbation theory, leading to the well--known $(2g)!$ growth of the genus expansion. This paper presents a detailed treatment of nonperturbative solutions in string theory, and their relation to the large--order behavior of perturbation theory, making use of transseries and resurgent analysis. These are powerful techniques addressing general nonperturbative contributions within non--linear systems, which are developed at length herein as they apply to string theory. The cases of topological strings, the Painlev\'e I equation describing 2d quantum gravity, and the quartic matrix model, are explicitly addressed. These results generalize to minimal strings and general matrix models. It is shown that, in order to completely understand string theory at a fully nonperturbative level, new sectors are required beyond the standard D--brane sector.
}

\keywords{Instantons, Resurgent Analysis, Topological and Minimal Strings, Matrix Models}

\begin{document}



\vfill

\eject

\allowdisplaybreaks

\section{Introduction and Summary}

String theory may be defined perturbatively, as a topological genus expansion, in terms of two couplings, $\a$ and $g_s$,
\be\label{stringgenus}
F \left( g_s; \left\{ t_i \right\} \right) \simeq \sum_{g=0}^{+\infty} g_s^{2g-2} F_g (t_i),
\ee
\noindent
where $F = \log Z$ is the string free energy and $Z$ the partition function. At fixed genus $g$ the free energies\footnote{In here the $t_i$ are geometric moduli: for instance, in topological string theory the $\left\{ t_i \right\}$ moduli are identified with K\"ahler parameters in the A--model and with complex structure parameters in the B--model.} $F_g (t_i)$ are themselves perturbatively expanded in $\a$. As it turns out this $\a$ expansion is the milder one, with finite convergence radius. What we address in this paper concerns the topological genus expansion where one is faced with the familiar string theoretic large--order behavior $F_g \sim (2g)!$ rendering the topological expansion as an asymptotic expansion \cite{s90}. How can one go beyond perturbation theory in $g_s$ and define nonperturbative string theory in general?

Stated in this form, this is a very broad and hard question. In order to actually be able to make progress on this front we shall need to specialize to a very concrete physical arena---large $N$ duality for topological strings, matrix models, and their double--scaling limits---where a set of mathematical tools, which go by the name of Borel and resurgent analysis, will allow for the construction of solutions to this problem. Let us thus start by reviewing the physical context.

\subsubsection*{Main Motivations}

Perhaps the most popular approach to the nonperturbative definition of string theory is within the context of large $N$ duality \cite{agmoo99}. In this framework, the partition function of some gauge theoretic system defines, nonperturbatively, a dual large--$N$ closed--string background. This background is, in turn, described by some geometrical construction which is itself determined by the particular asymptotic (large $N$) limit under consideration. Let us focus, for example, on the rather complete picture of \cite{dv02}. Here, one starts off on the gauge theoretic side with a matrix model with some potential, $V(z)$, and, given a classical vacuum---that is, a distribution of the matrix eigenvalues across the several distinct critical points of the potential---, the 't~Hooft large $N$ limit \cite{th74} yields a (holographic) closed string background which is described by the topological B--model on a specific non--compact Calabi--Yau geometry. It is important to notice that different choices of classical vacua will yield different large $N$ geometries and the same gauge theoretic system will thus allow for different large $N$ asymptotic expansions, represented by these distinct semi--classical geometrical backgrounds. Now, the construction of the large $N$ dual in \cite{dv02} is essentially achieved by comparing free energies. On the matrix model side, the $1/N$ 't~Hooft expansion of the free energy starts off by a choice of a semi--classical saddle--point, described by a spectral curve (see, \textit{e.g.}, \cite{fgz93} for a review\footnote{Let us further point out that this matrix model problem of constructing the $1/N$ expansion, spelled out in \cite{bipz78} and which gained an appealing geometrical flavor in \cite{ackm93}, has recently been exactly solved purely in terms of spectral geometry in \cite{eo07, eo08}, and that these results lie at the conceptual basis of our description above.}). Given this spectral geometry there then exists a well--defined procedure to compute the large $N$ expansion of the free energy \cite{eo07} which puts the results in \cite{dv02} on solid ground, with an explicit construction of the genus expansion (\ref{stringgenus}) of the dual closed string geometry\footnote{This procedure later allowed for very interesting extensions of the proposal in \cite{dv02} to more general topological string backgrounds, including duals of closed strings on mirrors of toric backgrounds \cite{m06, bkmp07}.} \cite{emo07} (which can of course be checked by explicit calculations strictly \textit{within} the topological B--model closed string theory).

We may now address our main question above within this physical set--up. In fact, it is also the case that the 't~Hooft expansion is of the form (\ref{stringgenus}), this time around in powers of $N^{2-2g}$ but still with $F_g \sim (2g)!$, \textit{i.e.}, the $1/N$ expansion is an asymptotic perturbative expansion with zero radius of convergence. As we shall discuss at length in this work, this means that there will be nonperturbative corrections of the form $\exp \left( -N \right)$ that still need to be taken into consideration. These are associated to instantons and from a dual closed string point of view, in the large $N$ limit, these corrections typically enjoy a semi--classical description corresponding to D--brane instanton effects. So we shall see that, given a gauge theoretic system and considering one of its possible large $N$ limits, one obtains a closed string dual from the semi--classical data of the gauge theory 't~Hooft limit, both at perturbative \textit{and} nonperturbative levels. This set--up indeed allows us to move beyond the perturbative 't~Hooft expansion. One question we address in this work is to which point the nonperturbative description is complete, and whether D--branes account for the full semi--classical nonperturbative data in such a complete description. That is, if the finite $N$ gauge theoretic partition function is the correct nonperturbative definition for closed strings in certain backgrounds, one must also understand how this finite $N$ system encodes, from a dual spacetime point of view, all semi--classical nonperturbative contributions.

In order to tackle the aforementioned problems, we shall need to resort to an extensive use of resurgent analysis. This is a framework which allows for the construction of exact nonperturbative solutions to rather general non--linear problems in terms of so--called transseries solutions (first introduced in the string theoretic context in \cite{m08}), and we shall further fully develop this framework as it applies to string theory. Transseries solutions account for all possible saddle--points of a given problem, and denoting them as resurgent essentially means that the asymptotic behavior of the perturbative expansion around some chosen saddle is dictated by contributions from all other saddles (we shall be precise about these ideas in the main body of the text). There are two different but complementary aspects to these solutions: on the one hand the specific construction of transseries solutions, and the check of their resurgent properties, amounts to the mathematical study of either differential or finite--difference equations (in the context of matrix models and their double--scaling limits). On the other hand, we also have a physical interpretation of these solutions: in particular, we shall find that these transseries solutions, encoding the complete nonperturbative content of the large $N$ description, have sectors which cannot be associated to D--branes, at least not in a straightforward fashion (as first anticipated in \cite{gikm10, kmr10}). Further, the resurgent nonperturbative solutions have a holographic flavor, in the sense that although one starts from the gauge theoretic (matrix model) side, these solutions may be understood in terms of dual large $N$ data. Setting up a nonperturbative large $N$ duality framework is of obvious relevance to many diverse issues. For example, a particularly interesting question is whether going beyond perturbation theory around some classical vacuum of the matrix model will allow, in the holographic dual, to ``see'' other closed string backgrounds (which are naturally included in the finite $N$ gauge theoretic system).

\subsubsection*{Literature Overview}

In order to place our results in perspective, let us now present an overview of the literature that led up to this work. The present research program started in \cite{m06, msw07}, which proposed to generalize many of the nonperturbative results previously obtained within minimal strings to the realm of matrix models off--criticality and topological strings (intimately related via \cite{dv02}, as mentioned before). Indeed, the double--scaled instantons uncovered in \cite{s90}, and studied from the matrix model point of view in \cite{d91, d92}, were instrumental for, \textit{e.g.}, the discovery of D--branes in critical string theory\footnote{Of course these instantons also played a decisive role in many nonperturbative questions addressed within minimal string theory \cite{fgz93, n04} and were later precisely identified as D--brane configurations in \cite{akk03}.} sparked in \cite{p94}. The approach of \cite{msw07} used saddle--point techniques to extend results such as \cite{d91, d92, hhikkmt04, iy05} away from criticality. This is an approach that relies on the matrix model spectral curve, identifying instantons with B--cycles in the spectral geometry, and which can also be extended to the study of multi--instanton corrections---developed in \cite{msw08, kmr10}, albeit not very explicitly on what concerns general multi--cut saddle--point configurations. These results were later extended in \cite{ps09} to further include instantons associated to A--cycles, which play a relevant role in many topological string theories (in the so--called $c=1$ class).

A complementary approach was introduced in \cite{m08}, this time around making use of orthogonal polynomial techniques \cite{biz80}, where transseries were first introduced to deal with string theoretic problems. One of the results in the present paper is to fully generalize these ideas to obtain complete nonperturbative solutions to matrix models. In some sense, as we shall make much more precise as we go along in this work, the transseries approach amounts to summing over \textit{all} possible backgrounds, \textit{i.e.}, all possible distributions of matrix eigenvalues across the many cuts, which correspond to all possible large $N$ saddle geometries. In particular, multi--instanton corrections within multi--cut geometries \cite{msw08, kmr10} amount to the exchange of matrix eigenvalues along the different cuts, which is effectively interpreted as a change of semi--classical background. This naturally led to the construction of a grand--canonical, manifestly background independent, partition function in \cite{em08} (building upon results in \cite{bde00, e08}) which was further proved to be both holomorphic and modular covariant. Summing over all possible backgrounds or over all possible nonperturbative instanton corrections amounts to the same effect. This grand--canonical partition function is built by making use of theta--functions, implying, in particular, that there will be regions in the gauge theory phase diagram where there are no large $N$ expansions (\textit{i.e.}, there is no $1/N$ expansion due to the oscillatory nature of the theta--functions). This rather important idea was later explored, from a large $N$ duality point of view, in \cite{mpp09}. Finally, most of the transseries results extend beyond the context of matrix models. All they require is the existence of a string equation \cite{biz80}, typically a finite difference equation in the context of off--critical matrix models, or a differential equation in the context of double--scaled minimal strings, which is known to also exist in other examples of topological strings without a very clean matrix model relation, \textit{e.g.}, \cite{msw07, cgmps06}. There may well be larger classes of examples where this is the case.

A very important role in all this analysis was played by the relation of instantons to the large--order behavior of the string perturbation theory \cite{m06, msw07}, \textit{i.e.}, to the fact that these instanton effects are testable via their connection to the large--order behavior of the $1/N$ asymptotic expansion \cite{z81}. Rather impressive agreement was found for many of the calculations in the previous references and this will also be a very important point in the present paper: the resurgent framework we uncover, from an analytical approach, is extensively---and extremely rigorously---tested by exploring the connections between the asymptotics of multi--instanton sectors as dictated by resurgence. As we shall explain, resurgence demands for a very tight web of relations in between all these distinct nonperturbative sectors, which is translated into their large--order behavior. These relations may be very thoroughly checked, and to very high precision, making use of numerical tests, a fact which will clearly justify the construction we shall propose.

It might be fair to say that the first truly unexpected result along this line of research appeared in \cite{gikm10}, which addressed the large--order asymptotic behavior of \textit{multi}--instanton sectors, rather than just focusing on the usual large--order behavior of perturbation theory. In particular, that work addressed the large--order behavior of the 2--instantons sector in the Painlev\'e I system (the $(2,3)$ minimal string) and found that \textit{new} nonperturbative sectors, besides the usual multi--instanton contributions, were required in order to properly describe the full asymptotic behavior of this sector. This was done at leading order, in the resurgent framework, and was in fact the main motivation behind the full construction we embrace in our present work: to understand the complete set of nonperturbative contributions demanded by resurgence, within the minimal string context, and further extend it to general matrix models and topological strings. At this stage the reader might complain that we have mentioned the word ``resurgence'' a lot but have been a bit vague about the nature of this framework. This is due to the fact that this formalism, a rather general framework introduced in \cite{e81} to address general solutions of non--linear systems in terms of multi--instanton data, is a bit involved. In here, we wanted to motivate the need for more general approaches to nonperturbative issues within large $N$ duality from a purely string theoretic point of view. In section \ref{borelrev} we introduce this formalism (alongside with some new results concerning multi--instanton asymptotics) and indicate how it may be used in string theory. In this way we recommend the reader to regard this section as an enlarged introduction to the ideas that are then explored at length in the rest of the paper.

\subsubsection*{Outline of the Paper}

This paper is organized as follows. As just mentioned, we begin in section \ref{borelrev} with an introduction to resurgence and the development of asymptotic formulae. Asymptotic expansions, with zero radius of convergence, need to be resummed if one is to extract any information out of them. There are, of course, many different possible ressumation techniques (see, \textit{e.g.}, \cite{cmrsj07}) but since our models deal with asymptotic series which diverge factorially, the natural procedure to use in this case turns out to be the Borel resummation framework. This leads in turn to the resurgent framework of \'Ecalle which we introduce in a physical context in this section. We also discuss the relation to the Stokes phenomenon; previously discussed in, \textit{e.g.}, \cite{ps09, mpp09}. Then, in section \ref{topstringGV}, we apply some of the ideas of resurgence to topological string theory in the Gopakumar--Vafa representation \cite{gv98a, gv98c}. This is, essentially, an extension of the work developed in the context of topological strings on the resolved conifold in \cite{ps09}. Section \ref{sec:4} starts developing the resurgent framework to more general string theoretic systems, in such a way that we can  apply it to minimal strings and matrix models. This is where we develop the main structure of our nonperturbative solutions, which will later materialize with explicit results in the following sections. In section \ref{PIsection} we discuss one of our main examples, the $(2,3)$ minimal string theory, which describes pure gravity in two dimensions. In this section we shall construct the full two--parameters transseries solution to the Painlev\'e I equation, generalizing the work of \cite{gikm10}. Do notice that, for the Painlev\'e I perturbative solution, leading asymptotic checks have been carried out in, \textit{e.g.}, \cite{k89, jk01, k04, msw07}. A partial transseries analysis was done in \cite{d05}. As for its multi--instantons solutions, as mentioned above, leading asymptotic checks have been carried out in \cite{gikm10}. Our present analysis extends all these partial results to a full general solution. Furthermore, by analysis of the resulting resurgent structure we show that this solution has complete nonperturbative information concerning the minimal model. More importantly, in this complete set of nonperturbative data, and besides the standard instanton or D--brane sector, we find new nonperturbative sectors with a ``generalized'' instanton structure. We perform high--precision numerical tests of \textit{all} nonperturbative sectors, including the new ``generalized'' instanton sectors, which clearly show the need for all these contributions in the full exact solution. We also compute many, previously unknown, Stokes constants of the Painlev\'e I equation and of the $(2,3)$ minimal string theory. In section \ref{sec:6} we analyze the full fledged quartic matrix model, starting around the one--cut saddle--point geometry. In a similar fashion to what we previously did for the Painlev\'e I equation, we construct the transseries solution which yields the complete nonperturbative solution to this matrix model. We further show how this solution relates back to the Painlev\'e I transseries solution in the double--scaling limit. This includes a discussion of the new nonperturbative sectors of the quartic matrix model, alongside with extensive numerical checks which use the resurgent relations to prove the validity of these sectors. We also show that the transseries of the quartic model may be set up in such a way that the Stokes constants of this problem are essentially given by the Stokes constants of the $(2,3)$ model. We close in section \ref{sec:7} with a discussion and some ideas for future work. Do notice that our analysis generated a rather large amount of data which, for reasons of space, cannot be all presented in the body of this paper. \textit{Mathematica} files with the relevant data are available from the authors upon request. We do however present some partial data in a few appendices, to indicate how the full set--up was constructed.

\section{Borel Analysis, Resurgence and Asymptotics}\label{borelrev}

One framework to address nonperturbative completions of rather general non--linear systems is the resurgent formalism of \'Ecalle \cite{e81}, building upon results of Borel analysis and Stokes phenomena, and we shall briefly review it in this section\footnote{The reason for the term ``resurgent''---roughly meaning ``reappearing''---will also be explained in what follows.}. In short, it amounts to a procedure which constructs solutions to non--linear problems by addressing all possible multi--instanton sectors, \textit{i.e.}, all possible saddle--point configurations in the path integral. Notice that this means that one constructs the full solution perturbatively as a power series in the string coupling \textit{and} also perturbatively in the instanton number, \textit{i.e.}, as a power series in the (exponential) instanton contribution---although each instanton contribution is itself nonperturbative. Besides allowing for the construction of nonperturbative solutions, the multi--instanton sectors also allow for a quantitative understanding of the large--order behavior of the corresponding perturbative expansions around a given, fixed multi--instanton sector (the large--order behavior of the zero--instanton sector being the simplest case to analyze), a subject with a long tradition in quantum mechanics and field theory, \textit{e.g.}, \cite{bw73, cs78, z81}. Some ideas of resurgence have also been partially addressed recently within the matrix model context, see, \textit{e.g.}, \cite{m06, msw07, m08, em08, msw08, ps09, gikm10, kmr10}. At least in principle, the multi--instanton information could provide for a reconstruction of the exact free energy, or partition function, in any region of the coupling--constant complex plane.

Let us begin with a rather general introduction to some of these ideas, by considering the free energy in the zero--instanton sector of any given model (stringy or not), $F(z)$, given as an asymptotic perturbative expansion\footnote{In the following we shall do perturbation theory around $z \sim \infty$, rather than $g_s \sim 0$ as usual.} in some coupling parameter $z$ (we will soon take $z \in \BC$),
\be\label{d}
F(z) \simeq \sum_{g=0}^{+\infty} \frac{F_g}{z^{g+1}}. 
\ee
\noindent
Let us assume that, at large $g$, the coefficients above behave as $F_g \sim g!$, rendering the series asymptotic with zero radius of convergence. In this case, while we are assuming that $F(z)$ exists as a function, one must still make sense out of the formal power series on the right--hand--side and we shall use the notation $\simeq$ to signal this fact. There are many quantum mechanical and quantum field theoretic examples where this is the typical behavior of the perturbative series and this is essentially due to the growth of Feynman diagrams in perturbation theory \cite{z81}. In the following we shall explain how resurgent analysis makes sense of asymptotic series. For the moment, let us just mention that the factorial growth of the $F_g$ is precisely controlled by nonperturbative instantons corrections, which behave as $\rme^{-n A z}$ with $A$ denoting the instanton action and $n$ the instanton number \cite{z81}. As we shall see in great detail, although each perturbative/multi--instanton sector is very different due to the non--analytic contribution $\rme^{-n A z}$ (at $z \sim \infty$), resurgence will relate the asymptotic growth of each sector to the leading coefficients of every other sector.

Let us now further perform a perturbative expansion around the (nonperturbative) contribution at a given fixed instanton number. One finds that the full $n$--instanton contribution is of the form (see, \textit{e.g.}, \cite{msw07, m08, gm08, gikm10} for discussions in the context of matrix models, and topological and minimal strings)
\be\label{multiseries}
F^{(n)}(z) \simeq z^{-n \beta}\, \rme^{-n A z}\, \sum_{g=1}^{+\infty} \frac{F^{(n)}_g}{z^g}.
\ee
\noindent
Here $\beta$ is an exponent which varies from example to example\footnote{As such, we shall be more explicit on how to find it when we actually address some examples.}, and $F^{(n)}_g$ is the $g$--loop contribution around the $n$--instanton configuration. Let us now further assume that, at large $g$, these coefficients \textit{also} behave as $F^{(n)}_g \sim g!$, rendering all multi--instanton contributions as (divergent) asymptotic series (just as above, this is a typical behavior in many quantum mechanical or quantum field theoretic examples \cite{z81}). As we shall see, it is possible to precisely understand the asymptotics, in $g$, of the multi--instanton contributions $F^{(n)}_g$, in terms of the coefficients $F^{(n')}_g$, with $n'$ close to $n$. This means that all these asymptotic expansions are resurgent \cite{gm08}, and we shall delve into this in the following.

As an approximation to the exact solution these asymptotic, \textit{divergent} formal power series must be truncated and one is consequently faced with the problem that the perturbative expansion has zero convergence radius. In particular, if we do not know the exact function, $F(z)$, but only its asymptotic series expansion, how do we associate a value to the divergent sum? One framework to address issues related to (factorially divergent) asymptotic series is Borel analysis. Introduce the Borel transform as the linear map\footnote{Notice that the Borel transform is not defined for $\alpha=-1$, \textit{i.e.}, for a constant term. Thus, in order to Borel transform an asymptotic power series with constant term (denoted the residual coefficient), one \textit{first} drops this constant term and \textit{then} performs the Borel transform by the rule presented above.} from (asymptotic) power series around $z \sim \infty$ to (convergent) power series around $s \sim 0$, defined by
\be
\CB \left[ \frac{1}{z^{\alpha+1}} \right](s) = \frac{s^{\alpha}}{\Gamma(\alpha+1)},
\ee
\noindent
so that the Borel transform of the asymptotic series (\ref{d}) is the function
\be\label{defborelt}
\CB[F](s) = \sum_{g=0}^{+\infty} \frac{F_g}{g!}\,s^g,
\ee
\noindent
which ``removed'' the divergent part of the coefficients $F_g$ and renders $\CB[F](s)$ with finite convergence radius around the origin in $\BC$. In general, however, $\CB[F](s)$ will have singularities and it is crucial to locate them in the complex plane. Indeed, if $\CB[F](s)$ has no singularities along a given direction in the complex $s$--plane, say $\arg s = \theta$, one may analytically continue this function on the ray $\rme^{\rmi\theta} \BR^+$ and thus define the \textit{inverse} Borel transform---or \textit{Borel resummation} of $F(z)$ along $\theta$---by means of a Laplace transform with a rotated contour as\footnote{If the original asymptotic series one started off with had a constant term, dropped in the Borel transform, one may now define the Borel resummation as shown, plus the addition of this constant term.}
\be\label{borelinttheta}
\CS_\theta F (z) = \int_0^{\rme^{\rmi\theta}\infty} \rmd s\, \CB[F](s)\, \rme^{-zs}.
\ee
\noindent
The function $\CS_\theta F (z)$ has, by construction, the same asymptotic expansion as $F(z)$ and may provide a solution to our original question; it associates a value to the divergent sum (\ref{d}). In the following we shall further define the \textit{lateral} Borel resummations along $\theta$, $\CS_{\theta^\pm} F(z)$, as the Borel resummations $\CS_{\theta\pm\epsilon} F(z)$ for $\epsilon \sim 0^+$.

Let us consider a simple example where we take as asymptotic series
\be\label{example}
F(z) \simeq \sum_{g=0}^{+\infty} \frac{\Gamma (g+a)}{\Gamma (a)}\, \frac{1}{A^g}\, \frac{1}{z^{g+1}}. 
\ee
\noindent
In this case the Borel transform immediately follows as
\be
\CB[F](s) = \frac{1}{\left( 1 - \frac{s}{A} \right)^a},
\ee
\noindent
and it has a singularity (either a pole or a branch--cut, depending on the value of $a$) at $s=A$.

Thus, if the function $\CB[F](s)$ has poles or branch cuts along its integration contour above, from $0$ to $\rme^{\rmi\theta}\infty$, things get a bit more subtle: in order to perform the integral (\ref{borelinttheta}) one needs to choose a contour which avoids such singularities. This choice of contour naturally introduces an ambiguity (a \textit{nonperturbative} ambiguity) in the reconstruction of the original function, which renders $F(z)$ \textit{non}--Borel summable. As it turns out, different integration paths produce functions with the same asymptotic behavior, but differing by (non--analytical) exponentially suppressed terms. It is precisely when there are such obstructions to Borel resummation along some direction $\theta$ that the lateral Borel resummations become relevant: for instance, in the presence of a simple pole singularity at a distance $A$ from the origin, along some direction $\theta$ in $\BC$, one may define the Borel resummation on contours $\CC_{\theta^\pm}$, either avoiding the singularity via the left (as moving towards infinity), and leading to $\CS_{\theta^+} F (z)$, or from the right, and leading to $\CS_{\theta^-} F (z)$ (see figure \ref{borelstokeshankel}). One finds that these two functions differ by a nonperturbative term \cite{z81}
\be\label{npa}
\CS_{\theta^+} F(z) - \CS_{\theta^-} F(z) \propto \oint_{(A)} \rmd s\, \frac{\rme^{-zs}}{s-A} \propto \rme^{-A z}.
\ee
\noindent
Further nonperturbative ambiguities arise as one reconstructs the original function along different directions (with singularities) in the complex $s$--plane. As such, different integration paths produce functions with the same asymptotic behavior, but differing by exponentially suppressed terms. To be fully precise about these, we shall need to delve into resurgence \cite{cnp93, ss03, gm08}.

\subsection{Alien Calculus and the Stokes Automorphism}

Let us return to our formal power series (\ref{d}). This asymptotic expansion is said to be a \textit{simple resurgent function} if its Borel transform, $\CB[F](s)$, only has simple poles or logarithmic branch cuts as singularities, \textit{i.e.}, near each singular point $\omega$
\be\label{borelomega}
\CB[F](s) = \frac{\alpha}{2\pi\rmi \left( s-\omega \right)} + \Psi (s-\omega)\, \frac{\log \left( s-\omega \right)}{2\pi\rmi} + \Phi (s-\omega),
\ee
\noindent
where $\alpha \in \BC$ and $\Psi$, $\Phi$ are analytic around the origin. It can be shown that simple resurgent functions allow for the resummation of formal power series along \textit{any} direction in the complex $s$--plane, thus leading to a family of sectorial analytic functions $\{\CS_\theta F(z)\}$. For rigorous details and the proof of this statement, we refer the reader to \cite{cnp93, ss03, gm08}. 

\FIGURE[ht]{
\label{borelstokeshankel}
\centering
\includegraphics[width=4cm]{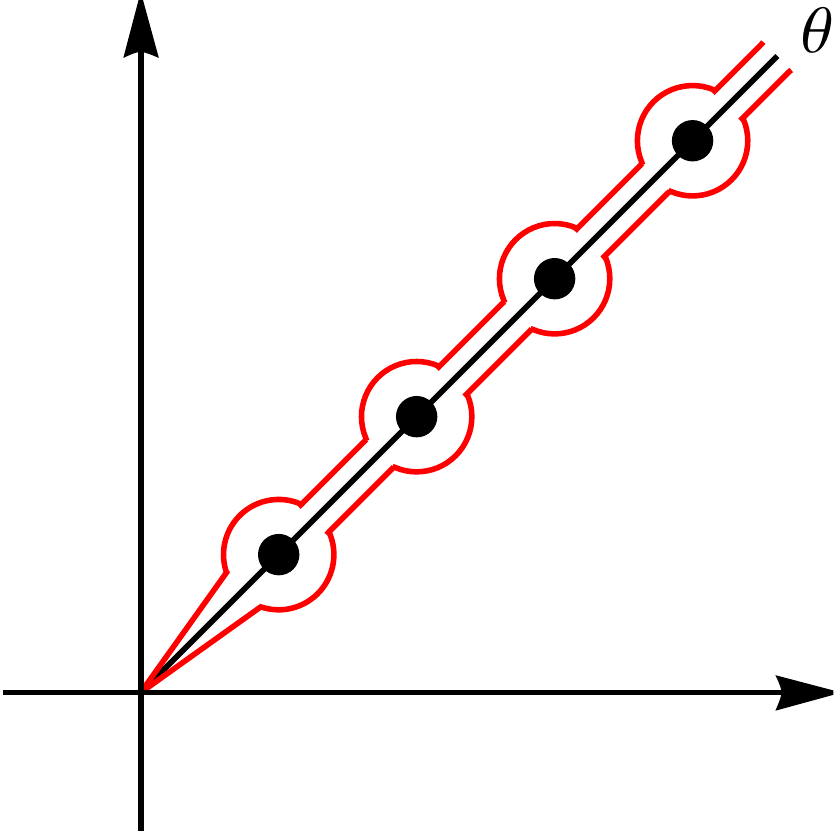}
$\qquad$
\includegraphics[width=4cm]{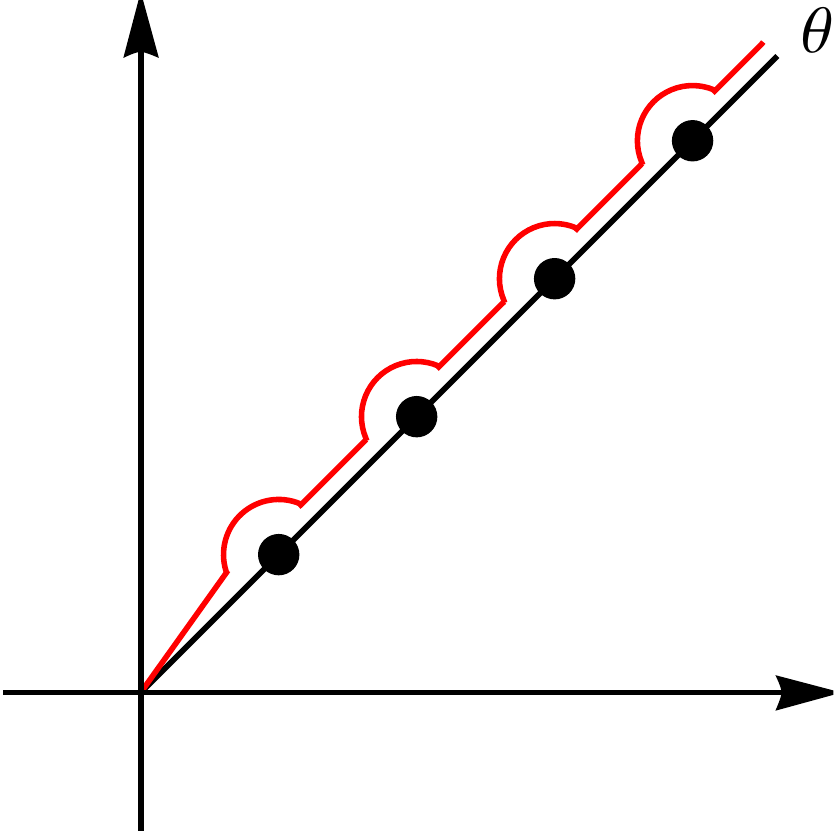}
$\qquad$
\includegraphics[width=4cm]{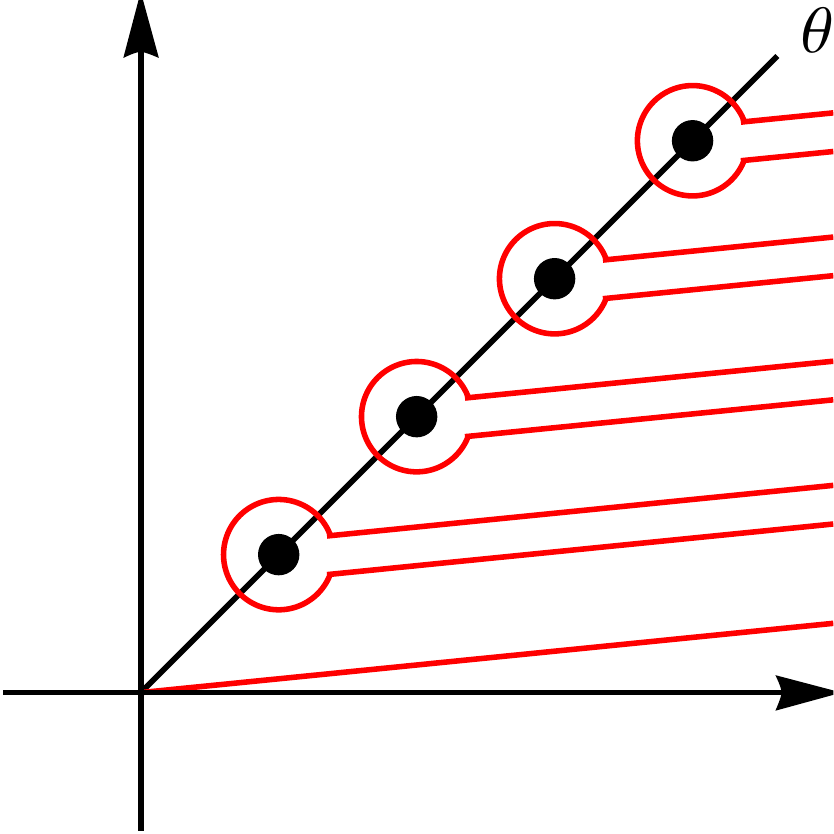}
\caption{The first image shows singularities along some direction $\theta$ in the Borel complex plane, and the contours corresponding to the left and right lateral Borel resummations along such direction. The second and third images show how to cross this singular direction, or Stokes line, via the Stokes automorphism: the left Borel resummation equals the right Borel resummation plus the discontinuity of the singular direction (given by the sum over Hankel contours around all singular points).}
}

As should be clear---and up to nonperturbative ambiguities---in different sectors one obtains different resummations and one needs to fully understand Borel singularities in order to ``connect'' these sectorial solutions together. The next step in order to analyze these Borel singularities in greater detail is to introduce \'Ecalle's alien calculus \cite{e81, cnp93}. At its basis lies a differential operator acting on resurgent functions, the alien derivative $\Delta_\omega$. Let us define it within the context of simple resurgent functions\footnote{The definition for general resurgent functions is more intricate; see, \textit{e.g.}, \cite{cnp93, ss03, gm08}.}: $\Delta_\omega$ is a linear differential operator from simple resurgent functions to simple resurgent functions, satisfying the Leibniz rule and the following two basic properties:
\begin{itemize}
\item If $\omega$ is not a singular point (a simple pole or a logarithmic cut), then $\Delta_\omega F(z) = 0$.
\item If $\omega$ is a singular point, let us first consider the Borel transform of our resurgent function \eqref{borelomega}, which we now conveniently write as
\be\label{alienborel}
\CB[F](\omega+s) = \frac{\alpha}{2\pi\rmi\, s} + \CB[G] (s)\, \frac{\log s}{2\pi\rmi} + \mathrm{holomorphic},
\ee
\noindent
with $G(z)$ the resurgent function whose Borel transform yields $\Psi(s)$ in \eqref{borelomega} (of course in practice it might be hard to find $G(z)$ explicitly). In this case, the alien derivative at a singular point $\omega$ is given by
\be\label{alienresurgent}
\CS \Delta_\omega F(z) = \alpha + \CS_{\arg \omega} G(z).
\ee
\end{itemize}
\noindent
To have a better grasp on the calculation of alien derivatives let us consider another example, slightly more involved than \eqref{example}, where we now take as asymptotic series
\be\label{secondexample}
F(z) \simeq \sum_{g=0}^{+\infty} \frac{\Gamma (g+1)}{\Gamma (1)}\, \frac{1}{g+1}\, \frac{1}{A^g}\, \frac{1}{z^{g+1}}. 
\ee
\noindent
It is again very simple to evaluate the Borel transform as
\be\label{borelsecondexample}
\CB[F](s) = - \frac{A}{s}\, \log \left( 1-\frac{s}{A} \right),
\ee
\noindent
with a branch cut in the complex $s$--plane running from $A$ to infinity. It is immediate to check that our asymptotic series \eqref{secondexample} defines a simple resurgent function. Further noticing that $\frac{A}{s} = \CB[G](s-A)$ with $G(z)$ a resurgent function closely related to our earlier example \eqref{example},
\be
G(z) \simeq \sum_{g=0}^{+\infty} \frac{\Gamma (g+1)}{\Gamma (1)}\, \frac{1}{\left( -A \right)^g}\, \frac{1}{z^{g+1}},
\ee
\noindent
it immediately follows by definition that
\be\label{alienexample}
\Delta_{A} F = - 2\pi\rmi\, G.
\ee

Alien derivatives thus encode the whole singular behavior of the Borel transform (they encode how much $\CB[F](s)$ ``jumps'' at a singularity) and allow for the aforementioned ``connection'' of sectorial solutions. Indeed, let us consider a \textit{singular direction} $\theta$, \textit{i.e.}, a direction along which there are singularities in the Borel complex plane. In the original complex $z$--plane such a direction is known as a \textit{Stokes line} (more on this later). Understanding how to connect the distinct sectorial solutions on both sides of such direction necessarily entails understanding their ``jump'' across this direction, and this is accomplished via the \textit{Stokes automorphism}, $\underline{\frak{S}}_\theta$, or its related discontinuity, $\disc_{\theta}$, acting on resurgent functions and satisfying \cite{cnp93}
\be
\CS_{\theta^+} = \CS_{\theta^-} \circ \underline{\frak{S}}_\theta \equiv \CS_{\theta^-} \circ \left( \mathbf{1} - \disc_{\theta^-} \right),
\ee
\noindent
in such a way that the action of $\underline{\frak{S}}_\theta$ on resurgent functions immediately translates into the required connection of distinct sectorial solutions, across a singular direction $\theta$. In particular,
\be\label{boreldisc}
\CS_{\theta^+} - \CS_{\theta^-} = - \CS_{\theta^-} \circ \disc_{\theta^-},
\ee
\noindent
such that $\disc_{\theta}$ precisely encodes the full discontinuity\footnote{A function $\phi$ satisfying $\underline{\frak{S}}_\theta \phi = \phi$, or, equivalently, $\disc_{\theta} \phi = 0$, has no Borel singularities along the $\theta$--direction and is known as a \textit{resurgence constant} along this direction. In particular, in this region its Borel transform is analytic and $\phi$ is thus given by a convergent power series.} of the resurgent function across $\theta$. Geometrically, one may think of $\disc_{\theta^-}$ as the sum over all Hankel contours which encircle each singular point in the $\theta$--direction, on the left, and part off to infinity, on the right (see figure \ref{borelstokeshankel}). The main point now is that, as it turns out \cite{cnp93, ss03}, one finds
\be\label{stokesauto}
\underline{\frak{S}}_\theta = \exp \left( \sum_{\{\omega_\theta\}} \rme^{-\omega_\theta z} \Delta_{\omega_\theta} \right),
\ee
\noindent
where $\{\omega_\theta\}$ denote all singular points along the $\theta$--direction. Explicitly, for singularities along the $\theta$--direction in an ordered sequence, one can write \cite{ss03}
\be\label{glue}
\CS_{\theta^+} F(z) = \CS_{\theta^-} F(z) + \sum_{r \ge 1; \{ n_i \ge 1 \}} \frac{1}{r!}\, \rme^{- \left( \omega_{n_1} + \omega_{n_2} + \cdots \omega_{n_r} \right) z}\, \CS_{\theta^-} \left( \Delta_{\omega_{n_1}} \Delta_{\omega_{n_2}} \cdots \Delta_{\omega_{n_r}} F(z) \right).
\ee
\noindent
One concludes that, given all possible alien derivatives, this result provides the necessary connection, and thus allows for a full construction of the exact nonperturbative solution alongside with its Riemann surface domain.

It is rather instructive to explicitly write the Stokes automorphism in our multi--instanton setting. Consider the positive real axis, where $\theta=0$, and where the Borel singularities are located at the multi--instanton points $n A$ with $n \in \mathbb{N}^*$. In this case:
\be\label{stokesdisc}
\underline{\frak{S}}_0 = \exp \left( \sum_{n=1}^{+\infty} \rme^{-n A z} \Delta_{n A} \right) = 1 + \rme^{-Az}\, \Delta_A + \rme^{-2Az} \left( \Delta_{2A} + \frac{1}{2} \Delta_A^2 \right) + \cdots.
\ee
\noindent
This expression will be rather important in what follows. For the moment let us just go back to our earlier example and compute the action of the Stokes automorphism $\underline{\frak{S}}_0$ on (\ref{secondexample}). Given the only non--vanishing alien derivative of $F(z)$, (\ref{alienexample}), and the fact that higher--order alien derivatives of $F(z)$ at $A$ also vanish (as $\Delta_A G = 0$), it is immediate to check from (\ref{stokesdisc}) above that
\be
\underline{\frak{S}}_0 F(z) = F(z) - 2\pi\rmi\, \rme^{-Az}\, G(z).
\ee

Computing alien derivatives straight from their definition is a hard task. Fortunately, as we shall see, there are much simpler ways to compute alien derivatives. In fact, it turns out that things will greatly simplify by introducing the \textit{pointed} alien derivative
\be
\dot{\Delta}_\omega \equiv \rme^{-\omega z} \Delta_\omega,
\ee
\noindent
as this operator commutes with the usual derivative \cite{ss03},
\be
\left[ \dot{\Delta}_\omega, \frac{\rmd}{\rmd z} \right] = 0.
\ee
\noindent
We shall now turn to explicit computations of alien derivatives in different settings.

\subsection{Transseries and the Bridge Equations}

Having understood the central role that alien derivatives play in the construction of nonperturbative solutions, the question remains: how to compute them in a---preferably simple---systematic fashion? The answer arises in the construction of the \textit{bridge equations}, constructing a ``bridge'' between ordinary and alien calculus.

Focusing on our familiar multi--instanton setting, with instanton action $A$ (one may allow $A$ to be complex, in which case we shall be addressing the $\arg A$ direction in the Borel complex plane), let us consider a \textit{transseries ansatz} for our resurgent function,
\be\label{Ftransseries}
F(z,\sigma) = \sum_{n=0}^{+\infty} \sigma^n F^{(n)} (z),
\ee
\noindent
where $F^{(0)} (z)$ is the formal asymptotic power series (\ref{d}), and where
\be
F^{(n)} (z) = \rme^{- n A z}\, \Phi_{n} (z), \qquad n \ge 1,
\ee
\noindent
are the $n$--instanton contributions (\ref{multiseries}), as discussed before---the $\Phi_{n} (z)$ being further formal asymptotic power series. In here, $\sigma$ is the nonperturbative ambiguity or transseries parameter, selecting, in specified wedges of the complex $z$--plane, distinct nonperturbative completions to our problem. In the resurgence framework, where transseries also go by the name of resurgent symbols along some wedge of the complex plane \cite{cnp93}, this is the most general solution to a given non--linear system. In this work we shall only be concerned with so--called log--free height--one transseries: this means that the \textit{ansatz} above is a formal sum of trans--monomials $z^\alpha\, \rme^{S(z)}$, with $\alpha \in \BR$ and, in our case, $S(z)$ a particularly simple convergent series. More general transseries may be constructed, with $S(z)$ a transseries itself possibly further involving compositions with exponentials or with logarithms, but we refer to \cite{e0801} for a complete discussion.

It is also important to realize that the transseries formalism is a rather powerful technology: when inserted in the non--linear equation satisfied by $F(z)$ (\textit{e.g.}, a finite--difference equation in the case of matrix models, or an ordinary differential equation in the case of minimal strings, as we shall see later), it will yield back the non--linear equation for $F^{(0)}(z)$---which is now to be solved perturbatively---; it will yield a \textit{linear} and \textit{homogeneous} equation for $F^{(1)}(z)$; and it will yield \textit{linear} but \textit{inhomogeneous} equations for $F^{(n)}(z)$, $n \ge 2$. It is thus feasible to solve for \textit{all} members of this hierarchy of equations and fully compute the transseries solution. Indeed, in all examples of interest to us, it will turn out that all perturbative coefficients $F_g^{(n)}$ appearing in the infinite hierarchy of formal asymptotic power series $\Phi_{n} (z)$ can be computed by means of (non--linear) recursions. It will be further the case that the \textit{asymptotics} of these transseries coefficients $F_g^{(n)}$ will be exactly determined in terms of neighboring coefficients $F_g^{(n')}$, with $n'$ close to $n$, and in terms of a finite number of Stokes constants (defined in the following).

Finally, notice that we have assumed the transseries \textit{ansatz} to depend on a \textit{single} parameter, $\sigma$, assuming that the resurgent function arises as a solution to some problem depending on a single ``boundary condition''. More complicated problems could lead to more general transseries \textit{ans\"atze}, and we shall see some such examples further down the line, but for the moment we just consider the simple case where we may power series expand the transseries \textit{ansatz} in a single parameter, \textit{i.e.}, the transseries is an expansion in $\BC[[z^{-1}, \sigma\, \rme^{-Az} ]]$. For simplicity, we shall further assume that the $\Phi_{n} (z)$ asymptotic series are simple resurgent functions.

Given the pointed alien derivative $\dot{\Delta}_{\ell A} = \rme^{-\ell A z} \Delta_{\ell A}$, $\ell \in \BN^*$, one may now compute
\be
\dot{\Delta}_{\ell A} F(z,\sigma) = \sum_{n=0}^{+\infty} \sigma^n\, \rme^{- \left( \ell+n \right) A z}\, \Delta_{\ell A} \Phi_{n} (z).
\ee
\noindent
The key point is the following: suppose the transseries $F(z,\sigma)$ is an \textit{ansatz} for the solution to some differential equation, in the variable $z$. Because the pointed alien derivative commutes with the usual derivative, it is straightforward to obtain the (linear) differential equation which $\dot{\Delta}_{\ell A} F(z,\sigma)$ satisfies. But, clearly, this will be the exact \textit{same} differential equation as the one that
\be
\frac{\partial F}{\partial \sigma} (z,\sigma)
\ee
\noindent
satisfies---simply because also this derivative commutes with the usual derivative\footnote{In full generality this is slightly more subtle: indeed, it will be often the case that the differential equation one is considering will depend on some other (``initial data'') functions. In this case, for the above reasoning to hold, these functions must either be entire functions, or their Borel transforms cannot have singularities at the points $\ell A$ (of course these functions will also have no dependence on $\sigma$ whatsoever; the transseries expression (\ref{Ftransseries}) is simply an \textit{ansatz} for the solution, introducing a new parameter).}. Assuming for simplicity that the differential equation is of first order, it must thus be the case that
\be\label{bridgeF}
\dot{\Delta}_{\ell A} F(z,\sigma) = S_\ell (\sigma)\, \frac{\partial F}{\partial \sigma} (z,\sigma);
\ee
\noindent
a relation known as \'Ecalle's \textit{bridge equation} \cite{ss03}, relating alien derivatives to familiar ones! In here $S_\ell (\sigma)$ is a proportionality factor which may, naturally, depend on $\sigma$. Recalling that alien derivatives encode the singular behavior of the Borel transform, the bridge equation tells us that, in some sense, at these singularities we find back the original asymptotic power series we started off with---hence its name as a ``resurgent'' function. Let us explore the implications of (\ref{bridgeF}). Spelling it out as formal power series, and given that $\Phi_n \not = 0$, $\forall_n$, this immediately implies
\be
S_\ell (\sigma) = 0, \quad \ell > 1 \qquad \Leftrightarrow \qquad \Delta_{\ell A} F(z,\sigma) = 0, \quad \ell > 1.
\ee
\noindent
While one may generically expect that the proportionality factor $S_\ell (\sigma)$ has a formal power series expansion as $S_\ell (\sigma) = \sum_{k=0}^{+\infty} S_{\ell}^{(k)}\, \sigma^k$, homogeneity in $\sigma$ of the bridge equation (\ref{bridgeF}) demands $k=1-\ell$. One may quickly realize this by introducing a notion of degree such that
\be
\deg \left( \sigma^n\, \rme^{m A z} \right) = n + m.
\ee
\noindent
In this case $\deg F (z,\sigma) = 0$ (which follows since $F (z,\sigma)$ only depends on $\sigma\, \rme^{-Az}$) immediately yields $\deg S_\ell (\sigma) = 1-\ell$, \textit{i.e.},
\be
S_\ell (\sigma) = S_\ell\, \sigma^{1-\ell}, \qquad \ell \le 1.
\ee
\noindent
Plugging this back into the power series expansion of the bridge equation one finally obtains a clearer expression for the bridge, or resurgence equations\footnote{One also obtains a clearer explanation for the name ``resurgent'': via the bridge equations the alien derivatives, encoding the singular behavior of the Borel transform, are given in terms of the original asymptotic power series one started off with (multiplied by suitable Stokes' constants).}
\be\label{bridgePhi}
\Delta_{\ell A} \Phi_n = 
\begin{cases}
0, & \ell > 1, \\
S_\ell \left( n+\ell \right) \Phi_{n+\ell}, & \ell \le 1,
\end{cases}
\ee
\noindent
where we have used conventions in which $\Phi_n$ vanishes if $n$ is less than zero. This expression yields \textit{all} alien derivatives, in terms of a (possibly) infinite sequence of unknowns $S_\ell \in \BC$, $\ell \in \left\{ 1, -1, -2, \cdots \right\}$, the so--called analytic invariants of the differential equation we started off with. Knowledge of these analytic invariants allows for a \textit{full} nonperturbative reconstruction of the original function $F(z)$, the problem we first set out to solve. However, generically, the analytic invariants are transcendental functions of the initial data (say, the differential equation one started off with) and quite hard to compute. 

For completeness, it is interesting to notice that the above resurgence equations (\ref{bridgePhi}) may be translated back to the structure of the Borel transform, at least near each singularity $\ell A$, by making use of the definition of alien derivative (\ref{alienresurgent}) for a simple resurgent function. Indeed, with
\be
\Phi_n (z) \simeq \sum_{g=1}^{+\infty} \frac{F^{(n)}_g}{z^{g+n\beta}} \qquad \mathrm{and} \qquad \CB [\Phi_n] (s) = \sum_{g=1}^{+\infty} \frac{F^{(n)}_g}{\Gamma(g+n\beta)}\, s^{g+n\beta-1},
\ee
\noindent
it simply follows via (\ref{alienborel})
\be
\CB \left[ \Phi_n \right] (s+\ell A) = S_\ell \left( n+\ell \right) \CB \left[ \Phi_{n+\ell} \right] (s)\, \frac{\log s}{2\pi\rmi}, \qquad \ell \leq 1.
\ee

Going back to the connection formulae (\ref{glue}) or to Stokes' automorphism (\ref{stokesdisc}) makes clear how important (\ref{bridgePhi}) is: it is telling us that the somewhat initial multi--instanton data is \textit{enough} for a full reconstruction of the nonperturbative solution. Consider first the positive real axis, where $\theta = 0$, and where the Stokes automorphism is
\be
\underline{\frak{S}}_0 = \exp \left( \sum_{\ell=1}^{+\infty} \rme^{-\ell A z} \Delta_{\ell A} \right) = 1 + \rme^{-Az}\, \Delta_A + \rme^{-2Az} \left( \Delta_{2A} + \frac{1}{2} \Delta_A^2 \right) + \cdots.
\ee
\noindent
Given the transseries \textit{ansatz}, the action of $\underline{\frak{S}}_0$ on $F(z,\sigma)$ is entirely encoded by the action of $\underline{\frak{S}}_0$ on the several $\Phi_n (z)$, and this can now be completely determined by the use of the bridge equations (\ref{bridgePhi}). But because these vanish whenever $\ell>1$, the Stokes automorphism immediately simplifies to
\be
\underline{\frak{S}}_0 = \exp \left(\rme^{-A z} \Delta_{A} \right) = 1 + \rme^{-Az}\, \Delta_A + \frac{1}{2} \rme^{-2Az}\, \Delta_A^2 + \frac{1}{3!} \rme^{-3Az}\, \Delta_A^3 + \cdots,
\ee
\noindent
where (just iterate (\ref{bridgePhi}))
\be
\Delta_A^N \Phi_n = \left( S_1 \right)^N \cdot \prod_{i=1}^N \left( n+i \right) \cdot \Phi_{n+N}.
\ee
\noindent
One may now simply compute
\be\label{disc@zero}
\underline{\frak{S}}_0 \Phi_n = \sum_{\ell=0}^{+\infty} \binom{n+\ell}{n}\, S_1^\ell\, \rme^{- \ell A z}\, \Phi_{n+\ell}.
\ee
\noindent
The interesting fact about the bridge equations (\ref{bridgePhi}) is that they contain much more information than just that concerning the positive real axis. Indeed, consider instead the negative real axis, where $\theta = \pi$, and where the Stokes automorphism becomes
\be
\underline{\frak{S}}_\pi = \exp \left( \sum_{\ell=1}^{+\infty} \rme^{\ell A z} \Delta_{-\ell A} \right) = 1 + \rme^{Az}\, \Delta_{-A} + \rme^{2Az} \left( \Delta_{-2A} + \frac{1}{2} \Delta_{-A}^2 \right) + \cdots.
\ee
\noindent
The action of $\underline{\frak{S}}_\pi$ on $F(z,\sigma)$ is again entirely encoded by the action of $\underline{\frak{S}}_\pi$ on the several $\Phi_n (z)$, and is determined by the use of the bridge equations (\ref{bridgePhi}). All one needs are formulae for multiple alien derivatives, which follow straightforwardly as\footnote{Of course this expression holds as long as $n - \sum_{i=1}^N \ell_i \not = 0$. As soon as this term vanishes, so does the multiple alien derivative, and consequently so will all subsequent ones.}
\be
\prod_{i=1}^N \Delta_{-\ell_{(N+1-i)} A} \Phi_n = \prod_{i=1}^N S_{-\ell_i} \cdot \prod_{i=1}^N \left( n - \sum_{j=1}^i \ell_j \right) \cdot \Phi_{n - \sum_{i=1}^N \ell_i}.
\ee
\noindent
Notice that the \textit{ordering} of the alien derivatives in the left--hand side of the expression above is rather fundamental, as alien derivatives computed at \textit{different} singular points do \textit{not} commute. For example, it is simple to check that $\left[ \Delta_{-n A}, \Delta_{-m A} \right] \propto \left( n-m \right)$. Also, because the alien derivatives vanish as soon as one considers $\Delta_{-n A} \Phi_n = 0$, this apparent series actually truncates to a finite sum, at each stage. One may simply compute
\bea\label{disc@pi}
\underline{\frak{S}}_{\pi} \Phi_{n} &=& \Phi_{n} + \sum_{\ell=1}^{n-1} \rme^{\ell A z}\, \sum_{k=1}^{\ell} \frac{1}{k!}\, \underset{\sum_i \ell_{i} = \ell}{\sum_{\ell_{1}, \dots, \ell_{k} \ge 1}} \left\{ \prod_{j=1}^{k} S_{-\ell_{j}} \cdot \prod_{j=1}^{k} \left( n - \sum_{m=1}^{j}\ell_{m} \right) \right\} \Phi_{n-\ell} \\
&=&
\Phi_{n} + \sum_{\ell=1}^{n-1} \rme^{\ell Az}\, \sum_{k=1}^{\ell} \frac{1}{k!}\, \sum_{0 = \gamma_{0} < \gamma_{1} < \dots < \gamma_{k} = \ell}\, \left( \prod_{j=1}^{k} \left( n-\gamma_{j} \right) S_{-\mathbf{d} \gamma_j} \right) \Phi_{n-\ell}. 
\eea
\noindent
In the last line, the sum over all possible partitions $\ell_{i} \ge 1$ was replaced by a sum over their consecutive sums $\gamma_{s} = \sum_{j=1}^{s} \ell_{j}$ and we defined $\mathbf{d} \gamma_j \equiv \gamma_{j}-\gamma_{j-1}$ (\textit{i.e.}, the partitions). A few examples of the Stokes automorphism at $\theta = \pi$ are given below:
\bea
\underline{\frak{S}}_\pi \Phi_0 &=& \Phi_0,\\
\underline{\frak{S}}_\pi \Phi_1 &=& \Phi_1, \\
\underline{\frak{S}}_\pi \Phi_2 &=& \Phi_2 + S_{-1}\, \rme^{Az}\, \Phi_1, \\
\underline{\frak{S}}_\pi \Phi_3 &=& \Phi_3 + 2 S_{-1}\, \rme^{Az}\, \Phi_2 + \left( S_{-2} + S_{-1}^2 \right) \rme^{2Az}\, \Phi_1. \label{stokes-at-pi-n-3}
\eea

Finally, making use of the Stokes automorphism (\ref{stokesauto}), one may directly apply the bridge equation (\ref{bridgeF}) in order to find, \textit{e.g.},
\be
\CS_{\theta^+} F(z,\sigma) = \CS_{\theta^-} \exp \left( \dot{\Delta}_\omega \right) F(z,\sigma) = \CS_{\theta^-} F \left( z, \sigma \left( 1 + \omega\, S_\omega\, \sigma^{-\omega}\right)^{\frac{1}{\omega}} \right).
\ee
\noindent
In particular, when $\omega=1$, $\arg \omega =0$, this is
\be\label{stokestransition@0}
\CS_{+} F(z,\sigma) = \CS_{-} F \left( z, \sigma + S_1 \right),
\ee
\noindent
in such a way that $S_1$ acts as a Stokes constant for the transseries expression. For this reason, we shall generally refer to the analytic invariants $S_\ell$ as ``Stokes constants''. Of course this exact same expression could be obtained by applying the Stokes automorphism at $\theta=0$, (\ref{disc@zero}), to the transseries (\ref{Ftransseries}) (trying the same at $\theta=\pi$, via (\ref{disc@pi}), would be much more complicated). In the original complex $z$--plane this Borel--plane singular--direction corresponds to a Stokes line and what the expression above describes is precisely the Stokes phenomena of classical asymptotics---here fully and naturally incorporated in the resurgence analysis. At a Stokes line, subleading exponentials start contributing to the asymptotics and this is accomplished in here by the ``jump'' of $\sigma$, the \textit{coefficient} associated to the transseries formal sum over (multi--instanton) solutions. In other words, the ``connection'' expression (\ref{glue}) yields a relation between the coefficient(s) in the transseries solution, in different parts of its domain, or, on different sides of the Stokes line.

\subsection{Stokes Constants and Asymptotics}

One may wonder why the long detour into resurgence and alien calculus. As it turns out, understanding the \textit{full} asymptotic behavior of \textit{all} multi--instanton sectors---which is to say, fully understanding the nonperturbative structure of the problem at hand---demands for this complete formalism. Let us first recall the standard large--order dispersion relation that follows from Cauchy's theorem \cite{z81}: if a function $F(z)$ has a branch--cut along some direction, $\theta$, in the complex plane, and is analytic elsewhere, it follows
\be\label{cauchy}
F(z) = \frac{1}{2\pi\rmi} \int_0^{\rme^{\rmi\theta} \cdot \infty} \rmd w\, \frac{\disc_\theta\, F(w)}{w-z} - \oint_{(\infty)} \frac{\rmd w}{2\pi\rmi}\, \frac{F(w)}{w-z}.
\ee
\noindent
In certain situations \cite{bw73, cs78} it is possible to show by scaling arguments that the integral around infinity does not contribute. In such cases Cauchy's theorem provides a remarkable connection between perturbative and nonperturbative expansions. Let us first consider our familiar perturbative expansion (\ref{d}), within the transseries set up (\ref{Ftransseries}), where $F^{(0)}(z) = \Phi_0(z)$. In this case, the bridge equations (\ref{bridgePhi}) tell us, via the Stokes automorphisms (\ref{disc@zero}) and (\ref{disc@pi}), that $F^{(0)}(z)$ has the following discontinuities:
\bea
\disc_0\, \Phi_0 &=& - \sum_{\ell=1}^{+\infty} S_1^\ell\, \rme^{- \ell A z}\, \Phi_{\ell}, \\
\disc_\pi\, \Phi_0 &=& 0,
\eea
\noindent
\textit{i.e.}, $F^{(0)}(z)$ has a single branch cut along the Stokes direction corresponding to the positive real axis in the Borel complex plane. From the perturbative expansion (\ref{d}) and using (\ref{cauchy}) above, it immediately follows
\be\label{zeroinstasymp}
F^{(0)}_g \simeq \sum_{k=1}^{+\infty} \frac{S_1^k}{2\pi\rmi}\, \frac{\Gamma \left( g-k\beta \right)}{\left( k A \right)^{g-k\beta}}\, \sum_{h=1}^{+\infty} \frac{\Gamma \left( g-k\beta-h+1 \right)}{\Gamma \left( g-k\beta \right)}\, F^{(k)}_h \left( k A \right)^{h-1},
\ee
\noindent
where we have used the asymptotic expansions for the multi--instanton contributions (\ref{multiseries}). It is instructive to explicitly write down the first terms in this double--series,
\bea
F^{(0)}_g &\simeq& \frac{S_1}{2\pi\rmi}\, \frac{\Gamma \left( g-\beta \right)}{A^{g-\beta}} \left( F^{(1)}_1 + \frac{A}{g-\beta-1}\, F^{(1)}_2 + \cdots \right) + \nonumber \\
&&
+ \frac{S_1^2}{2\pi\rmi}\, \frac{\Gamma \left( g-2\beta \right)}{\left( 2 A \right)^{g-2\beta}} \left( F^{(2)}_1 + \frac{2A}{g-2\beta-1}\, F^{(2)}_2 + \cdots \right) + \nonumber \\
&&
+ \frac{S_1^3}{2\pi\rmi}\, \frac{\Gamma \left( g-3\beta \right)}{\left( 3 A \right)^{g-3\beta}} \left( F^{(3)}_1 + \frac{3A}{g-3\beta-1}\, F^{(3)}_2 + \cdots \right) + \cdots.
\eea
\noindent
This is the multi--instanton generalization of a well--known result, also from previous work within the matrix model and topological string theory contexts, \textit{e.g.}, \cite{msw07, m08}. In particular, it relates the coefficients of the perturbative expansion around the zero--instanton sector with a sum over the coefficients of the perturbative expansions around all multi--instanton sectors, in an asymptotic expansion which holds for large $g$ (and positive real part of the instanton action). In particular, the computation of the one--loop one--instanton partition function determines the leading order of the asymptotic expansion for the perturbative coefficients of the zero--instanton partition function, up to the Stokes factor $S_1$. Higher loop contributions then yield the successive $\frac{1}{g}$ corrections. Furthermore, multi--instanton contributions with action $n A$ will yield corrections to the asymptotics of the $F^{(0)}_g$ coefficients which are exponentially suppressed as $n^{-g}$.

The novelty here arises due to the use of alien calculus, which allows for a straightforward incorporation of \textit{all} multi--instanton sectors in the asymptotic formulae, as well as a generalization of this procedure to \textit{all} multi--instanton sectors! Indeed, in terms of asymptotics of instanton series, we shall find that the bridge equations (\ref{bridgePhi}) essentially tell us that, given a fixed instanton sector, its \textit{leading} asymptotics are determined by both the \textit{next} and the \textit{previous} instanton contributions---at least in examples where a transseries \textit{ansatz} depending on a single parameter is enough (we shall later see examples where things get more complicated). In particular, at the level of Borel transforms, the singularities closest to the origin, of $\CB [\Phi_n] (s)$, are located at $s=\pm A$ (if $n=0,1$ there is a single closest--to--the--origin singularity located at $s=A$), and these singularities will necessarily control the large--order behavior of the multi--instanton sectors. 

Let us now address the full $n$--instanton sector. Consider our perturbative expansion (\ref{multiseries}) within the transseries set--up (\ref{Ftransseries}), where $F^{(n)} (z) = \rme^{-n A z}\, \Phi_n (z)$. In this case, the bridge equations (\ref{bridgePhi}) tell us, via the Stokes automorphisms (\ref{disc@zero}) and (\ref{disc@pi}), that $F^{(n)}(z)$ has the following discontinuities:
\begin{eqnarray}
\disc_{0}\, \Phi_{n} &=& - \sum_{\ell=1}^{+\infty} \binom{n+\ell}{n}\, S_{1}^{\ell}\, \rme^{-\ell A z}\, \Phi_{n+\ell}, \\
\disc_{\pi}\, \Phi_{n} &=& - \sum_{\ell=1}^{n-1} \rme^{\ell A z}\, \sum_{k=1}^{\ell} \frac{1}{k!} \sum_{0 = \gamma_{0} < \gamma_{1} < \dots < \gamma_{k} = \ell} \left( \prod_{j=1}^{k} \left( n - \gamma_{j} \right) S_{-\mathbf{d} \gamma_j} \right) \Phi_{n-\ell},
\end{eqnarray}
\noindent
\textit{i.e.}, $F^{(n)}(z)$ has branch cuts along the Stokes directions corresponding to both positive and negative real axes in the Borel complex plane. The contribution from the discontinuity at $\theta=\pi$ can also be rewritten as a sum over Young diagrams $\gamma_{i} \in \Gamma(k,\ell) \,:\, 0 \le \gamma_{1} \le \cdots \le \gamma_{k} = \ell$ of length $\ell(\Gamma) = k$, and with maximum number of boxes for each part $\gamma_{i}$ (also called the length of the transposed Young diagram) being $\ell(\Gamma^{T})=\ell$. This sum is only completely well--defined if we also set $S_{0},\gamma_{0}\equiv0$, in which case one finally obtains
\begin{equation}\label{disc-pi-one-param-young}
\disc_{\pi}\, \Phi_{n} = - \sum_{\ell=1}^{n-1} \rme^{\ell A z}\, \sum_{k=1}^{\ell} \frac{1}{k!} \sum_{\gamma_{i} \in \Gamma(k,\ell)} \left( \prod_{j=1}^{k} \left( n-\gamma_{j} \right) S_{-\mathbf{d} \gamma_j} \right) \Phi_{n-\ell}.
\end{equation}
\noindent
For example, consider once again the case $n=3$ in the notation above,
\be
\disc_{\pi}\, \Phi_{3} = - \rme^{Az} \sum_{\gamma_i \in \Gamma(1,1)} \Big( \left( 3-\gamma_{1} \right) S_{-\mathbf{d} \gamma_1} \Big)\, \Phi_{2} - \rme^{2Az}\, \sum_{k=1}^{2} \frac{1}{k!} \sum_{\gamma_i \in \Gamma(k,2)} \left( \prod_{j=1}^{k} \left( 3-\gamma_{j} \right) S_{-\mathbf{d} \gamma_j} \right) \Phi_{1}.
\ee
\noindent
Expanding the sums, there will be only one Young diagram corresponding to $\Gamma(1,1)$, $\tableau{1}$, for which $\gamma_{1}=1$. For $\Gamma(1,2)$ one can only find $\tableau{2}$ ($\gamma_{1}=2$), while for $\Gamma(2,2)$ there are two possible Young diagrams: $\tableau{21}$ ($\gamma_{1}=1,\,\gamma_{2}=2$) and $\tableau{22}$ ($\gamma_{1}=\gamma_{2}=2$; but this will not contribute because $S_{0}=0$). The expected result arising from (\ref{stokes-at-pi-n-3}) then simply follows.

From the perturbative expansion (\ref{multiseries}) and the dispersion relation (\ref{cauchy}), which now needs to account for both branch cuts, it finally follows
\begin{eqnarray}
F_{g}^{(n)} &\simeq& \sum_{k=1}^{+\infty} \binom{n+k}{n}\, \frac{S_{1}^{k}}{2\pi\rmi} \cdot \frac{\Gamma \left( g-k\beta \right)}{\left( k A \right)^{g-k\beta}}\, \sum_{h=1}^{+\infty} \frac{\Gamma \left( g-k\beta-h \right)}{\Gamma \left( g-k\beta \right)}\, 
F_{h}^{(n+k)} \left( k A \right)^{h} + \nonumber \\
&&
+ \sum_{k=1}^{n-1} \left\{ \frac{1}{2\pi\rmi} \sum_{m=1}^{k} \frac{1}{m!} \sum_{\gamma_{i} \in \Gamma(m,k)}  \left( \prod_{j=1}^{m} \left( n-\gamma_{j} \right) S_{-\mathbf{d} \gamma_j} \right) \right\} \times \nonumber \\
&&
\times \frac{\Gamma \left( g+k\beta \right)}{\left( - k A \right)^{g+k\beta}}\, \sum_{h=1}^{+\infty} \frac{\Gamma \left( g+k\beta-h \right)}{\Gamma \left( g+k\beta \right)}\, F_{h}^{(n-k)} \left( - k A \right)^{h}. \label{ninstasymp}
\end{eqnarray}
\noindent
This expression relates the coefficients of the perturbative expansion around the $n$--instanton sector with sums over the coefficients of the perturbative expansions around \textit{all} other multi--instanton sectors, in an asymptotic expansion which holds for large $g$. All Stokes factors are now needed for the general asymptotic problem, and this analysis has essentially boiled down the asymptotic problem to a problem of precisely computing these Stokes factors. These numbers are transcendental invariants of the problem one is addressing and generically hard to compute---although, as we shall see, a matrix model computation partially solves this issue. Notice that for instanton numbers $n=0,1$ the combinatorial factor associated to the Stokes factors $S_\ell$ at negative $\ell$ vanishes. As such, the contributions arising from the second and third lines in the expression above can only be seen at instanton number $n=2$ and above. Finally, let us note that the explicit treatment of the \textit{leading} contribution to this type of asymptotics was first presented, to the best of our knowledge, in \cite{gm08}.

\section{Topological Strings in the Gopakumar--Vafa Representation}\label{topstringGV}

The first concrete example we shall explore deals with topological string theory, where the free energy admits an integral Gopakumar--Vafa (GV) representation, see \cite{gv98a, gv98c}. Consider the free energy of the A--model, on a Calabi--Yau (CY) threefold $\CX$, with complexified K\"ahler parameters $\{ t_i \}$. As a string theory it satisfies the standard topological genus expansion ({\ref{stringgenus}) where, at genus $g$, for large values of the K\"ahler parameters (the large--radius phase), one finds \cite{bcov93}
\be
F_g (t_i) = \sum_{d_i = 1}^{+\infty} N_{g,d} \left(\CX\right) \rme^{- \boldsymbol{d} \cdot \boldsymbol{t}},
\ee
\noindent
where the sum is over K\"ahler classes\footnote{$\boldsymbol{d} = \left( d_1, \ldots, d_{b_2 (\CX)} \right)$ denotes the expansion of the two--homology class $d$ on a basis of $H_2 \left( \CX, \BZ \right)$; see, \textit{e.g.}, \cite{m05}.} and where the coefficients $N_{g,d} (\CX)$ are the Gromov--Witten (GW) invariants of $\CX$, counting world--sheet instantons, \textit{i.e.}, the number of curves of genus $g$ and degree $d$ in $\CX$. As we mentioned before, this $\a$ expansion is the milder one with finite convergence radius $t_{\mathrm{c}}$, where the conifold singularity is reached, which may be estimated from the asymptotic behavior of GW invariants at large degree (here $\gamma$ is a critical exponent; see, \textit{e.g.}, \cite{cgmps06}) \cite{bcov93}
\be
N_{g,d} \sim d^{(\gamma-2)(1-g)-1}\, \rme^{d\, t_{\mathrm{c}}}, \qquad d \to + \infty.
\ee

What we shall be interested in next is instead the asymptotic genus expansion. In this case, and as throughly investigated for the resolved conifold in \cite{ps09}, the GV integral representation for the free energy may be interpreted as a Borel resummation formula, immediately yielding, as we will see in the following, the ``leading'' part of the topological string resurgent data.

\subsection{Topological String Free Energy and Borel Resummation}

Let us thus consider the GV integral representation for the all--genus topological string free energy on a CY threefold $\CX$, including nonperturbative M--theory corrections via the type IIA $\leftrightarrow$ M$/\BS^1$ duality \cite{gv98c} (see also \cite{gv98, gv98a, gv98b} and \cite{ps09} for a discussion in the present context which further highlights the Schwinger--like nature of this result),
\be\label{gvrep}
F_\CX (g_s) \simeq \sum_{r=0}^{+\infty} \sum_{d_i=1}^{+\infty} n^{(d_i)}_r (\CX) \sum_{m \in \BZ} \int_0^{+\infty} \frac{\rmd s}{s} \left( 2 \sin \frac{s}{2} \right)^{2r-2}\, \exp \left( - \frac{2 \pi s}{g_s} \left( \boldsymbol{d} \cdot \boldsymbol{t} + \rmi\, m \right) \right).
\ee
\noindent
In here, the integers $n^{(d_i)}_r (\CX)$ are the GV invariants of the threefold $\CX$, depending both on the K\"ahler class $d_i$ and on a spin label $r$, and the combination $Z = \boldsymbol{d} \cdot \boldsymbol{t} + \rmi\, m$ represents the central charge of certain four--dimensional BPS states \cite{gv98c}. To be completely precise, notice that in order to obtain the full topological string free energy one still has to add to this expression the (alternating) constant map contribution \cite{mm98, fp98}
\be\label{constmap}
N_{g,0} = \frac{(-1)^{g} \left| B_{2g} B_{2g-2} \right|}{4 g \left( 2g-2 \right) \left( 2g-2 \right)!}\, \chi (\CX),
\ee
\noindent
where $\chi (\CX) = 2 \left( h^{1,1}-h^{2,1} \right)$ is the Euler characteristic of $\CX$. This term can also be written as a Borel--like resummation, where the result is\footnote{This may also be obtained directly from the GV representation by simply setting $d_i=0$ and $r=0$ in  (\ref{gvrep}) and properly identifying the ``degree zero'' and ``spin zero'' GV invariant with the Euler number of the CY threefold.} \cite{gv98a, ps09}
\be
F_{d=0} (g_s) \simeq \sum_{g=0}^{+\infty} g_s^{2g-2} N_{g,0} = \frac{1}{2}\, \chi (\CX) \sum_{m \in \BZ} \int_0^{+\infty} \frac{\rmd s}{s}\, \left( 2 \sin \frac{s}{2} \right)^{-2}\, \exp \left( - \frac{2\pi s}{g_s}\, \rmi\, m \right).
\ee

Apart from the overall multiplicative factor of the Euler characteristic, the universal constant map contribution has already been fully addressed in \cite{ps09} and we shall thus leave it aside for the moment. Let us focus on the GV contribution (\ref{gvrep}) instead. By rewriting the sum in $m \in \BZ$ as a sum over delta--functions it is simple to obtain the GV formula for the topological string free energy as \cite{gv98c}
\be
F_\CX (g_s) \simeq \sum_{r=0}^{+\infty} \sum_{d_i=1}^{+\infty} n^{(d_i)}_r \sum_{n=1}^{+\infty} \frac{1}{n} \left( 2 \sin \frac{n g_s}{2} \right)^{2r-2}\, \rme^{- 2 \pi n\, \boldsymbol{d} \cdot \boldsymbol{t}}.
\ee
\noindent
This makes it quite clear how the input data for a given CY threefold is simply its set of GV integer invariants (and its Euler number if one is also to write down the constant map contribution). Expressed as a topological genus expansion one finds, at genus $g$, (see, \textit{e.g.}, \cite{mm98, m05, glm08} for partial expressions)
\be\label{gvgenusformula}
F_g (t_i) = \sum_{d_i = 1}^{+\infty} \left\{ \frac{|B_{2g}|}{2g \left( 2g-2 \right)!}\, n_0^{(d_i)} + \sum_{h=1}^g (-1)^{g-h} \frac{\alpha^{(h-1)}_{g-h+1}}{\left( 2g-2 \right)!}\, n_h^{(d_i)} \right\} \mathrm{Li}_{3-2g} \left( \rme^{- 2\pi\, \boldsymbol{d} \cdot \boldsymbol{t}} \right),
\ee
\noindent
where the coefficients $\alpha^{(n)}_m$ are obtained from the generating function
\be
A_n (x) = \frac{(2n)!}{\prod_{k=1}^n \left( 1 - k^2 x \right)} \equiv \sum_{m=0}^{+\infty} \alpha^{(n)}_{m+1} x^m
\ee
\noindent
by power series expansion, and where $\mathrm{Li}_p (x)$ is the polylogarithm of order $p$, defined as 
\be
\mathrm{Li}_p (x) = \sum_{n=1}^{+\infty} \frac{x^n}{n^p}.
\ee
\noindent
Two things to notice are the following: at fixed genus $g$, only GV invariants $n_h^{(d_i)}$ with $h \leq g$ contribute to the free energy \cite{gv98c}; in particular the ``highest'' GV invariant at genus $g$ has $h=g$ and appears with coefficient one in (\ref{gvgenusformula}) as $\frac{\alpha^{(g-1)}_1}{\left( 2g-2 \right)!}=1$, $\forall g$. Furthermore, $\alpha^{(0)}_g$ is only non--vanishing when $g=1$, implying that $n_1^{(d_i)}$ only contributes to the genus one free energy.

What we want to understand in here is how or when the GV representation (\ref{gvrep}) may be understood as a nonperturbative completion of the free energy genus expansion (\ref{gvgenusformula}), in the sense of resurgent analysis. Furthermore, one would like to understand how to relate this nonperturbative completion to the large--order behavior of the genus expansion (\ref{gvgenusformula}) via the use of Stokes' automorphism. Following the approach in \cite{ps09}, we shall interpret the GV integral representation for the free energy (\ref{gvrep}) as a Borel resummation formula (\ref{borelinttheta}), for $\CS_\theta F_\CX (g_s)$, in such a way that, after a simple change of variables, one obtains
\be\label{gvborel}
\CB[F_\CX](s) = \sum_{r=0}^{+\infty} \sum_{d_i=1}^{+\infty} n^{(d_i)}_r \sum_{m \in \BZ} \frac{1}{s} \left( 2 \sin \frac{s}{4 \pi \left( \boldsymbol{d} \cdot \boldsymbol{t} + \rmi\, m \right)} \right)^{2r-2},
\ee
\noindent
with the GV representation (\ref{gvrep}) now amounting to the statement that
\be
\CS_\theta F_\CX (g_s) = \int_0^{\rme^{\rmi\theta}\infty} \rmd s\, \CB[F_\CX](s)\, \rme^{-\frac{s}{g_s}}.
\ee
\noindent
This rewriting, of course, required changing the integration with the (in general) infinite sums over GV invariants, a procedure which is only valid if there is uniform convergence of the partial sums in (\ref{gvborel}). As we have seen before, the sum in $m$ is the milder one. Furthermore, at fixed degree, the sum in $r$ will truncate, \textit{i.e.}, given a fixed two--homology class $\{d_i\}$, there is $r_*$ such that $n^{(d_i)}_r = 0$ for all $r>r_*$ \cite{k06}. The real issue concerning uniform convergence of the GV Borel transform thus arises when we fix genus and sum over degree. In this case one finds that the asymptotic behavior of, for example, the genus zero GV invariants at large degree is \cite{kz03}
\be
n^{(d)}_0 \sim \frac{\exp \left( 2\pi t_2(1) \cdot d \right)}{d^3 \left( \log d \right)^2}, \qquad d \to + \infty,
\ee
\noindent
where $2\pi t_2(1)$ is a critical exponent (for instance, in the example of local $\BP^2$ this would be  $2\pi t_2(1) \simeq 2.90759...$ \cite{kz03}). This is an exponential growth and, as such, in strict validity, the results that follow only hold for threefolds with a \textit{finite} number of GV invariants, \textit{i.e.}, without compact four--cycles. This is also in line with the general expectations briefly discussed in \cite{ps09}.

In this context, the only singularities of the GV Borel transform (\ref{gvborel}), with $s\not=0$, appear when $r=0$ as the zeroes of the sine (located at $\omega_n = (2\pi)^2\, n \left( \boldsymbol{d} \cdot \boldsymbol{t} + \rmi\, m \right)$, $n \in \BZ^*$). In this case one will only find pole singularities and the Borel transform (\ref{gvborel}) may be written as
\be\label{gvborelsing}
\CB[F_\CX](\omega_n+s) = \frac{1}{2\pi}\, n^{(d_i)}_0 \left( \frac{2\pi \left( \boldsymbol{d} \cdot \boldsymbol{t} + \rmi\, m \right)}{n\, s^2} - \frac{1}{2 \pi n^2\, s} \right) + \mathrm{holomorphic},
\ee
\noindent
near each singular point $\omega_n$. The (multiple) instanton action, $\omega_n = n A$, is further obtained as
\be\label{topstringinstaction}
A_m (t_i) = (2\pi)^2 \left( \boldsymbol{d} \cdot \boldsymbol{t} + \rmi\, m \right).
\ee
\noindent
In the following we shall make use of this information in order to explore, from a resurgent point of view, when does the Borel interpretation of the GV integral representation (\ref{gvborel}) provide for a nonperturbative completion of the topological string free energy.

\subsection{Simple Resurgence in Topological String Theory}

The first step in understanding the resurgence of topological strings is to compute alien derivatives. At first, this could seem non--trivial as the GV Borel transform (\ref{gvborelsing}) is not quite a simple resurgent function due to the second order pole. However, explicitly evaluating the difference of lateral Borel resummations as in (\ref{npa}), one notices that the contribution from this second order pole is simple to include in the alien derivative, which now becomes, for $n \in \BZ^*$,
\be
\Delta_{n A} F_\CX = - \frac{\rmi}{2\pi g_s}\, n^{(d_i)}_0 \left( \frac{(2\pi)^2 \left( \boldsymbol{d} \cdot \boldsymbol{t} + \rmi\, m \right)}{n} + \frac{g_s}{n^2} \right) \equiv \Lambda_n,
\ee
\noindent
with an added unusual dependence on the coupling constant. In spite of this, the right--hand side above is in fact a resurgent constant, in such a way that all multiple alien derivatives vanish. In this case, it is trivial to compute Stokes' automorphism, (\ref{stokesauto}). Denoting\footnote{Notice that in the original integration variable of (\ref{gvrep}), therein denoted $s$, this would correspond to $\theta=0$.} by $\theta = \arg A$, this is
\be
\underline{\frak{S}}_\theta F_\CX = F_\CX + \sum_{n=1}^{+\infty} \sum_{d_i=1}^{+\infty} \sum_{m \in \BZ} \Lambda_n \cdot \exp \left( - \frac{(2\pi)^2\, n \left( \boldsymbol{d} \cdot \boldsymbol{t} + \rmi\, m \right)}{g_s} \right),
\ee
\noindent
leading to the discontinuity
\be
\disc_\theta\, F_\CX = \frac{\rmi}{2\pi g_s} \sum_{n=1}^{+\infty} \sum_{d_i=1}^{+\infty} n^{(d_i)}_0 \sum_{m \in \BZ} \left( \frac{(2\pi)^2 \left( \boldsymbol{d} \cdot \boldsymbol{t} + \rmi\, m \right)}{n} + \frac{g_s}{n^2} \right)
\rme^{- \frac{(2\pi)^2\, n \left( \boldsymbol{d} \cdot \boldsymbol{t} + \rmi\, m \right)}{g_s}}.
\ee
\noindent
Finally, making use of the dispersion relation (\ref{cauchy}), where one further assumes that the contribution around infinity may be neglected, one may now compute all coefficients in the perturbative asymptotic expansion of $F_\CX$, which has the usual genus expansion form (\ref{stringgenus}). Focusing on the discontinuity naturally induced by the GV integral representation (\ref{gvrep}), namely $\arg s = 0$, and following a calculation very similar to the one in \cite{ps09} for the case of the resolved conifold, it follows
\be
F_\CX (g_s) \simeq \sum_{g=1}^{+\infty} g_s^{2g-2} \sum_{d_i = 1}^{+\infty} n_0^{(d_i)}\, \frac{|B_{2g}|}{2g \left( 2g-2 \right)!}\, \mathrm{Li}_{3-2g} \left( \rme^{- 2\pi\, \boldsymbol{d} \cdot \boldsymbol{t}} \right).
\ee

Some comments are in order concerning this result. The first obvious one is that this does not fully match against the GV result (\ref{gvgenusformula}), as it only captures the leading, dominant Bernoulli growth of the free energy. While this is certainly the correct expectation for an asymptotic formula in the case of a finite number of GV invariants, one may also ask if it is possible to do any better. Of course, if one is to start with the GV Borel transform (\ref{gvborel}), its singular part (\ref{gvborelsing}) will not include any GV invariant  $n^{(d_i)}_r$ with $r\not=0$ and, as such, will never be able to yield the subleading contributions in (\ref{gvgenusformula}) unless these should arise from the singularity at infinity in the Cauchy dispersion relation (\ref{cauchy}). While this is a possibility, it is also a notoriously difficult case to handle---the singularity at infinity is an essential singularity, leading us far from the realm of simple resurgent functions---further departing from the conventional set--up of resurgent asymptotics. At the end of the day this ``loss'' of GV invariants $n^{(d_i)}_r$ with $r\not=0$ arises from the exchange of integration and infinite sums in (\ref{gvrep}) to obtain (\ref{gvborel}) and all it says is that another procedure will be required in order to look beyond the Bernoulli growth in (\ref{gvgenusformula}), \textit{i.e.}, to study the \textit{full} nonperturbative information of topological string theory. In other words, while the GV integral representation is extremely useful in order to solve topological string theory at the perturbative level, (\ref{gvgenusformula}), one needs extra work if one wants, in general, to obtain a closed form expression for the topological string Borel transform---possibly in terms of GV invariants.

At this point it might be useful to make a bridge to the case of matrix models with polynomial potentials (a subject we shall study in detail later in this paper). For these, the Gaussian component of the polynomial potential will induce a contribution to the free energy which also leads to Bernoulli growth \cite{ps09}, rather similar to the one above arising from genus zero GV invariants. From a spectral curve point of view, both these contributions are associated to A--cycle\footnote{These are instantons whose action is given by the period of the spectral curve one--form around one of its A--cycles \cite{ps09}. They are simpler than B--cycle instantons (almost ``universal'' as they directly relate to the 't~Hooft moduli), whose action is given by the period of the spectral curve one--form around one of its B--cycles \cite{msw07}.} instantons \cite{ps09}. Instantons of this type are always very simple to handle. As described above, the alien derivative is essentially trivial (it equals a resurgent constant) and at the end of the day the asymptotics is somewhat universal---and certainly much simpler than the discussion in the previous section. All multi--instanton sectors have no non--trivial large--order behavior (their alien derivatives vanish, their series truncate and their structure is thus rather different from the one in (\ref{multiseries})) and the perturbative sector is essentially dominated by Bernoulli growth. For matrix models with polynomial potentials the truly non--trivial resurgent structure will then be associated to higher monomials in the potential which will induce different contributions to the free energy, this time around associated to B--cycle instantons \cite{msw07}. More realistic examples of this non--trivial resurgent structure associated to B--cycle instantons will be discussed next, as we move to the realm of minimal strings and matrix models in the following sections. For the moment, let us just notice that, in general, we still expect topological strings to display full non--trivial resurgence: if one wants to see beyond the Bernoulli growth in (\ref{gvgenusformula}) one will certainly need to find a proper Borel transform, leading to non--trivial alien derivatives and asymptotic growth of all multi--instanton sectors. Thus, in general, there will be both A and B--cycle instantons in topological string models, both contributing to the full instanton action, and controlling (in turns, depending on the absolute value of their corresponding actions) the large--order behavior of perturbation theory at different values of the 't~Hooft moduli, as recently discussed in \cite{dmp11}.

\section{The Resurgence of Two--Parameters Transseries}\label{sec:4}

In order to address broader string theoretic contexts, in particular those involving minimal string theory or matrix models, as we shall study later in this work, we now need to generalize the formalism introduced in section \ref{borelrev} in order to include transseries depending on multiple parameters. Let us start off with some words on the general transseries set--up (see, \textit{e.g.}, \cite{m10} for a recent review, or, \textit{e.g.}, \cite{c95, c98} for more technical accounts).

A rank--$n$ system of non--linear ordinary differential equations,
\be\label{nlode}
\frac{\rmd \boldsymbol{u}}{\rmd z} (z) = \boldsymbol{F} \big( z, \boldsymbol{u}(z) \big),
\ee
\noindent
may always be written, via a suitable change of variables, in the so--called prepared form \cite{m10}:
\be\label{preparednlode}
\frac{\rmd \boldsymbol{u}}{\rmd z} (z) = - \boldsymbol{A} \cdot \boldsymbol{u} (z) - \frac{1}{z} \boldsymbol{B} \cdot \boldsymbol{u} (z) + \boldsymbol{G} \big( z, \boldsymbol{u}(z) \big).
\ee
\noindent
Denoting by $\left\{ \alpha_i \right\}_{i=1\cdots n}$ the eigenvalues of the linearized system,
\be
\mathbb{A} = \left[ \frac{\partial F_i}{\partial u_j} \left( \infty, \boldsymbol{0} \right) \right]_{i,j=1\cdots n},
\ee
\noindent
then, in the expression above, $\boldsymbol{A} = \mathrm{diag} \left( \alpha_1, \ldots, \alpha_n \right)$ and $\boldsymbol{B} = \mathrm{diag} \left( \beta_1, \ldots, \beta_n \right)$ are diagonal matrices and one further insures that $\boldsymbol{G} \big( z, \boldsymbol{u}(z) \big) = \CO \left( \| \boldsymbol{u} \|^2, z^{-2} \boldsymbol{u} \right)$. It is also convenient to choose variables such that $\alpha_1 > 0$. Most cases addressed in the literature deal with the \textit{non--resonant} case, where the eigenvalues $\left\{ \alpha_i \right\}_{i=1\cdots n}$ are $\BZ$--linearly independent, in many cases with $\arg \alpha_i \not = \arg \alpha_j$. This will \textit{not} be the case in the present work, as the string theoretic systems we address \textit{resonate}. In the above set--up, a formal transseries solution to our system of differential equations (\ref{nlode}) is given by \cite{m10}
\be
\boldsymbol{u} (z, \boldsymbol{\sigma}) = \boldsymbol{u}^{(0)} (z) + \sum_{\boldsymbol{n}\in\BN^n\backslash\{\boldsymbol{0}\}} \boldsymbol{\sigma}^{\boldsymbol{n}}\, z^{-\boldsymbol{n} \cdot \boldsymbol{\beta}}\, \rme^{- \boldsymbol{n} \cdot \boldsymbol{\alpha}\, z}\, \boldsymbol{u}^{(\boldsymbol{n})} (z),
\ee
\noindent
where $\boldsymbol{\sigma} = \left( \sigma_1, \ldots, \sigma_n \right)$ are the transseries parameters, and where both the perturbative contribution, $\boldsymbol{u}^{(0)} (z)$, as well as instanton and multi--instanton\footnote{Linear systems have no multi--instanton sectors.} contributions, $\boldsymbol{u}^{(\boldsymbol{n})} (z)$, are formal asymptotic power series of the form
\be
\boldsymbol{u}^{(\boldsymbol{n})} (z) \simeq \sum_{g=0}^{+\infty} \frac{\boldsymbol{u}^{(\boldsymbol{n})}_{g}}{z^g}.
\ee
\noindent
The fact that the systems we shall address in the following resonate now translates to
\be
\exists_{\boldsymbol{n} \not = \boldsymbol{n}'} \,\, | \,\, \boldsymbol{n} \cdot \boldsymbol{\alpha} = \boldsymbol{n}' \cdot \boldsymbol{\alpha}.
\ee
\noindent
Furthermore, one often deals with \textit{proper} transseries, where only exponentially suppressed contributions appear: the eigenvalues $\boldsymbol{\alpha}$ are such that, for some chosen direction in the complex $z$--plane, all contributions along this direction with $\sigma_i \not = 0$ are exponentially suppressed; $\re \left( \boldsymbol{n} \cdot \boldsymbol{\alpha}\, z \right) > 0$. Again, as first pointed out in \cite{gikm10}, if one wishes to fully address the instanton series in a string theoretic context one will also have to allow for less studied non--proper transseries. We thus see that resurgence in string theory is more intricate than usual, with resonant non--proper transseries.

As we have reviewed in section \ref{borelrev}, asymptotic series need to be Borel resummed in order to extract sensible information from them. Naturally, this will also be a required step in the construction of a transseries solution to the non--linear differential equation (\ref{nlode}), and it follows that \cite{c95, c98}
\be
\CS_{\theta^{\pm}} \boldsymbol{u} (z, \boldsymbol{\sigma}_{\pm}) = \CS_{\theta^{\pm}} \boldsymbol{u}^{(0)} (z) + \sum_{\boldsymbol{n}\in\BN^n\backslash\{0\}} \boldsymbol{\sigma}^{\boldsymbol{n}}_{\pm}\, z^{-\boldsymbol{n} \cdot \boldsymbol{\beta}}\, \rme^{- \boldsymbol{n} \cdot \boldsymbol{\alpha}\, z}\, \CS_{\theta^{\pm}} \boldsymbol{u}^{(\boldsymbol{n})} (z),
\ee
\noindent
is a good solution to our problem along a proper direction (at least for sufficiently large $|z|$). Many of the concepts introduced in section \ref{borelrev} now have a straightforward generalization, for instance a simple extension of Stokes' automorphism (\ref{glue}) where this time around one may write
\be
\CS_{\theta^{+}} \boldsymbol{u} (z, \boldsymbol{\sigma}) = \CS_{\theta^{-}} \boldsymbol{u} (z, \boldsymbol{\sigma} + \boldsymbol{S})
\ee
\noindent
for the crossing of a Stokes line, with $\boldsymbol{S}$ the associated Stokes constants.

We shall now construct the resurgent formalism for the specific case of two--parameters transseries, which will turn out to be the required framework to address the instanton series in 2d quantum gravity (as first uncovered in \cite{gikm10} for the case of the Painlev\'e I equation) as well as the instanton series in the quartic matrix model, as we shall discuss in this work.

\subsection{The Bridge Equations Revisited}\label{sub:Bridge-Eqns-revisited}

We have seen in section \ref{borelrev} how the bridge equations allow for a simple evaluation of alien derivatives (up to the determination of the Stokes invariants), (\ref{bridgePhi}), and how this result then allows for an exact evaluation of the Stokes automorphism along a singular direction in the Borel complex plane, (\ref{disc@zero}) and (\ref{disc@pi}). We have further seen in section \ref{borelrev} how the discontinuities associated to these singular directions end up determining the full multi--instanton asymptotics (\ref{ninstasymp}) and, in essence, solve the nonperturbative problem via the use of transseries solutions.

In general one requires multi--parameter transseries in order to set up full nonperturbative solutions which completely encode the multi--instanton asymptotics. For the main examples we shall study in this work, the quartic matrix model and its double--scaling limit, the Painlev\'e I equation, it turns out that a two--parameters transseries is required, as we shall see later and as discussed in \cite{gikm10}. We shall now derive the bridge equations in this situation.

In particular, we consider the special case of two--parameters transseries where the prepared form eigenvalues are $\{\pm A\}$, with $A$ the instanton action\footnote{This will be the relevant case for both the Painlev\'e I equation and the quartic matrix model.}. The transseries \textit{ansatz} is now simply
\begin{equation}\label{2Ftransseries}
F(z,\sigma_{1},\sigma_{2}) = \sum_{n=0}^{+\infty} \sum_{m=0}^{+\infty} \sigma_{1}^{n} \sigma_{2}^{m} F^{(n|m)}(z),
\end{equation}
\noindent
where the perturbative asymptotic series is 
\begin{equation}\label{2Fpertseries}
F^{(0|0)}(z) \simeq \sum_{g=0}^{+\infty} \frac{F_{g}^{(0|0)}}{z^{g+1}} \equiv \Phi_{(0|0)} (z)
\end{equation}
\noindent
and where the generalized multi--instanton contributions take the form\footnote{In here we are simplifying things a bit: as we shall later discuss in the Painlev\'e I framework, $\Phi_{(n|m)}(z)$ is not always a plain formal power series in $z$ but may sometimes also include logarithmic powers, of the form $\log^k z$ multiplied by formal power series in $z$. For clarity of discussion, we shall proceed under this simpler assumption.}
\begin{equation}\label{2Fnmseries}
F^{(n|m)}(z) \simeq z^{-\beta_{nm}}\, \rme^{-n\left(+A\right)z}\, \rme^{-m\left(-A\right)z}\, \sum_{g=1}^{+\infty} \frac{F_{g}^{(n|m)}}{z^{g}} \equiv \rme^{-\left(n-m\right)Az}\, \Phi_{(n|m)}(z).
\end{equation}
\noindent
The characteristic exponent is often taken to be of the form $\beta_{nm} = n\beta_{1} + m\beta_{2}$, but we shall also allow for more general combinations. Everything else is a a straightforward generalization of the standard result (\ref{multiseries}) and a simple application of our discussion at the beginning of this section.

Because $\exists_{(n,m) \not= (n',m')}\,|\, n-m=n'-m'$ this transseries describes a resonant system and it is not too hard to see that one can make the ``instanton number'' explicit by slightly reorganizing the previous transseries representation, obtaining
\begin{equation}
F(z,\sigma_{1},\sigma_{2}) = \sum_{n=0}^{+\infty} \sigma_{1}^{n}\, \rme^{-nAz} \sum_{m=0}^{+\infty} \left( \sigma_{1}\sigma_{2} \right)^{m} \Phi_{(m+n|m)}(z) + \sum_{n=1}^{+\infty} \sigma_{2}^{n}\, \rme^{nAz} \sum_{m=0}^{+\infty} \left( \sigma_{1}\sigma_{2} \right)^{m} \Phi_{(m|m+n)}(z).
\end{equation}
\noindent
This also introduces a natural notion of degree, 
\begin{equation}
\deg \left( \sigma_{1}^{n} \sigma_{2}^{m} \rme^{kAz} \right) = n-m+k,
\end{equation}
\noindent
such that the transseries $F(z,\sigma_{1},\sigma_{2})$ has degree zero.

Let us now consider the pointed alien derivative $\dot{\Delta}_{\ell A} = \rme^{-\ell Az} \Delta_{\ell A}$, $\ell\in\BZ^{*}$, which, as we discussed earlier, commutes with the usual derivative. The reasoning of section \ref{borelrev} used in deriving the bridge equation also holds now, albeit in the two--parameters case the space of solutions to the differential, or finite difference, string equation becomes two--dimensional \cite{gikm10} (we shall see this very explicitly in the examples that follow). It must then be the case that 
\begin{equation}\label{2bridgeF}
\dot{\Delta}_{\ell A} F(z,\sigma_{1},\sigma_{2}) = S_{\ell} (\sigma_{1},\sigma_{2})\, \frac{\partial F}{\partial\sigma_{1}} (z,\sigma_{1},\sigma_{2}) + \widetilde{S}_{\ell} (\sigma_{1},\sigma_{2})\, \frac{\partial F}{\partial\sigma_{2}} (z,\sigma_{1},\sigma_{2});
\end{equation}
\noindent
the bridge equation in the two--parameters setting. Let us explore its implications. First of all, it is quite simple to notice that this immediately determines the degrees of the proportionality factors as 
\begin{equation}
\deg S_{\ell} (\sigma_{1},\sigma_{2}) = 1-\ell \qquad \mathrm{and} \qquad \deg \widetilde{S}_{\ell} (\sigma_{1},\sigma_{2}) = -1-\ell.
\end{equation}
\noindent
Because these should be expressed as formal power series expansions, this further implies 
\begin{equation}
S_{\ell} (\sigma_{1},\sigma_{2}) = \sum_{k=\max(0,-1+\ell)}^{+\infty} S_{\ell}^{(k+1-\ell,k)}\, \sigma_{1}^{k+1-\ell} \sigma_{2}^{k}
\end{equation}
\noindent
and 
\begin{equation}
\widetilde{S}_{\ell} (\sigma_{1},\sigma_{2}) = \sum_{k=\max(0,-1-\ell)}^{+\infty} \widetilde{S}_{\ell}^{(k,k+1+\ell)}\, \sigma_{1}^{k} \sigma_{2}^{k+1+\ell}.
\end{equation}
\noindent
Clearly, there are now a whole lot more Stokes constants than before. For simplicity of notation, and noticing that the Stokes constants depend only on the two parameters $k$ and $\ell$, we redefine them as 
\begin{equation}
S_{\ell}^{(k+1-\ell,k)} \equiv S_{\ell}^{(k+1-\ell)} \qquad \mathrm{and} \qquad\widetilde{S}_{\ell}^{(k,k+1+\ell)} \equiv \widetilde{S}_{\ell}^{(k+1+\ell)}.
\end{equation}
\noindent
Plugging these expressions back into the power series expansion of the bridge equation (\ref{2bridgeF}) one obtains, after a rather long but straightforward calculation, 
\begin{eqnarray}
\Delta_{\ell A} \Phi_{(n|m)} &=& \sum_{k=\max(0,\ell-1)}^{\min(m,n+\ell-1)} \left(n-k+\ell\right) S_{\ell}^{(k-\ell+1)}\, \Phi_{(n-k+\ell|m-k)} + \nonumber \\
&&
+ \sum_{k=\max(-\ell-1,0)}^{\min(m-\ell,n)} \left(m-k-\ell\right) \widetilde{S}_{\ell}^{(k+\ell+1)}\, \Phi_{(n-k|m-k-\ell)},
\label{eq:2paraliender}
\end{eqnarray}
\noindent
valid for all $\ell \ne 0$. Looking at the $\ell \ge 1$ case ($\ell \le -1$ is completely analogous), one finds
\begin{eqnarray}
\Delta_{\ell A}\Phi_{(n|m)} &=& \sum_{k=0}^{\min(m-\ell+1,n)} \left(n-k+1\right) S_{\ell}^{(k)}\, \Phi_{(n-k+1|m-k-\ell+1)} + \nonumber \\
&&
+ \sum_{k=0}^{\min(m-\ell,n)} \left(m-k-\ell\right) \widetilde{S}_{\ell}^{(k+\ell+1)}\, \Phi_{(n-k|m-k-\ell)},
\end{eqnarray}
\noindent
which can be directly compared with equivalent expressions from \cite{gikm10}. In these expressions we have used conventions in which $\Phi_{(n|m)}$ vanishes if either $n$ or $m$ are less than zero. As compared to the one--parameter case, (\ref{bridgePhi}), the increase in complexity is evident. Analyzing the bridge equations in the form (\ref{eq:2paraliender}), it is not difficult to notice that the cases $\Delta_{\ell A} \Phi_{(n|m)}$ and $\Delta_{-\ell A} \Phi_{(m|n)}$ with $\ell > 0$ are intimately related. In fact, one can go from one to the other by performing the simple changes  $( S_{\ell}^{a}, \widetilde{S}_{\ell}^{b} ) \leftrightarrow ( \widetilde{S}_{-\ell}^{a}, S_{-\ell}^{b} )$ and $\Phi_{(a|b)} \leftrightarrow \Phi_{(b|a)}$, where $a$, $b$ can be any combination of indices. The same relation can be seen to extend to the full Stokes automorphisms---changing between the $\underline{\frak{S}}_{0}\Phi_{(n|m)}$ and $\underline{\frak{S}}_{\pi}\Phi_{(m|n)}$ cases---which we shall further discuss in the following. In any case, the main focus of our concern deals with the instanton series, $\Phi_{(n|0)}$, where these formulae become
\begin{equation}\label{2bridgePhi}
\Delta_{\ell A}\Phi_{(n|0)} =
\begin{cases}
0, & \ell>1, \\
S_{1}^{(0)} \left( n+1 \right) \Phi_{(n+1|0)}, & \ell=1, \\
S_{\ell}^{(1-\ell)} \left( n+\ell \right) \Phi_{(n+\ell|0)} + \widetilde{S}_{\ell}^{(0)} \Phi_{(n+\ell+1|1)}, & \ell \le -1.
\end{cases}
\end{equation}
\noindent
This result clearly illustrates that in the present situation, unlike the one--parameter case, understanding the asymptotics of the \textit{physical} instanton series necessarily requires the use of the \textit{generalized} multi--instanton contributions, due to the appearance of the term in $\Phi_{(\bullet|1)}$ which, upon multiple alien derivation, will make materialize the full generalized instanton sector.

As we have further seen in section \ref{borelrev}, the bridge equations may also be translated back to the structure of the Borel transform, at least near each singularity in the Borel complex plane. In the present case we have to consider, for $\beta_{nm} = n\beta_{1} + m\beta_{2}$,
\begin{equation}
\Phi_{(n|0)}(z) \simeq \sum_{g=1}^{+\infty} \frac{F_{g}^{(n|0)}}{z^{g+n\beta_{1}}} \qquad \mathrm{and} \qquad \CB[\Phi_{(n|0)}](s) = \sum_{g=1}^{+\infty} \frac{F_{g}^{(n|0)}}{\Gamma(g+n\beta_{1})}\, s^{g+n\beta_{1}-1},
\end{equation}
\noindent
as well as\footnote{In the Painlev\'e I case there will also be logarithmic contributions to $\Phi_{(n|1)}(z)$, which we ignore for the moment.}
\begin{equation}
\Phi_{(n|1)}(z) \simeq \sum_{g=1}^{+\infty} \frac{F_{g}^{(n|1)}}{z^{g+n\beta_{1}+\beta_{2}}} \qquad \mathrm{and} \qquad \CB[\Phi_{(n|1)}](s) = \sum_{g=1}^{+\infty} \frac{F_{g}^{(n|1)}}{\Gamma(g+n\beta_{1}+\beta_{2})}\, s^{g+n\beta_{1}+\beta_{2}-1}.
\end{equation}
\noindent
Then, from (\ref{2bridgePhi}) above and the definition of alien derivative, it simply follows, \textit{e.g.},
\begin{equation}
\CB\left[\Phi_{(n|0)}\right](s+\ell A) = \left(S_{\ell}^{(1-\ell)} \left(n+\ell\right) \CB \left[ \Phi_{(n+\ell|0)} \right] (s) + \widetilde{S}_{\ell}^{(0)}\, \CB \left[ \Phi_{(n+\ell+1|1)} \right] (s) \right) \frac{\log s}{2\pi\rmi}, \quad \ell \leq -1.
\end{equation}

The next step is to use the alien derivatives in order to fully construct Stokes' automorphism, allowing for a full reconstruction of the nonperturbative solution. Consider first the positive real axis, where $\theta=0$, and where the Stokes automorphism is
\begin{equation}
\underline{\frak{S}}_{0} = \exp \left( \sum_{\ell=1}^{+\infty} \rme^{-\ell Az} \Delta_{\ell A} \right) = 1 + \rme^{-Az}\, \Delta_{A} + \rme^{-2Az} \left( \Delta_{2A} + \frac{1}{2} \Delta_{A}^{2} \right) + \cdots.
\end{equation}
\noindent
Just like in the one--parameter case of section \ref{borelrev}, given the transseries \textit{ansatz}, the action of $\underline{\frak{S}}_{0}$ on $F(z,\sigma_{1},\sigma_{2})$ is entirely encoded by the action of $\underline{\frak{S}}_{0}$ on the several $\Phi_{(n|m)}(z)$, and this can now be completely determined by the use of the bridge equations. When focusing on the physical instanton series, and again akin to the one--parameter case of section \ref{borelrev}, the bridge equations (\ref{2bridgePhi}) vanish whenever $\ell>1$, and when $\ell=1$ both one--parameter (\ref{bridgePhi}) and two--parameters (\ref{2bridgePhi}) cases are entirely analogous. Thus, the Stokes automorphism immediately simplifies to
\begin{equation}
\underline{\frak{S}}_{0} = \exp \left( \rme^{-Az} \Delta_{A} \right) = 1 + \rme^{-Az}\, \Delta_{A} + \frac{1}{2} \rme^{-2Az}\, \Delta_{A}^{2} + \frac{1}{3!} \rme^{-3Az}\, \Delta_{A}^{3} + \cdots,
\end{equation}
\noindent
where 
\begin{equation}
\Delta_{A}^{N} \Phi_{(n|0)} = \left( S_{1}^{(0)} \right)^{N} \cdot \prod_{i=1}^{N} \left( n+i \right) \cdot \Phi_{(n+N|0)}.
\end{equation}
\noindent
One may now simply compute 
\begin{equation}\label{2disc@zero}
\underline{\frak{S}}_{0} \Phi_{(n|0)} = \sum_{k=0}^{+\infty} \binom{n+k}{n}\, \left( S_{1}^{(0)} \right)^{k}\, \rme^{-kAz}\, \Phi_{(n+k|0)},
\end{equation}
\noindent
a completely straightforward generalization of the one--parameter case (\ref{disc@zero}). 

The novelties arise as we turn to the Borel negative real axis, where $\theta=\pi$, and where the Stokes automorphism becomes 
\begin{eqnarray}
\underline{\frak{S}}_{\pi} = \exp \left( \sum_{\ell=1}^{+\infty} \rme^{\ell Az} \Delta_{-\ell A} \right) &=& 1 + \sum_{\ell=1}^{+\infty} \rme^{\ell Az} \sum_{k=1}^{\ell} \frac{1}{k!} \underset{\sum \ell_{i} = \ell}{\sum_{\ell_{1},...,\ell_{k}\ge1}} \Delta_{-\ell_{k}A} \cdots \Delta_{-\ell_{1}A} = \label{Stokes-2param-disc@pi} \\
&=& 1 + \rme^{Az}\, \Delta_{-A} + \rme^{2Az} \left( \Delta_{-2A} + \frac{1}{2} \Delta_{-A}^{2} \right) + \cdots.
\end{eqnarray}
\noindent
Things are now much more complicated than in the simple one--parameter transseries case, as the different terms in $\underline{\frak{S}}_{\pi}$ will mix contributions arising from all $\Phi_{(n|m)}$. From the expression above for the Stokes automorphism it becomes obvious that, in order to find the final expression for $\underline{\frak{S}}_{\pi} \Phi_{(n|m)}$, one first needs to focus on determining $\Delta_{-\ell_{k}A} \cdots \Delta_{-\ell_{1}A} \Phi_{(n|0)}$, with $\ell_{j}\ge1$. For $k=1,2$, this calculation is pretty straightforward. Using (\ref{2bridgePhi}) one can write
\bea
\Delta_{-\ell_{1}A} \Phi_{(n|0)} &=& (n-\ell_{1})\, S_{-\ell_{1}}^{(1+\ell_{1})}\, \Phi_{(n-\ell_{1}|0)} + \widetilde{S}_{-\ell_{1}}^{(0)}\, \Phi_{(n-\ell_{1}+1|1)}, \\
\Delta_{-\ell_{2}A} \Delta_{-\ell_{1}A} \Phi_{(n|0)} &=& \left( n-\sum_{i=1}^{2}\ell_{i} \right) \left( \left( n-\ell_{1} \right) S_{-\ell_{1}}^{(1+\ell_{1})} S_{-\ell_{2}}^{(1+\ell_{2})} + \widetilde{S}_{-\ell_{1}}^{(0)} S_{-\ell_{2}}^{(2+\ell_{2})} \right) \Phi_{\left. \left( n-\sum_{i=1}^{2}\ell_{i} \right| 0 \right)} + \nonumber \\
&&
\hspace{-65pt}
+ \left((n-\ell_{1})\, S_{-\ell_{1}}^{(1+\ell_{1})} \widetilde{S}_{-\ell_{2}}^{(0)} + \left( n+1-\sum_{i=1}^{2}\ell_{i} \right) \widetilde{S}_{-\ell_{1}}^{(0)} S_{-\ell_{2}}^{(1+\ell_{2})} + \widetilde{S}_{-\ell_{1}}^{(0)} \widetilde{S}_{-\ell_{2}}^{(1)} \right) \Phi_{\left. \left( n+1-\sum_{i=1}^{2}\ell_{i} \right| 1 \right)} + \nonumber \\
&&
\hspace{-65pt}
+ 2\, \widetilde{S}_{-\ell_{1}}^{(0)} \widetilde{S}_{-\ell_{2}}^{(0)} \Phi_{\left. \left( n+2-\sum_{i=1}^{2}\ell_{i} \right| 2 \right)}.
\label{2alien-der-2-param-inst-series}
\eea
\noindent
In order to go further and generalize these cases to an arbitrary product of alien derivatives, one first needs to determine $\Delta_{-\ell_{k+1}A} \Phi_{\left. \left( n+m-\sum_{i=1}^{k}\ell_{i} \right| m \right)}$. After some effort one can find that
\bea
\Delta_{-\ell_{k+1}A} \Phi_{\left. \left( n+m-\sum_{i=1}^{k}\ell_{i} \right| m \right)} &=& \sum_{q=0}^{m+1} \left( \left( n+m+1-q-\sum_{i=0}^{k+1}\ell_{i} \right) S_{-\ell_{k+1}}^{(q+\ell_{k+1})} + \left( m+1-q \right) \widetilde{S}_{-\ell_{k+1}}^{(q)}\right) \times \nonumber \\
&&
\times\, \Phi_{\left. \left( n+m+1-q-\sum_{i=0}^{k+1}\ell_{i} \right| m+1-q \right)}, \label{extra-der-in-general-2par-term} 
\eea
\noindent
where we have set $S_{-\ell_{i}}^{(\ell_{i})} \equiv 0$, for any $\ell_{i} \ge 1$, in order to simplify the final result. The general case for the ordered product of $k$ alien derivatives of the form $\prod_{i=1}^{k} \Delta_{-\ell_{k+1-i}A} = \Delta_{-\ell_{k}A} \cdots \Delta_{-\ell_{1}A}$, acting on $\Phi_{(n|0)}$, is then given by 
\begin{eqnarray}
\prod_{i=1}^{k} \Delta_{-\ell_{k+1-i}A} \Phi_{(n|0)} &=& \sum_{m=0}^{k} \prod_{s=1}^{k} \left\{ \sum_{q_{s}=0}^{s} \left[ \left( s-\sum_{i=1}^{s}q_{i} \right) \widetilde{S}_{-\ell_{s}}^{(q_{s})} + \left( n-\sum_{i=1}^{s}\ell_{i}+s-\sum_{i=1}^{s}q_{i} \right) S_{-\ell_{s}}^{(\ell_{s}+q_{s})} \right] \right. \times \nonumber \\
&&
\left. \times\, \Theta \left( s-\sum_{i=1}^{s}q_{s} \right) \right\} \delta \left( \sum_{i=1}^{k}q_{i}\, ,\, k-m \right) \Phi_{\left. \left( n+m-\sum_{i=1}^{k}\ell_{i} \right| m \right)}. 
\end{eqnarray}
\noindent
In this expression $\delta(n,m) \equiv \delta_{nm}$ is the usual Kronecker--delta, the function $\Theta(x)$ is the usual Heaviside step--function
\begin{equation}
\Theta(x) =
\begin{cases}
1, & x \ge 0, \\
0, & x < 0,
\end{cases}
\end{equation}
\noindent
and once again we set $S_{-\ell_{i}}^{(\ell_{i})}\equiv0$. A proof of this result can be found in appendix \ref{app:2-param-stokes-autom}. In the same manner as we have done earlier in the one--parameter case for the discontinuity at $\theta=\pi$ (\ref{disc-pi-one-param-young}), this result can also be rewritten using a sum over Young diagrams. To do so, let us first  define $\delta_{s} = \sum_{i=1}^{s} q_{s}+1$, such that $0 < \delta_{1} \le \delta_{2} \le \cdots \le \delta_{k} = k-m+1$ and $0 < \delta_{s} \le s+1$\footnote{The reason for adding the one in the present definition of $\delta_{s}$ is to make all $\delta_{s}$ strictly positive, and thus naturally labeled by some Young diagram.}. As explained in section \ref{borelrev}, the set of $\delta_{s}$, with $s=1,\ldots,k$, form a Young diagram $\Gamma(k,k-m+1)$ of lengths $\ell(\Gamma)=k$ and $\ell(\Gamma^{T})=k-m+1$, with the extra constraint that each component $\delta_{s}\in\Gamma(k,k-m+1)$ has a maximum number of $s+1$ boxes. As such, one may finally rewrite the above result as
\begin{eqnarray}
\prod_{i=1}^{k} \Delta_{-\ell_{k+1-i}A} \Phi_{(n|0)} &=& \sum_{m=0}^{k}\, \sum_{\delta_{s}\in\Gamma(k,k-m+1)}\, \prod_{s=1}^{k} \bigg\{ \bigg[ \left( s+1-\delta_{s} \right) \widetilde{S}_{-\ell_{s}}^{(\mathbf{d}\delta_{s})} + \label{k-alien-der-2-param-youngT} \\
&&
\left. \left. + \left( n-\sum_{i=1}^{s}\ell_{i}+s+1-\delta_{s} \right) S_{-\ell_{s}}^{(\ell_{s}+\mathbf{d}\delta_{s})} \right] \Theta \left( s+1-\delta_{s} \right) \right\} \Phi_{\left. \left( n+m-\sum_{i=1}^{k}\ell_{i} \right| m \right)}. \nonumber
\end{eqnarray}
\noindent
For this expression to hold, one still needs to set $\delta_{0} \equiv 1$ and $\widetilde{S}_{0}^{(s)} = S_{0}^{(s)} = S_{-\ell_{s}}^{(\ell_{s})} = 0$ (notice that some of these conditions will only be needed in the following).

Due to the complexity of this expression, let us pause for an example. Let us choose the case of $k=2$, which we have also described in (\ref{2alien-der-2-param-inst-series}) above, and see what the sum over Young diagrams (\ref{k-alien-der-2-param-youngT}) above yields. One finds:
\bea
\Delta_{-\ell_{2}A} \Delta_{-\ell_{1}A} \Phi_{(n|0)} &=& \sum_{m=0}^{2} \sum_{\delta_{s}\in\Gamma(2,3-m)} 
\Theta \left( 2-\delta_{1} \right) \left( \left( 2-\delta_{1} \right) \widetilde{S}_{-\ell_{1}}^{(\mathbf{d}\delta_{1})} + \left( n-\ell_{1}+2-\delta_{1} \right) S_{-\ell_{1}}^{(\ell_{1}+\mathbf{d}\delta_{1})} \right) \times \nonumber \\
&&
\hspace{-65pt}
\times\, \Theta \left( 3-\delta_{2} \right) \left( \left( 3-\delta_{2} \right) \widetilde{S}_{-\ell_{2}}^{(\mathbf{d}\delta_{2})} + \left( n-\ell_{1}-\ell_{2}+3-\delta_{2} \right) S_{-\ell_{2}}^{(\ell_{2}+\mathbf{d}\delta_{2})} \right) \Phi_{\left. \left( n+m-\sum_{i=1}^{2}\ell_{i} \right| m \right)}.
\eea
\noindent
The sum over Young diagrams in this expression is over $\delta_{s} \in \Gamma(2,3-m)$, with $m=0,1,2$. For $m=0$, one sums over all diagrams $\delta_{s} \in \Gamma(2,3)$ and there are three possible diagrams: $\tableau{31}$, $\tableau{32}$ and $\tableau{33}$. But because $\delta_{1}\le2$ and $\delta_{2}=3$, only two will remain: $\tableau{31}$ (where $\delta_{1}=1$) and $\tableau{32}$ ($\delta_{1}=2$). For $m=1$, one sums over diagrams $\delta_{s} \in \Gamma(2,2)$ and there are now two possible diagrams with $\delta_{2}=2$: $\tableau{21}$ ($\delta_{1}=1$) and $\tableau{22}$ ($\delta_{1}=2$). Finally, for $m=2$, one sums over diagrams $\delta_{s} \in \Gamma(2,1)$, which corresponds to the single diagram: $\tableau{11}$ ($\delta_{1}=\delta_{2}=1$). Plugging these results back into the expression above, one easily finds (\ref{2alien-der-2-param-inst-series}) as expected.

There is now enough information in order to completely determine the Stokes automorphism, at $\theta=\pi$, of the instanton series $\Phi_{(n|0)}$. Going back to its definition (\ref{Stokes-2param-disc@pi}) and making use of our formulae for multiple alien derivatives (\ref{k-alien-der-2-param-youngT}) it follows
\bea
\underline{\frak{S}}_{\pi} \Phi_{(n|0)} &=& \Phi_{(n|0)} + \sum_{\ell=1}^{+\infty} \rme^{\ell Az} \sum_{k=1}^{\ell} \frac{1}{k!} \underset{\sum\ell_{i} = \ell}{\sum_{\ell_{1},...,\ell_{k} \ge 1}} \Delta_{-\ell_{k}A} \cdots \Delta_{-\ell_{1}A} \Phi_{(n|0)} = \\
&=& \Phi_{(n|0)} + \sum_{\ell=1}^{+\infty} \rme^{\ell Az} \sum_{k=1}^{\ell} \frac{1}{k!} \underset{\sum\ell_{i} = \ell}{\sum_{\ell_{1},...,\ell_{k} \ge 1}}\, \sum_{m=0}^{k}\, \sum_{\delta_{s} \in \Gamma(k,k-m+1)} \\
&&
\hspace{-60pt}
\prod_{s=1}^{k} \left\{ \left[ \left( s+1-\delta_{s} \right) \widetilde{S}_{-\ell_{s}}^{(\mathbf{d}\delta_{s})} + \left( n-\sum_{i=1}^{s}\ell_{i}+s+1-\delta_{s} \right) S_{-\ell_{s}}^{(\ell_{s}+\mathbf{d}\delta_{s})} \right] \Theta \left(s+1-\delta_{s}\right) \right\} \cdot\, \Phi_{(n-\ell+m|m)}. \nonumber
\eea
\noindent
Interestingly enough, if we further define $\gamma_{s} = \sum_{i=1}^{s} \ell_{i}$, then the sum over the $\ell_{i}$ can also be rewritten as a sum of Young diagrams $\gamma_{i} \in \Gamma(k,\ell) :\, 0 < \gamma_{1} \le \cdots \le \gamma_{k} = \ell$; as long as we set  $\widetilde{S}_{0}^{(s)} = S_{0}^{(s)} = 0$. In this case, one finally obtains the simpler expression
\begin{eqnarray}
\underline{\frak{S}}_{\pi} \Phi_{(n|0)} &=& \Phi_{(n|0)} + \sum_{\ell=1}^{+\infty}  \sum_{k=1}^{\ell} \frac{\rme^{\ell Az}}{k!} \sum_{\gamma_{i} \in \Gamma(k,\ell)} \, \sum_{m=0}^{k} \, \sum_{\delta_{s} \in \Gamma(k,k-m+1)} \label{Stokes-pi-2-param-inst-series} \\
&&
\hspace{-40pt}
\prod_{s=1}^{k} \left\{ \left[ \left( s+1-\delta_{s} \right) \widetilde{S}_{-\mathbf{d}\gamma_{s}}^{(\mathbf{d}\delta_{s})} + \left( n-\gamma_{s}+s+1-\delta_{s} \right) S_{-\mathbf{d}\gamma_{s}}^{(\mathbf{d}\gamma_{s}+\mathbf{d}\delta_{s})} \right] \Theta \left( s+1-\delta_{s} \right) \right\} \cdot\, \Phi_{(n-\ell+m|m)}. \nonumber
\end{eqnarray}
\noindent
Some comments are now in order. Comparing the Stokes automorphism of the instanton series at the $\theta=\pi$ discontinuity, for both the one--parameter (\ref{disc-pi-one-param-young})\footnote{Recall that $\underline{\frak{S}}_{\theta}=1-\disc_{\theta}$.} and the two--parameter cases (above), one can see the that the increased degree of complexity of the latter is translated in the fact that there is now a sum over \textit{two} independent sets of Young diagrams (instead of summing over just one set of diagrams as in the one--parameter case). It is thus natural to infer that for a general $\ell$--parameter transseries \textit{ansatz} such sums would be substituted by sums over $\ell$ independent sets of Young diagrams. It is also not too difficult to see that one can recover the one--parameter result (\ref{disc-pi-one-param-young}) starting from (\ref{Stokes-pi-2-param-inst-series}) above, by simply setting $\delta_{s}=s+1$ for all $\delta_{s}$. This corresponds to choosing the Young diagrams $\delta_{s} \in \Gamma(k,k+1)$ (with $m=0$ and consequently $\ell\le n$) where each row has one more box than the previous one, \textit{e.g.}, $\tableau{5432}$ for $k=4$.

\subsection{Stokes Constants and Asymptotics Revisited}

The main outcome of the above calculations are expressions for the discontinuities of the full, physical, multi--instanton series, encoded in the Stokes automorphism of $\Phi_{(n|0)}$, in both $\theta=0,\pi$, directions. As we have seen earlier, in section \ref{borelrev}, these discontinuities lie at the basis of understanding the full asymptotic behavior of all multi--instanton sectors and we shall next use these new Stokes' discontinuities in order to generalize our results on asymptotics, from the one--parameter to the two--parameters case. Recall that by making use of Cauchy's theorem a given function $F(z)$ with a branch--cut along some direction $\theta$ in the complex plane (and analytic elsewhere) can actually be fully described precisely by its discontinuity along that direction (\ref{cauchy}), at least as long as its behavior at infinity does not contribute. In the present case of interest, the multi--instanton free energies $F^{(n|0)}(z)$, which are the coefficients of the two--parameters transseries \textit{ansatz} (\ref{2Ftransseries}), have asymptotic expansions given by (\ref{2Fpertseries}) for the perturbative series and by (\ref{2Fnmseries}) for the generalized multi--instanton contributions. Their discontinuities are essentially given by the Stokes automorphisms of $\Phi_{(n|0)}(z)$ previously calculated.

Let us first look at the perturbative expansion (\ref{2Fpertseries}). The discontinuities of $F^{(0|0)}(z)$ arise directly from the bridge equations (\ref{2bridgePhi}), via the Stokes automorphisms (\ref{2disc@zero}) and (\ref{Stokes-pi-2-param-inst-series}),
\bea
\disc_{0}\, \Phi_{(0|0)} &=& - \sum_{k=1}^{+\infty} \left( S_{1}^{(0)} \right)^{k} \rme^{-kAz}\, \Phi_{(k|0)},\\
\disc_{\pi}\, \Phi_{(0|0)} &=& - \sum_{k=1}^{+\infty} \left( \widetilde{S}_{-1}^{(0)} \right)^{k} \rme^{kAz}\, \Phi_{(0|k)}.
\eea
\noindent
Note that $F^{(0|0)}(z)$ will now have two branch cuts in the Borel complex plane (instead of only one as in the one--parameter transseries case), along both positive and negative real axes. By using (\ref{2Fpertseries}), (\ref{2Fnmseries}) and (\ref{cauchy}) it is not difficult to find the asymptotic coefficients of the perturbative expansion to be given by
\bea
F_g^{(0|0)} &\simeq& \sum_{k=1}^{+\infty} \frac{\left( S_1^{(0)} \right)^k}{2\pi\rmi}\, \frac{\Gamma \left( g-\beta_{k,0} \right)}{\left( k A \right)^{g-\beta_{k,0}}}\, \sum_{h=1}^{+\infty} \frac{\Gamma \left( g-\beta_{k,0}-h+1 \right)}{\Gamma \left( g-\beta_{k,0} \right)}\, F_h^{(k|0)} \left( k A \right)^{h-1} + \nonumber \\
&&
+ \sum_{k=1}^{+\infty} \frac{\left( \widetilde{S}_{-1}^{(0)} \right)^k}{2\pi\rmi}\, \frac{\Gamma \left( g-\beta_{0,k} \right)}{\left( - k A \right)^{g-\beta_{0,k}}}\, \sum_{h=1}^{+\infty} \frac{\Gamma \left( g-\beta_{0,k}-h+1 \right)}{\Gamma \left( g-\beta_{0,k} \right)}\, F_h^{(0|k)} \left( - k A \right)^{h-1}.
\label{zerozerotwostokes}
\eea
\noindent
As should be by now expected, we find that the coefficients of the pertubative expansion around the zero--instanton sector are given by an asymptotic double--sum expansion, valid for large values of $g$, over the coefficients of the perturbative expansions around (some of) the generalized multi--instanton sectors. The novelty in here, as compared to the one--parameter case of section \ref{borelrev}, is that this expansion includes not only the coefficients of the physical instanton series $F_{g}^{(n|0)}$, associated with positive real part of the instanton action, but also the generalized coefficients $F_{g}^{(0|n)}$, associated with negative real part of the instanton action. In particular, the leading order of this zero--instanton asymptotic expansion is determined by the coefficients of the one--loop (generalized) one--instanton partition functions, but now up to two Stokes constants, namely $S_{1}^{(0)}$ and $\widetilde{S}_{-1}^{(0)}$. Higher loop contributions will arise as $\frac{1}{g}$ corrections, while other multi--instanton contributions, with action $\pm n A$, will yield corrections suppressed as $n^{-g}$.

Thus, what we have found in the present two--parameters transseries setting is that, such as in the one--parameter case, through the use of alien calculus and the bridge equations it is possible to include \textit{all} multi--instanton sectors in the asymptotics of the perturbative zero--instanton sector. Furthermore, through essentially the same methods it is also straightforward to generalize this asymptotic result to \textit{all} multi--instanton sectors. This is what we shall do next for the $n$--instanton sector, $F^{(n|0)} (z)$. Using the formulae for the Stokes automorphism in the directions $\theta=0,\pi$, of $\Phi_{(n|0)}$, given in (\ref{2disc@zero}) and (\ref{Stokes-pi-2-param-inst-series}), we can easily find the related discontinuities in the said directions. As usual, $F^{(n|0)}(z)$ has branch cuts in the Stokes directions corresponding to both positive and negative real axes in the Borel complex plane. Then, by means of (\ref{cauchy}) and (\ref{2Fnmseries}), in particular the identification $F^{(n|m)}(z) = \rme^{-(n-m)Az}\, \Phi_{(n|m)} (z)$, a lengthy but straighforward calculation leads to (it might be interesting for the reader to compare this expression against its one--parameter counterpart, (\ref{ninstasymp}))
\bea
F_{g}^{(n|0)} &\simeq& \sum_{k=1}^{+\infty} \binom{n+k}{n}\, \frac{(S_{1}^{(0)})^{k}}{2\pi \rmi} \cdot \frac{\Gamma \left( g+\beta_{n,0}-\beta_{n+k,0} \right)}{\left( k A \right)^{g+\beta_{n,0}-\beta_{n+k,0}}}\, \sum_{h=1}^{+\infty} \frac{\Gamma \left( g+\beta_{n,0}-\beta_{n+k,0}-h \right)}{\Gamma \left( g+\beta_{n,0}-\beta_{n+k,0} \right)}\, F_{h}^{(n+k|0)} \left( k A \right)^{h} \nonumber \\
&&
+\sum_{k=1}^{+\infty} \left\{ \frac{1}{2\pi\rmi} \sum_{m=1}^{k} \frac{1}{m!}\, \sum_{\ell=0}^{m}\, \sum_{\gamma_{i}\in\Gamma(m,k)}\, \sum_{\delta_{j}\in\Gamma(m, m-\ell+1)} \left( \prod_{j=1}^{m} \Sigma(n,j) \right) \right\} \times \nonumber \\
&&
\times\, \frac{\Gamma \left( g+\beta_{n,0}-\beta_{n+\ell-k,\ell} \right)}{\left( - k A \right)^{g+\beta_{n,0}-\beta_{n+\ell-k,\ell}}}\, \sum_{h=1}^{+\infty} \frac{\Gamma \left( g+\beta_{n,0}-\beta_{n+\ell-k,\ell}-h \right)}{\Gamma \left( g+\beta_{n,0}-\beta_{n+\ell-k,\ell} \right)}\, F_{h}^{(n+\ell-k|\ell)} \left( - k A \right)^{h},
\label{nzeroinstasymp}
\eea
\noindent
where we have introduced
\begin{equation}\label{sigmanj}
\Sigma(n,j) = \left( \left( j+1-\delta_{j} \right) \widetilde{S}_{-\mathbf{d}\gamma_{j}}^{(\mathbf{d}\delta_{j})} + \left( n-\gamma_{j}+j+1-\delta_{j} \right)  S_{-\mathbf{d}\gamma_{j}}^{(\mathbf{d}\gamma_{j}+\mathbf{d}\delta_{j})} \right) \Theta \left( j+1-\delta_{j} \right).
\end{equation}
\noindent
Recall that we have previously defined $S_{-\ell}^{(\ell)} = S_{0}^{(\ell)} = \widetilde{S}_{0}^{(\ell)} = 0$, with $\ell>0$, and $\gamma_{0}=0$, $\delta_{0}=1$, which are required to fully understand the formulae above. This result relates the coefficients of the perturbative expansion around the $n$--instanton sector with sums over the coefficients of the perturbative expansions around \textit{all} other \textit{generalized} multi--instanton sectors, in asymptotic expansions which hold for large $g$. All Stokes factors are needed to compute the general asymptotics of $F_{g}^{(n|0)}$, whose computation is, in general, quite hard to do from first principles, but which may, nonetheless, be explored numerically in specific examples as shall be seen in great detail in the following sections.

\subsection{Resurgence of the String Genus Expansion}\label{sub:String-genus-exp}

The results we obtained in the previous subsections are rather general and do not take into account any symmetries or properties of the physical system that one might have started from. If we now specialize to the cases of interest in this work, models with a topological genus expansion such as topological strings, minimal strings or matrix models, then it is well known that the corresponding free energy in the zero--instanton sector will have a genus expansion as (\ref{stringgenus}), \textit{i.e.}, an expansion in the closed string coupling $g_s^2$,
\begin{equation}\label{eq:genus-expansion-0-inst}
g_{s}^{2} F^{(0|0)}(g_s; \{t_i\}) \simeq \sum_{g=0}^{+\infty} g_{s}^{2g}\, \widehat{F}_{g}^{(0|0)} (t_i) \equiv \Phi^{(0|0)} (g_s; \{ t_i \}).
\end{equation}
\noindent
This expansion resembles (\ref{2Fpertseries}) if one sets $z=1/g_{s}$ and assumes a $t_i$ dependence for the coefficients $\widehat{F}_{g}^{(0|0)}(t_i)$ in the asymptotic expansion (and similarly for the instanton action, $A(t_i)$). These parameters, $t_i$, encode a possible dependence of the result on the 't~Hooft moduli, as will be the case of matrix models. We also need to consider a string theoretic version of the \textit{ansatz} (\ref{2Fnmseries}) for the generalized multi--instanton free energies, this time around as an expansion in the open string coupling $g_s$,
\be\label{eq:Def-Phi_(n|m)-with-all-modes}
F^{(n|m)} (g_s; t_i) \simeq \rme^{-(n-m) \frac{A(t_i)}{g_{s}}}\, \sum_{k=0}^{k_{nm}}\log^{k} g_{s} \sum_{g=0}^{+\infty} g_{s}^{g+\beta_{nm}^{[k]}}\, F_{g}^{(n|m)[k]}(t_i) \equiv \rme^{-(n-m) \frac{A(t_i)}{g_s}}\Phi_{(n|m)} (g_s; t_i).
\ee
\noindent
Notice that in this expression we have further included an expansion in logarithmic powers of the open string coupling (up to some \textit{finite} logarithmic power, $k_{nm}$) in order to account for resonant effects which will appear later in the Painlev\'e I case and in the quartic matrix model, and which we have already mentioned at the beginning of this section\footnote{Our discussion up to now solely focused on the ``$k=0$ sector'' of the logarithmic expansion.} (see \cite{gikm10} as well, for the logarithmic terms). The integer $\beta_{nm}^{[k]}$ will also be necessary in order to take into account possible different starting powers of our asymptotic expansions. For instance, in the case of the Painlev\'e I equation we shall later find $k_{nm}=\min(n,m)-m\, \delta_{nm}$ and $\beta_{nm}^{[k]}=\beta(m+n)-\left[(k_{nm}+k)/2\right]_{\mathrm{I}}$, where $[\bullet]_{\mathrm{I}}$ denotes the integer part of the argument, and where $\beta=1/2$. We shall also make the assumption that the resonant effects do not appear in the $n$--instanton sector, that is $k_{n,0}=0=k_{0,m}$. Finally we will focus on the cases where both $\beta_{nm}^{[k]}$ and $k_{nm}$ are symmetrical in $n$, $m$. As we shall see later, all these assumptions turn out to be properties of string theoretic systems.

Starting off with the zero--instanton sector, we have
\begin{equation}
F^{(0|0)} (g_s; \{t_i\}) \simeq \sum_{\ell\ge 0} g_{s}^{\ell+\beta_{0,0}^{[0]}}\, F_{\ell}^{(0|0)[0]}(t_i).
\end{equation}
If we compare this expansion with (\ref{eq:genus-expansion-0-inst}) above, one easily concludes that, in order to find a topological genus expansion, it must be the case that $F_{2\ell+1}^{(0|0)[0]} (t_i) \equiv 0$ with $\beta_{0,0}^{[0]} = 0$. Do notice that the free energy coefficients in the genus expansion (\ref{eq:genus-expansion-0-inst}) are given by $\widehat{F}_{g}^{(0|0)} \equiv F_{2g}^{(0|0)[0]}$, which will naturally include both even and odd powers of the genus, $g$, as expected. Via Cauchy's theorem (\ref{cauchy}), now applied in the complex $g_{s}$--plane\footnote{Notice that a blind application of Cauchy's theorem (\ref{cauchy}) in the $g_{s}$--variable leads to a large--order relation with an (incorrect) overall minus sign as compared to, \textit{e.g.}, (\ref{zeroinstasymp}). Instead, one should recall that the definition of the Stokes discontinuities in terms of the Stokes automorphism, (\ref{boreldisc}), depends on what one means by left and right Borel resummations. Under a change of variables of the type $x \to 1/x$ these orientations change and so does the sign of the discontinuity---thus leading to the correct result.}, one essentially recovers the result of the previous section for the $F_{\ell}^{(0|0)[0]}$ and, in particular, one finds for the asymptotics of $F_{2\ell+1}^{(0|0)[0]}$
\begin{eqnarray}
F_{2\ell+1}^{(0|0)[0]} &\simeq& \sum_{k=1}^{+\infty} \frac{\left(S_{1}^{(0)}\right)^{k}}{2\pi\rmi}\, \frac{\Gamma(2\ell+1-\beta_{k,0}^{[0]})}{\left( k A \right)^{2\ell+1-\beta_{k,0}^{[0]}}}\, \sum_{h=0}^{+\infty} \frac{\Gamma(2\ell+1-h-\beta_{k,0}^{[0]})}{\Gamma(2\ell+1-\beta_{k,0}^{[0]})}\, F_{h}^{(k|0)[0]} \left( k A \right)^{h} + \nonumber \\
&&
+ \sum_{k=1}^{+\infty} \frac{\left(\widetilde{S}_{-1}^{(0)}\right)^{k}}{2\pi\rmi}\, \frac{\Gamma(2\ell+1-\beta_{0,k}^{[0]})}{\left( - k A \right)^{2\ell+1-\beta_{0,k}^{[0]}}}\, \sum_{h=0}^{+\infty} \frac{\Gamma(2\ell+1-h-\beta_{0,k}^{[0]})}{\Gamma(2\ell+1-\beta_{0,k}^{[0]})}\, F_{h}^{(0|k)[0]} \left( - k A \right)^{h}.
 \end{eqnarray}
\noindent
The ``genus expansion condition'', that $F_{2\ell+1}^{(0|0)[0]}=0$, now becomes equivalent to a set of relations between $F_{g}^{(k|0)[0]}$, $F_{g}^{(0|k)[0]}$, $S_{1}^{(0)}$ and $\widetilde{S}_{-1}^{(0)}$. We find, for each $k$ and $g$,
\begin{equation}\label{sf-wtsf-relation}
\left( S_{1}^{(0)} \right)^{k} F_{g}^{(k|0)[0]} = (-1)^{g+\beta_{0,k}^{[0]}} \left( \widetilde{S}_{-1}^{(0)} \right)^{k} F_{g}^{(0|k)[0]}.
\end{equation}
\noindent
In the following sections we shall see in detail that by considering special properties of the systems we will address, such as 2d quantum gravity or the quartic matrix model, there are in fact more general relations between the $F_{g}^{(n|m)[k]}$, under exchange of $n$ and $m$. Furthermore this will also allow us to find relations between $S_{1}^{(0)}$ and $\widetilde{S}_{-1}^{(0)}$ (and, in fact, relations between other Stokes constants) effectively reducing the number of independent Stokes constants needed to account for the large--order behavior of all multi--instanton sectors.

The relation determined above can now be used to simplify the large--order behavior of the coefficients in the topological genus expansion (\ref{eq:genus-expansion-0-inst}), as
\be\label{eq:Free-En-zero-inst-coeff}
\widehat{F}_{g}^{(0|0)} \simeq \sum_{k=1}^{+\infty} \frac{\left( S_{1}^{(0)} \right)^{k}}{\rmi\pi}\, \frac{\Gamma(2g-\beta_{k,0}^{[0]})}{\left( k A \right)^{2g-\beta_{k,0}^{[0]}}}\, \sum_{h=0}^{+\infty} \frac{\Gamma(2g-h-\beta_{k,0}^{[0]})}{\Gamma(2g-\beta_{k,0}^{[0]})}\, F_{h}^{(k|0)[0]} \left( k A \right)^{h},
\ee
\noindent
which in fact, as just mentioned, reduced the number of Stokes constants one effectively needs to completely understand the asymptotics of the perturbative sector (comparing with the corresponding result in the previous subsection, (\ref{zerozerotwostokes}), we see that this final expression is much closer to its one--parameter counterpart, (\ref{zeroinstasymp})). Further notice that this expression coincides with the result in \cite{gikm10}, at leading order in $k$, if one takes into account that in our case we are considering a genus expansion in the variable $g_{s}$, instead of an expansion in $z=1/g_{s}$ as used in that paper. 

One can also use the string theoretic generalized multi--instanton expansion (\ref{eq:Def-Phi_(n|m)-with-all-modes}) to determine the large--order behavior of the physical $n$--instanton series $F^{(n|0)}(z)$. This follows by applying Cauchy's theorem to the string coupling, $g_{s}$, and using the discontinuities for $\Phi_{(n|0)}(z)$ determined in section \ref{sub:Bridge-Eqns-revisited}. The novelty now is that we are further considering logarithmic power contributions to the asymptotic series of $\Phi_{(n|m)}(z)$. Thus, in order to obtain the large--order coefficients $F_{g}^{(n|0)[0]}$ we shall apply Cauchy's theorem as before, but when making use of the expansion
(\ref{eq:Def-Phi_(n|m)-with-all-modes}) new integrals will have to be addressed:
\begin{flalign}
\mbox{Discontinuity at }\theta=0: & \quad \int_{0}^{+\infty} \rmd x\, x^{-g-1}\, \rme^{- \frac{kA}{x}}\, \log^r x \underset{z=\frac{1}{x}}{\rightarrow} (-1)^{r} \int_{0}^{+\infty} \rmd z\, z^{g-1}\, \rme^{-kAz}\, \log^r z, \\
\mbox{Discontinuity at }\theta=\pi: & \quad \int_{0}^{-\infty} \rmd x\, x^{-g-1}\, \rme^{\frac{kA}{x}}\, \log^r x \underset{z=\frac{1}{x}}{\rightarrow} (-1)^{r} \int_{0}^{-\infty} \rmd z\, z^{g-1}\, \rme^{kAz}\, \log^r z.
\end{flalign}
\noindent
The relevant quantity needed to perform these integral is the following Laplace transform
\begin{eqnarray}
\mathcal{L} \left[ z^{g} \log^{r}(z) \right] (s) &\equiv& \int_{0}^{+\infty} \rmd z\, z^{g}\, \rme^{-s\, z}\, \log^{r} z = \left( \frac{\partial}{\partial g} \right)^r \int_{0}^{+\infty} \rmd z\, z^{g}\, \rme^{-s\, z} = \left( \frac{\partial}{\partial g} \right)^{r} \left[ \frac{\Gamma(g+1)}{s^{g+1}} \right] = \nonumber \\
&=& \frac{\Gamma(g+1)}{s^{g+1}} \left( \delta_{r0} + \Theta(r-1) \left( \widetilde{B}_{s}(g) + \partial_{g} \right)^{r-1} \widetilde{B}_{s}(g) \right),
\end{eqnarray}
\noindent
and its analogous $\theta = \pi$ version
\begin{equation}
\mathcal{L} \left[ z^{g} \log^{r} (-z) \right] (s) = \frac{\Gamma(g+1)}{s^{g+1}} \left( \delta_{r0} + \Theta(r-1)\, \Big( B_{s}(g) + \partial_{g} \Big)^{r-1} B_{s}(g) \right),
\end{equation}
\noindent
where\footnote{In here $\psi(z) = \frac{\Gamma'(z)}{\Gamma(z)}$ is the digamma function; the logarithmic derivative of the gamma function.}
\begin{eqnarray}
\widetilde{B}_{s}(a) &=& \psi(a+1)-\log(s), \label{eq:Btildedef} \\
B_{s}(a) &=& \psi(a+1)-\log(-s) = \widetilde{B}_{s}(a) - \rmi\pi.
\end{eqnarray}
\noindent
Collecting all these results, one can now easily find the large--order behavior of $F_{g}^{(n|0)[0]}$ (again, it might be interesting for the reader to compare this expression against the two--parameters case without logarithms, (\ref{nzeroinstasymp}), or the one--parameter counterpart, (\ref{ninstasymp})),
\begin{eqnarray}
F_{g}^{(n|0)[0]} &\simeq& \sum_{k=1}^{+\infty} \binom{n+k}{n}\, \frac{(S_{1}^{(0)})^{k}}{2\pi\rmi}\, \frac{\Gamma (g+\beta_{n,0}^{[0]}-\beta_{n+k,0}^{[0]})}{\left( k A \right)^{g+\beta_{n,0}^{[0]}-\beta_{n+k,0}^{[0]}}}\, \sum_{h=1}^{+\infty} \frac{\Gamma (g+\beta_{n,0}^{[0]}-\beta_{n+k,0}^{[0]}-h)}{\Gamma (g+\beta_{n,0}^{[0]}-\beta_{n+k,0}^{[0]})}\, F_{h}^{(n+k|0)[0]} \left( k A \right)^{h} \nonumber \\
&&
\hspace{-10pt}
+ \sum_{k=1}^{+\infty} \left\{ \frac{1}{2\pi\rmi} \sum_{m=1}^{k} \frac{1}{m!}\, \sum_{\ell=0}^{m}\,\sum_{\gamma_{i}\in\Gamma(m,k)}\, \sum_{\delta_{j}\in\Gamma(m,m-\ell+1)} \left( \prod_{j=1}^{m} \Sigma(n,j) \right) \right\} \times \nonumber \\
&&
\hspace{-10pt}
\times\, \sum_{r=0}^{k_{n+\ell-k,\ell}} \frac{\Gamma (g+\beta_{n,0}^{[0]}-\beta_{n+\ell-k,\ell}^{[r]})}{\left( - k A \right)^{g+\beta_{n,0}^{[0]}-\beta_{n+\ell-k,\ell}^{[r]}}}\, \sum_{h=0}^{+\infty} \frac{\Gamma (g+\beta_{n,0}^{[0]}-\beta_{n+\ell-k,\ell}^{[r]}-h)}{\Gamma (g+\beta_{n,0}^{[0]}-\beta_{n+\ell-k,\ell}^{[r]})}\, F_{h}^{(n+\ell-k|\ell)[r]} \left( - k A \right)^{h} \times \nonumber \\
&&
\hspace{-10pt}
\times \left.\left\{ \delta_{r0} + \Theta(r-1)\, \Big( B_{kA}(a) + \partial_{a} \Big)^{r-1} B_{kA}(a) \right\} \right|_{a=g+\beta_{n,0}^{[0]}-\beta_{n+\ell-k,\ell}^{[r]}-h-1}.
\label{ninstasymp-log}
\end{eqnarray}
\noindent
The quantity $\Sigma(n,j)$ was previously defined in (\ref{sigmanj}) as
\begin{equation}
\Sigma(n,j) = \left( \left( j+1-\delta_{j} \right) \widetilde{S}_{-\mathbf{d}\gamma_{j}}^{(\mathbf{d}\delta_{j})} + \left( n-\gamma_{j}+j+1-\delta_{j} \right) S_{-\mathbf{d}\gamma_{j}}^{(\mathbf{d}\gamma_{j}+\mathbf{d}\delta_{j})} \right) \Theta \left( j+1-\delta_{j} \right).
\end{equation}

One thing to notice is that, due to the logarithmic contributions appearing in the generalized multi--instanton expansion of $\Phi_{(n|m)} (z)$, the large--order behavior now includes contributions depending on the function $B_s (a)$. The simplest possible contribution of this type in (\ref{ninstasymp-log}) is
\be
B_{kA} \left( g+\beta_{n,0}^{[0]}-\beta_{n+\ell-k,\ell}^{[r]}-h-1 \right) = \psi \left( g+\beta_{n,0}^{[0]}-\beta_{n+\ell-k,\ell}^{[r]}-h \right)-\log \left( k A \right) - \rmi\pi.
\ee
\noindent
When $g$ is very large (\textit{i.e.}, considering the large--order behavior) this expression may be expanded as
\be
B_{kA} \left( g \right) \simeq \psi \left( g \right) \simeq \log g - \CO \left( 1/g \right),
\ee
\noindent
where we made use of the asymptotic expansion of the digamma function around infinity. This shows that, in addition to the familiar $g!$ growth of the large--order coefficients, we now further find a large--order growth of the type $g! \log g$ in the instanton sectors (which was also noticed in \cite{gikm10} for Painlev\'e I) and generalizations thereof---as explicitly contained in (\ref{ninstasymp-log}). In particular, this is a leading growth when compared with $g!$ and will be clearly visible at large order.

As an application of the expression (\ref{ninstasymp-log}) above let us look at the case $n=1$ and $k=2$, that is, the 2--instantons contributions to $F_{g}^{(1|0)[0]}$, with particular focus on the ones which display a logarithmic behaviour. The contribution from the discontinuity at $\theta=0$ is straightforward so we shall focus instead on the contributions arising from $\theta=\pi$. The sums in $m$ and $\ell$ have to be such that $n+\ell-k\ge 0$, which implies $2\ge m\ge\ell\ge 1$. The cases with $\ell=1$ will not have any logarithmic contributions as $k_{n+\ell-k,\ell}\equiv k_{0,1}=0$. Thus, the only case of interest is $m=\ell=2$, which can have logarithmic contributions as long as $k_{n+\ell-k,\ell}\equiv k_{1,2}\ne 0$. In this case $\gamma_{i}\in\Gamma(2,2)$ will have contributions from the Young diagrams $\tableau{21}$ and $\tableau{22}$, and $\delta_{j}\in\Gamma(2,1)$ will have only one contributing diagram, $\tableau{11}$. Assuming that $k_{1,2}=1$ (as will be the case of Painlev\'e I) the 2--instantons contribution to $F_{g}^{(1|0)[0]}$ becomes
\begin{eqnarray}
\left. F_{g}^{(1|0)[0]}\right|_{m=\ell=2}^{\text{2-inst}} &\approx& \frac{(\widetilde{S}_{-1}^{(0)})^{2}}{2\pi\rmi}\, \frac{\Gamma(g+\beta_{1,0}^{[0]}-\beta_{1,2}^{[0]})}{\left( - 2 A \right)^{g+\beta_{1,0}^{[0]}-\beta_{1,2}^{[0]}}}\, \sum_{h \ge 0} \frac{F_{h}^{(1|2)[0]} \left( - 2 A \right)^{h}}{\prod_{m=1}^{h} \left( g+\beta_{1,0}^{[0]}-\beta_{1,2}^{[0]}-m \right)} + \\
&&
\hspace{-65pt}
+\frac{(\widetilde{S}_{-1}^{(0)})^{2}}{2\pi\rmi}\, \frac{\Gamma(g+\beta_{1,0}^{[0]}-\beta_{1,2}^{[1]})}{\left( - 2 A \right)^{g+\beta_{1,0}^{[0]}-\beta_{1,2}^{[1]}}}\, \sum_{h \ge 0} \frac{F_{h}^{(1|2)[1]} \left( - 2 A \right)^{h}}{\prod_{m=1}^{h} \left( g+\beta_{1,0}^{[0]}-\beta_{1,2}^{[1]}-m \right)}\, B_{2A}(g+\beta_{1,0}^{[0]}-\beta_{1,2}^{[1]}-h-1). \nonumber
\end{eqnarray}

The results obtained in this section can be extended to the generalized instanton series, such as, for example, the $(n,1)$--series. However, those generalizations yield extremely lengthy formulae. Consequently, we shall present those results only as they become needed in the following sections, and always in the specific form applicable to either of the particular cases of interest: the Painlev\'e I equation and the quartic matrix model.

\section{Minimal Models and the Painlev\'e I Equation}\label{PIsection}

We now want to apply the general theory of two--parameters resurgence developed in the previous section to some concrete examples appearing in string theory. The specific examples we have in mind are matrix models and minimal string theories, which, as is well known, are closely related: all minimal models can be obtained as double--scaling limits of matrix models \cite{fgz93}.

In this section we shall be mainly interested in the $(2,3)$ minimal string theory, which describes pure gravity in two dimensions, and whose free energy may be obtained from a solution of the Painlev\'e I differential equation. Later, in section \ref{sec:6}, we will turn to a similar resurgent treatment of the one--matrix model, where we shall see that, in the double--scaling limit, it exactly reproduces the minimal model results of this section.

\subsection{Minimal String Theory and the Double--Scaling Limit}\label{sec:PQfromMM}

Minimal models, labeled by two relatively prime integers, $p$ and $q$, are among the simplest two--dimensional conformal field theories (CFT) and, starting with the seminal work of \cite{bpz84}, they have been studied in great detail in the past (see, \textit{e.g.}, the excellent review \cite{fgz93}).

Strictly speaking, the models we are interested in are not the minimal CFTs \textit{per se}, but the string theories that they lead to. That is, we consider these models coupled to Liouville theory and ghosts and sum over all worldsheet topologies that the CFT can live on. The resulting genus expansion for the free energy is an asymptotic series, with the familiar large--order behavior $\sim (2g)!$ \cite{s90}, and it is the nonperturbative completion of this asymptotic series that we shall study. In particular, the simplest non--topological minimal string is the model with $(p,q)=(2,3)$. It has a single primary operator, which after coupling to Liouville theory can be thought of as the worldsheet cosmological constant, and the central charge of the CFT is $c=0$, meaning that the ``target space'' is a point: this minimal string theory describes pure gravity on the worldsheet.

We shall discuss one--matrix models and their double--scaling limits in some detail later in section \ref{sec:6}. For the moment, we only need one important result from the double--scaling analysis. The free energy $F(z)$ of the minimal string theory depends on a single parameter, $z$, which is essentially the string coupling constant\footnote{More precisely, as we shall see in what follows, the $c=0$ \textit{closed} string coupling constant equals $z^{-5/2}$.}. It is also convenient to define the function
\be
u(z) = - F'' (z). 
\ee
\noindent
Then, from the double--scaling limit of the string equations of the matrix model one can show that the function $u (z)$ satisfies a relatively simple ordinary differential equation which, for the $(2,3)$ minimal string describing two--dimensional pure gravity, is the famous Painlev\'e I equation,
\be
u^2 (z) - \frac{1}{6} u'' (z) = z.
\ee
\noindent
One can solve this equation perturbatively in the string coupling constant, and the resulting asymptotic series gives the genus expansion of the $(2,3)$ minimal string free energy. What we are interested in here is to describe the full \textit{nonperturbative} solution to this equation, in terms of a transseries. Since the differential equation is of second order, we expect such a solution to have two integration constants, and hence we should find a two--parameters transseries solution---exactly the type of transseries that we have discussed in the previous section.

The construction of the two--parameters transseries solution to the Painlev\'e I equation was started in \cite{gikm10}, where the structure of the full instanton series and of the contribution with a single ``generalized instanton'' were found. Here, we complete this analysis by describing the structure of the full, general nonperturbative contributions to the solution.

\subsection{The Transseries Structure of Painlev\'e I Solutions}\label{sec:PItransseries}

Let us now develop the transseries framework as applied to the Painlev\'e I equation.

\subsubsection*{Review of the One--Parameter Transseries Solution}

As explained above, our aim is to solve the Painlev\'e I equation,
\be
u^2 (z) - \frac{1}{6} u'' (z) = z,
\ee
\noindent
in terms of a two--parameters transseries, where the perturbative parameter of the solution is the string coupling constant. As it turns out, in the minimal string, small string coupling corresponds to large $z$ and hence the perturbative series in our solutions should be expansions around $z=\infty$. It is well known, and one can easily check, that there is indeed an asymptotic series solution around $z=\infty$ in terms of the parameter $z$. It is given by
\be\label{eq:upert}
u_{\text{pert}}(z) \simeq \sqrt{z} \left(1 - \frac{1}{48}\, z^{-5/2} - \frac{49}{4608}\, z^{-5} - \frac{1225}{55296}\, z^{-15/2} - \cdots \right).
\ee
Note that, apart from the leading factor of $z^{1/2}$, this solution is a power series in $z^{-5/2}$. This parameter is indeed known to be the coupling constant of the minimal string theory. However, $z^{-5/2}$ is not quite the perturbative parameter that we should choose for our transseries solution. The minimal string theory is a closed string theory, so indeed we expect its perturbative free energy to be a function of the closed string coupling constant. But nonperturbative effects in string theory, on the other hand, are associated to D--branes, and hence to open strings. As usual in string theory, the closed string coupling constant is the square of the open string coupling constant and, therefore, we may expect the nonperturbative contributions to the free energy to be expansions in 
\be
x = z^{-5/4}.
\ee
\noindent
We shall later see that this is indeed the case.

As a first step in finding a transseries solution to the Painlev\'e I equation, one may now try to find a one--parameter transseries solution of the form
\be\label{eq:u1par}
u(x) \simeq x^{-2/5} \sum_{n=0}^{+\infty} \sigma_1^n\, \rme^{-nA/x}\, x^{n\beta} \sum_{g=0}^{+\infty} u^{(n)}_{g}\, x^g,
\ee
\noindent
where $x$ is expressed in terms of $z$ by the relation above, and $A$ and $\beta$ are coefficients that still need to be determined. Plugging this \textit{ansatz} back into the Painlev\'e I equation (see, for example, \cite{sy90, msw07, gikm10}), one finds that a solution exists if one chooses
\be\label{eq:tspars}
A = \pm \frac{8 \sqrt{3}}{5}, \qquad \beta = \frac{1}{2}.
\ee
\noindent
The same result could be obtained by writing the Painlev\'e I equation in prepared form, (\ref{preparednlode}), where one would find
\be
\frac{\rmd \boldsymbol{u}}{\rmd z} (z) = -
\begin{bmatrix}
+\frac{8 \sqrt{3}}{5} & 0 \\
0 & -\frac{8 \sqrt{3}}{5}
\end{bmatrix}
\cdot \boldsymbol{u} (z) + \cdots.
\ee
\noindent
For the ``instanton action'', $A$, there is a choice of sign. In the one--parameter transseries one usually chooses the positive sign, since with that choice the instanton factor $\exp(-A/x)$ is exponentially suppressed as expected. Doing this one finds, for example, the one--instanton correction
\be
u_{\text{1-inst}}(x) \simeq \sigma_1\, x^{1/10}\, \rme^{-A/x} \left(1 - \frac{5}{64 \sqrt{3}}\, x + \frac{75}{8192}\, x^2 - \cdots \right).
\ee
\noindent
Note that indeed we now find a series in the \textit{open} string coupling $x = z^{-5/4}$, whereas the purely perturbative part (\ref{eq:upert}) of $u(z)$ is a series in the \textit{closed} string coupling $x^2 = z^{-5/2}$. The coefficients in this expression can be determined recursively by plugging the \textit{ansatz} (\ref{eq:u1par}) into the Painlev\'e I equation. One finds that this determines all coefficients except the leading one, $u^{(1)}_0$. Its (non--zero) value can in fact be chosen arbitrarily without loss of generality, since we can rescale it by choosing the nonperturbative ambiguity $\sigma_1$. For now, we adopt a normalization where $u^{(1)}_0=1$.

\subsubsection*{The Two--Parameters Transseries Solution}

So far, we have only considered the positive sign choice for the instanton action $A$ in (\ref{eq:tspars}). However, at the level of formal solutions, the negative sign choice is also required in order to obtain the most general solutions of the Painlev\'e I equation, \textit{i.e.}, we should really apply the machinery developed in section \ref{sec:4} and solve the Painlev\'e I equation using a \textit{two}--parameters transseries. To do this, it is very convenient to change variables once again. Recall that the $\beta$--parameter we found for the \textit{ansatz} (\ref{eq:u1par}) equals $\beta=1/2$. As we will see, the $x$--dependent prefactor in the two--parameters transseries will no longer be of the simple form $x^{n \beta}$. It is therefore no longer convenient to take it outside the perturbative sum over $g$, as we did in (\ref{eq:u1par}). The analogue of $x^{n \beta}$, on the other hand, will still be a half--integer power of $x$, so if we want to consider it as part of the perturbative series, it is more convenient to use the variable
\be
w = x^{1/2} = z^{-5/8}.
\ee
\noindent
Of course, up to a possible odd overall power in $w$, we still expect all perturbative series to be expansions in the \textit{open} string coupling constant, $w^2$, and we will find that this is indeed the case. Let us be a bit pedantic and stress this point once again, in order not to raise any confusions later on: the \textit{open} string coupling constant is $x=w^2$ and we shall mostly work in the $w$ variable.

It is also useful for calculational purposes to scale away the overall power of $z^{1/2}$ in $u(z)$, and set
\be\label{eq:urescale}
u(w) \equiv \left. \frac{u(z)}{\sqrt{z}}\right|_{z=w^{-8/5}}.
\ee
\noindent
Here, we slightly abuse notation; it would have been more precise to call the function on the left--hand side $\widehat{u}(w)$, but to avoid writing too many hats we will stick to the above notation and simply remember whether we use the rescaled $u$ or not by looking at the variable that we use. 

It is now a simple exercise to rewrite the Painlev\'e I equation in terms of the function $u(w)$; one finds
\be\label{eq:PIw}
u^2(w) + \frac{1}{24} w^4\, u(w) - \frac{25}{384} w^5\, u'(w) - \frac{25}{384} w^6\, u''(w) = 1,
\ee
\noindent
where we want to solve this equation using a two--parameters transseries \textit{ansatz},
\be\label{eq:u2par}
u(w,\sigma_1,\sigma_2) = \sum_{n=0}^{+\infty} \sum_{m=0}^{+\infty} \sigma_1^n \sigma_2^m\, \rme^{-(n-m)A/w^2}\, \Phi_{(n|m)}(w).
\ee
\noindent
Note that, as we have mentioned, we have not included a factor of $w^{\beta_{nm}}$ in the transseries expansion, but absorbed it in $\Phi_{(n|m)}(w)$. This means that the leading coefficients in $\Phi_{(n|m)}$ will in general not multiply the constant term. Conversely, we can find back the analogue of the prefactor $w^{\beta_{nm}}$ (as we will do below) by finding the first nonzero coefficient in $\Phi_{(n|m)}(w)$.

One may now be tempted to complete the \textit{ansatz} above by assuming that $\Phi_{(n|m)}(w)$ is a power series in $w$. However, an \textit{ansatz} of this form turns out \textit{not} to work, essentially since the Painlev\'e I equation is a \textit{resonant} equation (a property we have previously discussed in section \ref{sec:4} and to which we shall come back in a moment). It turns out that, for a correct \textit{ansatz}, one needs terms multiplying powers of $\log(w)$, a phenomenon first observed in \cite{gikm10}. In that paper, the authors calculated $\Phi_{(n|1)}(w)$, and found that it had the general form
\be
\Phi_{(n|1)}(w) = \sum_{g=0}^{+\infty} u^{(n|1)[0]}_g\, w^g + \log(w) \cdot \sum_{g=0}^{+\infty} u^{(n|1)[1]}_g\, w^g.
\ee
\noindent
In fact, for $n=0,1,$ the logarithmic terms are absent, but they are always present whenever $n>1$. One may now wonder what the general form of $\Phi_{(n|m)}$ is. From the $u^2$--term in the Painlev\'e I equation, one sees that $\Phi_{(n|m)}$ is determined recursively in terms of products $\Phi_{(n-p|m-q)} \Phi_{(p|q)}$. This means that, starting\footnote{We shall actually see below that, due to resonance, the $\log^2 w$ behaviour already sets in at $\Phi_{(3|2)}$.} at $\Phi_{(4|2)}$, we can expect to encounter $\log^2 w$ terms coming from terms such as $\Phi_{(2|1)}\Phi_{(2|1)}$. Extending this reasoning, we see that a natural \textit{ansatz} for the general $\Phi_{(n|m)}$ is
\be\label{eq:phiansatz}
\Phi_{(n|m)}(w) = \sum_{k=0}^{\min(n,m)} \log^k w \cdot \sum_{g=0}^{+\infty} u^{(n|m)[k]}_g\, w^g.
\ee
\noindent
Our job now is to determine if a solution for all coefficients $u^{(n|m)[k]}_g$ can be found. It is a tedious but straightforward exercise to plug the \textit{ans\"atze} (\ref{eq:u2par}) and (\ref{eq:phiansatz}) into the Painlev\'e I equation (\ref{eq:PIw}) and, in this process, to find that the coefficients $u^{(n|m)[k]}_g$ must satisfy the relation
\bea\label{eq:PIrec}
\delta^n_0\, \delta^m_0\, \delta^k_0\, \delta^g_0 &=& \sum_{\widehat{n}=0}^n \sum_{\widehat{m}=0}^m \sum_{\widehat{g}=0}^g \sum_{\widehat{k}=0}^k u^{(\widehat{n}|\widehat{m})[\widehat{k}]}_{\widehat{g}}\, u^{(n-\widehat{n}|m-\widehat{m})[k-\widehat{k}]}_{g-\widehat{g}} - \\
&&
- \frac{25}{96} \left(n-m\right)^2 A^2\, u^{(n|m)[k]}_g + \frac{25}{96} \left(m-n\right) \left(k+1\right) A\, u^{(n|m)[k+1]}_{g-2} + \nonumber \\
&&
+ \frac{25}{96} \left(m-n\right) \left(g-3\right) A\, u^{(n|m)[k]}_{g-2} - \frac{25}{384} \left(k+2\right) \left(k+1\right) u^{(n|m)[k+2]}_{g-4} - \nonumber \\
&&
- \frac{25}{192} \left(k+1\right) \left(g-4\right) u^{(n|m)[k+1]}_{g-4} - \frac{1}{384} \left(5g-16\right) \left(5g-24\right) u^{(n|m)[k]}_{g-4}. \nonumber
\eea
\noindent
This relation is valid for any 4--tuple $(n,m,k,g)$ if we assume that all non--existent coefficients---that is, the ones with $k$ larger than $\min(n,m)$ and the ones with $g<0$---are vanishing. The relation can be used to recursively determine $u^{(n|m)[k]}_g$ in terms of coefficients which have smaller $n,m,k$ or $g$. A \textit{Mathematica} notebook with the results is available from the authors upon request.

\subsubsection*{The Consequences of Resonance}

In using the relation (\ref{eq:PIrec}), one finds that something special happens whenever $|n-m|=1$. In this case, the first term on the second line of the relation equals
\be
- \frac{25 A^2}{96} u^{(n|m)[k]}_g = -2 u^{(n|m)[k]}_g,
\ee
\noindent
where we inserted the explicit value (\ref{eq:tspars}) for $A$. However, this is not the only term multiplying $u^{(n|m)[k]}_g$: the sum in the first line of (\ref{eq:PIrec}) also contains two terms with this factor, which add up to
\be
2 u^{(0|0)[0]}_0\, u^{(n|m)[k]}_g = 2 u^{(n|m)[k]}_g,
\ee
where we read off the leading coefficient $u^{(0|0)[0]}_0=1$ from (\ref{eq:upert}). Thus, we see that, whenever $|n-m|=1$, the leading terms in the recursion formula \textit{cancel}. This is precisely the phenomenon of resonance! The cancellation of the leading terms in itself is not a problem---it simply means that one should use our formula to determine $u^{(n|m)[k]}_{g-2}$ instead. However, it could potentially be a problem whenever $u^{(n|m)[k]}_{g-2}$ does not exist---that is, when we try to determine the leading term in $w$ for each perturbative series, given $n,m,k$. Here, two things can happen:
\begin{enumerate}
\item The recursion relation may reduce to $const = 0$, in which case it cannot be satisfied. This is what happens if one does not include the correct $\log w$ terms. For example, if we would include no logarithmic terms at all, the recursion for $n=2$, $m=1$ would lead to such an inconsistency. Thus, resonance \textit{forces} us to include the logarithmic terms. In a similar way, we will need $\log^2 w$ terms starting at $n=3$, $m=2$. Note that above we have already argued that such terms must appear for $n=4$, $m=2$; now we find that we also need to include them in $\Phi_{(3|2)}$, as we did in our \textit{ansatz}. Only at $n=m=2$ are the $\log^2 w$ terms absent. This pattern actually continues to higher $m$: $\Phi_{(m|m)}$ will \textit{never} contain any logarithmic terms; but the $\log^m w$ terms set in immediately at $n=m+1$ due to resonance.
\item The recursion relation may reduce to $0=0$. This is of course consistent, but it means that we have a leading coefficient which can be chosen arbitrarily. We already saw an example of this: $u^{(1)}_0$, the leading coefficient of the one--instanton series, can have an arbitrary value due to the choice in the normalization of the nonperturbative ambiguity $\sigma_1$. In our two--parameters transseries terminology, this coefficient is now denoted $u^{(1|0)[0]}_1$. The same thing now holds for $u^{(0|1)[0]}_1$, its value can be absorbed into $\sigma_2$. However, it turns out that the recursion relation allows for a whole lot more free parameters: for any $m\geq0$, the coefficients $u^{(m+1|m)[0]}_1$ and $u^{(m|m+1)[0]}_1$ are not fixed by our recursion relation.
\end{enumerate}
The second property above seems confusing at first sight. How can a two--parameters transseries, solving a second order differential equation, have infinitely many free parameters? The answer turns out to be that our \textit{ansatz} still has a large degree of reparametrization symmetry.

\subsubsection*{Reparametrization Invariance}

Recall that our general transseries \textit{ansatz} for the solution to the Painlev\'e I equation has the form
\be\label{eq:uC}
u(w) = \sum_{n=0}^{+\infty} \sum_{m=0}^{+\infty} \sigma_1^n \sigma_2^m\, \rme^{-(n-m)A/w^2}\, \Phi_{(n|m)}(w).
\ee
\noindent
It is important to note that the nonperturbative factor in each term only depends on the difference $n-m$. This means that when we make a (degree preserving) change of variables,
\be
\sigma_1 = \widehat{\sigma}_1 \sum_{p=0}^{+\infty} \alpha_p \left( \widehat{\sigma}_1 \widehat{\sigma}_2 \right)^p, \qquad \sigma_2 = \widehat{\sigma}_2 \sum_{q=0}^{+\infty} \beta_q \left( \widehat{\sigma}_1 \widehat{\sigma}_2 \right)^q,
\ee
\noindent
with arbitrary coefficients $\alpha_p$, $\beta_q$, we will find a new expression with exactly the same nonperturbative structure. Let us work this out in some detail. From the above change of variables, we get expansions of the form
\be
\sigma_1^n = \widehat{\sigma}_1^n \sum_{r=0}^{+\infty} \gamma^n_r \left( \widehat{\sigma}_1 \widehat{\sigma}_2 \right)^r, \qquad \sigma_2^m = \widehat{\sigma}_2^m \sum_{s=0}^{+\infty} \delta^m_s \left( \widehat{\sigma}_1 \widehat{\sigma}_2 \right)^s,
\ee
\noindent
where it is not too hard to find explicit formulae for the coefficients $\gamma^n_r$, $\delta^m_s$, given by
\be
\gamma^n_r = \sum_{\{ \lambda \}} \prod_{i=1}^{n} \alpha_{\lambda_i} \qquad \text{and} \qquad \delta^m_s = \sum_{\{ \mu \}} \prod_{i=1}^{m} \beta_{\mu_i}.
\ee
\noindent
In here, $\{\lambda\}$ and $\{\mu\}$ are ordered partitions, where ``ordered'' means that, for example, we consider $\{0,1,4\}$ and $\{4,1,0\}$ as different partitions of the integer $5$. In the first sum, $\{\lambda\}$ runs over all ordered partitions of $r$ with length $n$, and the analogous statement holds for the second sum. These formulae only hold for $n,m \geq 0$; for $n=0$ we have that $\gamma_0^0 = 1$ and all other $\gamma_r^0 = 0$. The same thing of course holds for $\delta^0_s$.

Inserting these results in (\ref{eq:uC}), it follows
\be
u(w) = \sum_{n=0}^{+\infty} \sum_{m=0}^{+\infty} \sum_{r=0}^{+\infty} \sum_{s=0}^{+\infty} \widehat{\sigma}_1^{n+r+s}\, \widehat{\sigma}_2^{m+r+s}\, \gamma^n_r\, \delta^m_s\, \rme^{-(n-m)A/w^2}\, \Phi_{(n|m)}(w).
\ee
\noindent
Changing the summation variables $(n,m)$ to $(\widehat{n},\widehat{m})=(n+r+s,m+r+s)$, one obtains
\be
u(w) = \sum_{\widehat{n}=0}^{+\infty} \sum_{\widehat{m}=0}^{+\infty} \widehat{\sigma}_1^{\widehat{n}} \widehat{\sigma}_2^{\widehat{m}}\, \rme^{-(\widehat{n}-\widehat{m})A/w^2} \sum_{r=0}^{r_0} \sum_{s=0}^{s_0} \gamma^{\widehat{n}-r-s}_r\,  \delta^{\widehat{m}-r-s}_s\, \Phi_{(\widehat{n}-r-s|\widehat{m}-r-s)}(w).
\ee
\noindent
In this expression, $r_0 = \min(\widehat{n}, \widehat{m})$ and $s_0 = \min(\widehat{n},\widehat{m})-r$. In other words, $r$ and $s$ run over the triangle given by
\be
r \geq 0, \qquad s \geq 0, \qquad r+s \leq \min(\widehat{n}, \widehat{m}).
\ee
\noindent
Thus, we have found that, after reparametrization, $u(w)$ can be written in exactly the same form albeit in terms of new functions,
\be
\widehat{\Phi}_{(n|m)}(w) = \sum_{r,s} \gamma^{n-r-s}_r\, \delta^{m-r-s}_s\, \Phi_{(n-r-s|m-r-s)}(w).
\ee
\noindent
Let us write out the first few of those:
\be
\widehat{\Phi}_{(n|0)} = \alpha_0^{n}\, \Phi_{(n|0)}, \qquad \widehat{\Phi}_{(0|m)} = \beta_0^{m}\, \Phi_{(0|m)}.
\ee
\noindent
Since we have already fixed the leading coefficients of $\Phi_{(1|0)}$ and $\Phi_{(0|1)}$ to equal one, this means that we cannot freely choose $\alpha_0$ and $\beta_0$: we have to set them equal to one as well. Using this, one finds for the next few $\widehat{\Phi}$,
\bea
\widehat{\Phi}_{(n|1)} &=& \Phi_{(n|1)} + \alpha_1 \left( n-1 \right) \Phi_{(n-1|0)}, \label{eq:phirepar1} \\
\widehat{\Phi}_{(1|m)} &=& \Phi_{(1|m)} + \beta_1 \left( m-1 \right) \Phi_{(0|m-1)},
\eea
\noindent
where $\widehat{n}, \widehat{m} > 1$. Thus, the $\Phi_{(n|1)}$ are only defined up to additions of $\Phi_{(n-1|0)}$. One can continue like this: after fixing $\alpha_1$ and $\beta_1$ it turns out that the free parameters $\alpha_2$ and $\beta_2$ show up for the first time in $\Phi_{(n|2)}$ and $\Phi_{(2|m)}$, and multiply possible additions of $\Phi_{(n-2|0)}$ and $\Phi_{(0|m-2)}$. 

This explains the fact that, in the previous subsection, we found that our recursive transseries solution had an infinite number of undetermined parameters. They are simply the parameters $\alpha_p$ and $\beta_q$ that determine the freedom in the parametrization of the coefficients $\sigma_1$ and $\sigma_2$. One will find a unique two--parameters transseries solution to the Painlev\'e I equation only after fixing these parameters by some sort of ``gauge condition''.

\subsubsection*{Two--Parameters Transseries: Results}

There is a rather natural condition\footnote{This condition is applied implicitly in the function $\Phi_{(2|1)}$ reported in \cite{gikm10}.} to fix the free parameters in our transseries \textit{ansatz}. Calculating $\Phi_{(m+1|m)}$ up to $m=10$ for arbitrary values of the free parameters, we find that these transseries components do not have a constant term. We have also seen that $\Phi_{(1|0)}$ starts at order $w^1$, and we now know that we can use reparametrization invariance to add an arbitrary multiple of $\Phi_{(1|0)}$ to $\Phi_{(m+1|m)}$. Thus, one can tune the free parameter $\alpha_m$ in such a way that the $w^1$--term in $\Phi_{(m+1|m)}$ vanishes. That is, one can fix half of the reparametrization invariance by simply setting
\be\label{um1m01}
u^{(m+1|m)[0]}_1 = 0, \qquad \forall m \geq 1.
\ee
\noindent
In the exact same way, one can use the $\beta_n$--parameters to set
\be\label{unn101}
u^{(n|n+1)[0]}_1 = 0, \qquad \forall n \geq 1,
\ee
\noindent
by adding the appropriate multiples of $\Phi_{(0|1)}$.

This fixing of the undetermined parameters is the last ingredient one needs in order to use the recursive formula (\ref{eq:PIrec}) and solve for the entire transseries. Using a computer, this can be efficiently done up to $n=m=10$ and $g=50$ in a matter of minutes, and we have tabulated some of the $\Phi_{(n|m)}(w)$ in appendix \ref{app:PI}. One thing the reader should note from those expressions is that the resulting functions are always, up to an overall factor, indeed expansions in the \textit{open} string coupling constant $x=w^2$.

\begin{table}[t]
\centering
\begin{tabular}{c|ccccccc}
\begin{picture}(20,20)(0,0)
\put(17,8){$n$}
\put(3.6,18){\line(1,-1){22.3}}
\put(3,0){$m$}
\end{picture}
& 0 & 1 & 2 & 3 & 4 & 5 & 6 \\
\hline
0 & 0 & 1 & 2 & 3 & 4 &  5 &  6 \\
1 & 1 & 2 & 3 & 4 & 5 &  6 &  7 \\
2 & 2 & 3 & 4 & 3 & 4 &  5 &  6 \\
3 & 3 & 4 & 3 & 6 & 5 &  6 &  7 \\
4 & 4 & 5 & 4 & 5 & 8 &  5 &  6 \\
5 & 5 & 6 & 5 & 6 & 5 & 10 &  7 \\
6 & 6 & 7 & 6 & 7 & 6 &  7 & 12 
\end{tabular}
\qquad
\begin{tabular}{c|ccccccc}
\begin{picture}(20,20)(0,0)
\put(17,8){$n$}
\put(3.6,18){\line(1,-1){22.3}}
\put(3,0){$m$}
\end{picture}
& 0 & 1 & 2 & 3 & 4 & 5 & 6 \\
\hline
0 & $\ast$ & $\ast$ & $\ast$ & $\ast$ & $\ast$ & $\ast$ & $\ast$ \\
1 & $\ast$ & $\ast$ & 1 & 2 & 3 & 4 & 5 \\
2 & $\ast$ & 1 & $\ast$ & 3 & 4 & 5 & 6 \\
3 & $\ast$ & 2 & 3 & $\ast$ & 3 & 4 & 5 \\
4 & $\ast$ & 3 & 4 & 3 & $\ast$ & 5 & 6 \\
5 & $\ast$ & 4 & 5 & 4 & 5 & $\ast$ & 5 \\
6 & $\ast$ & 5 & 6 & 5 & 6 & 5 & $\ast$
\end{tabular}
\caption{Values for $2\beta_{nm}^{[0]}$ (left) and $2\beta_{nm}^{[1]}$ (right). An asterisk in the second table means that there are no logarithmic terms in $\Phi_{(n|m)}$.}
\label{table:beta}
\end{table}

The choices (\ref{um1m01}) and (\ref{unn101}) simplify our results a lot, and sets many more of the leading coefficients to zero. Let us fix $n$, $m$ and $k$, and ask ourselves what the lowest index $g$ is for which $u^{(n|m)[k]}_g$ is nonzero. We will call this index $2\beta_{nm}^{[k]}$ (the factor of $2$ is essentially due to the fact that we are now working with the $w$ variable rather than $x$); it is the analogue of the $\beta_{nm}$ in the general logarithm--free two--parameters transseries (\ref{2Fnmseries}). Whereas in that case $\beta_{nm}$ is usually of the form $(n+m)\beta$ for a fixed $\beta$, in the Painlev\'e I case we find a more complicated structure. We tabulate $2\beta_{nm}^{[0]}$ and $2\beta_{nm}^{[1]}$ in table \ref{table:beta}. This table clearly has some structure and, in fact, it is not too hard to find a general formula for $2\beta_{nm}^{[k]}$. When $n=m$, none of the contributions have logarithms, and we have that
\be
2\beta_{nn}^{[0]} = 2n.
\ee
\noindent
For $n \neq m$, it is easiest to write separate formulae for the cases $n > m$ and $m > n$. When either $n$ or $m$ is smaller than $k$, we have no $\log^k$ corrections. When $n > m \geq k$, one finds
\be
2\beta_{nm}^{[k]} = n - k + (m+k \mod 2).
\ee
\noindent
For $m > n \geq k$, the formula is the same, but with $n$ and $m$ interchanged. This can be summarized by defining, for all $n$ and $m$, 
\begin{equation}\label{eq:beta_nmk-Painleve}
2\beta_{nm}^{[k]} \equiv n+m-2\left[ \frac{k_{nm}+k}{2} \right]_{\text{I}},
\end{equation}
\noindent
where $\left[\bullet\right]_{\text{I}}$ represents the integer part, and
\begin{equation}\label{eq:k_nm-max-log-power-Painleve}
k_{nm}=\min(n,m)-m\,\delta_{nm}
\end{equation}
\noindent
is just the maximum power of the logarithm appearing in the expansion of $\Phi_{(n|m)} (w)$.

\subsection{The String Genus Expansion Revisited}

We now have enough information to address the string genus expansion of the Painlev\'e I solution, applying the general formulae previously obtained in section \ref{sub:String-genus-exp}. Let us start by re--writing the asymptotic expansion for the $\Phi_{(n|m)}$, given in (\ref{eq:Def-Phi_(n|m)-with-all-modes}), as\footnote{Notice that the $F_{g}^{(n|m)[k]}$ coefficients in the following just denote coefficients of a general transseries solution, in the abstract setting of section \ref{sec:4}, and \textit{not} the free energy of the $(2,3)$ model. We shall discuss the relation between the Painlev\'e I solution and the $(2,3)$ free energy at the end of this section.}
\be
\Phi_{(n|m)} (x) \simeq \sum_{k=0}^{k_{nm}} \log^{k} x\, \sum_{g=0}^{+\infty} F_{g}^{(n|m)[k]}\, x^{g+\beta_{nm}^{[k]}} = \sum_{k=0}^{k_{nm}} \log^{k} w\, \sum_{g'=0}^{+\infty} 2^{k}\, F_{\frac{g'}{2}}^{(n|m)[k]}\, w^{g'+2\beta_{nm}^{[k]}},
\ee
\noindent
\textit{i.e.}, as an expansion in $w$ rather than as an expansion in $x$. Recall that our formulae in section \ref{sub:String-genus-exp} were written in terms of the open string coupling $g_s=x=w^{2}$, while in here we find it more convenient to work directly in the $w$--variable. Furthermore, in this expression it is understood that all the $F_{\frac{g}{2}}^{(n|m)[k]}$ with $g$ odd vanish (in order to have an expansion in integer powers of $x$). We can now directly compare with the expansion (\ref{eq:phiansatz}) for the Painlev\'e I solution, and easily find that $k_{nm}$ is given by (\ref{eq:k_nm-max-log-power-Painleve}) above, and\footnote{More precisely, the relation between $F_{g}^{(n|m)[k]}$ and $u_{g}^{(n|m)[k]}$ is given by 
$$
2^{k}\, F_{g}^{(n|m)[k]} = u_{g'}^{(n|m)[k]},
$$
\noindent
where $g' = 2 \left( g+\beta_{nm}^{[k]} \right)$ and $g$ starts at $0$. To write the expansion of $\Phi_{(n|m)}(x)$ we performed a shift on the variable $g$ such that $u_{g'}^{(n|m)[k]} \rightarrow u_{2g}^{(n|m)[k]}$ where now the expansion starts at $u_{0}^{(n|m)[k]}\, x^{\beta_{nm}^{[k]}}$.}
\begin{equation}\label{eq:Painleve-relation-ug-and-Fg}
u_{g}^{(n|m)[k]} = 2^{k}\, F_{\frac{g}{2}}^{(n|m)[k]} \qquad \Leftrightarrow \qquad u_{2g}^{(n|m)[k]} = 2^{k}\, F_{g}^{(n|m)[k]}.
\end{equation}
\noindent
In particular, this implies that the $u_{g}^{(n|m)[k]}$ vanish for odd $g$. Moreover, the lowest index in $g$ for which $u_{g}^{(n|m)[k]}$ is non--zero, $\beta_{nm}^{[k]}$, can also be obtained via a comparison with the results of the previous section, being given by (\ref{eq:beta_nmk-Painleve}) above. 

In this way, we can rewrite the expansion of $\Phi_{(n|m)}(x)$ for the Painlev\'e I solution as 
\begin{equation}\label{eq:Full-exp-Phi(nm)-Painleve}
\Phi_{(n|m)} (x) \simeq \sum_{k=0}^{k_{nm}} \frac{\log^{k} x}{2^k}\, \sum_{g=0}^{+\infty} u_{2g}^{(n|m)[k]}\, x^{g+\beta_{nm}^{[k]}} \equiv \sum_{k=0}^{k_{nm}} \frac{\log^{k} x}{2^k}\, \Phi_{(n|m)}^{[k]}(x),
\end{equation}
\noindent
with the $k_{nm}$ and $\beta_{nm}^{[k]}$ given earlier. It is now straightforward to apply the results of section \ref{sub:String-genus-exp} to the current case. But, before that, let us address two important properties arising from the Painlev\'e I recursion relations (\ref{eq:PIrec}), \textit{i.e.}, from the physics of the $(2,3)$ model, which will refine our results even further (also see appendix \ref{app:PI}). The first of these properties relates the coefficients $\Phi_{(n|m)}^{[k]}$, at the $k$--th logarithmic power, with $\Phi_{(n|m)}^{[0]}$, the contribution without logarithms, as
\begin{equation}
\Phi_{(n|m)}^{[k]} = \frac{1}{k!} \left( \frac{4 \left( m-n \right)}{\sqrt{3}} \right)^k \Phi_{(n-k|m-k)}^{[0]}.
\end{equation}
\noindent
This is a rather important relation; it amounts to saying that the logarithmic terms in (\ref{eq:Full-exp-Phi(nm)-Painleve}) are actually \textit{not independent} of each other, as their coefficients are all related to the coefficients of the logarithm--free term. In other words, these logarithmic contributions simply amount to a useful arrangement of the resonant transseries solution. The previous relation can be written in terms of the $u_{g}^{(n|m)[k]}$ by noting that $\beta_{nm}^{[k]} = \beta_{n-k,m-k}^{[0]}$ and thus
\begin{equation}\label{eq:Painleve-relating-logk-to-log0}
u_{g}^{(n|m)[k]} = \frac{1}{k!} \left( \frac{4 \left( m-n \right)}{\sqrt{3}} \right)^k u_{g}^{(n-k|m-k)[0]}.
\end{equation}

The second property we shall be using relates the different $u_{2g}^{(n|m)[k]}$ under  interchange of $n \leftrightarrow m$. This relation can be found in appendix \ref{app:PI} and is given by\footnote{Recall that we previously performed the change $u_{g'}^{(n|m)[k]} \rightarrow u_{2g}^{(n|m)[k]}$, with $g'=2 \left( g+\beta_{nm}^{[k]} \right)$.}
\begin{equation}\label{eq:Painleve-relating-mn-to-nm}
u_{2g}^{(n|m)[k]} = (-1)^{g+\beta_{nm}^{[k]}-(n+m)/2}\, u_{2g}^{(m|n)[k]} = (-1)^{g-\left[(k_{nm}+k)/2\right]_{\text{I}}}\, u_{2g}^{(m|n)[k]}.
\end{equation}
\noindent
Note that the exponent of $(-1)$ in the above expression is always an integer. In the case where $n=m$ (and consequently $k=0$) we find that this relation returns
\begin{equation}\label{eq:Painleve-relating-nn-to-nn}
u_{2g}^{(n|n)[0]} = (-1)^{g}\, u_{2g}^{(n|n)[0]} \qquad \Rightarrow \qquad u_{2(2g+1)}^{(n|n)[0]} = 0.
\end{equation}
\noindent
Consequentially, the $(n|n)$--instanton series will always have a topological genus expansion
\begin{equation}
\Phi_{(n|n)} (x) \simeq x^{n}\, \sum_{g=0}^{+\infty} \widehat{u}_{g}^{(n|n)}\, x^{2g} \equiv x^{n}\, \sum_{g=0}^{+\infty} u_{4g}^{(n|n)[0]}\, x^{2g}.
\end{equation}

Looking back at the zero--instanton series from section \ref{sub:String-genus-exp}, we have the genus expansion 
\begin{equation}
\Phi_{(0|0)}(x)\simeq\sum_{g=0}^{+\infty}\widehat{u}_{g}^{(0|0)[0]}\, x^{2g},
\end{equation}
\noindent
where $x \equiv g_{s}$ and where the large--order behavior follows from
\be
\widehat{u}_{g}^{(0|0)} = u_{4g}^{(0|0)[0]} \simeq \sum_{k=1}^{+\infty} \frac{\left(S_{1}^{(0)}\right)^{k}}{\rmi\pi}\, \frac{\Gamma (2g-\beta_{k,0}^{[0]})}{\left( k A \right)^{2g-\beta_{k,0}^{[0]}}}\, \sum_{h=0}^{+\infty} \frac{\Gamma (2g-h-\beta_{k,0}^{[0]})}{\Gamma (2g-\beta_{k,0}^{[0]})}\, u_{2h}^{(k|0)[0]} \left( k A \right)^{h}.
\ee
\noindent
One can also write large--order formulae for the asymptotics of the Painlev\'e I multi--instanton coefficients in the current language. This amounts to inserting these coefficients, written as (\ref{eq:Painleve-relation-ug-and-Fg}), back in (\ref{ninstasymp-log}). The condition $u_{2(2m+1)}^{(0|0)[0]}=0$ was studied in equation (\ref{sf-wtsf-relation}), which, when applied to the present case and by further using (\ref{eq:Painleve-relating-mn-to-nm}), yields
\begin{equation}
\left( S_{1}^{(0)} \right)^{k} u_{2h}^{(k|0)[0]} = (-1)^{h+\beta_{0,k}^{[0]}} \left( \widetilde{S}_{-1}^{(0)} \right)^{k} u_{2h}^{(0|k)[0]} = (-1)^{\beta_{0,k}^{[0]}} \left( \widetilde{S}_{-1}^{(0)} \right)^{k} u_{2h}^{(k|0)[0]}.
\end{equation}
\noindent
This immediately implies the following relation between $S_{1}^{(0)}$ and $\widetilde{S}_{-1}^{(0)}$
\begin{equation}
S_{1}^{(0)} = (-1)^{\frac{1}{2}}\, \widetilde{S}_{-1}^{(0)},
\end{equation}
\noindent
which coincides with a result found in \cite{gikm10}. 

The aforementioned properties (\ref{eq:Painleve-relating-mn-to-nm}) and (\ref{eq:Painleve-relating-nn-to-nn}) for the Painlev\'e I coefficients can, in principle, allow us to find many possible relations between the Stokes coefficients $S_{k}^{(n)}$ and $\widetilde{S}_{\ell}^{(m)}$. We will present one more such example in the following, with the study of the $(n,1)$--instanton series. First, using the same tools as in section \ref{sub:Bridge-Eqns-revisited}, we can find the Stokes automorphism for the series $\Phi_{(n|1)}(z)$, both at $\theta=0$,
\begin{eqnarray}
\underline{\frak{S}}_{0} \Phi_{(n|1)}(z) &=& \Phi_{(n|1)}(z) + \sum_{k=1}^{+\infty} \binom{n+k}{n}\, \left( S_{1}^{(0)} \right)^{k-2}\, \rme^{-kAz} \times \\
&&
\hspace{-15pt}
\times\, \left\{ \left(S_{1}^{(0)}\right)^{2} \Phi_{(n+k|1)}(z) + \left( \frac{k(k-1)}{k+n}\, S_{2}^{(0)} + \frac{k(2n+k-1)}{2(n+k)}\, S_{1}^{(1)}\, S_{1}^{(0)} \right) \Phi_{(n+k-1|0)}(z) \right\}, \nonumber
\end{eqnarray}
\noindent
and at $\theta=\pi$,
\be
\underline{\frak{S}}_{\pi}\Phi_{(n|1)}(z) = \Phi_{(n|1)}(z) + \sum_{k=1}^{+\infty} \sum_{m=1}^{k} \frac{\rme^{k A z}}{m!} \sum_{\ell=0}^{m+1}\, \sum_{\gamma_{i}\in\Gamma(m,k)}\, \sum_{\delta_{j}\in\Gamma(m,m-\ell+2)}\, \prod_{j=1}^{m} \Sigma^{(1)}(n,j) \cdot \Phi_{(n-1-k+\ell|\ell)}, 
\ee
\noindent
where this time around we find\footnote{Comparing against the $(n,0)$ case, (\ref{sigmanj}), the reader may want to guess a solution for the arbitrary $(n,m)$--instanton series.}
\begin{equation}
\Sigma^{(1)} (n,j) = \left( \left(j+2-\delta_{j}\right) \widetilde{S}_{-\mathbf{d}\gamma_{j}}^{(\mathbf{d}\delta_{j})} + \left(n-1-\gamma_{j}+j+2-\delta_{j}\right) S_{-\mathbf{d}\gamma_{j}}^{(\mathbf{d}\gamma_{j}+\mathbf{d}\delta_{j})} \right) \Theta \left( j+2-\delta_{j} \right).
\end{equation}

With these results in hand, one can use the asymptotic expansion (\ref{eq:Full-exp-Phi(nm)-Painleve}) and Cauchy's theorem to obtain the large--order behavior of the coefficients $u_{2g}^{(n|1)[r]}$, with $n \ge 1$ and $r=0,1$. In order to simplify this calculation we shall now make use of the property (\ref{eq:Painleve-relating-logk-to-log0}), relating the logarithmic sectors, and thus write the expansion of $\Phi_{(n|1)}(x)$ as
\begin{eqnarray}
\Phi_{(n|1)}(x) &\simeq& \sum_{g=0}^{+\infty} u_{2g}^{(n|1)[0]}\, x^{g+\beta_{n,1}^{[0]}} + \frac{1}{2} \log x\, \sum_{g=0}^{+\infty} u_{2g}^{(n|1)[1]}\, x^{g+\beta_{n,1}^{[1]}} \\
&=&
\Phi_{(n|1)}^{[0]} (x) + \frac{2(1-n)}{\sqrt{3}}\, \log x \cdot \Phi_{(n-1|0)}^{[0]}(x).
\label{phi(n1)phi0logphi0}
\end{eqnarray}
\noindent
At this stage, we already know the asymptotic behavior of $u_{2g}^{(n-1|0)[0]}$ and now want to determine the asymptotics of $u_{2g}^{(n|1)[0]}$. Furthermore, we know the discontinuities of $\Phi_{(n|1)}(x)$ given the Stokes automorphisms above. Thus, applying the Cauchy formula to the function $\Phi_{(n|1)}^{[0]}(x)$, and making use of the relation (\ref{phi(n1)phi0logphi0}), one obtains
\be
\Phi_{(n|1)}^{[0]}(x) = \sum_{\theta=0,\pi} \left\{ \int_{0}^{\rme^{\rmi \theta}\infty} \frac{\rmd w}{2\pi\rmi}\, \frac{\disc_{\theta} \Phi_{(n|1)}(w)}{w-x} + \frac{2(n-1)}{\sqrt{3}}\, \int_{0}^{\rme^{\rmi\theta}\infty} \frac{\rmd w}{2\pi\rmi}\, \log w\, \frac{\disc_{\theta}\Phi_{(n-1|0)}^{[0]}(w)}{w-x}\right\}.
\ee
\noindent
The asymptotics of $u_{2g}^{(n|1)[0]}$ will have a contribution from each of these integrals, except in the case when $n=1$, where only the first integral is present. In this case we have already seen that $\Phi_{(1|1)}(x)$ will have a genus expansion, as a consequence of the condition that $u_{2(2m+1)}^{(1|1)[0]}=0$. Solving this condition, using (\ref{eq:Painleve-relating-mn-to-nm}), we find more relations between the Stokes coefficients. Summarizing, these relations are
\begin{eqnarray}
S_{1}^{(0)} &=& (-1)^{\frac{1}{2}}\, \widetilde{S}_{-1}^{(0)}, \label{eq:stokesrel} \\
S_{2}^{(0)} &=& \widetilde{S}_{-2}^{(0)}, \\
S_{1}^{(1)} &=& -(-1)^{\frac{1}{2}}\, \widetilde{S}_{-1}^{(1)} - \frac{4\pi\rmi}{\sqrt{3}}\, S_{1}^{(0)}. \label{eq:stokesrel3}
\end{eqnarray}
\noindent
As discussed before, requiring a genus expansion of $\Phi_{(n|n)}(x)$ for $n>1$, which is equivalent to setting $u_{2(2m+1)}^{(n|n)[0]}=0$, will then yield a tower of relations between different Stokes coefficients, effectively reducing the number of independent coefficients needed to account for both the full multi--instanton asymptotics as well as any possible Stokes transition one might wish to consider.

\subsection{Resurgence of Instantons in Minimal Strings}\label{sec:PIresurgence}

The recursion formula (\ref{eq:PIrec}) provides us with a tool to calculate the two--parameters transseries solution of the Painlev\'e I equation, to arbitrary precision. In particular, this allows us to do high--precision tests of the resurgent properties that were discussed in general terms in section \ref{sec:4} and that were discussed in the specific Painlev\'e I case in the preceding paragraphs.

\subsubsection*{Resurgence of the Perturbative Series}

One of the main new phenomena that our resurgence analysis uncovers is the fact that the large--order behavior of transseries coefficients is itself subject to nonperturbative corrections. This phenomenon is already present in the simplest case: the large--order behavior of $u^{(0|0)[0]}_g$, the zero--instanton, perturbative expansion coefficients of the Painlev\'e I transseries solution.

Recall that, in our normalizations, these coefficients are only nonzero when $g$ is a multiple of four. To avoid writing unnecesary factors, let us rescale
\be
\label{eq:uredef}
\widetilde{u}_{4g} = \frac{\rmi \pi A^{2g-\frac{1}{2}}}{S_1^{(0)}\, \Gamma \left( 2g- \frac{1}{2} \right)}\, u^{(0|0)[0]}_{4g}.
\ee
\noindent
We can then write the large--order formula (\ref{eq:Free-En-zero-inst-coeff}) as
\bea
\widetilde{u}_{4g} &\simeq& \sum_{h=0}^{+\infty} u^{(1|0)[0]}_{2h+1} \cdot A^{h}\, \frac{\Gamma \left( 2g-h-\frac{1}{2} \right)}{\Gamma \left( 2g-\frac{1}{2} \right)} + \sum_{h=0}^{+\infty} S_1^{(0)}\, u^{(2|0)[0]}_{2h+2} \cdot 2^{h-2g+1} \cdot A^{h+\frac{1}{2}}\, \frac{\Gamma \left( 2g-h-1 \right)}{\Gamma \left( 2g-\frac{1}{2} \right)} + \nonumber \\
&&
+ \sum_{h=0}^{+\infty} \left( S_1^{(0)} \right)^2 u^{(3|0)[0]}_{2h+3} \cdot 3^{h-2g+\frac{3}{2}} \cdot A^{h+1}\, \frac{\Gamma \left( 2g-h-\frac{3}{2} \right)}{\Gamma \left( 2g-\frac{1}{2} \right)} + \cdots.
\label{eq:PI0largeorder}
\eea
\noindent
The ratios of gamma functions in this expression should be thought of as perturbative $1/g$ expansions. For example, we can rewrite the ratio of gamma functions in the first sum of the first line above as
\be
\frac{\Gamma \left( 2g-h-\frac{1}{2} \right)}{\Gamma \left( 2g-\frac{1}{2} \right)}\, =\,\, \prod_{k=1}^h \frac{1}{2g-k-\frac{1}{2}}\, =\,\, \frac{1}{2^h}\, g^{-h} + \frac{h^2+2h}{2^{h+2}}\, g^{-h-1} + \cdots.
\ee
\noindent
In this way, we can define these ratios as (possibly asymptotic) series for any values of $g$ and $h$. In particular, this allows us to work with expressions such as, for instance, the factor of $\Gamma(2g-h-1)$ in the second sum in (\ref{eq:PI0largeorder}), even when $2g-h-1$ is a negative integer for which the actual gamma function would have had a pole. 

Thus, the first sum in (\ref{eq:PI0largeorder}) gives a purely perturbative description of the large $g$ behavior of the $\widetilde{u}_{4g}$ coefficients, as a series in $1/g$. This perturbative large--order series has been studied in detail in \cite{msw07, gikm10} and was found to give correct results up to high precision. What we see now is that, nevertheless, the perturbative large--order behavior is \textit{not} the full story. For example, the second sum in (\ref{eq:PI0largeorder}) contains further corrections that come with a factor $2^{-2g}$, and therefore are invisible in a perturbative study. The sum in the second line of (\ref{eq:PI0largeorder}) gives $3^{-2g}$ corrections, and so on; one keeps finding subleading multi--instanton corrections in this way.

The question is: can we actually \textit{see} those nonperturbative corrections to the large--order behavior? It should be intuitively clear that in order to see an effect as small as $2^{-2g}$ at large $g$, we first need to subtract the leading perturbative series to very high order. Here one actually runs into a problem since the perturbative series in $1/g$, the first sum appearing in (\ref{eq:PI0largeorder}), is \textit{not}  convergent---it is an asymptotic series. This should not come as a great surprise: we know that the presence of nonperturbative effects in a quantity is closely related to the nonconvergence of its perturbation series. This phenomenon pops up again in the large--order formula.

\subsubsection*{Optimal Truncation}

\FIGURE[ht]{
\label{fig:PIc000ot}
\centering
\includegraphics[width=10cm]{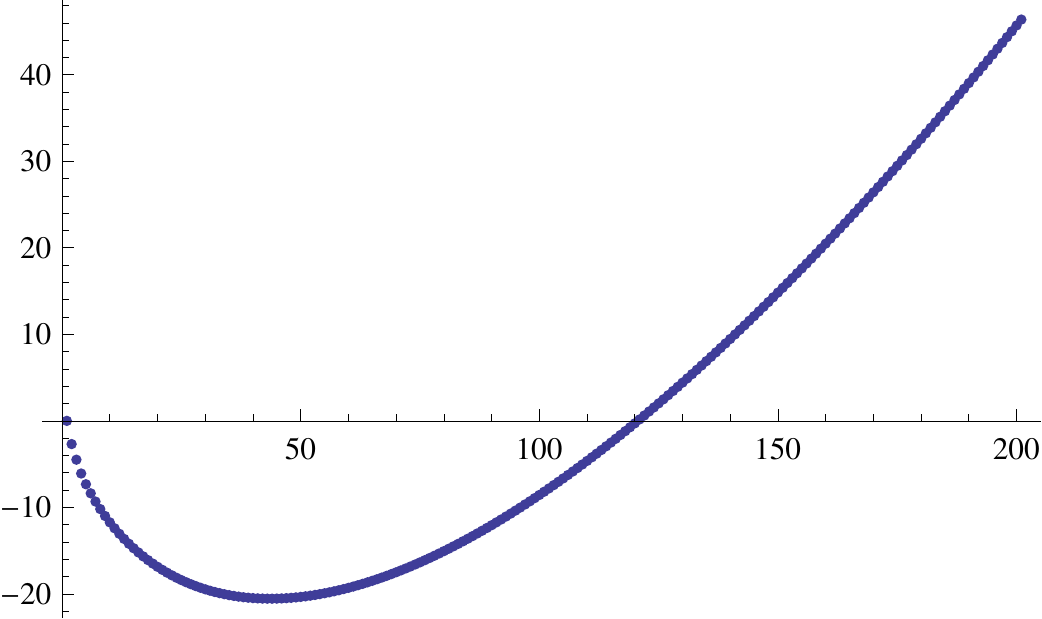}
\caption{The $\log_{10}$ of the absolute value of the first 200 coefficients in the $1/g$--expansion associated to the first sum appearing in (\ref{eq:PI0largeorder}), for the case where $g=30$.}
}

The simplest way to deal with asymptotic series is to do a so--called \textit{optimal truncation}: one simply sums the terms in the series for as long as their absolute value decreases, and cuts off the sum at this point. As an example, let us look at the case where $g=30$. In figure \ref{fig:PIc000ot}, we have plotted the $\log_{10}$ of the absolute value of the first 200 terms in the $1/g$--expansion associated to the first sum in (\ref{eq:PI0largeorder}). The smallest term in the series occurs at order $g^{-43}$ and equals, approximately, $-2.8 \times 10^{-21}$. We see from the figure that, after this term, the terms in the asymptotic expansion start growing again. Thus, optimal truncation instructs us to cut off the sum after the order $g^{-43}$ term. We expect that the size of the final term gives a good indication of the precision of the calculation. This is indeed true: one finds that
\bea
\label{eq:g30pert}
\widetilde{u}_{4 \cdot 30} &=& 0.9978832395689425456292\ldots, \\
\widetilde{u}^{\text{ot}}_{4 \cdot 30} &=& 0.9978832395689425456257\ldots,
\eea
\noindent
where ``ot'' stands for ``optimal truncation'', and, in the first line, we have calculated the exact value using (\ref{eq:uredef}). Thus, we get the correct result within an error of $3.5 \times 10^{-21}$---indeed of the order of magnitude of the last term in the optimally truncated sum.

The problem with this method is that it is only barely sufficient to distinguish the $2^{-2g}$ effects associated to the second sum appearing in (\ref{eq:PI0largeorder}). For our example value of $g=30$, the leading term in this sum is
\be
\label{eq:g30inst}
\frac{S_1^{(0)}\, u^{(2|0)[0]}_2\, A^{\frac{1}{2}}\, \Gamma(59)}{2^{59}\, \Gamma \left( \frac{119}{2} \right)} = 2.33\ldots \times 10^{-20}\, \rmi.
\ee
\noindent
We see that this leading term in the $2^{-2g}$ corrections is roughly of the same order of magnitude as the error in the optimal truncation\footnote{We will soon also explain the perhaps surprising fact that this term is imaginary.}. In other words, this term is only just within the ``resolution'' that optimal truncation allows us, and any $1/g$ corrections to it (let alone the $3^{-2g}$ corrections) will be completely washed out by the error due to optimal truncation. This is not just an unlucky coincidence: one can show using general arguments (see, \textit{e.g.}, \cite{m10}) that optimal truncation always leads to an error which is of the same order of magnitude as the first nonperturbative contribution.

\subsubsection*{Borel--Pad\'e Approximation}

Since optimal truncation is not powerful enough, we need a better method to approximate the asymptotic series associated to the first sum appearing in (\ref{eq:PI0largeorder}). That is, we actually need to \textit{resum} this series. Of course we already know of a very powerful method to resum asymptotic series: the method of Borel resummation, discussed at length in section \ref{borelrev}. To employ this method, we would in principle need to find the Borel transform (\ref{defborelt}) of the first sum in (\ref{eq:PI0largeorder}), and then do the Laplace transform (\ref{borelinttheta}) that inverts the Borel transform. The problem with this procedure is that we only have a recursive definition of the coefficients in the asymptotic series and, as a result, it seems impossible to find an exact expression for the Borel transform. Note that approximating the Borel transform by a Taylor series will not do: the inverse Borel transform will then simply give back our original divergent series.

The solution to this problem lies in the method of \textit{Borel--Pad\'e approximations}. Let us write the $1/g$ expansion associated to the first sum of (\ref{eq:PI0largeorder}) as
\be
P(g) \simeq \sum_{n=0}^{+\infty} a_n\, g^{-n}.
\ee
\noindent
The Borel transform (\ref{defborelt}) of this asymptotic series is
\be
\CB [P] (s) = \sum_{n=0}^{+\infty} \frac{a_n}{n!}\, s^{-n}.
\ee
\noindent
One can check that the $a_n$ grow factorially with $n$, so that this new series has a finite radius of convergence. However, we can only calculate the $a_n$ recursively, so in numerical calculations we will actually have to cut off the above sum at some large order. For convenience, we choose this order to be an even number, $2N$,
\be
\CB [P] (s) \approx \sum_{n=0}^{2N} \frac{a_n}{n!}\, s^{-n}.
\ee
\noindent
Instead of directly performing the inverse Borel transform (which, as we mentioned, would give back the original asymptotic series), we now further approximate this function by an \textit{order N Pad\'e approximant\footnote{More precisely, this is the order $(N,N)$ Pad\'e approximant. One could, in principle, choose different orders of $g^{-	1}$ for the numerator and the denominator, but in numerical approximations this so--called \textit{diagonal} choice often leads to the best results. As we shall see, in our case it indeed leads to very precise numerics.}}
\be
\CB^{[N]} [P] (s) = \frac{\sum_{n=0}^{N} b_n\, s^{-n}}{\sum_{n=0}^{N} c_n\, s^{-n}}.
\ee
\noindent
That is, the degree $2N$ polynomial in $1/g$ is replaced by a rational function which is the ratio of two degree $N$ polynomials in $1/g$. The coefficients in this approximation are chosen in such a way that the first $2N+1$ terms in a $1/g$--expansion of $\CB^{[N]} [P] (s)$ reproduce $\CB [P] (s)$. When one furthermore chooses $c_0 = 1$, to remove the invariance under homogeneous rescalings of all coefficients, this requirement can be shown to lead to a unique set of $(b_n, c_n)$. There exist fast algorithms to determine Pad\'e approximants; for instance in \textit{Mathematica} such an algorithm is implemented under the name \texttt{PadeApproximant}.

The virtue of replacing the polyomial by this rational function is that, for small $1/g$, both functions look very similar, but for large $1/g$, the rational function approaches a constant and is therefore much better behaved. As a result, one can now calculate the inverse Borel transform, or Borel resummation, 
\be
\label{eq:bpint}
\CS^{[N]}_0 P (g) = \int_0^{+\infty} \rmd s\, \CB^{[N]} [P] ( s/g )\, \rme^{-s}.
\ee
\noindent
Contrary to our original asymptotic series, this result will indeed converge in the limit where $N \to \infty$. Note that the subject of Borel--Pad\'e approximations has been studied intensively from a mathematical point of view, and has been applied to several physical problems in the past---the reader can find further details, for example, in \cite{bo78, m08}.

One thing one needs to be careful about when doing a Borel--Pad\'e approximation is that the rational function $\CB^{[N]} [P] (s)$ will, in general, have poles on the positive $s$--axis, making the integral (\ref{eq:bpint}) ill--defined. This problem is precisely the same as the one we encountered earlier for the ordinary Borel resummation in section \ref{borelrev}, and we now know how to solve it: instead of integrating (\ref{eq:bpint}) along the real $s$--axis, we need to integrate around the poles using a $+\rmi \epsilon$ prescription\footnote{This sign is a matter of convention; integrating using a $-\rmi \epsilon$ prescription will lead to the same large--order formulae, but with the imaginary Stokes constant $S_1^{(0)}$ replaced by $-S_1^{(0)}$.}. As a result, the resummed approximation $\CS^{[N]}_+ P (g)$ will no longer be purely real, but will have a small imaginary part. For example, using a Borel--Pad\'e approximation for the first sum appearing in (\ref{eq:PI0largeorder}), in our example case of $g=30$, we find the value
\be
\widetilde{u}^{\text{BP} \langle 1 \rangle}_{4 \cdot 30} = 0.9978832395689425456292\ldots - 2.26\ldots \times 10^{-20}\,  \rmi,
\ee
\noindent
where the $\langle 1 \rangle$ indicates that we only resummed the first sum in (\ref{eq:PI0largeorder}). Comparing this to (\ref{eq:g30pert}) and (\ref{eq:g30inst}), we notice two very important facts. First of all, the Borel--Pad\'e approximation indeed gives \textit{better} results than optimal truncation: at the precision to which we are presently calculating, the real part of the above expression exactly reproduces (\ref{eq:g30pert}). Moreover, shedding light on our previous evaluation, the imaginary part of the above result is of the same order of magnitude as (\ref{eq:g30inst}), albeit of \textit{opposite} sign. That is, it is largely canceled by the leading $2^{-2g}$ term in (\ref{eq:PI0largeorder}) which, as we now understand, indeed \textit{needs} to be imaginary. The fact that the cancellation is not precise is because in (\ref{eq:g30inst}) we only calculated the leading term in the $2^{-2g}$ corrections; adding further terms will give more precise results.

\subsubsection*{Testing the $2^{-2g}$ Corrections using Richardson Transforms}

We shall see in a moment how incredibly precise these results can be made, but first we want to perform an additional test on the validity of our large--order formula (\ref{eq:PI0largeorder}). Traditionally (see \cite{msw07, m08, gikm10} for many examples), large--order formulae are tested as follows. One finds a $g$--dependent quantity, $X_g$, such that the ratio
\be
R_g = \frac{X_g}{X_{g+1}}
\ee
\noindent
approaches a certain coefficient, $R_\infty$, at large $g$, and such that the corrections to this large--order value take, at least to a good approximation, the form of a $1/g$ expansion. One then calculates $R_g$ for a sequence of low values of $g$, and finds $R_\infty$ using the numerical method of Richardson transforms (see, \textit{e.g.}, \cite{bo78, msw07}).

Let us work this out for a particular example. Since the perturbative large--order behavior associated to the first sum in (\ref{eq:PI0largeorder}) has already been tested extensively in \cite{msw07, gikm10}, the first interesting thing we can test is whether the corrections to this term scale as $2^{-2g}$ as $g \to \infty$. To this end, let us define
\be
X_g \equiv \widetilde{u}_{4g} - \widetilde{u}^{\text{ot}}_{4g} = \frac{S_1^{(0)}\, A^{\frac{1}{2}}\, \Gamma(2g-1)}{3 \times 2^g\, \Gamma \left( 2g-\frac{1}{2} \right)} + \CO \left( 1/g \right),
\ee
\noindent
where the right hand side is the result we expect if our formula (\ref{eq:PI0largeorder}) is correct. From this, we find the expectation that the ratio $R_g$ should go like
\be
R_g = 4 + \CO \left( 1/g \right),
\ee
\noindent
as $g \to \infty$. To check this expectation, we have plotted the values of $R_g$ for the first 100 values of $g$ in figure \ref{fig:PIc000instact} (top blue line). We see that the sequence $R_g$ indeed approaches the numerical value 4, albeit slowly. To increase convergence, we can remove $1/g$ effects by calculating the Richardson transforms of this sequence. This method is explained in some detail in \cite{bo78}, for example. The figure shows, from top to bottom, the first three Richardson transforms of the sequence $R_g$. We see that the sequences accurately approach the value $R_\infty = 4$. The best convergence happens after seven Richardson transforms, and in this way we find a limiting value of
\be
R_\infty \approx 4.000000000038.
\ee
\noindent
The fact that we numerically find the expected answer up to one part in $10^{11}$ gives us a lot of confidence that the $2^{-2g}$ behavior in our large order formula (\ref{eq:PI0largeorder}) is correct.

We could continue and define a new $X_g$ which tests the prefactor of the $2^{-2g}$ corrections, and then the subleading terms in $1/g$, and so on. We will not do this here, since we shall now see that there are other tests that check the coefficients in our formula to even higher precision.

\FIGURE[ht]{
\label{fig:PIc000instact}
\centering
\includegraphics[width=10cm]{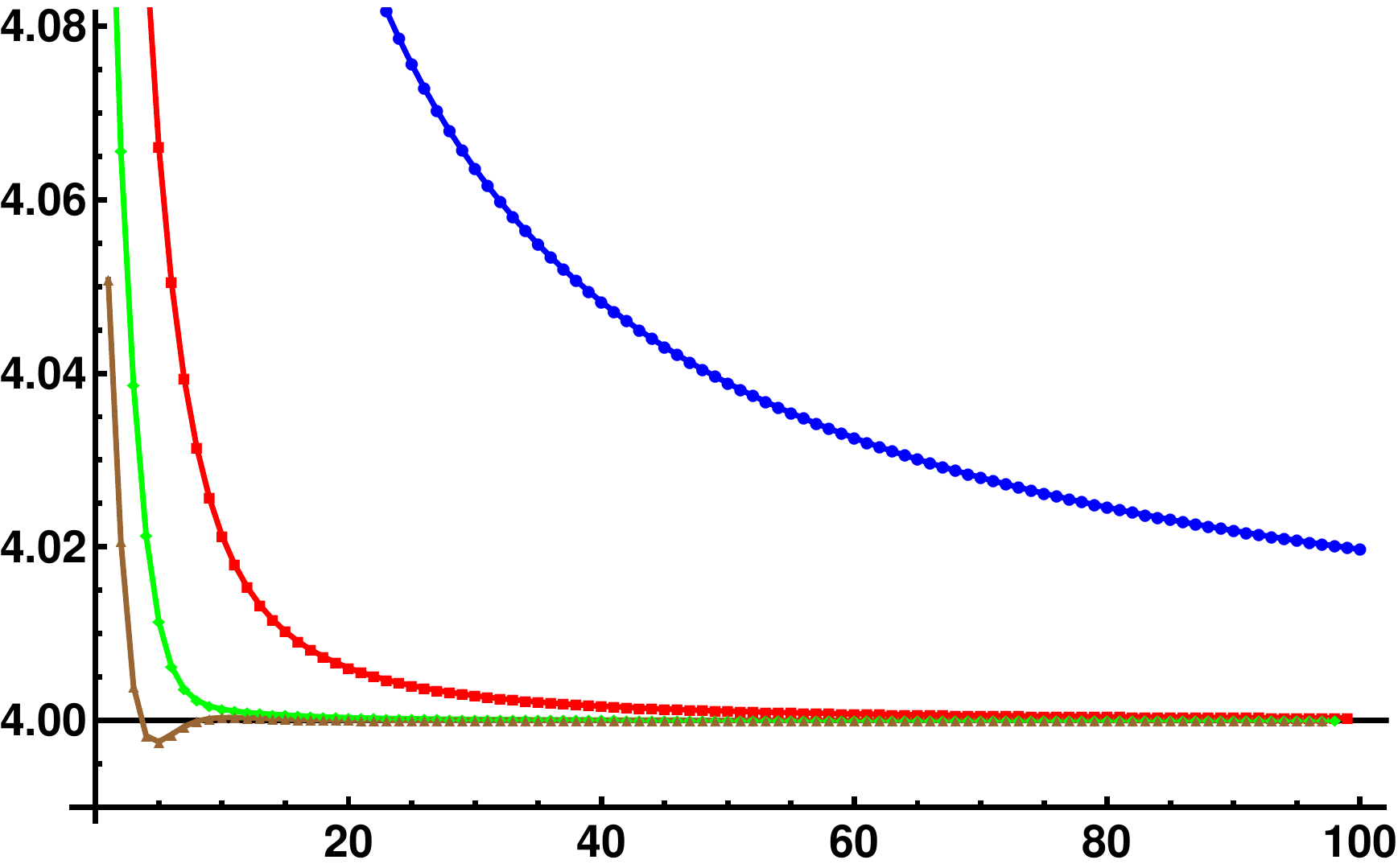}
\caption{The first 100 values of the sequence $R_g$, and the first three Richardson transforms of this sequence. The sequences accurately approach the numerical value 4, as expected.}
}

\subsubsection*{Direct Numerical Evaluation}

One may jump ahead and wonder: can we also see the $3^{-2g}$ corrections in our large--order formulae numerically, and perhaps even go beyond those? It is clear what needs to be done for this: to see $3^{-2g}$ effects, we not only need to resum the perturbative asymptotic series, but also the asymptotic series multiplying the factor $2^{-2g}$ in the first line of (\ref{eq:PI0largeorder}). As we shall now see, the method of Borel--Pad\'e approximations is powerful enough for this to be done.

Let us begin by again focusing on an example. To make sure the numbers that follow fit on a single line, let us now look at the case $g=10$. We will denote the Borel--Pad\'e resummation of the first $n$ distinct sums in (\ref{eq:PI0largeorder}) by $\widetilde{u}_{4 \cdot 10}^{\langle n \rangle}$. A numerical evaluation of the Pad\'e approximant and the consecutive Laplace transform gives the following results:
\bea
\widetilde{u}_{4 \cdot 10} &\approx& \phantom{-} 0.995695607481681532429, \\
\widetilde{u}_{4 \cdot 10} - \widetilde{u}_{4 \cdot 10}^{\langle 1 \rangle} &\approx& \phantom{-} 0.000000000000249496840 + 0.000000041490689176523 \, \rmi, \\
\widetilde{u}_{4 \cdot 10} - \widetilde{u}_{4 \cdot 10}^{\langle 2 \rangle} &\approx& -0.000000000000498993666 + 0.000000000000000063033 \, \rmi, \\
\widetilde{u}_{4 \cdot 10} - \widetilde{u}_{4 \cdot 10}^{\langle 3\rangle} &\approx& -0.000000000000000000043 - 0.000000000000000063033 \; \rmi.
\eea
\noindent
From these numbers, we learn the following. First of all, we see again that already the leading Borel--Pad\'e approximant $\widetilde{u}_{4 \cdot 10}^{\langle 1 \rangle}$ gives a very good approximation to the actual value $\widetilde{u}_{4 \cdot 10}$. It is off by a term of order $10^{-8}$ in the imaginary direction, and only by a term of order $10^{-13}$ in the real direction. This imaginary error is then canceled to very high precision by the order $2^{-2g}$ terms, leaving an imaginary error of order $10^{-17}$. Meanwhile, the real error is not further corrected at this level. The reason for this last fact is that the real error in both the perturbative terms and in the order $2^{-2g}$ terms come from $3^{-2g}$ effects, and are therefore of the same order of magnitude. We see even more: they are not only of the same order of magnitude, but actually related by a simple rational factor: the real error in the $2^{-2g}$ terms is $-3$ times the real error in the perturbative terms, thus giving the overall real error a factor of $-2$, as seen above. That these errors are so simply related could have been anticipated: both come from the 3--instantons series in the transseries solution to the Painlev\'e I equation. 

The remaining real error is then canceled to order $10^{-20}$ by the 3--instantons effects and, at this order, the imaginary error stays of the same magnitude, again being related by a simple rational factor to the imaginary error at the previous level. One can continue like this: the remaining imaginary error will now be canceled by $4^{-2g}$ effects, the next improvement in the real error will occur at order $5^{-2g}$, and so on\footnote{Since we know that the precise value of $\widetilde{u}_{4g}$ is real, we could actually have ignored all imaginary errors. This reduces the number of Borel--Pad\'e approximations that one needs to make by a factor of two. As the simplest example, we could take the real part of $\widetilde{u}_{4g} - \widetilde{u}_{4g}^{\langle 1 \rangle}$, and find a result which is correct up to $3^{-2g}$ corrections instead of the expected $2^{-2g}$ corrections. In the explicit calculations, however, to make sure that our methods are correct in more general cases, we have not used this simplification.}.

\FIGURE[ht]{
\label{fig:PIc000mia}
\centering
\includegraphics[width=10cm]{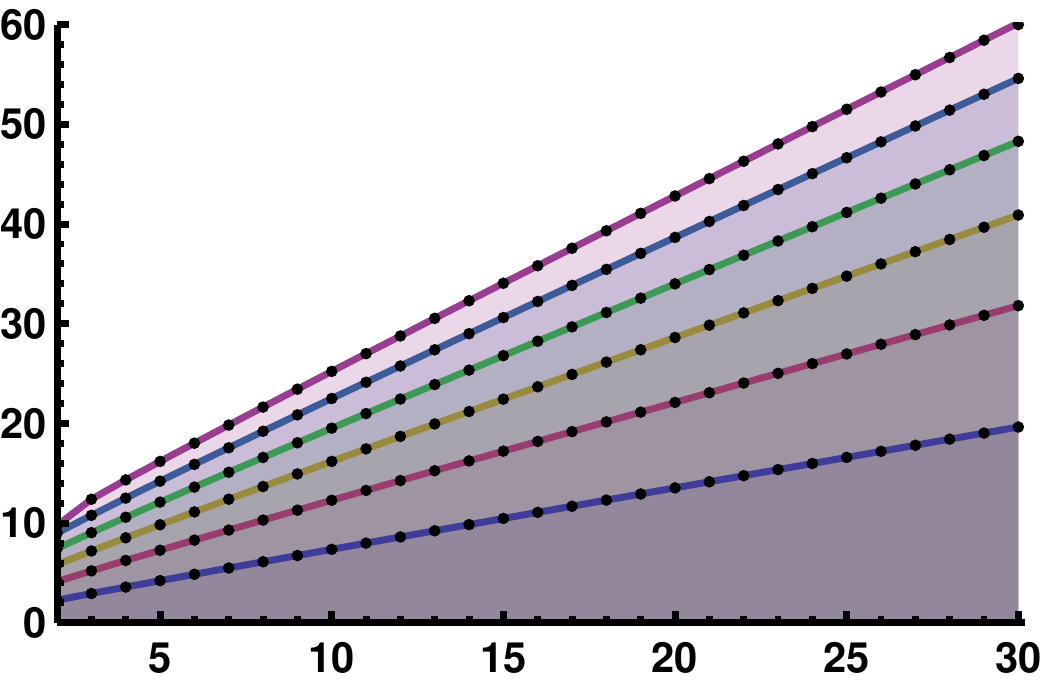}
\caption{The precision of $\widetilde{u}_{4g}^{\langle n \rangle}$ for $g$ ranging from 2 to 30 and $n$ ranging from 1 to 6.}
}

To see how well this method works, we show in figure \ref{fig:PIc000mia} the precision of $\widetilde{u}_{4g}^{\langle n \rangle}$ for $g$ ranging from 2 to 30 and $n$ ranging from 1 to 6 (\textit{i.e.}, we have tested our results up to six instantons). To obtain these numbers, we have done the appropriate Borel--Pad\'e resummations up to orders $(200, 180, 160, 120, 80, 80)$ for the $(1, 2^{-2g}, 3^{-2g}, 4^{-2g}, 5^{-2g}, 6^{-2g})$ corrections, respectively. Along the vertical axis, we have plotted the precision, which is defined as
\be
\log_{10} \left| \frac{\widetilde{u}_{4g}}{\widetilde{u}_{4g} - \widetilde{u}_{4g}^{\langle n\rangle}} \right|,
\ee
\noindent
that is, the number of decimal places to which $\widetilde{u}_{4g}^{\langle n \rangle}$ gives the correct result.

Thus, we see that our large--order formulae lead to \textit{extremely} accurate results. We saw before that, for $g=30$, optimal truncation of the perturbative $1/g$ large--order series gave the correct result up to approximately 20 decimal places. Now we see that using Borel--Pad\'e approximants the nonperturbative $n^{-2g}$ effects play a crucial role in getting higher precision, and that by including up to $6^{-2g}$ corrections we can get results that are correct up to 60 decimal places.

Two final remarks about these results are in order. First of all, even though we are speaking of ``large--order behavior'', we see from figure \ref{fig:PIc000mia} that already at $g=2$ we get results which are accurate up to 10 decimal places. There is still, however, a limit to this procedure. The reason for this is that, in our normalization, the $n$--instanton series at genus $g$ comes with a factor of
\be
\frac{\Gamma \left( 2g-h-\frac{n}{2} \right)}{\Gamma \left( 2g-\frac{1}{2} \right)}.
\ee
\noindent
The leading ($h=0$) terms for $n=8$ will therefore blow up when $g=2$. This is also the reason why we have not included $g=1$ in our graph: there, already the $n=4$ and $n=6$ contributions blow up. It would be interesting to know if this is indeed a fundamental problem or whether it is simply a matter of normalizations, and can be solved in a similar way to how we circumvented the analogous gamma function singularities for nonleading values of $h$.

A second remark is that this test can be viewed as a much more accurate test of certain coefficients than the traditional tests using Richardson transforms. For example, at $g=30$ we have seen that the $2^{-2g}$ effects set in at order $10^{-20}$. However, we have now checked formulae for $g=30$ up to order $10^{-60}$. If the base 2 in $2^{-2g}$ would have had an error, $\delta$, this would have shifted the $2^{-2g}$ effects to effects of order
\be
\left( 2+ \delta \right)^{-2g} = 2^{-2g} \left( 1 - \frac{\delta}{2g} + \CO \left( \delta^2 \right) \right),
\ee
\noindent
so from the fact that we get correct results up to order $10^{-60}$, we see that $\delta$ can be no larger than of order $10^{-40}$. This is a \textit{huge} improvement compared to the accuracy of order $10^{-11}$ that we found using Richardson transforms.

This does not mean that the method of Richardson transforms has become useless. Note that, in the above tests, we have essentially ``reversed the burden of proof'': we have assumed that the $n^{-2g}$ corrections (coming from higher instanton coefficients in the transseries) were correct, and checked these against the known perturbative coefficients. In the method of Richardson transforms, one starts from the known perturbative coefficients, and reproduces the expected coefficients in the $n^{-2g}$ corrections. Whereas the first method is more powerful as a test, in practical cases one is more likely to know the perturbative coefficients in a transseries than to know all nonperturbative coefficients, as we do in this example. Thus, in those cases, Richardson transforms can be used to learn something about the nonperturbative data, starting from perturbative data. This approach can be useful for example when studying topological string theories, where detailed nonperturbative information is often unknown.

\subsubsection*{Resurgence of the $(n|m)$ Instanton Series}

Now that we have gained some confidence in our resurgent techniques from studying the perturbative series $\Phi_{(0|0)}^{[0]}$, we can apply these techniques to the $n$--instanton perturbative series $\Phi_{(n|0)}^{[0]}$ and, more generally, to the generalized instanton series $\Phi_{(n|m)}^{[k]}$. 

A new phenomenon occurs here: the large--order behavior of the series coefficients, $u^{(n|m)[k]}_g$, no longer depends only on the single Stokes constant $S_1^{(0)}$, and further Stokes constants will appear. For example, applying (\ref{ninstasymp-log}) to the one--instanton series one finds that, up to order $2^{-g}$, its large--order behavior has the following six contributions
\bea
u^{(1|0)[0]}_{2g+1} &\simeq& \frac{2 S^{(0)}_1}{2\pi\rmi}\, \sum_{h=0}^{+\infty} u^{(2|0)[0]}_{2h+2} \cdot \frac{\Gamma \left( g-h-\frac{1}{2} \right)}{A^{g-h-\frac{1}{2}}} + \frac{(-1)^g\, S^{(0)}_1}{2\pi\rmi}\, \sum_{h=0}^{+\infty} u^{(1|1)[0]}_{4h+2} \cdot  \frac{\Gamma \left( g-2h-\frac{1}{2} \right)}{A^{g-2h-\frac{1}{2}}} + \nonumber \\
&& + \frac{3 \left( S^{(0)}_1 \right)^2}{2\pi\rmi}\, \sum_{h=0}^{+\infty} u^{(3|0)[0]}_{2h+3} \cdot \frac{\Gamma \left( g-h-1 \right)}{\left( 2A \right)^{g-h-1}} + \frac{(-1)^g \left( S^{(0)}_{1} \right)^2}{2\pi\rmi}\, \sum_{h=0}^{+\infty} u^{(2|1)[0]}_{2h+3} \cdot  \frac{\Gamma \left( g-h-1 \right)}{\left( 2A \right)^{g-h-1}} - \nonumber \\
&& - \frac{(-1)^g \left( S^{(0)}_{1} \right)^2}{4\pi\rmi}\, \sum_{h=0}^{+\infty} u^{(2|1)[1]}_{2h+1} \cdot \frac{\Gamma \left( g-h \right) \cdot B_{2A} (g-h)}{\left( 2A \right)^{g-h}} + \nonumber \\
&& + \frac{(-1)^g \left( \widetilde{S}^{(0)}_{-2} + \frac{1}{2} \widetilde{S}^{(0)}_{-1} \widetilde{S}_{-1}^{(1)} \right)}{2\pi\rmi}\, \sum_{h=0}^{+\infty} u^{(1|0)[0]}_{2h+1} \cdot \frac{\Gamma \left( g-h \right)}{\left( 2A \right)^{g-h}}.
\label{eq:u100LO}
\eea
\noindent
Several facts should be noted about this expansion:
\begin{itemize}
\item The first two sums determine the perturbative large--order behavior of the one--instanton coefficients, as a series in $1/g$. In the zero--instanton case, (\ref{eq:PI0largeorder}), we saw that this behavior was determined by the next instanton series---in that case, the one--instanton series. In the first sum above, we see this ``forward resurgence'' again: the large--order behavior of the one--instanton series is partly determined by the two--instantons series. However, we see from the second sum that there is also ``sideways resurgence'': the large--order behavior of the one--instanton series also depends on the $(1|1)$ generalized instanton coefficients. Thus, even though the physical interpretation of these generalized sectors is somewhat mysterious, they \textit{do} influence the physical instanton sectors in a very important way.
\item Even though it does not happen in the above example, from the structure (\ref{eq:2paraliender}) of alien derivatives, one can easily see that, in general, also ``backward resurgence'' will occur. For example, the large--order formulae for the two--instantons series will contain contributions coming from the previous, one--instanton series. Thus, already at the perturbative level in $1/g$, we find a very intricate pattern of relations between the different generalized instanton series. This pattern gets even more intricate at higher nonperturbative orders. For example, in the last four sums of the above formula, we see that, at order $2^{-g}$, the large--order behavior of the one--instanton series is determined by the 3--instanton series, by the generalized $(2|1)$--instanton series (including its logarithmic contributions $u^{(2|1)[1]}_{2h+1}$), and even recursively by the 1--instanton series itself\footnote{This behavior is a consequence of the symmetries of the problem: it is really $u^{(0|1)[0]}_{2h+1}$ that appears in the last sum of the large--order formula, but we have used equation (\ref{eq:PIsign}) to rewrite these coefficients in terms of $u^{(1|0)[0]}_{2h+1}$.}.
\item In the last sum above, two new Stokes constants appear: $\widetilde{S}^{(0)}_{-2}$ and $\widetilde{S}_{-1}^{(1)}$ (recall that $\widetilde{S}^{(0)}_{-1} = \rmi S_1^{(0)}$, so it is not a new constant). The new constants appear in the combination
\be
T = \widetilde{S}^{(0)}_{-2} + \frac{1}{2} \widetilde{S}^{(0)}_{-1} \widetilde{S}_{-1}^{(1)},
\ee
\noindent
so that by matching the right--hand side of (\ref{eq:u100LO}) to the left--hand side for large values of $g$, we can determine $T$ up to corrections coming from $3^{-g}$ terms. Note that to do this, we need to calculate Borel--Pad\'e approximations to the infinite sums in (\ref{eq:u100LO}). This is a procedure which takes some (computer) time, but other than that is relatively straightforward.
\item Finally, recall that $B_{2A} (g-h) = \psi \left( g-h+1 \right) - \log \left( 2A \right) - \rmi\pi$, where $\psi(z)$ is the digamma function. At large $g$ the digamma function has the asymptotic expansion
\be
\psi(z) = \log(z) - \frac{1}{2z} - \sum_{n=1}^{+\infty} \frac{B_{2n}}{2n\, z^{2n}},
\ee
\noindent
where in here $B_{2n}$ stands for the Bernoulli numbers. The leading term in this expansion implies that, at large order, $B_{2A}(g-h) \sim \log g$, \textit{i.e.}, we find at the 2--instantons level a growth of type $g! \log g$, leading as compared to $g!$. In calculating $T$, the easiest way to deal with this behavior is to gather all terms multiplying $\log g$, divide out the $\log g$, do a Borel--Pad\'e approximation and then multiply with $\log g$ again. The further terms coming from the above asymptotic expansion can then be treated as before, using Borel--Pad\'e approximation to resum all of them directly.
\end{itemize}

\FIGURE[ht]{
\label{fig:PIc100mia}
\centering
\includegraphics[width=10cm]{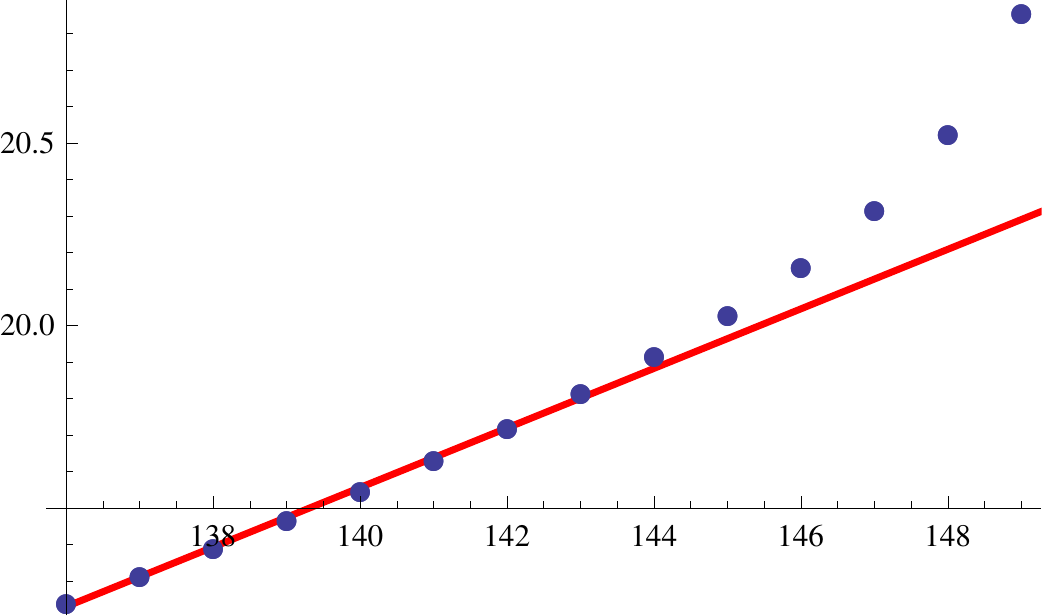}
\caption{The precision of $T_g$ with respect to $T_{151}$, and the resulting linear extrapolation to $g=151$.}
}

Carrying out the Borel--Pad\'e approximations, we have found that
\be
T = -0.90573009110532780736\ldots.
\ee
\noindent
The precision of this number can be determined as follows. Note that, for any $g$, we can determine $T_g$ from (\ref{eq:u100LO}), and we expect the result to become better as $g$ becomes larger. In fact, the true value of $T$ should be $T_\infty$. We have calculated $T_g$ for values of $g$ up to 151. In figure \ref{fig:PIc100mia}, we plot the number of decimal places to which $T_g$ agrees with $T_{151}$. When $g\ll151$, this is essentially the number of decimal places to which $T_g$ agrees with $T_{\infty}$. At $g \sim 151$, this is no longer true, since we are not really comparing with $T_\infty$, but with $T_{151}$. One finds that for $g\ll151$, the precision increases linearly, so by extrapolating this linear behavior, we find the expected precision of $T_{151}$, which in this case is a bit more than 20 decimal places.

Of course, it is really the separate values of $\widetilde{S}^{(0)}_{-2}$ and $\widetilde{S}_{-1}^{(1)}$ that we want to calculate, and not the value of the particular linear combination $T$. This can be achieved, for instance, by looking at the large--order behavior of the series $u^{(1|1)[0]}_g$ which, at the perturbative level, only depends on $\widetilde{S}_{-1}^{(1)}$ and on the known constant $S_{1}^{(0)}$. In exactly the same way, we can then calculate $\widetilde{S}_{-1}^{(1)}$ and from that constant and $T$ determine $\widetilde{S}^{(0)}_{-2}$. Applying this procedure to several generalized instanton series, we have calculated a series of Stokes coefficients that are tabulated in table \ref{table:PIstokescoeff}. Note that, to calculate these numbers, we have tested the resurgence of several of the generalized instanton series up to three instantons. In this table we have also indicated the number of decimal places to which we have calculated the answer. In the case of $S_{1}^{(0)}$, an analytic answer is known---see for example \cite{k89, d92, hhikkmt04, msw07} for derivations. One has
\be\label{eq:S10PI}
S_{1}^{(0)} = - \rmi \frac{3^{1/4}}{2 \sqrt{\pi}}.
\ee
\noindent
For $S_{3}^{(0)}$, we have actually listed more decimal places than we have calculated; we will see in a moment why we are able to conjecture some further digits. For readability, we have only listed about 20 decimal places for each Stokes constant; the authors will of course provide further data to the interested reader, upon request. In the table, we also list which type of large--order behavior the Stokes constants determine. For example, the constant $S_{3}^{(0)}$ appears for the first time in the large--order expansion of $\Phi_{(2|0)}^{[0]}$, where it multiplies the terms of order $3^{-g}$. Apart from $S_1^{(0)}$, and to the best of our knowledge, the only other number in this table which has been calculated before is $\widetilde{S}_1^{(2)}$ \cite{gikm10}. This number, called $S_{-1}$ in equation (5.38) of that paper, was calculated numerically in there up to 19 decimal places, and our result agrees with this up to 17 decimal places (\textit{i.e.}, the final 2 decimal places that are reported in \cite{gikm10} are incorrect---this is possibly a result of the onset of $2^{-g}$ effects that were not taken into account in that paper).

\begin{table}[htb]
\begin{center}
\begin{tabular}{|l|r|r|l|l|}
\hline
& & Precision & From & Order \\
\hline
$S_{1}^{(0)}$ & $- 0.371257624642845568... \; \rmi$ & $\infty$ & $\Phi_{(0|0)}^{[0]}$ & $1^{-g}$ \\
\hline 
$S_{2}^{(0)}$ & $0.500000000000000000... \; \rmi$ & 20 & $\Phi_{(1|0)}^{[0]}$ & $2^{-g}$ \\
\hline
$S_{3}^{(0)}$ & $-0.897849124725732240... \; \rmi$ & 13 & $\Phi_{(2|0)}^{[0]}$ & $3^{-g}$ \\
\hline
$S_{1}^{(1)}$ & $- 4.879253817220057751... \; \rmi$ & 81 & $\Phi_{(1|1)}^{[0]}$ & $1^{-g}$ \\
\hline 
$S_{2}^{(1)}$ & $9.856875980487862735... \; \rmi$ & 19 & $\Phi_{(2|1)}^{[0]}$ & $2^{-g}$ \\
\hline 
$S_{1}^{(2)}$ & $-22.825711248125715287... \; \rmi$ & 36 & $\Phi_{(2|2)}^{[0]}$ & $1^{-g}$ \\
\hline
$\widetilde{S}_{1}^{(2)}$ & $2.439626908610028875... \; \rmi$ & 112 & $\Phi_{(2|0)}^{[0]}$ & $1^{-g}$ \\
\hline
$\widetilde{S}_{1}^{(3)}$ & $15.217140832083810191... \; \rmi$ & 108 & $\Phi_{(3|1)}^{[0]}$ & $1^{-g}$ \\
\hline
$\widetilde{S}_{1}^{(4)}$ & $45.334204678679729580... \; \rmi$ & 108 & $\Phi_{(4|2)}^{[0]}$ & $1^{-g}$ \\
\hline
\end{tabular}
\end{center}
\caption{The Stokes constants that we have calculated. The third column gives the number of decimal places to which the answer is explicitly calculated. The fourth column lists the generalized instanton series for which the Stokes constant appears for the first time, and the fifth column lists what type of large--order behavior this constant determines.}
\label{table:PIstokescoeff}
\end{table}

Note that we have only listed Stokes constants $S_{\ell}^{(n)}$ and $\widetilde{S}_{\ell}^{(n)}$ with $\ell>0$. The reason is that all of these are purely imaginary, but, from them, one can then  easily calculate the corresponding set of Stokes constants with $\ell<0$ using relations such as (\ref{eq:stokesrel}) and the ones that follow it.

More interestingly, we find that the Stokes constants with $\ell>0$ also satisfy several (at this stage, unexpected) relations amongst themselves. The first thing one notices is that it seems extremely likely that
\be
S_{2}^{(0)} = \frac{\rmi}{2}.
\ee
\noindent
Studying table \ref{table:PIstokescoeff} some more, one also finds that
\bea
\label{eq:stokesrel2first}
\widetilde{S}_{1}^{(2)} &=& - \frac{1}{2}\, S_1^{(1)}, \\
\widetilde{S}_{1}^{(3)} &=& - \frac{2}{3}\, S_1^{(2)}, \\
S_{ 3}^{(0)} &=& -\frac{1}{3S_{ 1}^{(0)}}, \\
S_{1}^{(0)} S_2^{(1)} &=& \frac{3\rmi}{4}\, S_{1}^{(1)},
\label{eq:stokesrel2}
\eea
\noindent
are satisfied, at least up to the order to which we have calculated the relevant constants. Of course, we conjecture these results to be exact, even though we have no clear idea on how to prove these relations. Proving these relations and generalizing them to arbitrary Stokes constants\footnote{\label{fn:Stokesguesses}Using the limited amount of available data, one may make further bold guesses such as $n\, \widetilde{S}_{1}^{(n)} = - \left( n-1 \right) S_1^{(n-1)}$ and $n\, S_{n}^{(0)} = \rmi^{n-1}\, (S_1^{(0)})^{2-n}$. Also, it seems natural to write  (\ref{eq:stokesrel2}) as $2\, S_{1}^{(0)} S_2^{(1)} = 3\, S_{1}^{(1)} S_{2}^{(0)}$, since these two products often occur in the same alien derivatives. It is further tempting to guess that, in general, every Stokes constant can be expressed as a rational function of the $S_1^{(n)}$ alone.} is a very interesting problem whose solution will very likely give us a much deeper understanding of Stokes phenomena and resurgence in the Painlev\'e I framework.

\subsection{The Nonperturbative Free Energy of the $(2,3)$ Model}\label{sec:PIfreeenergy}

As discussed at the beginning of this section, we know that the free energy, $F(z)$, of the $(2,3)$ minimal string theory is related to the solution, $u(z)$, of the Painlev\'e I equation by
\be
u (z) = - F'' (z).
\ee
\noindent
We now want to investigate how our results for the Painlev\'e I solution translate into results for this free energy. Let us begin by studying the perturbative contribution to $u(z)$,
\be
\Phi_{(0|0)} (z) \simeq \sum_{g=0}^{+\infty} u^{(0|0)[0]}_{4g}\, z^{-(5g-1)/2}.
\ee
\noindent
Here, the reader should note that we have re--inserted the factor of $\sqrt{z}$ that we had removed earlier in (\ref{eq:urescale}). Two integrations then lead to
\be
F^{(0|0)} (z) \equiv - \sum_{g=0}^{+\infty} u^{(0|0)[0]}_{4g}\, \iint \rmd z\, z^{-(5g-1)/2}\, \simeq\, \sum_{g=0}^{+\infty} F^{(0|0)}_{g}\, x^{2g-2},
\ee
\noindent
where $x=z^{-5/4}$ is the string coupling constant, and where we defined the perturbative expansion coefficients for $F^{(0|0)} (z)$ as
\be\label{eq:Fpertupert}
F^{(0|0)}_{g} = - \frac{4}{(5g-3)(5g-5)}\, u^{(0|0)[0]}_{4g}.
\ee
\noindent
Our reason for not including a ``log index'' $[0]$ in the free energy coefficients $F^{(0|0)}_{g}$ will become clear in a moment. This asymptotic series is, once again, the perturbative part\footnote{Here and in what follows, we will not explicitly include any integration constants. In principle, these lead to undetermined terms in $F (z)$ which are constant and linear in $z$, and which cannot be fixed by using the Painlev\'e I analysis alone; they must be derived from the minimal string theory directly. It turns out that naively setting these terms to zero actually leads to the correct string theory result.} of a transseries expansion for the free energy $F(z)$. To see what form the one--instanton contribution takes, let us integrate the leading one--instanton term in the $u(z)$ transseries,
\be\label{eq:F10}
- \sigma_1\, u^{(1|0)[0]}_1 \iint \rmd z\, z^{-1/8} \rme^{-Az^{5/4}} = - \sigma_1\, \frac{u^{(1|0)[0]}_1}{12}\, z^{-5/8} \rme^{-Az^{5/4}} + \cdots.
\ee
\noindent
In this expression we have explicitly written the leading coefficient $u^{(1|0)[0]}_1$. Recall from our discussion in section \ref{sec:PItransseries} that the value of this constant can be absorbed by a rescaling of $\sigma_1$ and, for this reason, we have so far worked in a convention where $u^{(1|0)[0]}_1=1$. This was a very useful normalization for constructing the two--parameters transseries solution for $u(z)$ but, to discuss the free energy $F(z)$, we now actually want to change to a different convention.

The reason for this new choice of normalization is that we would like our one--instanton contribution to the free energy to agree with the equivalent result that was computed in \cite{d92, msw07}, for the free energy around the one--instanton configuration, straight out of a matrix model calculation associated to eigenvalue tunneling. That is, we want our one--instanton contribution to have the normalization (compare with, \textit{e.g.}, formula (4.35) in \cite{msw07})
\be
\sigma_1^F\, \frac{\rmi}{8 \cdot 3^{3/4} \sqrt{\pi}}\, z^{-5/8} \rme^{-Az^{5/4}} + \cdots.
\ee
\noindent
Notice that this coefficient is computed directly from the $(2,3)$ minimal model spectral curve \cite{msw07}. To find this answer, one simply has to rescale
\be
\sigma_1 = - \rmi \frac{3^{1/4}}{2 \sqrt{\pi}}\, \sigma_1^F.
\ee
\noindent
In order to keep the symmetry between instantons and generalized instantons, which we discussed before, we shall also choose to rescale $\sigma_2$ with this exact same factor.

The reader may have noticed that the above result may be equivalently written as
\be\label{eq:scalesigma}
\sigma_1 = S_1^{(0)}\, \sigma_1^F.
\ee
\noindent
The appearance of the Stokes constant $S_1^{(0)}$ in this formula turns out to be quite natural. A rescaling of the variables $\sigma_i$ does not only rescale the transseries components, but also the Stokes constants. A quick calculation shows that, under a general rescaling of the $\sigma_i$, these quantities scale as follows (recall (\ref{2bridgeF}) for example)
\bea
\sigma_1 &=& c_1\, \widehat{\sigma}_1, \\
\sigma_2 &=& c_2\, \widehat{\sigma}_2, \\
\Phi_{(n|m)} &=& c_1^{-n}\, c_2^{-m}\, \widehat{\Phi}_{(n|m)}, \\
S_\ell^{(k)} &=& c_1^{1-k}\, c_2^{1-k-\ell}\, \widehat{S}_\ell^{(k)}, \label{eq:Srescale1} \\
\widetilde{S}_\ell^{(k)} &=& c_1^{1+\ell-k}\, c_2^{1-k}\, \widehat{\widetilde{S}}_\ell^{(k)}. \label{eq:Srescale2}
\eea
\noindent
In particular, our rescaling sets $\widehat{S}_1^{(0)}=1$. Of course, \textit{physical} quantities cannot depend on arbitrary normalization choices, so any physical quantity must be a scale invariant combination of the above quantities. As we shall see in the following, we will be particularly interested in quantities which can be made scale invariant by multiplying with powers of the first Stokes contstant. When this Stokes constant equals 1, this means that the scale invariant quantity is numerically equal to the ``bare'' quantity.

\subsubsection*{Stokes Constants for the Free Energy}

Recall from (\ref{eq:PI0largeorder}) that the large--order behavior of $u^{(0|0)[0]}_{4g}$ has a leading term
\be\label{eq:uLO}
u^{(0|0)[0]}_{4g} \sim \frac{2 S_1^{(0)}}{2 \pi \rmi}\, \frac{\Gamma \left( 2g-\frac{1}{2} \right)}{A^{2g-\frac{1}{2}}}\, u^{(1|0)[0]}_1.
\ee
\noindent
For $F^{(0|0)}_{g}$ we can do the exact same large--order calculation and the result is very similar
\be\label{eq:FLO}
F^{(0|0)}_{g} \sim \frac{2 S_{1}^{(0)F}}{2 \pi \rmi}\, \frac{\Gamma \left( 2g-\frac{5}{2} \right)}{A^{2g-\frac{5}{2}}}\, F^{(1|0)}_0,
\ee
\noindent
where we denoted the leading one--instanton coefficient\footnote{For the free energy, and in order to avoid fractional indices, we will use a convention where all perturbative series start with a coefficient $F^{(n|m)}_0$, with lower index 0.} in the free energy transseries by $F^{(1|0)}_0$. We see that the only difference between the above two equations is in the argument of the gamma function and the power of the instanton action, $A$. Both of these are shifted by $-2$, as a result of the double integration involved in going from $u(z)$ to $F(z)$. One can see quite easily that this is a general property: all large--order formulae for $u(z)$ and $F(z)$ are the same up to these shifts.

Notice that in (\ref{eq:FLO}) we have denoted the Stokes constant as $S_{1}^{(0)F}$. Indeed, nothing guarantees that the Stokes constants for the transseries $F(z)$ equal those for the transseries $u(z)$---and in fact we shall see that, in general, they are different. However, the Stokes constants for $F(z)$ can easily be obtained from those for $u(z)$. For example, comparing (\ref{eq:uLO}) and (\ref{eq:FLO}), we can calculate the Stokes constant $S_{1}^{(0)F}$ for the free energy. First of all, note that we can rewrite (\ref{eq:Fpertupert}) as
\be
F^{(0|0)}_{g} = - \frac{16}{25}\, \frac{u^{(0|0)[0]}_{4g}}{\left( 2g-\frac{5}{2} \right) \left( 2g-\frac{3}{2} \right)} \left( 1 + \CO\left( \frac{1}{g} \right) \right).
\ee
\noindent
Moreover, we know from (\ref{eq:F10}) and (\ref{eq:scalesigma}) that
\be\label{eq:F100u100}
F^{(1|0)}_0 = - \frac{S_1^{(0)}}{12}\, u^{(1|0)[0]}_1.
\ee
\noindent
Inserting both of these in (\ref{eq:FLO}) gives
\be
u^{(0|0)[0]}_{4g} \sim \frac{2 S_{1}^{(0)F} S_1^{(0)}}{2 \pi \rmi}\, \frac{\Gamma \left( 2g-\frac{1}{2} \right)}{A^{2g-\frac{1}{2}}}\, u^{(1|0)[0]}_1.
\ee
\noindent
Comparing this to (\ref{eq:uLO}), we find that 
\be
S_{1}^{(0)F} = 1.
\ee
\noindent
This once again indicates why the rescaling (\ref{eq:scalesigma}) was a useful choice. In general, quantities such as $F^{(1|0)}_0$ or $S_{1}^{(0)F}$ cannot be physically meaningful quantities: only ``scale invariant'' quantities such as
\be
S_{1}^{(0)F} \cdot F^{(1|0)}_0
\ee
\noindent
can carry physical information. Having chosen our present normalization in such a way that $S_{1}^{(0)F} = 1$, we have shifted the full physical information into $F^{(1|0)}_0$. We could of course just as well have done the opposite thing, \textit{i.e.}, choosing a normalization where $F^{(1|0)}_0=1$ and absorbing all physical information into $S_{1}^{(0)F}$. The reason we have chosen the present normalization is that it agrees with the one usually chosen in the litarature. For example, in \cite{msw07} the above normalization is chosen and the resulting physical quantity $F^{(1|0)}_0$ (called $\mu_1$ in that paper) is calculated \textit{directly} from the spectral curve of a matrix model.

The above calculation, relating $S_{1}^{(0)F}$ to $S_{1}^{(0)}$, can be repeated for any Stokes constant. One simply finds a term in a large--order formula in which the Stokes constant appears, calculates the normalization of this term for both $u(z)$ and $F(z)$, and then compares the two formulae. Doing this carefully one finds the following relations between the Stokes constants for $F(z)$ and for $u(z)$
\bea
S_\ell^{(k)F} &=& \ell^2\, S_\ell^{(k)} \left( S_1^{(0)} \right)^{2k+\ell-2}, \label{eq:StokesFu1} \\
\widetilde{S}_\ell^{(k)F} &=& \ell^2\, \widetilde{S}_\ell^{(k)} \left( S_1^{(0)} \right)^{2k-\ell-2}. \label{eq:StokesFu}
\eea
\noindent
Note that the right--hand side in these equations consists of scale invariant quantities; the left--hand side consists of quantities which are implicitly scale invariant as well, due to the analogous powers of $S_{1}^{(0)F}=1$. The factor of $\ell^2$ comes from taking a second derivative of the instanton factor $\exp (\pm\ell A z^{5/4})$ in the free energy transseries.

\begin{table}[ht]
\begin{center}
\begin{tabular}{|l|r|r|}
\hline
$S_{1}^{(0)F}$ & $1.000000000000000000... \; \phantom{\rmi}$ & $\infty$ \\
\hline 
$S_{2}^{(0)F}$ & $2.000000000000000000... \; \rmi $ & 20 \\
\hline
$S_{3}^{(0)F}$ & $-3.000000000000000000... \; \phantom{\rmi}$ & 13 \\
\hline
$S_{1}^{(1)F}$ & $-1.811460182210655615... \; \phantom{\rmi}$ & 81 \\
\hline 
$S_{2}^{(1)F}$ & $-5.434380546631966844... \; \rmi $ & 19 \\
\hline 
$S_{1}^{(2)F}$ & $1.168020496900498115... \; \phantom{\rmi}$ & 36 \\
\hline
$\widetilde{S}_{1}^{(2)F}$ & $0.905730091105327807... \; \phantom{\rmi}$ & 112 \\
\hline
$\widetilde{S}_{1}^{(3)F}$ & $-0.778680331266998743... \; \phantom{\rmi}$ & 108 \\
\hline
$\widetilde{S}_{1}^{(4)F}$ & $0.319744372344502079... \; \phantom{\rmi}$ & 108 \\
\hline
\end{tabular}
\end{center}
\caption{Stokes constants for the $(2,3)$ minimal string free energy. The third column gives the number of decimal places to which the answer was computed. See table 2 for the corresponding quantities for the Painlev\'e I solution $u(z)$, from which the above numbers are derived using (5.111).}
\label{table:FEStokes}
\end{table}

We have listed the numerical values for the free energy Stokes constants in table \ref{table:FEStokes}. Our main reason for listing the free energy Stokes constants separately is that we expect those numbers to be the ones that can eventually be calculated from minimal string theory or spectral curve considerations, similar to the way in which one can calculate $S_1^{(0)}$. Of course, to actually carry out such calculations, one needs a physical understanding of what the generalized instantons are.

As was the case for $u(z)$, not all of the free energy Stokes constants are independent: using equation (\ref{eq:StokesFu}), the relations (\ref{eq:stokesrel2first}--\ref{eq:stokesrel2}) directly translate into relations between these numbers
\bea
\widetilde{S}_{1}^{(2)F} &=& - \frac{1}{2}\, S_{1}^{(1)F}, \\
\widetilde{S}_{1}^{(3)F} &=& - \frac{2}{3}\, S_{1}^{(2)F}, \\
S_{2}^{(1)F} &=& 3 \rmi\, S_{1}^{(1)F}.
\label{eq:stokesrelF}
\eea
\noindent
As in footnote \ref{fn:Stokesguesses}, one can then conjecture analogous further relations for the free energy Stokes constants that have not been calculated yet, such as, \textit{e.g.}, ${S}_{n}^{(0)F} = \rmi^{n-1} n$.

\subsubsection*{The Free Energy Transseries Coefficients}

We now want to calculate the explicit form of some of the $(n|m)$--instantons contributions to the free energy transseries. In the case where $n=m$, there are no logarithmic contributions to the $u$--transseries, and the double integration is easily carried out as we did for $n=m=0$ in (\ref{eq:Fpertupert}). 

Let us therefore study the ``off--diagonal'' $(n|m)$--instantons contribution to $u(z)$
\be
\sigma_1^n \sigma_2^m\, \rme^{-(n-m)A/w^2}\, \Phi_{(n|m)}^{[0]} (w) = \sigma_1^n \sigma_2^m\, \rme^{-(n-m)A/w^2} \sum_{g=0}^{+\infty} u^{(n|m)[0]}_{2g+2\beta_{nm}^{[0]}}\, w^{2g+2\beta_{nm}^{[0]}},
\ee
\noindent
where $n \neq m$. It is convenient to add to this term all the logarithmic terms that are proportional to it by (\ref{eq:Painleve-relating-logk-to-log0}), \textit{i.e.}, all terms of the form
\be
\sigma_1^{n+k} \sigma_2^{m+k}\, \rme^{-(n-m)A/w^2}\, \log^k (w) \cdot \Phi_{(n+k|m+k)}^{[k]} (w),
\ee
\noindent
with $k \geq 0$. Using (\ref{eq:Painleve-relating-logk-to-log0}), we can rewrite these terms as
\be
\frac{1}{k!} \left( \frac{4}{\sqrt{3}} \left( m-n \right) \sigma_1 \sigma_2\, \log w \right)^k \sigma_1^n \sigma_2^m\, \rme^{-(n-m)A/w^2}\, \Phi_{(n|m)}^{[0]} (w),
\ee
\noindent
and summing all of them over $k$ we find that we can incorporate all of those terms by simply replacing $\Phi_{(n|m)}^{[0]}$ by
\be
\Phi_{(n|m)}^{[\text{sum}]} (w) = \exp \left( \frac{4}{\sqrt{3}} \left( m-n \right) \sigma_1 \sigma_2\, \log w \right) \Phi_{(n|m)}^{[0]} (w).
\ee
\noindent
Formally, we can write this as\footnote{This also illustrates in a rather clear way, and as explained before, that the logarithmic sectors do not seem to represent any new nonperturbative sectors. Herein, they simply encode an irrational power function.}
\be
\Phi_{(n|m)}^{[\text{sum}]} (w) = w^{\frac{4}{\sqrt{3}} \left( m-n \right) \sigma_1 \sigma_2}\, \Phi_{(n|m)}^{[0]} (w).
\ee
\noindent
Rewriting the result in terms of $z=w^{-8/5}$ and reintroducing the scale factor $z^{1/2}$, we get the following $(n|m)$--contribution to the free energy transseries
\be\label{eq:logsummed}
\sigma_1^n \sigma_2^m\, \rme^{-(n-m)Az^{5/4}} \sum_{g=0}^{+\infty} u^{(n|m)[0]}_{2g+\beta_{nm}^{[0]}}\, z^{-\frac{10g+5\beta_{nm}^{[0]}-4}{8}+\frac{4 \left( n-m \right)\sigma_1 \sigma_2}{A}}.
\ee
\noindent
To integrate this part of the transseries, we use the fact that
\be
\label{eq:tsint}
- \iint \rmd z\, z^{\gamma} \rme^{- \ell A z^{5/4}} = \frac{4}{5\ell A}\, z^{\gamma+3/4}\, \rme^{- \ell A z^{5/4}} \sum_{k=1}^{+\infty} a_k (\gamma) \cdot \left( -\ell Az^{5/4} \right)^{-k},
\ee
\noindent
where
\be
a_k(\gamma) = \frac{\Gamma \left( k-\frac{4\gamma-1}{5} \right)}{\Gamma \left( -\frac{4\gamma-1}{5} \right)} - \frac{\Gamma \left( k-\frac{4\gamma+3}{5} \right)}{\Gamma \left( -\frac{4\gamma+3}{5} \right)}
\ee
\noindent
is a polynomial of degree $k-1$ in $\gamma$. It is important to notice that, in the components of our logarithmically summed transseries (\ref{eq:logsummed}), the coefficient $\gamma$ is linear in $\sigma_1 \sigma_2$ and thus $a_k(\gamma)$ in (\ref{eq:tsint}) above will be a polynomial of degree $k-1$ in $\sigma_1 \sigma_2$. This means that integrating the $(n|m)$ transseries component in $u(z)$ will not only contribute to the $(n|m)$ transseries component in $F(z)$, but also to all $(n+\alpha|m+\alpha)$ components with $\alpha>0$.

Using (\ref{eq:tsint}), the double integration of the $u$--transseries is now easily carried out in a computer. We find the result that the free energy has the following transseries structure
\be
\label{eq:Ftransseries}
F (z, \sigma_1^F, \sigma_2^F ) = \sum_{n=0}^{+\infty} \sum_{m=0}^{+\infty} \left( S_1^{(0)} \right)^{n+m} \left( \sigma_1^F \right)^n \left( \sigma_2^F \right)^m \rme^{-(n-m)Az^{5/4}}\, z^{\frac{5}{8\pi} \left( m-n \right) \sigma_1^F \sigma_2^F}\, F^{(n|m)}(z),
\ee
\noindent
where the $F^{(n|m)}(z)$ are perturbative expansions\footnote{We use this term with a bit of hand--waving since these expansions contain half--integral overall powers of the string coupling constant and two logarithmic terms actually appear in the lowest $F^{(n|m)}(z)$.} in the string coupling $z^{-5/4}$. The formal power of $z$ should once again be interpreted as
\be\label{transseriesexponentiation}
z^{\frac{5}{8\pi} \left( m-n \right) \sigma_1^F \sigma_2^F} = \exp \left( \frac{5}{8\pi} \left( m-n \right) \sigma_1^F \sigma_2^F\, \log z \right),
\ee
\noindent
which can be expanded to give $\log z$--dependent contributions exactly analogous to the ones we found for the $u$--transseries. That is, we could leave out this factor in (\ref{eq:Ftransseries}) and instead replace
\be
\sum_{n=0}^{+\infty} \sum_{m=0}^{+\infty} F^{(n|m)}(z) \longrightarrow \sum_{n=0}^{+\infty} \sum_{m=0}^{+\infty} \sum_{k=0}^{\min(n,m)} \log^k (z) \cdot F_{(n|m)}^{[k]}(z),
\ee
\noindent
with
\bea
F_{(n|m)}^{[0]}(z) &=& F^{(n|m)}(z), \\
F_{(n|m)}^{[k]}(z) &=& \frac{1}{k!} \left( \frac{5 \left( n-m \right)}{2 \sqrt{3}} \right)^k F_{(n-k|m-k)}^{[0]}(z).
\eea
\noindent
Keeping the (\ref{eq:Ftransseries}) transseries structure for the free energy, the first few of the $F^{(n|m)}(z)$ are
\bea
F^{(0|0)}(z) &=& - \frac{4}{15} z^{\frac{5}{2}} - \frac{1}{48} \log z + \frac{7}{5760} z^{-\frac{5}{2}} + \frac{245}{331776} z^{-5} + \cdots, \label{eq:F00z} \\
F^{(1|0)}(z) &=& - \frac{1}{12} z^{-\frac{5}{8}} + \frac{37}{768 \sqrt{3}} z^{-\frac{15}{8}} - \frac{6433}{294912} z^{-\frac{25}{8}} + \frac{12741169}{283115520 \sqrt{3}} z^{-\frac{35}{8}} - \cdots, \\
F^{(2|0)}(z) &=& - \frac{1}{288} z^{-\frac{5}{4}} + \frac{109}{27648 \sqrt{3}} z^{-\frac{5}{2}} - \frac{11179}{5308416} z^{-\frac{15}{4}} + \frac{11258183}{2548039680 \sqrt{3}} z^{-\frac{10}{2}} - \cdots, \\
F^{(1|1)}(z) &=& + \frac{16}{5} z^{\frac{5}{4}} + \frac{5}{96} z^{-\frac{5}{4}} + \frac{15827}{1474560} z^{-\frac{15}{4}} + \frac{6630865}{452984832} z^{-\frac{25}{4}} + \cdots, \\
F^{(2|1)}(z) &=& - \frac{71}{864} z^{-\frac{15}{8}} + \frac{2999}{18432 \sqrt{3}} z^{-\frac{25}{8}} - \frac{25073507}{191102976} z^{-\frac{35}{8}} + \cdots, \\
F^{(3|1)}(z) &=& - \frac{47}{6912} z^{-\frac{5}{2}} + \frac{16957}{995328 \sqrt{3}} z^{-\frac{15}{4}} - \frac{1843303}{127401984} z^{-\frac{10}{2}} + \cdots, \\
F^{(2|2)}(z) &=& - \frac{5}{6} \log z + \frac{1555}{20736} z^{-\frac{5}{2}} + \frac{5288521}{95551488} z^{-5} - \frac{1886134925}{13759414272} z^{-\frac{15}{2}} + \cdots, \\
F^{(3|2)}(z) &=& + \frac{47}{288 \sqrt{3}} z^{-\frac{15}{8}} - \frac{41341}{248832} z^{-\frac{25}{8}} + \frac{11044831}{21233664 \sqrt{3}} z^{-\frac{35}{8}} - \cdots, \\
F^{(4|2)}(z) &=& + \frac{47}{3456 \sqrt{3}} z^{-\frac{5}{2}} - \frac{116803}{5971968} z^{-\frac{15}{4}} + \frac{4714205}{71663616 \sqrt{3}} z^{-5} - \cdots. \label{eq:F42z}
\eea
\noindent
One easily checks that inserting these expansions in (\ref{eq:Ftransseries}), and taking minus its second derivative, reproduces the results for the $u(z)$ transseries that we listed in appendix \ref{app:PI}. We only listed the $F^{(n|m)}(z)$ with $n\geq m$ here; the ones with $n<m$ can be obtained by the rule
\be
F^{(m|n)}_g = (-1)^{g+[n/2]_{\text{I}}} F^{(n|m)}_g, \qquad n>m.
\ee
\noindent
The starting exponent of $F^{(n|m)}$ follows straightforwardly from that of $u^{(n|m)}$. One has
\bea
F^{(n|n)} & \sim & z^{-\frac{5}{4} \beta_{nn}^{[0]} + \frac{5}{2}}, \\
F^{(n|m)} & \sim & z^{-\frac{5}{4} \beta_{nm}^{[0]}},
\eea
\noindent
where $\beta_{nm}^{[0]}$ is defined in (\ref{eq:beta_nmk-Painleve}) and the second line above is valid for $n \neq m$. This concludes the nonperturbative solution to the $(2,3)$ minimal string.

\section{Matrix Models with Polynomial Potentials}\label{sec:6}

While there are many examples of exactly solvable matrix models (see, \textit{e.g.}, a few such examples within the context of nonperturbative completions in \cite{ps09}), it is certainly the case that in most situations one does not have access to anything more than perturbative techniques, most notoriously those introduced a long time ago \cite{bipz78, biz80, ackm93}. Enlarging these old techniques by the use of resurgent analysis naturally becomes of critical importance for the extraction of nonperturbative information out of a rather large class of string theoretic examples \cite{m08}. In here, we shall focus upon matrix models with polynomial potentials, mostly on the quartic one--matrix model, developing the two--parameters resurgent framework as it applies to this example. Notice that in the large $N$ limit all matrix model quantities will now depend upon 't~Hooft moduli, an additional complication as compared to the case of minimal strings. However, we shall further see how to make the bridge back to Painlev\'e I via a natural double--scaling limit of the quartic model.

The resurgent analysis of matrix models has another added feature, as compared to minimal string models. Within this context, perturbative techniques construct asymptotic expansions which are formal power series around a given \textit{saddle--point} of the partition function of the theory. In other words, one performs perturbation theory around a chosen \textit{background}---where one expects that a full nonperturbative solution should be background \textit{independent}, \textit{i.e.}, it should include all possible backgrounds \cite{em08}. This is where the full transseries framework comes into play: only by properly considering the correct multiple--parameters transseries (a \textit{two}--parameters transseries in the quartic example) can we expect to construct fully nonperturbative, background independent solutions. In fact, while it is possible to consider a one--parameter transseries \textit{ansatz} for the quartic matrix model, still yielding a rather interesting amount of nonperturbative information, this is not the most general multi--instanton expansion required and, as such, cannot possibly see all other backgrounds \cite{m08}. In the following we shall construct the full two--parameters transseries solution to the quartic matrix model around the so--called \textit{one--cut} large $N$ saddle--point. Because this is the most general solution to this problem, it is naturally applicable to the problem of changing of background: one can envisage starting off in the one--cut phase and, via Stokes transitions, reach other stable saddle--points of the quartic matrix models such as, \textit{e.g.}, its well--known two--cut phase. We hope to report on these issues in upcoming work.

\subsection{Matrix Models: Spectral Geometry and Orthogonal Polynomials}

For the purpose of completeness on what follows, let us begin with a lightening review of matrix models, both in the spectral geometry and orthogonal polynomial frameworks (for more complete accounts we refer the reader to, \textit{e.g.}, the excellent reviews \cite{fgz93, m04}).

The one--matrix model partition function for the hermitian ensemble is
\be\label{matrixZ}
Z (N,g_s) = \frac{1}{{\mathrm{vol}} \left( \mathrm{U}(N) \right)} \int \rmd M\, \exp \left( - \frac{1}{g_s}\, \tr\, V(M) \right),
\ee
\noindent
with 't~Hooft coupling $t=Ng_s$ (fixed in the 't~Hooft limit). In standard diagonal gauge one has
\be\label{diagonalgauge}
Z (N,g_s) = \frac{1}{N!} \int \prod_{i=1}^N \left( \frac{\rmd \lambda_i}{2\pi} \right) \Delta^2(\lambda_i)\, \exp \left( - \frac{1}{g_s} \sum_{i=1}^N V(\lambda_i) \right),
\ee
\noindent
where $\Delta(\lambda_i)$ is the Vandermonde determinant. The simplest possible saddle point for this integral is the one--cut solution, characterized by an eigenvalue density normalized to one, and where the cut is simply $\CC = [a,b]$. A rather convenient description of this saddle point is given by the Riemann surface which corresponds to a double--sheet covering of the complex plane with precisely the above cut. This geometry is described by the corresponding spectral curve\footnote{Where the imaginary part of the spectral curve simply relates to the eigenvalue density.}
\be
y(z) = M(z)\, \sqrt{(z-a)(z-b)},
\ee
\noindent
where\footnote{This particular expression only holds for polynomial potentials.}
\be
M(z) = \oint_{(0)} \frac{\rmd w}{2\pi\rmi}\, \frac{V'(1/w)}{1-wz}\, \frac{1}{\sqrt{(1 - a w)(1 - b w)}}.
\ee
\noindent
For future reference, it is also useful to define the holomorphic effective potential $V_{\mathrm{h;eff}}'(z) = y(z)$, which appears at leading order in the large $N$ expansion of the matrix integral as
\be
Z \sim \int \prod_{i=1}^N \rmd \lambda_i\, \exp \left( - \frac{1}{g_s} \sum_{i=1}^N V_{\mathrm{h;eff}} (\lambda_i) + \cdots \right).
\ee

There are many ways to solve matrix models. A recursive method, sometimes denoted by the \textit{topological recursion}, was recently introduced for computing connected correlation functions and genus $g$ free energies, entirely in terms of the spectral curve \cite{eo07, eo08}. However, for our purposes of computing the genus expansion of the free energy, one of the most efficient and simple methods is still that of orthogonal polynomials \cite{biz80}, which we now briefly introduce. Considering again the one--matrix model partition function in diagonal gauge (\ref{diagonalgauge}) it is natural to regard
\be
\rmd\mu (z) = \rme^{- \frac{1}{g_s} V(z)}\, \frac{\rmd z}{2\pi}
\ee
\noindent
as a positive--definite measure on $\BR$, and it is immediate to introduce orthogonal polynomials, $\{ p_n (z) \}$, with respect to this measure as
\be\label{op}
\int_{\BR} \rmd\mu(z)\, p_n(z) p_m(z) = h_n \delta_{nm}, \qquad n \ge 0,
\ee
\noindent
where one further normalizes $p_n (z)$ such that $p_n (z) = z^n + \cdots$. Further noticing that the Vandermonde determinant is $\Delta(\lambda_i) = \det p_{j-1} (\lambda_i)$, the one--matrix model partition function above may be computed as
\be\label{zop}
Z = \prod_{n=0}^{N-1} h_n = h_0^N \prod_{n=1}^N r_n^{N-n},
\ee
\noindent
where we have defined $r_n = \frac{h_n}{h_{n-1}}$ for $n \ge 1$, and where one may explicitly write
\be
h_0 = \int_{\BR} \rmd \mu(z) = \frac{1}{2\pi} \int_{-\infty}^{+\infty} \rmd z\, \rme^{- \frac{1}{g_s} V(z)}.
\ee
\noindent
The $r_n$ coefficients also appear in the recursion relations of the orthogonal polynomials,
\be\label{oprecursion}
p_{n+1} (z) = \left( z+s_n \right) p_n (z) - r_n\, p_{n-1} (z),
\ee
\noindent
together with the new coefficients $\{ s_n \}$, which actually vanish for an even potential.

The key point that follows is that once one has a precise form of the coefficients in the recursion (\ref{oprecursion}), one may then simply compute the partition function of the matrix model (and, in fact, all quantities in a large $N$ topological expansion). In the example of main interest to us in the following, that of the quartic potential $V(z) = \frac{1}{2} z^2 - \frac{\lambda}{24}\, z^4$, it is simple to find that $s_n=0$ and \cite{biz80}
\be\label{4stringeq}
r_n \left( 1 - \frac{\lambda}{6}\, \big( r_{n-1} + r_n + r_{n+1} \big) \right) = n g_s.
\ee
\noindent
This recursion sets up a perturbative expansion around the one--cut solution of the quartic matrix model which, as briefly outlined above, is described by a single cut $\CC = \left[ -2\alpha, 2\alpha \right]$ where
\be\label{eq:qmmcut}
\alpha^2 = \frac{1}{\lambda} \left( 1 - \sqrt{1 - 2 \lambda t} \right)
\ee
\noindent
and the spectral curve is
\be
y(z) = \left( 1 - \frac{\lambda}{6} \left( z^2+2\alpha^2 \right) \right) \sqrt{z^2-4\alpha^2}.
\ee
\noindent
Before attempting a nonperturbative transseries solution to the quartic matrix model, let us briefly consider its perturbative solution \cite{biz80} and what it implies towards resurgence.

\subsection{Resurgence of the Euler--MacLaurin Formula}\label{sec:resEM}

In the 't~Hooft limit, where $N \to + \infty$ with $t=g_s N$ held fixed, the perturbative, large $N$, topological expansion of the free energy $F = \log Z$ of the matrix model (\ref{matrixZ}) is precisely given by a standard string theoretic genus expansion (\ref{stringgenus}). This is usually normalized against the Gaussian weight, where $V_{\mathrm{G}} (z) = \frac{1}{2} z^2$, thus following from (\ref{zop})
\be\label{prefree}
\CF \equiv F-F_{\mathrm{G}} = \sum_{g=0}^{+\infty} g_s^{2g-2} \CF_g (t) = \frac{t}{g_s} \log \frac{h_0}{h^{\mathrm{G}}_0} + \frac{t^2}{g_s^2}\, \frac{1}{N} \sum_{n=1}^N \left( 1-\frac{n}{N} \right) \log \frac{r_n}{r^{\mathrm{G}}_n}.
\ee
\noindent
In this expression one first needs to understand the large $N$ expansion of the recursion coefficients, $\{ r_n \}$. Given the Gaussian solution $r^{\mathrm{G}}_n = n g_s$ it is natural to change variables\footnote{The $x$ variable in this section should not be confused with the $x$ variable of section \ref{PIsection}.} as $x \equiv n g_s$, where $x \in [0,t]$ in the 't~Hooft limit, and define the function
\be
\CR(x) = r_n, \qquad \mathrm{with} \qquad \CR^{\mathrm{G}}(x) = x.
\ee
\noindent
In the example of the quartic potential, (\ref{4stringeq}) is then rewritten as
\be\label{largeN4stringeq}
\CR(x) \left\{ 1 - \frac{\lambda}{6}\, \big( \CR(x-g_s) + \CR(x) + \CR(x+g_s) \big) \right\} = x.
\ee
\noindent
Noticing that this equation is invariant under $g_s \leftrightarrow - g_s$ it follows that $\CR(x)$ is an even function of the string coupling and thus admits an asymptotic large $N$ expansion of the form
\be
\CR(x) \simeq \sum_{g=0}^{+\infty} g_s^{2g} R_{2g} (x),
\ee
\noindent
which allows one to solve for the $R_{2g} (x)$ in a recursive fashion, given $R_{0} (0)=0$. Further noticing that in the 't~Hooft limit, where $x$ becomes a continuous variable, the sum in (\ref{prefree}) may be computed by making use of the Euler--MacLaurin formula\footnote{In here the $B_{2k}$ are the Bernoulli numbers and $x = t\, \xi$.}
\be\label{euler-maclaurin}
\lim_{N \to +\infty} \frac{1}{N} \sum_{n=1}^N \Phi \left( \frac{n}{N} \right) = \int_0^1 \rmd \xi\, \Phi(\xi) + \left. \frac{1}{2N} \Phi(\xi) \right|_{\xi=0}^{\xi=1} + \sum_{k=1}^{+\infty} \left. \frac{1}{N^{2k}}\, \frac{B_{2k}}{(2k)!}\, \Phi^{(2k-1)} (\xi) \right|_{\xi=0}^{\xi=1},
\ee
\noindent
we finally obtain
\bea
\CF (t,g_s) &=& \frac{t}{2g_s} \left( 2 \log \frac{h_0}{h^{\mathrm{G}}_0} - \left. \log \frac{\CR(x)}{x} \right|_{x=0} \right) + \frac{1}{g_s^2} \int_0^t \rmd x \left( t-x \right) \log \frac{\CR(x)}{x} + \nonumber \\
&&
+ \sum_{g=1}^{+\infty} \left. g_s^{2g-2}\, \frac{B_{2g}}{(2g)!}\, \frac{\rmd^{2g-1}}{\rmd x^{2g-1}} \left[ \left( t-x \right) \log \frac{\CR(x)}{x} \right] \right|_{x=0}^{x=t},
\label{eq:EMqmm}
\eea
\noindent
or, explicitly, using the expansion of $\CR(x)$ in powers of the string coupling \cite{biz80}, \textit{e.g.},
\bea
\CF_0 (t) &=& \int_0^t \rmd x \left( t-x \right) \log \frac{R_0 (x)}{x}, \\
\CF_1 (t) &=& \int_0^t \rmd x \left( t-x \right) \frac{R_2 (x)}{R_0 (x)} + \frac{1}{12}\, \frac{\rmd}{\rmd x} \left. \left[ \left( t-x \right) \log \frac{R_0(x)}{x} \right] \right|_{x=0}^{x=t} + \frac{1}{8}\, t\, \lambda.
\eea
\noindent
It is worth making some comments concerning these expressions. First notice that the Euler--MacLaurin formula is an asymptotic expansion, thus only capturing perturbative contributions to the matrix model free energy. These perturbative contributions to the free energy at genus $g$ then arise from a recursive solution to the string equation, \textit{i.e.}, out of the coefficients $R_{2g} (x)$, computed recursively in the quartic potential example out of the large $N$ string equation (\ref{largeN4stringeq}), and similarly for different potentials. For instance, in our main example, it is simple to obtain out of (\ref{largeN4stringeq}) that
\be
R_0 (x) = \frac{1}{\lambda} \left( 1 - \sqrt{1 - 2 \lambda x} \right),
\ee
\noindent
which is the one solution satisfying the initial condition $R_0 (0) = 0$. Finally, when computing $\CF_g (t)$ above, we have partially restricted our result to the quartic potential, as we have made use of the fact that in this case one has \cite{biz80}
\be
2 \log \frac{h_0}{h^{\mathrm{G}}_0} - \left. \log \frac{\CR(x)}{x} \right|_{x=0} \equiv \log \frac{h_0 \left( +|\lambda| \right)}{h_0 \left( -|\lambda| \right)} = \frac{1}{4}\, \lambda g_s + \frac{11}{48}\, \lambda^3 g_s^3 - \cdots.
\ee

We now arrive at the main point concerning the construction of perturbative solutions to matrix models, in the orthogonal polynomial framework, and its relation to resurgence. It can be shown that the asymptotic expansion (\ref{euler-maclaurin}), defining the Euler--MacLaurin formula, may also be written as a finite difference operator of Toda type \cite{m08},
\be\label{toda-eml}
\CF(t+g_s) - 2 \CF(t) + \CF(t-g_s) = \log \frac{\CR(t)}{t}.
\ee
\noindent
This expression encodes the relation between $\CR(t,g_s)$ and $\CF(t,g_s)$, expressed by the Euler--MacLaurin asymptotic formula, and it is essentially the large $N$ version of the identity
\be
\frac{Z_{N+1} Z_{N-1}}{Z_{N}^2} = r_N;
\ee
\noindent
which is in itself an immediate consequence of (\ref{zop}). The above finite--difference equation makes it clear that if the recursion function $\CR(t,g_s)$ has a non--trivial resurgent structure, arising via a transseries solution to the string equation (\ref{largeN4stringeq}), then, (\ref{toda-eml}) will immediately induce a non--trivial resurgent structure to the matrix model free energy $\CF(t,g_s)$, of the \textit{exact same form} \cite{m08}. This is essentially a statement concerning the particular solution to the non--homogeneous Toda--type relation above. One has, however, to check the general solution to the homogeneous version of (\ref{toda-eml}), \textit{i.e.}, check whether the Euler--MacLaurin formula induces any other new resurgent effects before further ado! But all such homogeneous ``Toda'' resurgent effects have already been studied in \cite{ps09}. Furthermore, it was shown in \cite{cg08} that, essentially because the homogeneous Euler--MacLaurin relation (\ref{toda-eml}) is linear with constant coefficients, it only has Borel singularities associated to A--cycle instantons, of the type discussed in \cite{ps09}. In this scenario, B--cycle instantons, displaying fully non--trivial resurgence, originate in transseries solutions to the string equation (\ref{largeN4stringeq}). These translate to the free energy as the non--homogeneous contribution to the solution of (\ref{toda-eml}) (relating back to our discussion on A and B--cycle instantons in section \ref{topstringGV}).

The bottom line is thus that the nonperturbative resurgent analysis can be all done at the level of the non--linear recursion, or string equation (\ref{largeN4stringeq}), alone. This will capture the full non--trivial resurgent structure of the matrix model free energy; the addition of ``Toda'' or A--cycle instantons then being completely straightforward to implement, following the results in \cite{ps09}.

\subsection{The Transseries Structure of the Quartic Matrix Model}\label{sec:tsqmm}

As just discussed, the solution $\CR(x)$ to the string equation (\ref{largeN4stringeq}) completely determines the free energy of the one--cut solution to the quartic matrix model. In order to nonperturbatively solve this model, our aim is now to construct $\CR(x)$ as a transseries solution.

The string equation is, in this case, the finite--difference analogue of a second--order differential equation. For this reason, one expects the full transseries solution to contain two free parameters, which is further consistent with the fact that the double--scaling limit of the quartic matrix model reproduces the $(2,3)$ minimal string theory. As we have seen, the free energy of that theory is described by the Painlev\'e I equation, which is also solved by a transseries with two free parameters. The one--parameter transseries solution to the string equation (\ref{largeN4stringeq}) was first discussed in \cite{m08}, building upon the perturbative results obtained in \cite{biz80}. Below, we review those results, and then continue to describe the full two--parameters transseries solution.

\subsubsection*{Review of the One--Parameter Transseries Solution}

In \cite{m08} the one--parameter transseries solution to the string equation
\be\label{largeN4stringeq2}
\CR(x) \left\{ 1 - \frac{\lambda}{6} \big( \CR(x-g_s) + \CR(x) + \CR(x+g_s) \big) \right\} = x
\ee
\noindent
was investigated. It was found that such a solution can indeed be constructed, having the form
\bea
\CR(x) &=& \sum_{n=0}^{+\infty} \sigma^n\, R^{(n)} (x), \\
R^{(n)}(x) &\simeq& \rme^{- n A(x) /g_s} \sum_{g=0}^{+\infty} g_s^g\, R^{(n)}_g (x),
\label{eq:QMM1par}
\eea
\noindent
where we used a notation which slightly differs from that in \cite{m08} but which is more convenient for our purposes. Note that, as in the Painlev\'e I case, the nonperturbative answer is an expansion in the ``open string coupling constant'', $g_s$, and not in the ``closed string coupling constant'', $g_s^2$.

To find expressions for the $R^{(n)}_g (x)$, one simply plugs (\ref{eq:QMM1par}) into (\ref{largeN4stringeq2}), and solves the resulting equation order by order in $\sigma$ and $g_s$. For example, at order $\sigma^0 g_s^0$, one finds the equation
\be
r \left( 1-\frac{\lambda r}{2} \right) = x,
\ee
\noindent
where we have introduced the shorthand
\be
r \equiv R^{(0)}_0 (x).
\ee
\noindent
Solving this quadratic equation leads to the answer we have already mentioned,
\be\label{eq:rres}
r = \frac{1}{\lambda} \left( 1-\sqrt{1-2\lambda x} \right).
\ee
\noindent
Here the square root is defined to be positive on real and positive arguments and we chose the sign in front of it in such a way that $r$ has a finite $\lambda \to 0$ limit.

At order $\sigma^0 g_s^2$, (\ref{largeN4stringeq2}) gives the equation
\be
R^{(0)}_2(x) \left( 1 - \lambda r \right) - \frac{\lambda r r''}{6} = 0.
\ee
\noindent
Using (\ref{eq:rres}), one can now solve for $R^{(0)}_2 (x)$ in terms of $x$. In fact, it will turn out to be useful to write this answer, as well as all other answers that will follow, in terms of $r$. Doing this, one obtains
\be
R^{(0)}_2(x) = \frac{1}{6}\, \frac{\lambda^2 r}{(1 - \lambda r)^4}.
\ee
\noindent
This procedure is easily continued to order $\sigma^0 g_s^{2g}$, which then determines all coefficients $R^{(0)}_{2g}(x)$. In this way, one reproduces the perturbative results that were first obtained in \cite{biz80}. Note that, at order $\sigma^0$, we are skipping all odd orders in $g_s$ since our answer should be an expansion in the closed string coupling constant $g_s^2$. As aforementioned, since equation (\ref{largeN4stringeq2}) is itself even in $g_s$, it is indeed possible to find a perturbative solution $\CR_{\text{pert}}(x)$ which is also even in $g_s$.

The next step is to calculate the one--instanton contributions, which appear at order $\sigma^1$. Expanding (\ref{largeN4stringeq2}) at order $\sigma^1 g_s^0$, one finds
\be\label{eq:R10}
R^{(1)}_0(x) \left( \rme^{+ A'(x)} + \rme^{- A'(x)} + 4 - \frac{6}{\lambda r} \right) = 0.
\ee
\noindent
One sees that the overall factor $R^{(1)}_0(x)$, which we will soon find to be nonzero, drops out. Hence, this equation determines the possible values for the instanton action $A(x)$. Expressed in terms of the variable $r$, these values are
\bea
A(x) &=& \pm \frac{r}{2} \left( 2-\lambda r \right) \text{arccosh} \left( \frac{3}{\lambda  r} - 2 \right) \mp \frac{1}{2\lambda}\, \sqrt{3 \left( 1-\lambda r \right) \left( 3 - \lambda r \right)} + \nonumber \\
&&
+ \pi \rmi\, p r (2-\lambda r) + c_{\text{int}},
\eea
\noindent
where the branch cuts are chosen such that both the arccosh and the square root are positive when $\lambda r \to 1^-$, and $p \in \BZ$. Furthermore, notice that in the first line there is only one single sign ambiguity: one can either choose both upper signs or both lower ones. The integration constant $c_{\text{int}}$ and the integer ambiguity $p$ were fixed in \cite{m08} by requiring that this expression reproduces the Painlev\'e I instanton action in the corresponding double--scaling limit. It turns out that, for this, both constants need to vanish. The sign in the first line was also fixed in \cite{m08}; to obtain the positive Painlev\'e I instanton action, one needs to choose
\be\label{eq:QMMinstaction}
A(x) = - \frac{r}{2} \left( 2-\lambda r \right) \text{arccosh} \left( \frac{3-2\lambda r}{\lambda r} \right) + \frac{1}{2\lambda}\, \sqrt{\left( 3-3\lambda r \right) \left( 3 - \lambda r \right)}.
\ee
\noindent
In our two--parameters case, we shall eventually be interested in both choices of sign. We simply take the above expression as the definition of $A(x)$ and, once we move on to the two--parameters transseries, one will see that both $A(x)$ and $-A(x)$ appear symmetrically.

Akin to the Painlev\'e I case, essentially the same results may be obtained by writing the string equation (\ref{largeN4stringeq2}) in prepared form. Indeed, also for finite difference equations there is a very similar story to the one we described in section \ref{sec:4} for ordinary differential equations, and which we shall now mention very briefly \cite{b01}. This time around one can show that, via a suitable change of variables, a rank--$n$ system of non--linear finite difference equations 
\be
\boldsymbol{\CR} (x+1) = \boldsymbol{F} \big(x, \boldsymbol{\CR} (x) \big),
\ee
\noindent
may always be written in prepared form as \cite{b01}
\be
\boldsymbol{\CR} (x+1) = \boldsymbol{\Lambda} (x)\, \boldsymbol{\CR} (x) + \boldsymbol{G} \big(x, \boldsymbol{\CR} (x) \big),
\ee
\noindent
with $\boldsymbol{G} \big( x, \boldsymbol{\CR} (x) \big) = \CO \left( \left\| \boldsymbol{\CR} \right\|^2, x^{-2}\, \boldsymbol{\CR} \right)$ and where
\be
\boldsymbol{\Lambda} (x) = \mathrm{diag} \left( \rme^{-\alpha_1} \left( 1 + x^{-1} \right)^{\beta_1}, \ldots, \rme^{-\alpha_n} \left( 1 + x^{-1} \right)^{\beta_n} \right).
\ee
\noindent
Within this setting formal transseries solutions to our system of non--linear finite difference equations essentially have the same form and properties as those discussed in section \ref{sec:4}.

Once we have fixed the instanton action $A$ (to keep the notation readable, we shall many times suppress the $x$--dependence of all our functions), one can continue to higer orders in $g_s$. At order $\sigma^1 g_s^1$ (\ref{largeN4stringeq2}) gives terms involving two unknown functions, $R^{(1)}_0$ and $R^{(1)}_1$. However, it turns out that the terms proportional to $R^{(1)}_1$ actually are
\be
R^{(1)}_1 \left( \rme^{+A'} + \rme^{-A'} + 4 - \frac{6}{\lambda r} \right),
\ee
\noindent
and hence vanish by (\ref{eq:R10}). One is left with the equation
\be
\frac{\rmd R^{(1)}_0}{\rmd x} \left( \rme^{-A'} - \rme^{+A'} \right) - R^{(1)}_0 \frac{A''}{2} \left( \rme^{-A'} + \rme^{+A'} \right) = 0.
\ee
\noindent
This differential equation is not too hard to solve; where the multiplicative integration constant is once again fixed by requiring that the double--scaling limit yields the Painlev\'e I solution \cite{m08}. Its solution is thus
\be\label{eq:R10sol}
R^{(1)}_0 = \frac{\sqrt{\lambda r}}{\left( 3-\lambda r \right)^{1/4} \left( 3-3\lambda r \right)^{1/4}},
\ee
\noindent
where the quartic roots are defined to be positive as $\lambda r \to 1^-$. In fact, we shall use this convention for any of the fractional powers that will appear in what follows.

Proceeding in this way, one finds a similar pattern: at order $\sigma^1 g_s^g$ both $R^{(1)}_{g-1}$ and $R^{(1)}_g$ appear as unknown functions, but $R^{(1)}_g$ multiplies the same terms as $R^{(1)}_0$ in (\ref{eq:R10}) and hence drops out. This is nothing but the phenomenon of resonance that we have also encountered in the Painlev\'e I case. What is left is a linear first--order differential equation for $R^{(1)}_{g-1}$, which can then be easily solved. The integration constant in this solution can be fixed by the requirement of a good double--scaling limit. In \cite{m08}, the answers for $R^{(1)}_1$ and $R^{(1)}_2$ were calculated in this way. Using a \textit{Mathematica} script, we have calculated the one--instanton contributions up to $R^{(1)}_{30}$. The general structure of these solutions will be described below.

In principle, one could now go on in the same way and calculate the higher instanton contributions $R^{(n)}_g$, for $n>1$. Instead of doing this in the one--parameter formalism, we shall now move on to the two--parameters case, and calculate the higher instanton contributions as part of this more general setting.

\subsubsection*{The Two--Parameters Transseries Solution}

In the framework of the present paper, one should not restrict to a single sign choice for the instanton action. Rather, we would like to find the general two--parameters transseries solution
\bea
\CR(x) &=& \sum_{n=0}^{+\infty} \sum_{m=0}^{+\infty} \sigma_1^n \sigma_2^m\, R^{(n|m)}(x), \\
R^{(n|m)}(x) &\simeq& \rme^{-(n-m) A(x) /g_s} \sum_{g=\beta_{nm}}^{+\infty} g_s^g\, R^{(n|m)}_g (x), \label{eq:QMM2paransatz}
\eea
\noindent
to the quartic model string equation. Note that, apart from the obvious changes in this \textit{ansatz}, as going from one parameter $\sigma$ to two parameters $\sigma_1$, $\sigma_2$, we have also included a ``starting genus'' $\beta_{nm}$, which plays the same role as the $\beta_{nm}$ in our previous examples. The reader may also wonder if it is not necessary, as in the Painlev\'e I case, to introduce $\log g_s$ terms in our \textit{ansatz}. As we shall see below, there is in fact \textit{no need} for such terms in the present context\footnote{As we will see in section \ref{sec:resurinstmm}, however, it may be useful to change variables in such a way that $\log g_s$ terms do appear. This will turn out to be especially useful when we want to study the double--scaling limit.}.

Once we have made this \textit{ansatz}, solving the string equation (\ref{largeN4stringeq2}) order by order in $n$, $m$ and $g$ is a tedious but relatively straightforward exercise. As in the one--parameter case, one simply inserts (\ref{eq:QMM2paransatz}) into the string equation, isolates the terms multiplying a certain power of $\sigma_1$, $\sigma_2$ and $g_s$, and solves the resulting equations inductively for $R^{(n|m)}_g (x)$.

In the case of the ordinary instanton series one finds algebraic equations for $R^{(n|0)}_g (x)$, with $n>1$. For example, at order $\sigma_1^2 \sigma_2^0 g_s^0$, one finds the equation
\be
R^{(2|0)}_0 \left( \rme^{+2A'} + \rme^{-2A'} + 4 - \frac{6}{\lambda r} \right) + \frac{R^{(1|0)}_0\, R^{(1|0)}_0}{r} \left( 1 + \rme^{+A'} + \rme^{-A'} \right) = 0,
\ee
\noindent
which, after inserting (\ref{eq:R10sol}) and (\ref{eq:QMMinstaction}), is solved by
\be\label{eq:R200}
R^{(2|0)}_0 = - \frac{\lambda^2 r}{2 \left( 3-\lambda r \right)^{1/2} \left( 3-3\lambda r \right)^{3/2}}.
\ee
\noindent
Going beyond the instanton series, we can now also calculate the ``generalized instanton contributions'' $R^{(n|m)}_g (x)$, with nonzero $m$. At order $\sigma_1^1 \sigma_2^1 g_s^0$, for example, one finds an algebraic equation that is solved by
\be\label{eq:R110}
R^{(1|1)}_0 = \frac{3\lambda \left( 2 - \lambda r \right)}{\left( 3 - \lambda r \right)^{1/2} \left( 3 - 3 \lambda r \right)^{3/2}}.
\ee
\noindent
Continuing to higher genus, one finds that all $R^{(1|1)}_g$ with odd $g$ vanish, so that the resulting perturbative series is a series in the closed string coupling constant $g_s^2$. The same holds for \textit{all} other functions $R^{(n|n)}_g$, with as many instantons as ``generalized anti--instantons''.

At generic order $\sigma_1^n \sigma_2^m g_s^g$ one has to solve an algebraic equation to find $R^{(n|m)}_g$. Generically, \textit{i.e.}, when $n \neq m$, the answers also contain ``open string'' contributions with $g$ odd. When $n=m \pm 1$, we again encounter the phenomenon of resonance: the terms multiplying $R^{(n|m)}_g$ drop out, and we actually need to solve a differential equation to obtain $R^{(n|m)}_{g-1}$. Some of the integration constants that appear in the solutions to these differential equations are equivalent to the ambiguities we found in the Painlev\'e I case: they parameterize the choices we have in rearranging $\sigma_1$ and $\sigma_2$ into new nonperturbative parameters. We fix those integration constants as for Painlev\'e I, by requiring that $\beta_{nm}$ is as large as possible. Other integration constants do not have this interpretation, and need to be fixed by requiring the correct double--scaling limit.

The solutions to the differential equations for $n = m \pm 1$ are not all of the form that we have encountered so far. Starting at $n=2$, $m=1$, we also have logarithms entering the game. For example, for $R^{(2|1)}_0$, we find that
\be\label{eq:R210}
R^{(2|1)}_0 = \frac{\lambda \sqrt{\lambda r} \left( 54 - 45 \lambda r - 6 \lambda^2 r^2 + 8 \lambda^3 r^3 \right)}{4 r \left( 3 - 3 \lambda r \right)^{11/4} \left( 3 - \lambda r \right)^{7/4}} - \frac{3 \lambda \sqrt{\lambda r} \left( 6 + 3 \lambda r - 6 \lambda^2 r^2 + 2 \lambda^3 r^3 \right)}{32 r \left( 3 - 3 \lambda r \right)^{11/4} \left( 3 - \lambda r \right)^{7/4}}\, \log f(x),
\ee
\noindent
with
\be\label{eq:deff}
f(x) = \frac{\left( 3 - \lambda r \right)^3 \left( 3 - 3 \lambda r \right)^5}{3 \lambda^4 r^4}.
\ee
\noindent
Note that, once again, we see logarithms appearing as was previously the case for the Painlev\'e I equation. The big difference as compared to the aforementioned situation is that now the logarithmic factors are functions of $x$, and \textit{not} of the perturbative parameter $g_s$. As it turns out, all instanton corrections still take the form of open string theory perturbation series. Only in the double--scaling limit (where, as we shall see shortly, $x$ becomes a function of $g_s$) do we find back the logarithmic coupling constant dependence of the Painlev\'e I solution.

Another interesting result is that, generically, the ``starting genus'' $\beta_{nm}$ in (\ref{eq:QMM2paransatz}) is nonzero. In fact, it is usually \textit{negative}: for example, one finds that the series for $n=2$, $m=1$, does not start with the above function but with
\be\label{eq:R21m1}
R^{(2|1)}_{-1} = \frac{\lambda \sqrt{\lambda r}}{12 \left( 3 - \lambda r \right)^{1/4} \left( 3 - 3 \lambda r \right)^{1/4}}\, \log f(x),
\ee
\noindent
so that $\beta_{2,1} = -1$. We find that the non--logarithmic terms have a true genus expansion in $g_s$, but that the expansion for the logarithmic terms actually starts at ``genus $-1/2$''. At higher generalized instanton numbers, the non--logarithmic terms will in general also appear with negative powers of $g_s$. While this may seem surprising, it is not a big problem: as we shall see in section \ref{sec:QMMFE} the transseries solution for the \textit{free energy} of the quartic matrix model still only has nonnegative genus contributions.

\subsubsection*{Two--Parameters Transseries: Results}

We have written a \textit{Mathematica} script to solve the equations for $R^{(n|m)}_g(x)$ to high orders in $n$, $m$ and $g$. In appendix \ref{app:QMM}, we present some further explicit results. Here, let us write down a formula for the generic structure of the answer:
\be\label{eq:structureQMM}
R^{(n|m)}_g (x) = \frac{\left( \lambda r \right)^{p_1}}{r^{p_2} \left( 3-3\lambda r \right)^{p_3} \left( 3-\lambda r \right)^{p_4}}\, P^{(n|m)}_g (x),
\ee
\noindent
where the powers in the prefactor are the following functions of $n$, $m$ and $g$,
\bea
p_1 &=& \frac{1}{2} \left( 3n-m-2 \right), \\
p_2 &=& n+m+g-1, \\
p_3 &=& \frac{1}{4} \left( 5n+5m+10g-4 \right), \\
p_4 &=& \frac{1}{4} \left( 3n+3m+6g+2\delta-4 \right),
\eea
\noindent
with $\delta = (n+m) \mod 2$. In general, the $g_s$ expansion starts at $g = \beta_{nm} = -\min(n,m)$, whereas $n$ and $m$ only take on nonnegative values. Finally, at each order $(n,m,g)$ we find a finite expansion in logarithms,
\be
P^{(n|m)}_g (x) = \sum_{k=0}^{\min(n,m)} P^{(n|m)[k]}_g (x) \cdot \log^k f(x),
\ee
\noindent
with $f(x)$ the function defined in (\ref{eq:deff}). The resulting components $P^{(n|m)[k]}_g$ are now \textit{polynomials in} $\lambda r$, of degree $(6g+n+5m+\delta-2)/2$. 

These formulae look somewhat complicated, but the crucial point is that all the information about the two--parameters transseries is now contained in a set of simple polynomials. Moreover, up to an overall rational factor consisting of powers of some small prime factors, the coefficients of these polynomials are \textit{integers}. Thus, we have reduced the full nonperturbative solution of the quartic matrix model to the determination of a list of $(n+m+6g-\delta+2)/2$ integers for every $n$, $m$, $k$ and $g$. This result makes one wonder if these integers have any further relations between them, and whether they contain any geometrical information, as for example in the case for GV invariants we have discussed in section \ref{topstringGV}. We have no concrete suggestions in this direction, but it would be very interesting if such an interpretation could be found.

The reader may have observed that both the power $p_1$ and the degree of the polynomials are not symmetric under the exchange of $n$ and $m$. The reason for this is that we wrote (\ref{eq:structureQMM}) in such a way that, in general, when $n>m$, the $P^{(n|m)[k]}_g$ are \textit{irreducible} polynomials\footnote{There are a few low--index exceptions to this rule, for example, $P^{(3|1)[0]}_1$ has an overall factor $\lambda r$ and $P^{(4|1)[0]}_0$ and all $P^{(5|2)[k]}_{-1}$ contain a factor of $(2-\lambda r)$.}. When $n<m$ the structure formula is still valid but the polynomials are no longer irreducible. In fact, the symmetry of the string equation dictates that
\be
R^{(n|m)}_g = \left(-1\right)^g R^{(m|n)}_g,
\ee
\noindent
and as a result there is a relation
\be
P^{(n|m)[k]}_g = \left(-1\right)^g \left(\lambda r\right)^{2m-2n} P^{(m|n)[k]}_g,
\ee
\noindent
when $n<m$. After inserting this back in (\ref{eq:structureQMM}), the symmetry in $n$ and $m$ is indeed restored.

When $n=m$, the polynomials $P^{(n|m)[0]}_g$ are highly reducible. It turns out that in this case these polynomials factorize as
\be\label{eq:defQ}
P^{(n|n)[0]}_g = \left(\lambda r\right)^{p_2-p_1} Q^{(n)}_g,
\ee
\noindent
with $Q^{(n)}_g$ a polynomial of degree $2g+2n-1$ in $\lambda r$. Thus, in these cases, one can rewrite (\ref{eq:structureQMM}) as follows
\be
R^{(n|n)}_g (x) = \frac{\lambda^{p_2}}{\left( 3-3\lambda r \right)^{p_3} \left( 3-\lambda r \right)^{p_4}}\, Q^{(n)}_g (x).
\ee
\noindent
When $n=0$, $Q^{(n)}_g$ factorizes even further, and can be written as
\be\label{eq:defS}
Q^{(0)}_g = \lambda r \left( 3-\lambda r \right)^{p_4} S_g,
\ee
\noindent
with $S_g$ a polynomial of degree $(g-2)/2$ in $\lambda r$ (recall that $S_g$ is only nonzero for $g$ even, so that this degree is always an integer). Thus, we now have
\be
R^{(0|0)}_g = \frac{\lambda^{g} r}{\left( 3-3\lambda r \right)^{p_3}}\, S_g.
\ee
\noindent
This expression is only truly valid for $g>0$, although formally we can use it for $g=0$ as well if we choose
\be
S_0 = \frac{1}{3-3\lambda r}
\ee
\noindent
as the ``degree $-1$ polynomial''.

\subsection{Resurgence of Instantons in Matrix Models and String Theory}\label{sec:resurinstmm}

Now that we know the full structure of the one--cut two--parameters transseries solution to the quartic matrix model, we can test the theory of resurgence as described earlier in this paper. Before we do this for the full solution, let us discuss the double--scaling limit, in which the string equation reduces to the Painlev\'e I equation that we studied in section \ref{PIsection}.

\subsubsection*{Double--Scaling Limit}

It is well--known that there is a double--scaling limit in which the double--line Feynman diagrams of the quartic matrix model reproduce the worldsheets of the $(2,3)$ minimal string theory (for details on the physical aspects of this relation, we refer the reader to the review \cite{fgz93}). At the level of equations, it is not too hard to see directly that this limit exists. To this end, we first change variables from $(x,g_s)$ to
\be
(z, g_s) = \left( \frac{1-2\lambda x}{\left(8 \lambda^2 g_s^2\right)^{\frac{2}{5}}}, \; g_s \right),
\ee
\noindent
and replace $\CR(x,g_s)$ by a function $u(z, g_s)$ using the substitution
\be\label{eq:dsR}
\CR(x,g_s) = \frac{1}{\lambda}\left( 1 - \left( 8 \lambda^2 g_s^2 \right)^{\frac{1}{5}}  u(z,g_s) \right).
\ee
\noindent
A little algebra then shows that, in the limit where $g_s \to 0$ and $z$ is held fixed, the string equation (\ref{largeN4stringeq2}) indeed reduces to the Painlev\'e I equation
\be
u^2 (z) - \frac{1}{6} u'' (z) = z.
\ee
\noindent
Note that this result is true for any value of $\lambda$. This is a consequence of the fact that we have a redundancy of variables: the coupling constant $\lambda$ in the quartic matrix model potential $V(M)=\frac{1}{2} M^2 - \frac{\lambda}{24} M^4$ can essentially be absorbed into $g_s$ (or the 't~Hooft coupling $t = g_s N$) by a rescaling of $M$. We will encounter this redundancy of variables a few times in what follows.

Of course, the fact that the string equation reduces to the Painlev\'{e} I equation does not automatically imply that the same is true for the particular {\em  solutions} $\CR(x,g_s)$ and $u(z)$ that we have constructed. It is well known (see, \textit{e.g.}, \cite{fgz93}) that this is nevertheless the case at the perturbative level;
\be
\CR_{\text{pert}}(x,g_s) \; \to \; u_{\text{pert}}(z)
\ee
\noindent
in the double--scaling limit. One might therefore hope that the same holds true for the full two--parameters transseries solutions. It turns out that this is indeed the case, but not in a completely straightforward way. As we shall see, the correct double--scaling limit  also requires a subtle transformation between the nonperturbative ambiguities $(\sigma_1, \sigma_2)$ for the two solutions.

To further understand this limit, let us look at the full two--parameters transseries solution $\CR (x)$. It turns out to be useful to make some shifts in the summation indices, and write the transseries in the form\footnote{Recall our convention that $R^{(n|m)}_{g} \equiv 0$ if $g<\beta_{nm}$.}
\be
\CR (x) = \sum_{n=0}^{+\infty} \sum_{m=0}^{+\infty} \sum_{g=\beta_{nm}}^{+\infty} \sum_{k=0}^{+\infty} \sigma_1^{n+k} \sigma_2^{m+k}\, \rme^{- (n-m) A(x)/g_s}\, g_s^{g-k}\,  \log^k \left( f (x) \right) R^{(n+k|m+k)[k]}_{g-k} (x).
\ee
\noindent
Here, we have used the shorthand (\ref{eq:deff})
\be
f (x) = \frac{\left( 3 - \lambda r \right)^3 \left( 3 - 3 \lambda r \right)^5}{3 \lambda^4 r^4},
\ee
\noindent
and split the $R^{(n|m)}_g$ components into logarithmic contributions in the obvious way
\be
R^{(n|m)}_g (x) = \sum_{k=0}^{\min(n,m)} \log^k \left( f(x) \right) \cdot R^{(n|m)[k]}_g (x).
\ee
\noindent
The reason for writing $\CR (x)$ in the above form is that we may now apply the same trick as we did for the Painlev\'e I solution: from (\ref{eq:logrelP}) and (\ref{eq:structureQMM}) one easily deduces that
\be
R^{(n+k|m+k)[k]}_{g-k} = \frac{1}{k!} \left( \frac{\lambda \left( n-m \right)}{12} \right)^k R^{(n|m)[0]}_{g},
\ee
\noindent
so that we can sum the full logarithmic sector in order to find
\be
\CR (x) = \sum_{n=0}^{+\infty} \sum_{m=0}^{+\infty} \sum_{g=\beta_{nm}}^{+\infty} \sigma_1^{n} \sigma_2^{m}\, \rme^{-(n-m) A(x)/g_s}\, g_s^{g}\, R^{(n|m)[0]}_{g} (x) \cdot \left( f(x) \right)^{\frac{\lambda}{12 g_s} (n-m) \sigma_1 \sigma_2}.
\label{eq:Rlogsummed}
\ee
\noindent
Next, we want to manipulate this expression in such a way that it gives the correct double--scaling limit, $u(z)$. To this end, we note that, in this double--scaling limit\footnote{In here we have scaled $R^{(n|m)[0]}_{g}$ with the same overall factor of $-\lambda^{-1} \left( 8 \lambda^2 g_s^2 \right)^{\frac15}$ that was present for $R^{(0|0)[0]}_{g}$.}, one finds
\bea
\label{eq:Rds1}
\left( C \sqrt{g_s} \right)^{n+m} g_s^g\, R^{(n|m)[0]}_{g} &\longrightarrow& z^{-\frac{10g+5(n+m)-4}{8}}\, u^{(n|m)[0]}_{2g+n+m}, \\
\label{eq:Rds2}
f &\longrightarrow& 5184 \, \lambda^2 g_s^2 \, z^{5/2}, \\
\label{eq:Rds3}
\frac{1}{g_s}\, A_{\text{QMM}} &\longrightarrow& A_{\text{PI}}\, z^{5/4},
\eea
\noindent
where we denoted the quartic matrix model instanton action by $A_{\text{QMM}}$ and the Painlev\'e I instanton action by $A_{\text{PI}}$. In what follows, whenever there is danger of confusion, we will label Painlev\'e I quantities with a subscript PI and the analogous quartic matrix model quantities with a subscript QMM. When no subscript is present, we always refer to the quartic matrix model quantitiy. In the above formulae, we have also introduced the constant
\be\label{eq:QMMC}
C \equiv - \frac{2 \cdot 3^{1/4}}{\sqrt{\lambda}}.
\ee
\noindent
Of these three double--scaling formulae, the last two can be simply derived from their definitions. However, since we have no closed form expression for $R^{(n|m)[0]}_{g}$, we cannot derive the first---we shall see nonetheless that it is  necessary for the double--scaling limit to work. Moreover, we have explicitly checked its validity on all of the (more than 100) $R^{(n|m)[0]}_{g}$ that we have calculated. 

As an example, consider the expressions for $P^{(2|0)}_g$ (the polynomial components of $R^{(2|0)[0]}_g$) in (\ref{eq:P200}--\ref{eq:P203}). In the double--scaling limit we find that they yield the following terms:
\be
\frac{1}{6} z^{-3/4} - \frac{55}{576\sqrt{3}} z^{-2} + \frac{1325}{36864} z^{-13/4} - \frac{3363653}{53084160\sqrt{3}} z^{-9/2} + \cdots.
\ee
\noindent
After removing the overall normalization of $\sqrt{z}$ and substituting $z=w^{-8/5}$, this  reproduces the $u(z)$--component $\Phi^{[0]}_{(2|0)}$ in (\ref{eq:Phi200}), as should be expected. More generally, for the polynomial components $P^{(n|m)}_g$ of the $R^{(n|m)}_g$ coefficients that we present in appendix \ref{app:QMM}, one easily derives from (\ref{eq:Rds1}) that the double--scaling limit gives\footnote{Recall that $\delta = (n+m) \mod 2$.}
\be\label{eq:QMMds}
\Phi_{(n|m)}(z) = - \sum_{g=\beta_{nm}}^{+\infty} \left( - \frac{1}{3 \sqrt{2}} \right)^{n+m} \frac{6\, z^{-\frac{10g+5(n+m)}{8}}}{2^{3g}\, 2^{\delta/2}\, 3^{5g/2}}\, P^{(n|m)}_g (1).
\ee
\noindent
We thus conclude that the double--scaling limit works nicely at the component level. However, when inserting (\ref{eq:Rds1}--\ref{eq:Rds3}) into (\ref{eq:Rlogsummed}), we see that for the full $\CR (x)$ the naive double--scaling limit has two problems:
\begin{enumerate}
\item The factors of $C$ and $\sqrt{g_s}$ in (\ref{eq:Rds1}) are not present in (\ref{eq:Rlogsummed}). The factors of $C$ can be absorbed into a redefinition of the $\sigma_i$, but the absence of the factors of $\sqrt{g_s}$ will make the $(n|m) \neq (0|0)$ terms blow up in the double--scaling limit.
\item The power of $f$ in (\ref{eq:Rlogsummed}) should reproduce the power of $z$ in (\ref{eq:logsummed}) in the double--scaling limit. We see from (\ref{eq:Rds2}) that this is essentially what happens but that, in the present form, the double--scaling limit of $f$ also has an unwanted $g_s$--dependence.
\end{enumerate}
\noindent
Both of these problems can be solved by the following somewhat unconventional change of variables:
\bea
\label{eq:covsigma1}
\sigma_1 &=& \sqrt{g_s} \, \widehat{\sigma}_1 \cdot \left(72 \lambda g_s\right)^{-\frac{\lambda}{6} \, \widehat{\sigma}_1\widehat{\sigma}_2}, \\
\label{eq:covsigma2}
\sigma_2 &=& \sqrt{g_s} \, \widehat{\sigma}_2 \cdot \left(72 \lambda g_s\right)^{+\frac{\lambda}{6} \, \widehat{\sigma}_1\widehat{\sigma}_2}.
\eea
\noindent
We have discussed in section \ref{sec:PItransseries} that one is allowed to make $\sigma_1 \sigma_2$--dependent changes of variables in a two--parameters transseries. The somewhat surprising fact in here is that we now find a transformation which is also $g_s$--dependent. In the Painlev\'e I case, the expansion parameter in the transseries was $z$. Thus, in that case, one was not allowed to make $z$--dependent changes of $\sigma_i$ for the simple reason that this would spoil the Painlev\'e I equation, itself a differential equation in $z$. However, in the present quartic matrix model case, although the expansion parameter in the transseries is $g_s$, the string equation is \textit{not} an equation in $g_s$---it is an equation in $x$, for which $g_s$ is a parameter. For this reason, a $g_s$--dependent change in $\sigma_i$ does not spoil the string equation, and we are in fact allowed to make the above change of variables.

Inserting the new variables into (\ref{eq:Rlogsummed}), we find that
\be
\CR (x) = \sum_{n=0}^{+\infty} \sum_{m=0}^{+\infty} \sum_{g=\beta_{nm}}^{+\infty} \widehat{\sigma}_1^n \widehat{\sigma}_2^m\, \rme^{-(n-m) A(x)/g_s} \left( \sqrt{g_s} \right)^{n+m} g_s^{g}\, R^{(n|m)[0]}_{g} (x) \cdot \left( \frac{f(x)}{5184 \lambda^2 g_s^2} \right)^{\frac{\lambda}{12} (n-m) \widehat{\sigma}_1 \widehat{\sigma}_2},
\ee
\noindent
and we see from (\ref{eq:Rds1}--\ref{eq:Rds3}) that if we define the transseries parameters for the Painlev\'e I equation as
\be\label{eq:sigmares}
\sigma_{i,\text{PI}} = \frac{\widehat{\sigma}_i}{C},
\ee
\noindent
we indeed get the correct double--scaling limit, $u(z)$,
\be
\CR (x) \to \sum_{n=0}^{+\infty} \sum_{m=0}^{+\infty} \sum_{g=\beta_{nm}}^{+\infty} \sigma_{1,\text{PI}}^{n} \sigma_{1,\text{PI}}^{m}\, \rme^{-(n-m) A_{\text{PI}} z^{5/4}}\, u^{(n|m)[0]}_{2g+n+m} \cdot z^{-\frac{10g+5(n+m)-4}{8}+\frac{4(n-m) \sigma_{1,\text{PI}} \sigma_{2,\text{PI}}}{A}}.
\ee
\noindent
Notice that the only difference between this expression and (\ref{eq:logsummed}) is that the present formula has a lower starting genus: the first term in the $g$--sum is the one with $u^{(n|m)[0]}_{2\beta_{nm}+n+m}$. However, as we have defined all coefficients with genus smaller than the starting genus to be identically zero, this is not a problem (in principle, we could have started all $g$--sums at $-\infty$).

\subsubsection*{Choice of Resurgent Variables}

Having indentified the correct double--scaling limit of the transseries $\CR (x)$, we can now test its resurgent properties. Recall that also for non--linear difference equations there exists a suitable transseries framework \cite{b01} for which one may develop resurgent analysis in a fashion similar to what we have worked out in section \ref{sec:4} (although the literature on this class of equations is considerably smaller than the one on non--linear differential equations). However, the difference equation we address in this problem, the string equation, arises from a matrix model set--up and, in particular, has very sharp physical requirements on what concerns double--scaling limits. In other words, our difference equation \textit{must} relate to a differential equation, in a prescribed way, also at the level of resurgence. This will introduce some new features as we shall now see.

Indeed, and as discussed previously, the transseries resurgent structure of the string equation is highly dependent upon a judicious choice of variables (the ones which properly implement the Painlev\'e I double--scaling limit). As we discussed above, the naive choice of variables for $\CR (x)$, \textit{i.e.}, the choice of variables that one would consider natural from a purely finite--difference string equation point of view, is \textit{not} the one that leads to the correct double--scaling limit---for this, one further needs to make the $g_s$--dependent change of variables, from $(\sigma_1, \sigma_2)$ to $(\widehat{\sigma}_1, \widehat{\sigma}_2)$, defined in (\ref{eq:covsigma1}--\ref{eq:covsigma2}). As a result, we get a new transseries representation for $\CR (x)$; schematically
\be\label{eq:hattedrep}
\sum_{n,m,g} \sigma_1^n \sigma_2^m\, g_s^g\, R^{(n|m)}_g = \sum_{n,m,g,k} \widehat{\sigma}_1^n \widehat{\sigma}_2^m\, g_s^{g-k+\frac{n+m}{2}}\, \log^k \left( 72 \lambda g_s \right) \widehat{R}^{(n|m) \langle k \rangle}_g.
\ee
\noindent
In this new representation, different powers of $g_s$ and entirely new powers of $\log g_s$ appear\footnote{We have now labeled the coefficients of the $\log g_s$ terms with $\langle k \rangle$ to avoid confusion with the (still present) coefficients of the $\log z$ terms, which we are labeling with a $[k]$ index.}. Applying the resurgent formalism using the standard expressions for the alien derivatives can only give correct large--order formulae in one of these cases. We shall thus make the obvious assumption: we will assume that the correct representation is the one on the right--hand side above, which is the one leading directly to $u(z)$ in the double--scaling limit. In the following, we shall find ample evidence supporting this assumption.

\subsubsection*{Tests of Resurgence: Perturbative Sector}

As a first test of resurgence, let us study the large--order behavior of $R^{(0|0)}_g$. Since
\be\label{eq:RRhatnolog}
\widehat{R}^{(n|0)\langle 0 \rangle}_g = R^{(n|0)}_g, \qquad \widehat{R}^{(0|m)\langle 0 \rangle}_g = R^{(0|m)}_g,
\ee
\noindent
the result takes essentially the same form for either hatted or unhatted components. Applying our resurgent formalism to the $\widehat{R}$--transseries, and making the above substitution, one finds the large--order prediction
\be\label{eq:R00lo}
R^{(0|0)}_{g} (x) \simeq \sum_{k=1}^{+\infty} \frac{\left( S_1^{(0)} \right)^k}{\rmi\pi}\, \frac{\Gamma(g-k/2)}{\left( k A(x) \right)^{g-k/2}}\, \sum_{h=0}^{+\infty} \frac{\Gamma(g-h-k/2)}{\Gamma(g-k/2)}\, R^{(k|0)}_h (x) \left( k A(x) \right)^h.
\ee
\noindent
This result is valid for even values of $g$, so that $R^{(0|0)}_{g}$ is defined. In here, our change to the hatted components has still played a role: if we had applied the resurgent formalism directly to the $R$--transseries, we would not have found the terms of $k/2$ in the gamma function and in the power of $A$. Notice that this issue was already present in \cite{m08}, albeit implicitly: in there, this was solved by leaving a $g_s$--dependent factor in the $R^{(k|0)}_g$, leading to the somewhat counterintuitive result (equation (3.50) in that paper) of a $g_s$--dependent $S_1^{(0)}$ Stokes factor. To the contrary, our present formalism leads to large--order formulae which are $g_s$--independent---a more natural form for a quantity describing the \textit{coefficients} in a $g_s$--expansion. Moreover, as we shall see, this procedure can be straightforwardly applied to \textit{all} generalized instanton sectors, including the ones where the relation between the $R$ and $\widehat{R}$--coefficients is more complicated.

We now wish to test the large--order formula (\ref{eq:R00lo}). The first prediction we get from it is that the leading large--order behavior of $R^{(0|0)}_g$ is
\be
R^{(0|0)}_{g} (x) \sim \frac{S_1^{(0)}}{\rmi\pi}\, \frac{\Gamma \left( g-\frac{1}{2} \right)}{\left( A(x) \right)^{g-\frac{1}{2}}}\, R^{(1|0)}_0 (x).
\ee
\noindent
We have tested this behavior in a computer for a large range of $x$ (or, equivalently, $r$) and $\lambda$, and found that it was completely consistent (up to at least 10 decimal places in all cases) with a value of $S_1^{(0)}$ equal to
\be\label{eq:S10QMM}
S_1^{(0)} = \rmi \sqrt{\frac{3}{\pi\lambda}}.
\ee
\noindent
This formula equals (3.50) in \cite{m08} if we take into account the removal of $\sqrt{g_s}$ that was discussed above, as well as the definition of $\lambda$ in that paper which differs from ours by a factor of 2. 

To illustrate these tests, let us set $\lambda=1/2$ and plot the large--$g$ values of
\be\label{eq:R00lo2}
\frac{\left( A(x) \right)^{g-\frac{1}{2}}}{\Gamma \left( g-\frac{1}{2} \right)}\, R^{(0|0)}_{g}(x)
\ee
\noindent
for a sequence of equally spaced values of $r$, defined as a function of $x$ in (\ref{eq:rres}), between 0 and its double--scaling value $r_{\text{ds}} = 1/\lambda = 2$. As before, we obtain very precise large--$g$ values by calculating the above expression for values up to $g=50$, and then applying a large number of Richardson transforms (10 in this case) to remove $g^{-n}$--effects. The result is given by the blue dots in figure \ref{fig:QMM_R00_1}; the red line in that graph represents the expected result of
\be
\frac{S_1^{(0)}}{\rmi\pi}\, R^{(1|0)}_0 (x) = \frac{\sqrt{3r}}{\pi^{3/2} \left( 3-\lambda r \right)^{1/4} \left( 3-3\lambda r \right)^{1/4}},
\ee
\noindent
where we have inserted the explicit expression for $R^{(1|0)}_0$ given in (\ref{eq:R10sol}). We see that the large--order results perfectly match the predicted values. At the smallest value of $r$, the error is 0.002\%. This error is mainly due to the fact that, for small $r$, a very large amount of $R_g^{(0|0)}$ data is required to get good Richardson transforms. The error quickly decreases as $r$ increases; from $r=0.18$ onward, it becomes stable at around $10^{-12}$\%.

\FIGURE[ht]{
\label{fig:QMM_R00_1}
\centering
\includegraphics[width=10cm]{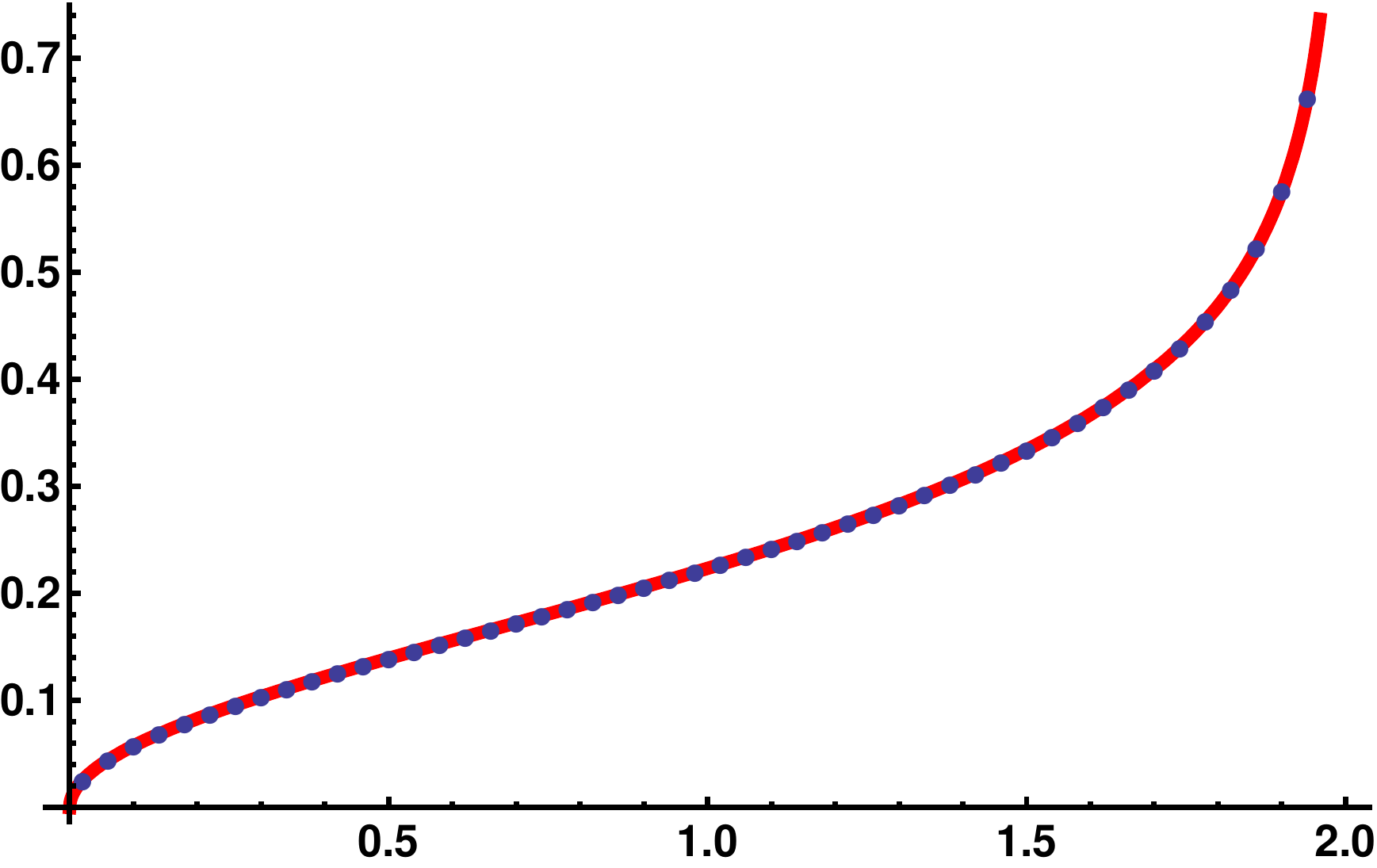}
\caption{The large--$g$ behavior of the $R^{(0|0)}_{g}$ (blue dots) compared to the predicted behavior arising from $R^{(1|0)}_0$ (red line). In this plot, we have set $\lambda=1/2$. The variable along the horizontal axis is $r$; along the vertical axis we plot the large $g$ value of the quantity in (\ref{eq:R00lo2}).}
}

As a further test of the large--order formula (\ref{eq:R00lo}), we could now study the next--to--leading order behavior in $g^{-1}$, arising from $R^{(1|0)}_1$, and so on. However, as discussed earlier in section \ref{sec:PIresurgence}, for the Painlev\'e I case, we can actually test all perturbative corrections at once by Borel--Pad\'e resumming them and going straight to the $2^{-g}$ corrections. That is, we calculate the quantity
\be
X_g (x) = R^{(0|0)}_{g} (x) - \frac{S_1^{(0)}}{\rmi\pi}\, \sum_{h=0}^{+\infty} \frac{\Gamma \left( g-h-\frac{1}{2} \right)}{\left( A(x) \right)^{g-h-\frac{1}{2}}}\, R^{(1|0)}_h (x),
\ee
\noindent
by Borel--Pad\'e resumming the second term as an expansion in $g^{-1}$, and then test the prediction that we get from (\ref{eq:R00lo}): that the large--order behavior of this quantity is
\be\label{eq:R00lo3}
- \rmi \frac{\left( 2 A(x) \right)^{g-1}}{\Gamma(g-1)}\, X_g (x) \sim -\frac{1}{\pi} \left( S_1^{(0)} \right)^2 R^{(2|0)}_0 (x) = - \frac{3\lambda r}{2 \pi^2 \left( 3-\lambda r \right)^{1/2} \left( 3-3\lambda r \right)^{3/2}},
\ee
\noindent
with $R^{(2|0)}_0$ given in (\ref{eq:R200}). Note that here we have also included a factor of $-\rmi$ to pick out the imaginary part of $X_g$: as in the Painlev\'e I case, the $2^{-g}$ correction in the large--order formula is purely imaginary, due to the fact that it comes from integrating around poles in the Borel plane with a given choice of $\pm \rmi \epsilon$--prescription.

In figure \ref{fig:QMM_R00_2}, we plot the large--order quantity on the left--hand side of (\ref{eq:R00lo3}), calculated using the usual Richardson transform method, as well as the expected result on the right--hand side of that equation (the red line in the plot). We have once again set $\lambda=1/2$ and varied $r$. The large--order data starts at a value of $r=0.22$; for smaller values, the amount of data required to get a good large--order approximation is too large to be calculated in a reasonable amount of time. The upper bound on $r$ is again its double--scaling value $r_{\text{ds}} = 1/\lambda = 2$. Akin to before, we find a very good match between the data and the prediction. For the smallest value of $r$, where the amount of data is barely sufficient, we find an error of 20\%. The error reduces quickly as the value of $X_g$ becomes larger: when $r=0.34$ the error is already less than 1\%, and it becomes as small as 0.007\% near the double--scaling limit.

\FIGURE[ht]{
\label{fig:QMM_R00_2}
\centering
\includegraphics[width=10cm]{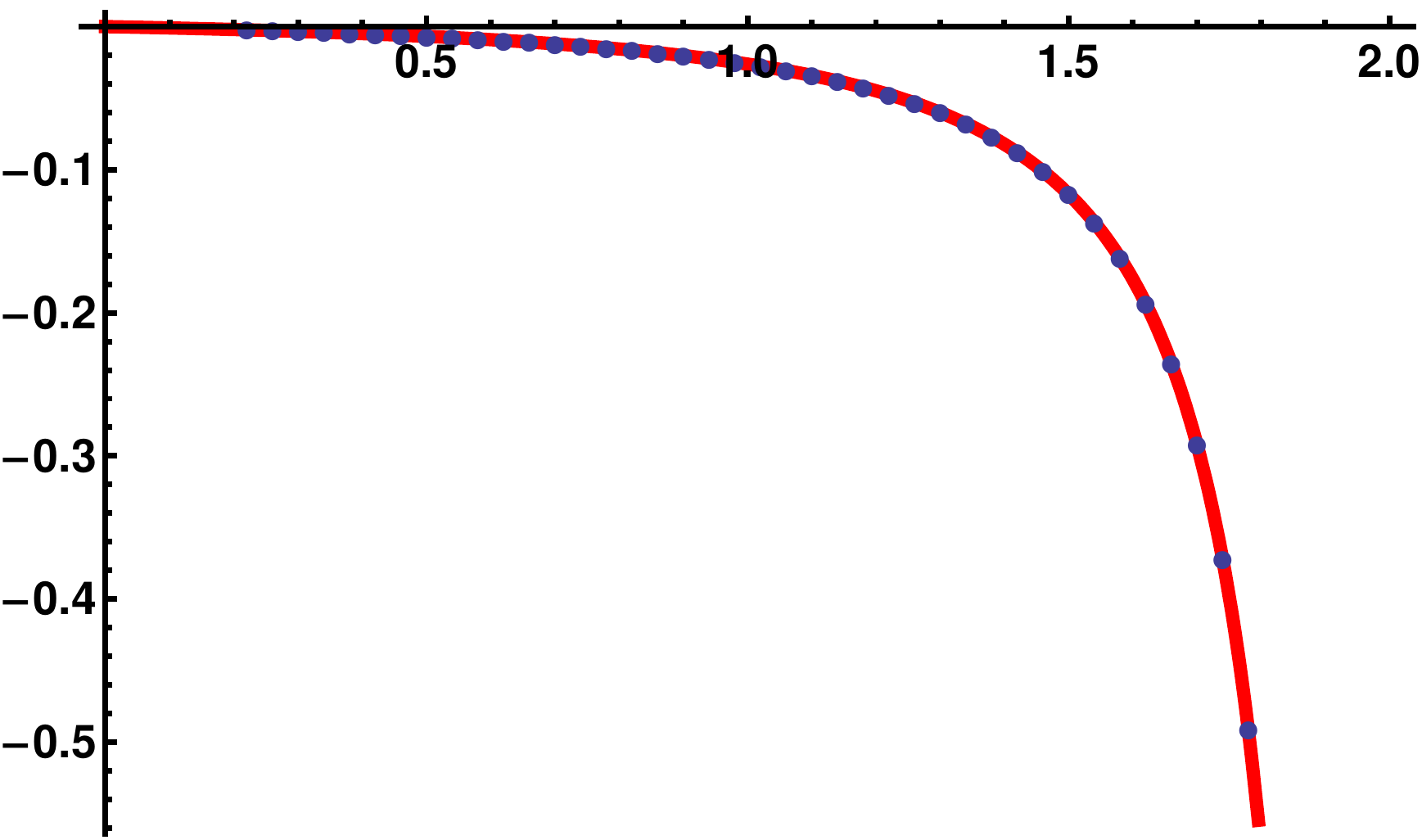}
\caption{The $2^{-g}$ corrections to the large--$g$ behavior of the $R^{(0|0)}_{g}$, expressed in terms of the quantity $X_g$ on the left--hand side of (\ref{eq:R00lo3}) (blue dots). The red line indicates the predicted value from the right--hand side of that equation. We have set $\lambda=1/2$; the variable along the horizontal axis is $r$.}
}

As a final remark on the validity of the large--order formula (\ref{eq:R00lo}), let us take its double--scaling limit using (\ref{eq:Rds1}) and (\ref{eq:Rds3}). After some straightforward algebra, one finds
\be
u^{(0|0)}_{2g} \simeq \frac{1}{\rmi\pi} \sum_{k=1}^{+\infty} \left( S_1^{(0)} C^{-1} \right)^k \frac{\Gamma(g-k/2)}{\left( k A_{\text{PI}} \right)^{g-k/2}}\, \sum_{h=0}^{+\infty} \frac{\Gamma(g-h-k/2)}{\Gamma(g-k/2)}\, u^{(k|0)}_{2h+k} \left( k A_{\text{PI}} \right)^{h}.
\ee
\noindent
This formula agrees with the Painlev\'e I large--order formula (\ref{eq:PI0largeorder}), provided that the Stokes constants for the quartic matrix model and for Painlev\'e I are related by
\be
S_{1,\text{QMM}}^{(0)} = C\, S_{1,\text{PI}}^{(0)}.
\ee
\noindent
Inserting the values (\ref{eq:S10QMM}), (\ref{eq:QMMC}), (\ref{eq:S10PI}) for these constants, we see that this is indeed the case.

\subsubsection*{Tests of Resurgence: Instanton Sectors}

Having tested the large--order behavior of the perturbative part of $\CR (x)$, we now want to switch to its (generalized) instanton components, as this is where new Stokes constants and ``backwards/sideways resurgence'' appear. The large--order behavior of the one--instanton coefficients $R^{(1|0)}_g$ still only depends on $S_1^{(0)}$ (at least perturbatively in $g^{-1}$), so the simplest coefficients to study for our purposes are the two--instantons coefficients, $R^{(2|0)}_g$. 

Thus, our first task is to derive a large--order formula for these coefficients. For this, it turns out to be essential to use the hatted representation of the transseries given in (\ref{eq:hattedrep}). The reason is that the large--order behavior of the $(2|0)$--component of any transseries depends, through ``sideways resurgence'', on its $(2|1)$--components. The latter components contain logarithms, and so it is essential that we correctly include the $\log g_s$ terms to get the correct large--order formula. After calculating the resulting large--order expression for the $\widehat{R}$--transseries, we can then translate the result back to the $R$--components using the relation (\ref{eq:RRhatnolog}), as well as the relation
\be
\widehat{R}^{(2|1)\langle 1 \rangle}_g = - \frac{\lambda}{6}\, R^{(1|0)}_g,
\ee
\noindent
that can be read off after expanding both sides of (\ref{eq:hattedrep}). Doing all of this carefully, one finds the following large--order expression
\bea
R^{(2|0)}_g (x) &\simeq& \frac{3 S_{1}^{(0)}}{2\pi\rmi}\, \sum_{h=0}^{+\infty} R^{(3|0)}_h (x) \cdot \frac{\Gamma \left( g-h-\frac{1}{2} \right)}{\left( A(x) \right)^{g-h-\frac{1}{2}}} + \nonumber \\
&&
+ \frac{(-1)^g S_{1}^{(0)}}{2\pi\rmi}\, \sum_{h=-1}^{+\infty} (-1)^h\, R^{(2|1)}_h (x) \cdot \frac{\Gamma \left( g-h-\frac{1}{2} \right)}{\left( A(x) \right)^{g-h-\frac{1}{2}}} - \nonumber \\
&&
- \frac{(-1)^g \lambda\, S_{1}^{(0)}}{12\pi\rmi}\, \sum_{h=0}^{+\infty} (-1)^h\, R^{(1|0)}_h (x) \cdot \frac{\Gamma \left( g-h+\frac{1}{2} \right) \cdot \widetilde{B}_{72\lambda\, A(x)} \left( g-h+\frac{1}{2} \right)}{\left( A(x) \right)^{g-h+\frac{1}{2}}} + \nonumber \\
&&
+ \frac{(-1)^g \widetilde{S}_1^{(2)}}{2\pi\rmi}\, \sum_{h=0}^{+\infty} (-1)^h\, R^{(1|0)}_h (x) \cdot \frac{\Gamma \left( g-h+\frac{1}{2} \right)}{\left( A(x) \right)^{g-h+\frac{1}{2}}},
\label{eq:R20lo}
\eea
\noindent
where $\widetilde{B}_{s}(a)$ is the shifted digamma function defined in (\ref{eq:Btildedef}), and we wrote the answer in terms of the purely imaginary combination
\be
\widetilde{S}_1^{(2)} = \rmi S_{-1}^{(2)} + \frac{\rmi\pi\lambda}{6}\, S_1^{(0)},
\ee
\noindent
which is also (compare against expressions such as (\ref{eq:stokesrel3})) the coefficient determining the large--order behavior of the ``conjugate'' coefficients $R^{(0|2)}_g$.

As we did several times before, (\ref{eq:R20lo}) can now be tested on a computer. Doing this, we found that the above large--order formula holds and that, for a wide range of $\lambda$ and $r$, up to 8 decimal places it is the case that
\be\label{eq:S12QMMPI}
\widetilde{S}_{1,\text{QMM}}^{(2)} = \frac{\widetilde{S}_{1,\text{PI}}^{(2)}}{C},
\ee
\noindent
with the numerical value of $\widetilde{S}_{1,\text{PI}}^{(2)}$ given in table \ref{table:PIstokescoeff}. As an illustrative example, we once again set $\lambda = 1/2$ and evaluate the quantity
\be
X_g (x) = R^{(2|0)}_g (x) - R^{(2|0)\{\text{T1-T3}\}}_g (x),
\ee
\noindent
where $R^{(2|0)\{\text{T1-T3}\}}_g (x)$ is the optimal truncation of the first three terms on the right--hand side of (\ref{eq:R20lo}). To leading order, we expect this quantity to grow as
\be\label{eq:R20lo2}
(-1)^g\, \frac{\left( A(x) \right)^{g-\frac{1}{2}}}{\Gamma \left( g-\frac{1}{2} \right)}\, X_g (x) \sim \frac{\widetilde{S}_1^{(2)}}{2\pi\rmi}\, R^{(1|0)}_0 (x).
\ee
\noindent
In figure \ref{fig:QMM_R20_1}, we plot the large--order quantity on the left--hand side of this equation as blue dots and the prediction on the right--hand side as a red line, for values of $r$ between $r=0.22$ (where we can generate just enough data) and the double--scaling value $r_{\text{ds}} = 1/\lambda = 2$. We see that the results once again match the prediction very nicely. We have included explicit error bars (estimated by comparing the results for two consecutive values of $g$) to indicate that the results for the lowest values of $r$ are still within the expectation. From $r=0.5$ onwards, the error due to lack of data is negligible, and we get results which are correct up to 8 decimal places.

\FIGURE[ht]{
\label{fig:QMM_R20_1}
\centering
\includegraphics[width=10cm]{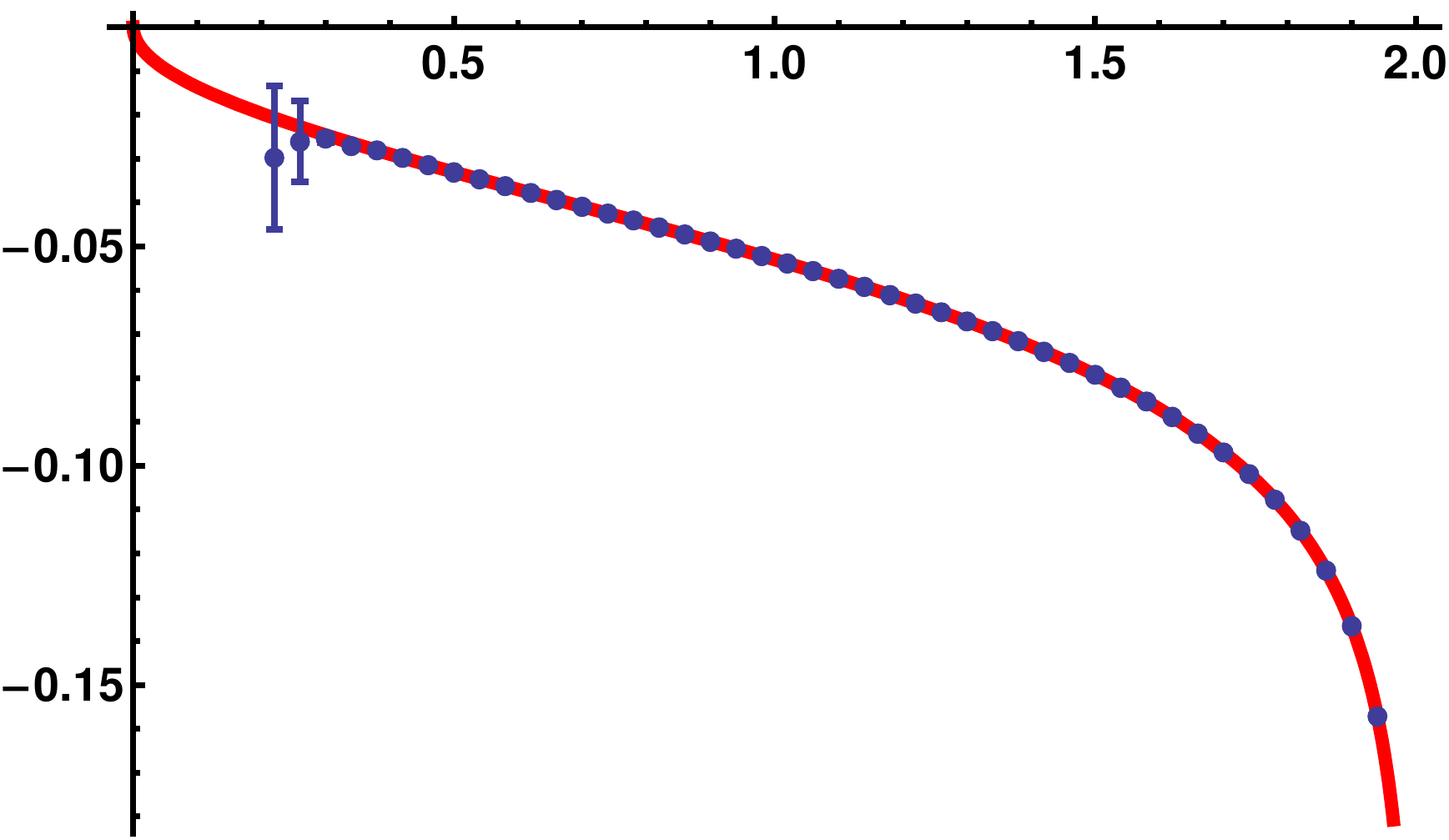}
\caption{The large--$g$ behavior of the $R^{(2|0)}_{g}$, expressed in terms of the quantity $X_g$ on the left--hand side of (\ref{eq:R20lo2}) (blue dots). The red line indicates the predicted value from the right--hand side of that equation. We have set $\lambda=1/2$; the variable along the horizontal axis is $r$.}
}

As an extra check on the validity of the large--order formula (\ref{eq:R20lo}), we can calculate its double--scaling limit using (\ref{eq:Rds1}--\ref{eq:Rds3}). It turns out that most of the logarithmic terms coming from $R^{(2|1)}_h$ and $\widetilde{B}_{72\lambda A}$ cancel, leaving a single term proportional to $\log{A_{\text{PI}}}$. All other terms reduce straightforwardly to terms involving the Painlev\'e I coefficients, and in the end one finds
\bea
u^{(2|0)[0]}_{2g+2} &\simeq& \frac{3 S_{1,\text{PI}}^{(0)}}{2\pi\rmi}\, \sum_{h=0}^{+\infty} u^{(3|0)[0]}_{2h+3} \cdot \frac{\Gamma \left( g-h-\frac{1}{2} \right)}{A_{\text{PI}}^{g-h-\frac{1}{2}}} + \frac{(-1)^g S_{1,\text{PI}}^{(0)}}{2\pi\rmi}\, \sum_{h=0}^{+\infty} (-1)^h\, u^{(2|1)[0]}_{2h+3} \cdot \frac{\Gamma \left( g-h-\frac{1}{2} \right)}{A_{\text{PI}}^{g-h-\frac{1}{2}}} - \nonumber \\
&&
- \frac{(-1)^g S_{1,\text{PI}}^{(0)}}{\sqrt{3}\pi\rmi}\, \sum_{h=0}^{+\infty} (-1)^h\, u^{(1|0)[0]}_{2h+1} \cdot \frac{\Gamma \left( g-h+\frac{1}{2} \right) \cdot  \widetilde{B}_{A_{\text{PI}}} \left( g-h+\frac{1}{2} \right)}{A_{\text{PI}}^{g-h+\frac{1}{2}}} + \nonumber \\
&&
+ \frac{(-1)^g\, C\, \widetilde{S}_{1,\text{QMM}}^{(2)}}{2\pi\rmi}\, \sum_{h=0}^{+\infty} (-1)^h\, u^{(1|0)[0]}_{2h+1} \cdot \frac{\Gamma \left( g-h+\frac{1}{2} \right)}{A_{\text{PI}}^{g-h+\frac{1}{2}}}.
\eea
\noindent
In this expression, everything is written in terms of Painlev\'e I quantities, except for the combination $C \widetilde{S}_{1,\text{QMM}}^{(2)}$ in the last term. If we now directly apply the resurgence formalism to the 2--instantons component of the Painlev\'e I transseries, we find precisely the same large--order formula, but with $C \widetilde{S}_{1,\text{QMM}}^{(2)}$ replaced by $\widetilde{S}_{1,\text{PI}}^{(2)}$. The two large--order formulae thus exactly coincide when (\ref{eq:S12QMMPI}) is valid, providing a good extra check on the validity of that equation.

As a final test, we study the large--order behavior of the generalized $(1|1)$--instanton coefficients, $R^{(1|1)}_g$. Applying the same techniques as above, we find the large--order formula
\bea
R^{(1|1)}_g (x) &\simeq& \frac{2 S_{1}^{(0)}}{\rmi\pi}\, \sum_{h=-1}^{\infty} R^{(2|1)}_h (x) \cdot \frac{\Gamma \left( g-h-\frac{1}{2} \right)}{\left( A(x) \right)^{g-h-\frac{1}{2}}} + \nonumber \\
&&
+ \frac{\lambda\, S_{1}^{(0)}}{3\pi\rmi}\, \sum_{h=0}^{\infty} R^{(1|0)}_h (x) \cdot \frac{\Gamma \left( g-h+\frac{1}{2} \right) \cdot \widetilde{B}_{72\lambda\, A(x)} \left( g-h+\frac{1}{2} \right)}{\left( A(x) \right)^{g-h+\frac{1}{2}}} + \nonumber \\
&&
+ \frac{S_{1}^{(1)}}{\rmi\pi}\, \sum_{h=0}^{\infty} R^{(1|0)}_h (x) \cdot \frac{\Gamma \left( g-h+\frac{1}{2} \right)}{\left( A(x) \right)^{g-h+\frac{1}{2}}}.
\label{eq:R11lo}
\eea
\noindent
In this formula, a new Stokes constant appears, $S_{ 1}^{(1)}$. We have checked by computer that, up to 4 decimal places, it equals
\be
S_{1,\text{QMM}}^{(1)} = \frac{S_{1,\text{PI}}^{(1)}}{C}.
\ee
\noindent
Furthermore, as we did before, one can also check that this result precisely leads to the correct Painlev\'e I large--order formula in the double--scaling limit. 

For a graphical illustration of the $S_1^{(1)}$ tests, let us choose $\lambda=1/2$ as usual and calculate the quantity
\be
X_g (x) = R^{(1|1)}_g (x) - R^{(1|1)\{\text{T1-T2}\}}_g (x),
\ee
\noindent
where $R^{(1|1)\{\text{T1-T2}\}}_g (x)$ is the optimal truncation of the first two terms on the right--hand side of (\ref{eq:R11lo}). To leading order, we expect this quantity to grow as
\be\label{eq:R11lo2}
(-1)^g\, \frac{\left( A(x) \right)^{g+\frac{1}{2}}}{\Gamma \left( g+\frac{1}{2} \right)}\, X_g (x) \sim \frac{S_1^{(1)}}{\rmi\pi}\, R^{(1|0)}_0 (x).
\ee
\noindent
Figure \ref{fig:QMM_R11_1} shows the large--order quantity on the left--hand side of the above equation as blue dots, and the prediction on the right--hand side as a red line. The variable $r$ ranges between $r=0.10$ and the double--scaling value $r = 1/\lambda = 2$. In spite of the fact that the coincidence is not perfect (due to a lack of $R^{(1|1)}_g$ data, the production of which consumes large amounts of computer time), the results still match the prediction within a few percent\footnote{The reason that we can actually calculate $S_1^{(1)}$ itself to higher precision is that, for that calculation, we can also take the optimal truncation of the third term in (\ref{eq:R11lo}).}.

\FIGURE[ht]{
\label{fig:QMM_R11_1}
\centering
\includegraphics[width=10cm]{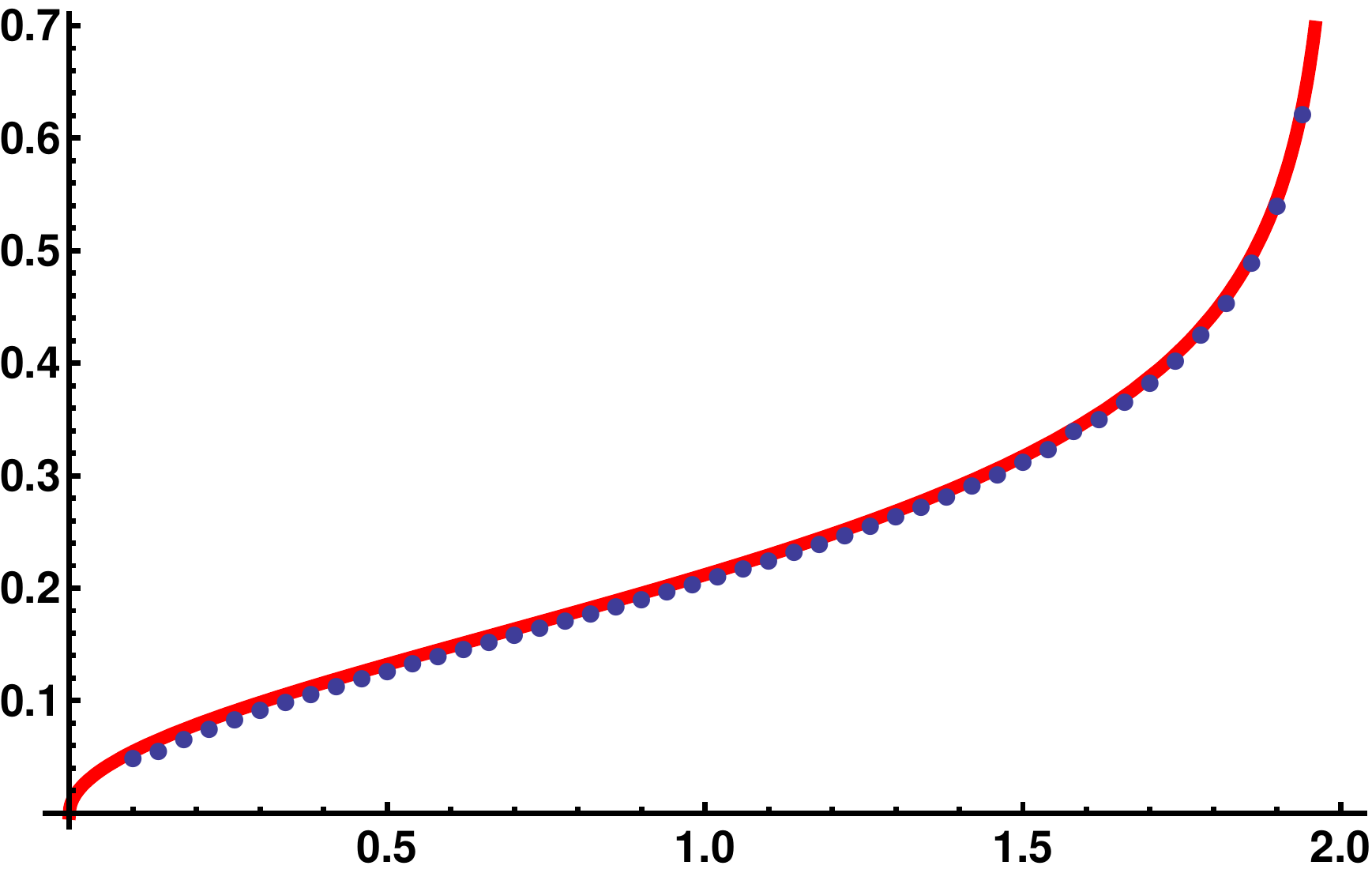}
\caption{The large--$g$ behavior of the $R^{(1|1)}_{g}$, expressed in terms of the quantity $X_g$ on the left--hand side of (\ref{eq:R11lo2}) (blue dots). The red line indicates the predicted value from the right--hand side of that equation. We have set $\lambda=1/2$; the variable along the horizontal axis is $r$.}
}

\subsubsection*{Moduli Independence of Stokes Factors}

We have now explicitly calculated three Stokes factors for the quartic matrix model and we have seen that, up to high accuracy, they satisfy
\be\label{eq:QMMStokes}
S_{1,\text{QMM}}^{(0)} = C\, S_{1,\text{PI}}^{(0)}, \qquad S_{1,\text{QMM}}^{(1)} = C^{-1}\, S_{1,\text{PI}}^{(1)}, \qquad \widetilde{S}_{1,\text{QMM}}^{(2)} = C^{-1}\, \widetilde{S}_{1,\text{PI}}^{(2)}.
\ee
\noindent
In particular, since $C \sim \lambda^{-1/2}$, the above quartic matrix model Stokes factors depend on the parameter $\lambda$. But this $C$--dependence is somewhat artificial: as we saw in (\ref{eq:Srescale1}--\ref{eq:Srescale2}), one can rescale Stokes factors by simply rescaling the parameters $\sigma_i$ with some factor, $c$, resulting in
\be\label{eq:Stokesrescale}
S_\ell^{(k)} \to c^{2-2k-\ell}\, S_\ell^{(k)}, \qquad \widetilde{S}_\ell^{(k)} \to c^{2-2k+\ell}\, \widetilde{S}_\ell^{(k)}.
\ee
\noindent 
Thus, by choosing $c=C^{-1}$, we can actually make all three Stokes factors $\lambda$--independent, and exactly \textit{equal} to their Painlev\'e I counterparts. Note that this is nothing but the scaling (\ref{eq:sigmares}) that produces the Painlev\'e I transseries solution $u(z)$ out of the quartic matrix model transseries solution $\CR(x)$, in the double--scaling limit\footnote{We could of course have chosen to absorb this scaling already in (\ref{eq:covsigma1}--\ref{eq:covsigma2}). The reason for not doing this was that it would have spoiled the simple relation (\ref{eq:RRhatnolog}) between the hatted and unhatted representation of our transseries.}.

The statement (\ref{eq:QMMStokes}) is much stronger than a statement just about the double--scaling limit: it says that, up to a trivial $C$--dependent rescaling, the quartic matrix model Stokes constants we have calculated are \textit{independent} of the parameters of the model. That is, their value at any point in parameter space equals their value in the double--scaling limit---and hence the value of the Painlev\'e I Stokes constants. That this is the case for $S_1^{(0)}$ alone is not too surprising: one can always choose a $c$ in (\ref{eq:Stokesrescale}) in such a way that $S_1^{(0)}$ becomes independent of the parameters. But that this is also the case for the other Stokes constants is indeed quite interesting. Nothing in the resurgence formalism seems to prevent these Stokes constants from depending on $\lambda$, or---as we shall see in more detail in the next section---on some combination of $\lambda$ and the 't~Hooft coupling $t$. The only consistency requirement is that the Painlev\'e I Stokes constants are reproduced when taking the double--scaling limit, which is expressed in here by choosing appropriate resurgent variables that allow for the matching of transseries solutions off--criticality and at criticality, as in (\ref{eq:hattedrep}). Here, we find that this requirement is fulfilled in the simplest possible way: by having off--critical Stokes constants which are fully independent of the parameters.

It would be very interesting to understand why the quartic matrix model Stokes constants that we have found are parameter--independent in the above sense. We have not been able to find a compelling argument for this, but of course it is very natural to conjecture that the above is not a coincidence, but that it is actually true for \textit{all} Stokes constants. That is, we conjecture
\be
S_{\ell,\text{QMM}}^{(k)} = C^{2-2k-\ell}\, S_{\ell,\text{PI}}^{(k)}, \qquad \widetilde{S}_{\ell,\text{QMM}}^{(k)} = C^{2-2k+\ell}\, \widetilde{S}_{\ell,\text{PI}}^{(k)}.
\ee
\noindent
This gives us conjectured values for many new quartic matrix model constants: up to the above $\lambda$--dependent rescalings, they should be equal to the Painlev\'e I values that we reported in table \ref{table:PIstokescoeff}. Combined with the further conjectures in footnote \ref{fn:Stokesguesses}, this gives us for example the conjecture that the exact value of the index $(0)$ Stokes constants is
\be
S_{n,\text{QMM}}^{(0)} = \frac{\rmi}{n} \left( \frac{3}{\pi\lambda} \right)^{\frac{2-n}{2}}.
\ee
\noindent
It would be very interesting to further test these conjectures, fully understand them from a physical point of view, and put them on a firm resurgent analysis mathematical footing.

\subsection{The Nonperturbative Free Energy of the Quartic Model}\label{sec:QMMFE}

Having fully constructed the two--parameters transseries solution for $\CR(x)$, our final task is to translate this solution into an expression for the free energy $\CF (t,g_s)$. In section \ref{sec:resEM}, we already briefly discussed how to do this. We saw that the Euler--MacLaurin formula leads to the expression (\ref{eq:EMqmm}), which we repeat in here for convenience:
\bea
\CF (t,g_s) &=& \frac{t}{2g_s} \left( 2 \log \frac{h_0}{h^{\mathrm{G}}_0} - \left. \log \frac{\CR(x)}{x} \right|_{x=0} \right) + \frac{1}{g_s^2} \int_0^t \rmd x \left( t-x \right) \log \frac{\CR(x)}{x} + \nonumber \\
&& + \sum_{g=1}^{+\infty} \left. g_s^{2g-2}\, \frac{B_{2g}}{(2g)!}\, \frac{\rmd^{2g-1}}{\rmd x^{2g-1}} \left[ \left( t-x \right) \log \frac{\CR(x)}{x} \right] \right|_{x=0}^{x=t}.
\label{eq:EM2}
\eea
\noindent
In applying this formula, we can of course choose any parametrization of $\CR(x)$ we wish, and one will thus end up with the corresponding parametrization of the free energy $\CF (t,g_s)$. To get a good double--scaling limit, in this section we shall once again work with the ``hatted representation'' that was introduced in (\ref{eq:hattedrep}). But do notice that, in order to avoid cluttering the notation too much, we will not put any hats on the corresponding coefficients of $\CF(t,g_s)$. 

As we saw in section \ref{sec:resEM}, the above expression is valid for the full two--parameters transseries, meaning that we can apply it both in the perturbative sector and in the (generalized) instanton sectors. We shall next discuss its application in these different sectors.

\subsubsection*{The Perturbative Sector}

In the zero--instanton sector, the above formula was already used in \cite{biz80, msw07} to compute the first few genus--$g$ free energies. There, it was found that the result takes its nicest form when expressed in terms of the 't~Hooft coupling constant $t=g_s N$ and a variable denoted by $\alpha^2$. This variable was also introduced in (\ref{eq:qmmcut}); it determines the end--points of the eigenvalue cut and is defined as
\be
\label{eq:alphadef}
\alpha^2 = \frac{1}{\lambda} \left( 1 - \sqrt{1 - 2 \lambda t} \right).
\ee
\noindent
Note that $\alpha^2$ is also essentially equal (up to an exchange $t \leftrightarrow x$) to the variable $r$ introduced in (\ref{eq:rres}). In terms of $t$ and $\alpha^2$, it was conjectured in \cite{biz80} and confirmed up to genus 10 in \cite{msw07} that the perturbative expansion coefficients of the free energy are of the form
\be\label{eq:F00struct}
\CF^{(0|0)}_g (t) = \frac{\left( t-\alpha^2 \right)^{g+1}}{t^g \left( 2t-\alpha^2 \right)^{5g/2}}\, \CS_g (t),
\ee
\noindent
where $\CS_g (t)$ is a homogeneous polynomial\footnote{As usual, there are exceptions at low genus, $g=-2,0$ in this case, where logarithmic contributions appear.} of degree $\frac{3g-2}{2}$ in $\alpha^2$ and $t$. 

In the above expression, the factor of $t^{-g}$ can be combined with the prefator $g_s^{g}$ to recover the original factor of $N^{-g}$ appearing in the large $N$ expansion of the matrix model free energy. Apart from this factor, we see that $\CF^{(0|0)}_g (t)$ in (\ref{eq:F00struct}) above, which could apparently be thought of as a natural function of $t$ and $\alpha^2$, is actually just a function of $t/\alpha^2$. In other words, the perturbative free energy components do not depend on the two separate parameters $\alpha^2$ and $t$ (or, equivalently, $\lambda$ and $t$), but only on a single combination of the two. As we mentioned in the previous subsection, this result could have been anticipated: the coupling constant $\lambda$ in the quartic matrix model potential $V(M)$ can be absorbed into $t$ by rescaling the variable $M$. We shall see that this pattern naturally extends to the full transseries solution.

Specifically, the first few $\CF^{(0|0)}_g (t)$ are\footnote{In our present conventions, these results differ by an overall minus sign from those in \cite{msw07}.}
\bea
\CF^{(0|0)}_{-2} (t) &=& \frac{1}{24} \left( 9 t^2 - 10 t \alpha^2 + \alpha^4 + 12 t^2\,  \log \left( \frac{\alpha^2}{t} \right) \right), \label{eq:FQMM1} \\
\CF^{(0|0)}_{0} (t) &=& -\frac{1}{12} \log \left( \frac{\alpha^2-2t}{t} \right), \\
\CF^{(0|0)}_{2} (t) &=& -\frac{\left( t - \alpha^2 \right)^3 \left( 82 t^2 + 21 t \alpha^2 - 3 \alpha^4 \right)}{720 t^2 \left( 2 t - \alpha^2 \right)^5}, \\
\CF^{(0|0)}_{4} (t) &=& \frac{\left( t - \alpha^2 \right)^5 \left( 17260 t^5 - 32704 t^4 \alpha^2 - 2925 t^3 \alpha^4 + 855 t^2 \alpha^6 - 135 t \alpha^8 + 9 \alpha^{10} \right)}{9072 t^4 \left( 2 t - \alpha^2 \right)^{10}}. \label{eq:FQMM2}
\eea
\noindent
We next want to investigate how these results extend to the (generalized) instanton sectors.

\subsubsection*{The Nonperturbative $(n|n)$--Sector}

To calculate the higher (generalized) instanton contributions to $\CF(t,g_s)$, it is more convenient to use the result of the Euler--MacLaurin formula in the form (\ref{toda-eml}),
\be\label{toda-eml2}
\CF(t+g_s) - 2 \CF(t) + \CF(t-g_s) = \log \frac{\CR(t)}{t}.
\ee
\noindent
A first consequence of this equation is that the instanton action $A(t)$ of the free energy equals the instanton action for $\CR(t)$ \cite{m08}. We constructed this action as a function of $r$ and $\lambda$ in (\ref{eq:QMMinstaction}); expressed in terms of $\alpha^2$ and $t$ it takes the form
\be
A(t) = - t\, \text{arccosh} \left( \frac{4t-\alpha^2}{2\alpha^2-2t} \right) + \frac{\alpha^2}{4\alpha^2-4t}\, \sqrt{12 t^2-3\alpha^4}.
\ee
\noindent
Note that, once again, $A(t)/g_s$ is a function of the single combination of variables $\alpha^2/t$.

For the free energy, we therefore make the following two--parameters transseries \textit{ansatz}\footnote{Our conventions differ from the usual ``perturbative'' ones, where $\CF_g(t)$ denotes the function multiplying $g_s^{2g-2}$. When including instanton sectors, it becomes more convenient when the subscript of $\CF^{(n|m)}_g(t)$ simply indicates the power of $g_s$ that it multiplies. Thus, our $g$ should be thought of as an Euler number, not a genus.}
\be\label{eq:qmmFansatz}
\CF(t,g_s) = \sum_{n=0}^{+\infty} \sum_{m=0}^{+\infty} \sigma_1^n \sigma_2^m\, \rme^{- (n-m) A(t)/g_s} \sum_{g=\beta^{\CF}_{nm}}^{+\infty} g_s^g\, \CF^{(n|m)}_g(t),
\ee
\noindent
where, as usual, $\beta^{\CF}_{nm}$ is the lowest $g$ for which a nonvanishing term is present. From the calculations we shall present below, it is a straightforward exercise to calculate that
\bea
n=m: && \beta^{\CF}_{nm} = \frac{n+m-4}{2}, \\
m=0, n>0: && \beta^{\CF}_{nm} = \frac{n+m}{2}, \\
m>0, n>m: && \beta^{\CF}_{nm} = \frac{n-m+2}{2},
\eea
\noindent
and by symmetry $\beta^{\CF}_{nm} = \beta^{\CF}_{mn}$. Now, it is a matter of plugging (\ref{eq:qmmFansatz}) into (\ref{toda-eml2}) and expanding in $(g_s, \sigma_1, \sigma_2)$ to obtain equations for the $\CF^{(n|m)}_g (t)$. When $n=m$, these are differential equations: one obtains
\be\label{eq:Fnnsol}
\frac{\rmd^2}{\rmd t^2} \CF^{(n|n)}_{g} (t) = \CL^{(n|n)}_{g+2} (t) - \frac{1}{12} \frac{\rmd^2}{\rmd t^2} \CL^{(n|n)}_{g} (t) + \frac{1}{240} \frac{\rmd^4}{\rmd t^4} \CL^{(n|n)}_{g-2} (t) + \cdots.
\ee
\noindent
Here, we have denoted the two--parameters transseries representation of the right--hand side of (\ref{toda-eml2}) by
\be
\log \frac{\CR(t)}{t} \equiv \CL(t),
\ee
\noindent
and $\CL(t)$ has a transseries expansion completely analogous to (\ref{eq:qmmFansatz}). Expressing the $\CL^{(n|m)}_{g} (t)$ in terms of the $\widehat{R}^{(n|m)}_{g} (t)$ is once again a straightforward exercise in Taylor expanding functions of transseries. The sum on the right--hand side of (\ref{eq:Fnnsol}) is infinite, but only a finite number of terms contribute for any given choice of $n$, as $g$ in $\CL^{(n|n)}_{g} (t)$ is bounded from below.

Solving the above equations for $n=1$, one obtains for the lowest two genera,
\bea
\label{eq:11exc}
\CF^{(1|1)}_{-1} (t) &=& \frac{\sqrt{2t-\alpha^2}\sqrt{2t+\alpha^2}}{2\sqrt{3}\, \alpha^2} + \frac{t \left( t-\alpha^2 \right)}{6\alpha^4} \log \left( \frac{2\sqrt{3} \left( 4t-\alpha^2 \right) + 6 \sqrt{2t-\alpha^2}\sqrt{2t+\alpha^2}}{\alpha^2} \right), \\
\CF^{(1|1)}_{ 1} (t) &=& - \frac{\left( t-\alpha^2 \right) \left( 8t^3-3t\alpha^4-2\alpha^6 \right)}{6\sqrt{3}\, \alpha^{4} \left( 2t-\alpha^2 \right)^{5/2} \left( 2t+\alpha^2 \right)^{3/2}}.
\eea
\noindent
Similarly, for $n=2$ one finds
\bea
\label{eq:22exc}
\CF^{(2|2)}_{0} (t) &=& \frac{\left( t-\alpha^2 \right)^2}{18\alpha^8} \log \left( \frac{\alpha^{8} \left( t-\alpha^2 \right)^4}{\left( 2t-\alpha^2 \right)^{5} \left( 2t+\alpha^2 \right)^{3}} \right), \\
\CF^{(2|2)}_{2} (t) &=& \frac{\left( t-\alpha^2 \right)^2 \left( 1696 t^6-816 t^5 \alpha^2+1896 t^4 \alpha^4-5408 t^3 \alpha^6+2229 t^2 \alpha^8+516 t \alpha^{10}+130 \alpha^{12} \right)}{486\alpha^{8} \left( 2t-\alpha^2 \right)^{5} \left( 2t+\alpha^2 \right)^{3}}.
\eea
\noindent
In both cases, the $g\leq0$ results are exceptional, with logarithmic contributions. For all strictly positive $g$, one finds the following general structure of the solution:
\be\label{eq:Fnnstruct}
\CF^{(n|n)}_g (t) = \frac{\left( t-\alpha^2 \right)^n}{\alpha^{4n} \left( 2t-\alpha^2 \right)^{5g/2} \left( 2t+\alpha^2 \right)^{3g/2}}\, \CP^{(n|n)}_g (t),
\ee
\noindent
where $\CP^{(n|n)}_g(t)$ is a homogeneous polynomial of degree $3g$. Note that, formally, $\CF^{(0|0)}_g (t)$ in (\ref{eq:F00struct}) is also of this form if we take the corresponding function $\CP^{(0|0)}_g (t)$ (which is now no longer a polynomial) to be
\be
\CP^{(0|0)}_g (t) = \frac{\left( t-\alpha^2 \right)^{g+1} \left( 2t+\alpha^2 \right)^{3g/2}}{t^g}\, \CS_g (t).
\ee
\noindent
In appendix \ref{app:QMMFE}, we present some higher--genus examples of $\CP^{(n|n)}_g (t)$ for $n=1,2$.

\subsubsection*{The Nonperturbative $(n|m)$--Sector}

When $n \neq m$, the Euler--MacLaurin formula in the form (\ref{toda-eml2}),
\be
\CF(t+g_s) - 2 \CF(t) + \CF(t-g_s) = \log \frac{\CR(t)}{t},
\ee
\noindent
gives, upon expansion in $(g_s, \sigma_1, \sigma_2)$, a set of {\em algebraic} equations\footnote{By ``algebraic'', we mean that $\CF^{(n|m)}_g (t)$ itself occurs algebraically (and even linearly), so that no integrations are needed to solve the equation. Derivatives of lower $\CF^{(n'|m')}_{g'} (t)$ still appear.} for $\CF^{(n|m)}_g (t)$. For example, for the lowest two orders, one finds
\bea
\CF^{(n|m)}_{\beta^{\CF}_{nm}} &=& \frac{1}{4} \sinh^{-2} \left( \frac{\ell A'}{2} \right) \CL^{(n|m)}_{\beta^{\CF}_{nm}}, \\
\CF^{(n|m)}_{\beta^{\CF}_{nm}+1} &=& \frac{1}{4} \sinh^{-2} \left( \frac{\ell A'}{2} \right) \left( \CL^{(n|m)}_{\beta^{\CF}_{nm}+1} + \frac{1}{4} \ell A'' \cosh \left( \frac{\ell A'}{2} \right) \CF^{(n|m)}_{\beta^{\CF}_{nm}} + \frac{1}{2} \sinh \left( \ell A'\right) \frac{\rmd}{\rmd t} \CF^{(n|m)}_{\beta^{\CF}_{nm}} \right),
\eea
\noindent
where $\ell = n-m$. Solving these equations is now straightforward (see also \cite{m08} where this was already done for the $(1|0)$--sector), and we find for example the one--instanton results
\bea
\label{eq:QMMF10l}
\CF^{(1|0)}_{1/2} (t) &=& \frac{\sqrt{2} \left( t-\alpha^2 \right)^{3/2}}{3^{5/4} \alpha^2 \left( 2t-\alpha^2 \right)^{5/4} \left( 2t+\alpha^2 \right)^{1/4}}, \\
\CF^{(1|0)}_{3/2} (t) &=& \frac{\left( t-\alpha^2 \right)^{3/2} \left( 40 t^3-12 t^2 \alpha^2-21 t \alpha^4-10 \alpha^6 \right)}{6 \sqrt{2}\, 3^{3/4}\, \alpha^2 \left( 2t-\alpha^2 \right)^{15/4} \left( 2t+\alpha^2 \right)^{7/4}},
\eea
\noindent
which agree with the results in \cite{m08}, and the two--instantons results
\bea
\CF^{(2|0)}_1 (t) &=& - \frac{4\left( t-\alpha^2 \right)^{3} \left( 4t-\alpha^2 \right)}{9 \sqrt{3}\, \alpha^4 \left( 2t-\alpha^2 \right)^{5/2} \left( 2t+\alpha^2 \right)^{3/2}}, \\
\CF^{(2|0)}_2 (t) &=& - \frac{\left( t-\alpha^2 \right)^{3} \left( 736 t^4-1096 t^3 \alpha^2+564 t^2 \alpha^4-253 t \alpha^6+22 \alpha^8 \right)}{162 \alpha^4 \left( 2t-\alpha^2 \right)^{5} \left( 2t+\alpha^2 \right)^{3}}.
\eea
\noindent
In appendix \ref{app:QMMFE}, we present some higher--genus results, as well as some results for the generalized instanton sectors $(2|1)$, $(3|1)$, $(3|2)$ and $(4|2)$. Their logarithm--free part (we will discuss the logarithmic terms in a moment) satisfies the general structure formula
\be\label{eq:Fnmstruct}
\CF^{(n|m)[0]}_g (t) = \frac{\left( t-\alpha^2 \right)^{(3n-m)/2}}{\left( \alpha^2 \right)^{n+m} \left( 2t-\alpha^2 \right)^{5g/2} \left( 2t+\alpha^2 \right)^{(3g-\delta)/2}}\, \CP^{(n|m)}_g (t),
\ee
\noindent
where $\CP^{(n|m)}_g (t)$ is a homogeneous polynomial of degree $(6g+\delta-4)/2$ (recall that here, as usual, $\delta=(n+m) \mod 2$). This expression should be compared to the very similar result (\ref{eq:structureQMM}) for $R^{(n|m)}_g(x)$. Also note that, apart from the different degree of the polynomial, the result (\ref{eq:Fnnstruct}) for $n=m$ is nothing but a specific case of the above equation.

\subsubsection*{The Logarithmic Sectors}

As is familiar by now, whenever $n>0$ and $m>0$, the $\CF^{(n|m)}_g (t)$ contain logarithmic terms. Once again, these logarithmic sectors do not contain any new information: one finds that when $n \neq m$, $\CF^{(n|m)}_g (t)$ is of the form
\be
\CF^{(n|m)}_g (t) = \sum_{k=0}^{\min(n,m)} \CF^{(n|m)[k]}_g (t) \cdot \log^k \left( \frac{f(t)}{5184 \lambda^2 g_s^2} \right),
\ee
\noindent
with
\be
\CF^{(n|m)[k]}_g (t) = \frac{1}{k!} \left( \frac{\lambda \left( n-m \right)}{12} \right)^k \CF^{(n-k|m-k)[0]}_g (t).
\ee
\noindent
The function $f(t)$ is essentially the same function as before (see (\ref{eq:deff})), but now conveniently written in the variables $\alpha^2$ and $t$,
\be
f(t) = \frac{81 \left( \alpha^2-2t \right)^5 \left( \alpha^2+2t \right)^3}{16\, \alpha^8 \left( \alpha^2-t \right)^4}.
\ee
\noindent
For readability reasons, we have left some factors of $\lambda$ explicit in the above expressions, but in principle, they should also be rewritten in terms of these variables, that is
\be
\lambda = \frac{2 \left( \alpha^2 - t \right)}{\alpha^4},
\ee
\noindent
which is the inverse of (\ref{eq:alphadef}). As before, one can also choose to sum all the logarithmic sectors resulting in the closed form
\be\label{eq:Ftotres}
\CF (t) = \sum_{n=0}^{+\infty} \sum_{m=0}^{+\infty} \sigma_1^n \sigma_2^m\, \rme^{- (n-m) A(t)/g_s} \sum_{g=\beta'^{\CF}_{nm}}^{+\infty} g_s^g\, \CF^{(n|m)[0]}_g (t) \cdot \left( \frac{f(t)}{5184 \lambda^2 g_s^2} \right)^{\frac{\lambda}{12}(n-m)\sigma_1 \sigma_2},
\ee
\noindent
for the two--parameters transseries. In here, we have introduced the shifted starting exponent
\be
\beta'^{\CF}_{nm} = \beta^{\CF}_{nm} \quad (n \geq m = 0), \qquad \beta'^{\CF}_{nm} = \beta^{\CF}_{nm}+1 \quad (n \geq m > 0),
\ee
extended by symmetry to the cases where $n<m$. The reason for the shifted exponent in the cases where $m>0$ is that in these cases, $\CF^{(n|m)}_g (t)$ starts off with a purely logarithmic term.

\subsubsection*{Double--Scaling Limit}

Because we started with the ``hatted representation'' for $\CR(x)$, which gives the Painlev\'e I solution $u(z)$ in the double--scaling limit, it is very natural to expect that the corresponding free energy $\CF(t, g_s)$ also gives the $(2,3)$ minimal model free energy $F(z)$ in the double--scaling limit. Indeed, as we discussed in detail for $\CR(x)$, the factor
\be
\left( \frac{f(t)}{5184 \lambda^2 g_s^2} \right)^{\frac{\lambda}{12}(n-m)\sigma_1 \sigma_2}
\ee
\noindent
nicely reproduces the structure of the $\log z$ terms in the Painlev\'e I solution. Thus, all we need to check is that the coefficients $\CF^{(n|m)[0]}_g (t)$ have the correct double--scaling limit. Indeed, we have checked that in this limit, and for all of the examples presented in appendix \ref{app:QMMFE},
\be
\sum_{g=\beta^{\CF}_{nm}}^\infty g_s^g\, \CF^{(n|m)[0]}_g (t) \,\to\, F^{(n|m)}(z),
\ee
\noindent
with $F^{(n|m)}(z)$ given in (\ref{eq:F00z}--\ref{eq:F42z}). This once again underlines the fact that the hatted transseries representation is the correct representation to study when one is interested in the double--scaling limit.

\subsubsection*{Stokes Constants for the Free Energy}

In the case of the $(2,3)$ minimal string, we found simple proportionality relations (\ref{eq:StokesFu1}--\ref{eq:StokesFu}) between the Stokes constants for the free energy $F(z)$ and those for the solution, $u(z)$, of the Painlev\'e I equation. We were further able to derive these relations analytically, because the map between $u(z)$ and $F(z)$ (a double integration) is a very simple and linear map.

Unfortunately, for the quartic matrix model, the situation is a whole lot more complicated. The Euler--MacLaurin formula (\ref{eq:EM2}) is very involved and it is difficult to deduce from it a direct relation between the large--order behavior of the $\widehat{R}^{(n|m)[k]}_g (x)$ and that of the $\CF^{(n|m)[k]}_g (t)$. Moreover, the computer generated data we have in this situation is insufficient to check or derive such a relation numerically, beyond the first Stokes constant.

Nevertheless, one can still make an educated guess as to what the result could be. It was found in \cite{msw07} (see equation (4.15) of that paper), both from a spectral curve analysis and using numerical results, that the large--order behavior of the perturbative series $\CF^{(0|0)}_g (t)$ is determined by the function
\be\label{eq:MSWmu1}
\mu_1 (t) = - \frac{\left( t-\alpha^2 \right)}{3^{3/4} \sqrt{\pi} \left( 2t-\alpha^2 \right)^{5/4} \left( 2t+\alpha^2 \right)^{1/4}}.
\ee
\noindent
In our notation, this function corresponds to the combination
\be
\mu_1 (t) = S_{1,\text{QMM}}^{(0)F} \cdot \CF^{(1|0)}_{1/2} (t).
\ee
\noindent
Thus, comparing (\ref{eq:MSWmu1}) to (\ref{eq:QMMF10l}), we find that
\be
S_{1,\text{QMM}}^{(0)F} = \sqrt{\frac{6 \alpha^4}{\pi \left( t-\alpha^2 \right)}} = \rmi \sqrt{\frac{3}{\pi\lambda}}.
\ee
\noindent
We see from this that $S_{1,\text{QMM}}^{(0)F}$ is exactly \textit{equal} to the Stokes constant $S_{1,\text{QMM}}^{(0)R}$ for the $\CR$--transseries, presented in (\ref{eq:S10QMM}). This is very similar to what we found in the Painlev\'e I case: in the correct parametrization, the Stokes constants for the free energy $F(z)$ are equal, up to a factor of $\ell^2$, to the Stokes constants for the corresponding solution $u(z)$. Thus, we may make the natural guess that the same pattern holds for \textit{all} Stokes constants of the quartic matrix model free energy,
\be
S_{\ell,\text{QMM}}^{(0)F} = \ell^2 S_{\ell,\text{QMM}}^{(0)R}, \qquad \widetilde{S}_{\ell,\text{QMM}}^{(0)F} = \ell^2 \widetilde{S}_{\ell,\text{QMM}}^{(0)R}.
\ee
\noindent
Note that this guess can also be viewed as extending the parameter--independence of the quartic matrix model Stokes constants for $\CR$, to the corresponding Stokes constants for $\CF$: it essentially states that, up to a trivial reparametrization, the quartic matrix model Stokes constants equal the Painlev\'e I Stokes constants. It would be quite interesting to prove (or disprove) this statement.

\section{Conclusions and Outlook}\label{sec:7}

In this paper we have hopefully made a strong case for the existence of new, previously unnoticed, nonperturbative sectors in string theory. The full structure we have uncovered was first anticipated in \cite{gikm10}, by studying the asymptotics of instantons of the Painlev\'e I equation, and first discussed, from a physical point of view, in \cite{kmr10}. But what exactly are these sectors? We hope to report on this question in upcoming work, but let us also make a few remarks herein.

The physical instanton series is simple to understand: it corresponds to standard matrix model instantons \cite{d91, d92, msw07, msw08} which, in the double--scaling limit,  become ZZ--brane amplitudes in Liouville gravity \cite{akk03}. As we shift our attention to the remaining sectors the first thing one notices is that the structure of the transseries solutions we have addressed, where purely ``generalized'' instantons have an overall minus sign in front of the instanton action as compared to standard instantons\footnote{Of course this is only the case in our present setting of a resonant two--parameters transseries. When dealing with general multi--parameter transseries, required in the solution of matrix models with more complicated potentials, or in the solutions of the minimal series coupled to gravity, this simple scenario will no longer be true.}, could seem to point towards understanding these new sectors as ghost D--branes \cite{ot06} (or, in the matrix model context, their counterpart of topological anti--D--branes \cite{v01} as dictated by the correspondence in \cite{dv02}). Indeed, these ghost D--brane sectors display this exact same feature as they have an overall minus sign in front of the Born--Infeld action \cite{ot06} (also see the discussion in \cite{kmr10}). This is an appealing picture: for instance, in the examples we have studied the free energies $F^{(n|n)}$, with as many instantons as purely ``generalized'' instantons, were found to have a resulting perturbative series which is a series in the \textit{closed} string coupling constant $g_s^2$. However, both ghost D--branes or topological anti--D--branes have one further property \cite{ot06, v01}, which is that their free energies must satisfy
\be
F^{(n|m)} = F^{(n-m|0)}, \qquad n>m.
\ee
\noindent
But this is a property we may explicitly check within our examples, and it is a property which is certainly \textit{not} satisfied. To illustrate, let us recall in here the case of the Painlev\'e I equation where we found
\bea
F^{(2|1)}(z) &=& - \frac{71}{864} z^{-\frac{15}{8}} + \frac{2999}{18432 \sqrt{3}} z^{-\frac{25}{8}} - \frac{25073507}{191102976} z^{-\frac{35}{8}} + \frac{2705576503}{6794772480 \sqrt{3}} z^{-\frac{45}{8}} - \cdots, \\
F^{(1|0)}(z) &=& - \frac{1}{12} z^{-\frac{5}{8}} + \frac{37}{768 \sqrt{3}} z^{-\frac{15}{8}} - \frac{6433}{294912} z^{-\frac{25}{8}} + \frac{12741169}{283115520 \sqrt{3}} z^{-\frac{35}{8}} - \cdots. 
\eea
\noindent
It is simple to see that these two sectors are \textit{not} proportional to each other. Furthermore, one can also show that there is \textit{no} reparametrization transformation that can achieve such proportionality. This is a straightforward consequence of (\ref{eq:phirepar1}) which states that, upon reparametrization, the only possible change of $F^{(2|1)}(z)$ is by a multiple of $F^{(1|0)}(z)$. Thus, if $F^{(2|1)}(z)$ is not a multiple of $F^{(1|0)}(z)$ in one representation, that statement is automatically true for any other reparametrization. Further notice that using the transseries structure of the free energy as in (\ref{eq:Ftransseries}), where the transseries parameters also appear exponentiated, does not change this conclusion. Indeed, the exponentiation (\ref{transseriesexponentiation}) is just a convenient way to rearrange the logarithmic sectors, which can always be reversed (by expanding the exponential). In this case one would then apply the aforementioned argument to each separate logarithmic sector with the exact same conclusion. As such, although we cannot at this stage state what the new nonperturbative sectors are, it seems we can state what they are not.

Another pertinent question is: why have we never seen these sectors before? The short answer is, of course, that two--parameters transseries were never addressed in a string theoretic context prior to \cite{gikm10}. Only by addressing the question of what controls the asymptotic behavior of \textit{multi}--instanton sectors can one realize that indeed the familiar physical instanton series \textit{cannot} be the full story. In fact, most large--order analyses have always been concentrated upon the leading asymptotics of the perturbative sector \cite{z81}. But, as we have shown at length in this paper, if we want to address harder questions than that, in the string theoretic nonperturbative realm, then the full multi--parameter transseries framework is indeed required.

On the other hand there are examples of exactly solvable models, where full nonperturbative answers have been computed. Should any of these expressions have shown these new sectors? Of course in order to see them one would have to know what to look for. But when one rewrites one of these exact nonperturbative solutions in terms of semi--classical data, one usually does so only for \textit{real} solutions around positive, \textit{real} coupling, and in the \textit{one}--parameter transseries framework! Let us briefly discuss the construction of real solutions, trivially generalizing a discussion in \cite{m08} to an arbitrary one--parameter transseries of the type (\ref{Ftransseries}),
\be
F(z,\sigma) = \sum_{n=0}^{+\infty} \sigma^n\, \rme^{- n A z}\, \Phi_{n} (z).
\ee
\noindent
A real solution starts around positive real coupling $z \in \BR^+$. But this is a Stokes line and we need to be careful in constructing such real solution. For instance, upon Borel resummation, either $\CS_{+} F$ or $\CS_{-} F$, will display an ambiguous \textit{imaginary} contribution to the solution which needs to be canceled, \textit{i.e.}, one needs to set\footnote{Notice that around the $\theta=0$ Stokes line one has $\im_0 = \frac{1}{2\rmi} \left( \CS_{+} - \CS_{-} \right)$ and $\re_0 = \frac{1}{2} \left( \CS_{+} + \CS_{-} \right)$.} $\im\, F(z,\sigma) = 0$. As it turns out \cite{m08}, $\im\, F(z,\sigma) = 0$ if and only if $\im\, \sigma = \frac{\rmi}{2} S_1$. As such, and as long as the instanton action is real, a real solution can be constructed by considering \cite{m08}
\be
F_\BR (z,\sigma) = \CS_{+} F \left( z,\sigma - \frac{1}{2} S_1 \right) = \CS_{-} F \left( z,\sigma + \frac{1}{2} S_1 \right),
\ee
\noindent
where the transseries parameter in the expression above is now $\sigma \in \BR$, and where the second equality follows trivially from the Stokes transition (\ref{stokestransition@0})
\be
\CS_{+} F(z,\sigma) = \CS_{-} F \left( z, \sigma + S_1 \right).
\ee
\noindent
Expanding, it immediately follows
\be
F_\BR (z,\sigma) = \re\, F^{(0)} (z) +  \sigma\, \re\, F^{(1)} (z) + \left( \sigma^2 - \frac{1}{4} S_1^2 \right) \re\, F^{(2)} (z) + \cdots.
\ee
\noindent
Two things are to be noticed. The first is that indeed real solutions display instanton corrections (even if $\sigma=0$). This is simply because the string equation (be it the Painlev\'e I equation or the quartic string equation or any other) is non--linear and, although $\CS_{+} F$ or $\CS_{-} F$ may be solutions, their sum is, consequentially, \textit{not} a solution. Indeed, their sum can only become a solution once we correct it appropriately by accounting for higher instanton corrections. The second point, however, is that this instanton expansion only includes information concerning $S_1$, not about any of the other Stokes constants. This is to say, as long as we consider the expansion in semi--classical data around the (natural) $\theta=0$ Stokes lines, we shall find no indication of the multi--parameter transseries sectors. Searching for signs of these new generalized instanton sectors within nonperturbative answers must thus start by properly addressing what type of expansion one wants to do---as shown, the standard one will not do.

In summary, we believe the most pressing question begging to be addressed is to fully understand, from a physical string theoretic point of view, the generalized instanton series. As discussed, D--branes only yield information on a limited set of Stokes constants and, if one is to address nonperturbative questions where all Stokes constants play a role, some information is missing. Examples where all Stokes constants would be required involve general Stokes transitions---even if we are just addressing the perturbative series, Stokes transitions along $\theta = \pi$ will require Stokes constants which, at this stage, have no first principles derivation. For instance, within the setting of the quartic model, one could imagine rotating the string coupling in the complex plane from the positive to the negative real axis. The saddle configuration would then change, from the one--cut spectral geometry we addressed in this paper to a two--cuts spectral curve. This change of background may be implemented  within our framework---the transseries does provide the complete nonperturbative answer---via a Stokes transition, but in order to explicitly construct the perturbative free energy around the new background, given the original one, we are still missing analytic expressions for the Stokes constants. This is a problem we hope to report upon soon. Furthermore, as one considers the two--cuts solution to the quartic matrix model, another double--scaling limit naturally appears: that of the Painlev\'e II equation describing 2d supergravity. Given that our off--critical transseries construction was very much attached to implementing correct double--scaling limits, this is certainly an interesting problem to address. Finally, we have just started uncovering what we believe is a very general method towards the construction of explicit nonperturbative solutions in string theory. Still within the matrix model realm, addressing two--matrix models and their associated minimal series seems to be a direction of great interest. We hope to return to many of these ideas in the near future.

\acknowledgments
We would like to thank Hirotaka Irie, Alexander Its, Marcos Mari\~no and Ricardo Vaz for useful discussions and comments. The authors would further like to thank CERN TH--Division for hospitality, where a part of this work was conducted.

\newpage

\appendix

\section{The Painlev\'e I Equation: Structural Data}\label{app:PI}

The general two--parameters transseries solution of the Painlev\'e I equation has the form
\be
u(w,\sigma_1,\sigma_2) = \sum_{n=0}^{+\infty} \sum_{m=0}^{+\infty} \sigma_1^n \sigma_2^m\, \rme^{-(n-m)A/w^2}\, \Phi_{(n|m)}(w),
\ee
\noindent
with
\be
\Phi_{(n|m)}(w) = \sum_{k=0}^{\min(n,m)} \log^k (w) \cdot \Phi_{(n|m)}^{[k]} (w).
\ee
\noindent
Table \ref{table:PIdata} shows up to which order in $w$ we have calculated $\Phi_{(n|m)}^{[k]}$. The table is for $k=0$; as we will see, the results for nonzero $k$ are directly proportional to those. Moreover, we only list the entries for $n\geq m$; as we shall see in a moment, the coefficients for $n<m$ can be easily obtained from those with $n>m$. It would go too far to reproduce all the data in this appendix---the interested reader may request a \textit{Mathematica} notebook from the authors containing all calculated coefficients. Below, we reproduce part of the expansions for some small values of $n$ and $m$.

\begin{table}[ht]
\centering
\begin{tabular}{c|rrrrrrrrrrr}
\begin{picture}(20,20)(0,0)
\put(17,8){$n$}
\put(3.6,18){\line(1,-1){22.3}}
\put(3,0){$m$}
\end{picture}
& 0 & 1 & 2 & 3 & 4 & 5 & 6 & 7 & 8 & 9 & 10 \\
\hline
0  & 1000 & 300 & 300 & 300 & 300 & 300 & 300 & 25 & 10 & 10 & 10 \\
1  &      & 300 & 300 & 300 & 300 & 300 &  23 & 24 \\
2  &      &     & 300 & 300 & 300 & 300 &  22 & 23 \\
3  &      &     &     & 300 & 300 &  20 &  21 & 22 \\
4  &      &     &     &     &  20 &  19 &  20 & 21 \\
5  &      &     &     &     &     &  20 &  19 & 20 \\
6  &      &     &     &     &     &     &  20 & 19 \\
7  &      &     &     &     &     &     &     & 20
\end{tabular}
\caption{Order in $w$ up to which we have calculated the $\Phi_{(n|m)}^{[k]}$.}
\label{table:PIdata}
\end{table}

The first few $\Phi_{(n|0)}^{[0]}$ are:
\bea
\Phi_{(0|0)}^{[0]} &=& 1 - \frac{1}{48}\, w^4 - \frac{49}{4608}\, w^8 - \frac{1225}{55296}\, w^{12} - \cdots, \\
\Phi_{(1|0)}^{[0]} &=& w - \frac{5}{64 \sqrt{3}}\, w^3 + \frac{75}{8192}\, w^5 - \frac{341329}{23592960 \sqrt{3}}\, w^7 + \cdots, \\
\Phi_{(2|0)}^{[0]} &=& \frac{1}{6}\, w^2 - \frac{55}{576 \sqrt{3}}\, w^4 + \frac{1325}{36864}\, w^6 - \frac{3363653}{53084160 \sqrt{3}}\, w^8 + \cdots. \label{eq:Phi200}
\eea
\noindent
The first few $\Phi_{(n|1)}^{[0]}$ are:
\bea
\Phi_{(1|1)}^{[0]} &=& - w^2 - \frac{75}{512}\, w^6 - \frac{300713}{1572864}\, w^{10} - \cdots, \\
\Phi_{(2|1)}^{[0]} &=& \frac{11}{72}\, w^3 - \frac{985}{4608 \sqrt{3}}\, w^5 + \frac{597575}{15925248}\, w^7 - \cdots, \label{phi210} \\
\Phi_{(3|1)}^{[0]} &=& \frac{3}{16}\, w^4 - \frac{3455}{10368 \sqrt{3}}\, w^6 + \frac{1712825}{7962624}\, w^8 - \cdots.
\eea
\noindent
In $\Phi_{(n|1)}$, one sees the first logarithms appearing. One finds that $\Phi_{(1|1)}$ has no logarithmic terms, and that
\bea
\Phi_{(2|1)}^{[1]} &=& - \frac{4}{\sqrt{3}}\, w + \frac{5}{48}\, w^3 - \frac{75}{2048 \sqrt{3}}\, w^5 + \cdots, \label{phi211} \\
\Phi_{(3|1)}^{[1]} &=& - \frac{4}{3 \sqrt{3}}\, w^2 + \frac{55}{216}\, w^4 - \frac{1325}{4608 \sqrt{3}}\, w^6 + \cdots.
\eea
\noindent
The reader may notice that these functions are very similar to the $\Phi_{(n|0)}^{[0]}$ listed above: in fact, using (\ref{eq:PIrec}), one can easily show that the recursion relations for the coefficients of the two power series are the same, and so they are equal up to an overal multiplicative constant. To be precise, one finds that
\be\label{eq:propPI}
\Phi_{(n|1)}^{[1]} = - \frac{4(n-1)}{\sqrt{3}}\, \Phi_{(n-1|0)}^{[0]}.
\ee
\noindent
This relation was first noted in \cite{gikm10}, and all the formulae we have tabulated so far can in fact be derived from the formulae in that paper. However, with our methods one can easily go beyond the results of \cite{gikm10}. At the next level, $\Phi_{(n|2)}$, we find for example that
\bea
\Phi_{(2|2)}^{[0]} &=& -\frac{5}{6}\, w^4 + \frac{54425}{82944}\, w^8 - \frac{26442605}{15925248}\, w^{12} + \cdots, \\
\Phi_{(3|2)}^{[0]} &=& -\frac{47}{24 \sqrt{3}}\, w^3 + \frac{4213}{20736}\, w^5 - \frac{1043455}{1769472 \sqrt{3}}\, w^7 + \cdots, \\
\Phi_{(4|2)}^{[0]} &=& -\frac{47}{72 \sqrt{3}}\, w^4 + \frac{54415}{124416}\, w^6 - \frac{6750359}{5971968 \sqrt{3}}\, w^8 + \cdots.
\eea
\noindent
These functions, except for the diagonal one $\Phi_{(2|2)}$, also have parts proportional to $\log w$. They are
\bea
\Phi_{(3|2)}^{[1]} &=& -\frac{11}{18 \sqrt{3}}\, w^3 + \frac{985}{3456}\, w^5 - \frac{597575}{3981312 \sqrt{3}}\, w^7 + \cdots, \\
\Phi_{(4|2)}^{[1]} &=& -\frac{3}{2 \sqrt{3}}\, w^4 + \frac{3455}{3888}\, w^6 - \frac{1712825}{995328 \sqrt{3}}\, w^8 + \cdots.
\eea
\noindent
The new phenomenon at this level is that we now also have $\log^2 w$ contributions. These are found to be
\bea
\Phi_{(3|2)}^{[2]} &=& \frac{8}{3}\, w - \frac{5}{24 \sqrt{3}}\, w^3 + \frac{25}{1024}\, w^5 - \cdots, \\
\Phi_{(4|2)}^{[2]} &=& \frac{16}{9}\, w^2 - \frac{55}{54 \sqrt{3}}\, w^4 + \frac{1325}{3456}\, w^6 - \cdots.
\eea
\noindent
Again, these functions have a close relation to the functions $\Phi_{(n|1)}^{[1]}$. In fact, with a bit of work, one can show from the recursion relation (\ref{eq:PIrec}) that terms with a given power of $\log w$ are always proportional to similar terms with lower $n$ and $m$, as well as lower logarithmic power,
\be
\Phi_{(n|m)}^{[k]} = \frac{4(m-n)}{k \sqrt{3}}\, \Phi_{(n-1|m-1)}^{[k-1]},
\ee
\noindent
where in this expression we have assumed that $n>m$. Applying this formula $k$ times, one can further express these coefficients in terms of log--free coefficients as
\be\label{eq:logprop}
\Phi_{(n|m)}^{[k]} = \frac{1}{k!} \left( \frac{4 \left( m-n \right)}{\sqrt{3}} \right)^k \Phi_{(n-k|m-k)}^{[0]}.
\ee
\noindent
This immediately implies that the logarithmic sectors are, from a certain point of view, artifacts of the resonant transseries solution---they do not contain any new physical content. Finally, we remark that we have only listed $\Phi_{(n|m)}^{[k]}$ above with $n \geq m$. The expansions for $n<m$ are very similar\footnote{Notice that the naive observation that all Painlev\'e I coefficients with $n<m$ are positive is, in fact, not true (even though the examples we have shown could seem to point in that way). This is only noticed for the first time when $n=3$, $m=4$ and at genus $11$, so it is indeed an assumption which is hard to falsify!}. In fact, one finds that
\be
\label{eq:PIsign}
u_g^{(n|m)[k]} = (-1)^{(g-n-m)/2} u_g^{(m|n)[k]}
\ee
\noindent
for $n \neq m$. This again generalizes a similar relation found in \cite{gikm10}.

\section{The Quartic Matrix Model: Structural Data}\label{app:QMM}

In this appendix, we present some of the explicit polynomials that determine the full nonperturbative solution (\ref{eq:structureQMM}) to the one--cut quartic matrix model. Recall from section \ref{sec:tsqmm} that this solution has the form
\be
\CR(x) = \sum_{n=0}^{+\infty} \sum_{m=0}^{+\infty} \sigma_1^n \sigma_2^m\, R^{(n|m)}(x)
\ee
\noindent
with
\be
R^{(n|m)}(x) \simeq \rme^{-(n-m) A(x) /g_s} \sum_{g=\beta_{nm}}^{+\infty} g_s^g\, R^{(n|m)}_g (x),
\ee
\noindent
and that the expansion coefficients $R^{(n|m)}_g (x)$ can be expressed in terms of polynomials $P^{(n|m)[k]}_g (x)$ as
\be
R^{(n|m)}_g (x) = \frac{\left( \lambda r \right)^{p_1}}{r^{p_2} \left( 3-3\lambda r \right)^{p_3} \left( 3-\lambda r \right)^{p_4}} \sum_{k=0}^{\min(n,m)} \log^k \left( f(x) \right) \cdot P^{(n|m)[k]}_g (x).
\ee
\noindent
The following table \ref{tablequartic} shows to which order in $g_s$ we have calculated the $P^{(n|m)}_g (x)$ polynomials:

\begin{table}[ht]
\centering
\begin{tabular}{c|ccccccccccc}
\begin{picture}(20,20)(0,0)
\put(17,8){$n$}
\put(3.6,18){\line(1,-1){22.3}}
\put(3,0){$m$}
\end{picture}
& 0 & 1 & 2 & 3 & 4 & 5 & 6 & 7 & 8 & 9 & 10 \\
\hline
0  & 100 & 30 & 30 & 30 & 10 &  10 & 10 & 10 & 10 & 10 & 10 \\
1  & & 12 & 4 & 4 & 4 & 3 \\
2  & & & 4 & 2 & 2 & 2
\end{tabular}
\caption{Values for the highest $g$ for which we have calculated $P^{(n|m)}_g$.}
\label{tablequartic}
\end{table}

\noindent
Note that the numbers in this table are actually smaller than the actual number of calculated polynomials. For example, at $n=5$ and $m=2$, $g$ starts at $\beta_{nm} = -2$. Therefore, the entry of $2$ means that we have calculated the five leading orders. At each of these orders (except for the leading one), the expression contains three polynomials multiplying different powers of the logarithm. Therefore, this entry of $2$ corresponds to a total of $13$ polynomials.

In the table, we have only mentioned the calculated polynomials for $n \geq m$. The ones with $n<m$ differ from those only by a sign,
\be
P^{(n|m)[k]}_g = \left(-1\right)^g P^{(m|n)[k]}_g.
\ee
\noindent
For reasons of space, in this appendix we only reproduce a very small sample of the calculated polynomials. A \textit{Mathematica} file containing all the calculated data is available from the authors.

Let us begin with the perturbative results---that is, $n=m=0$. At this order, the data is most easily reproduced in terms of the polynomials $S_g$ introduced in (\ref{eq:defS}). For the first three of those, we have
\bea
S_2 &=& \frac{27}{2}, \\
S_4 &=& \frac{15309}{8} \left( 5 + 2 X \right), \\
S_6 &=& \frac{177147}{16} \left( 1925 + 2864 X + 111 X^2 \right),
\eea
\noindent
where we substituted $X=\lambda r$. These results exactly match the results that were found in \cite{biz80, msw07}.

For the one--instanton contributions, appearing at $n=1$ and $m=0$, we list the first four of the polynomials $P^{(1|0)}_g$,
\bea\label{eq:Qpol}
P^{(1|0)}_{0} &=& 1, \\
P^{(1|0)}_{1} &=& -\frac{9}{8} \left( 6 + 3 X - 6 X^2 + 2 X^3 \right), \\
P^{(1|0)}_{2} &=& \frac{81}{128} \left( 36 + 36 X + 1665 X^2 - 2844 X^3 + 1800 X^4 - 536 X^5 + 68 X^6 \right), \\
P^{(1|0)}_{3} &=& \frac{243}{5120} \left( 30024 - 234900 X + 608958 X^2 - 3803895 X^3 + 6142554 X^4 - \right. \nonumber \\
&&
\left. - 4634370 X^5 + 2034360 X^6 - 588060 X^7 + 116520 X^8 - 12520 X^9 \right).
\eea
\noindent
These expressions agree with the one--instanton results presented in \cite{m08}.

We now turn to some of the new results. For the two--instanton case, $n=2$ and $m=0$,
\bea
P_{0}^{(2|0)} &=& -\frac{1}{2}, \label{eq:P200} \\
P_{1}^{(2|0)} &=& \frac{3}{8} \left( 18 + 117 X - 102 X^2 + 22 X^3 \right), \\
P_{2}^{(2|0)} &=& -\frac{81}{64} \left( 36 + 468 X + 5577 X^2 - 8204 X^3 + 4460 X^4 - 1128 X^5 + 116 X^6 \right), \\
P_{3}^{(2|0)} &=& \frac{81}{1280} \left( -20088 + 238140 X + 989334 X^2 + 23247945 X^3 - 41702958 X^4 + \right. \nonumber \\
&&
\left. + 29306340 X^5 - 10628280 X^6 + 2188980 X^7 - 276120 X^8 + 20360 X^9 \right). \label{eq:P203}
\eea
\noindent
The main novelty of our method is that we can also calculate contributions with generalized instantons, having the ``wrong sign'' of the instanton action. For example, for $n=m=1$, we have, using the notation introduced in (\ref{eq:defQ}),
\bea
Q^{(1)}_{0} &=& 3 \left( 2-X \right), \\
Q^{(1)}_{2} &=& \frac{729}{8} \left( 72 + 220 X - 380 X^2 + 207 X^3 - 48 X^4 + 4 X^5 \right), \\
Q^{(1)}_{4} &=& \frac{59049}{128} \left( 272160 + 2748816 X - 5760432 X^2 + 4023324 X^3 - \right. \nonumber \\
&&
\left. - 724722 X^4 - 548049 X^5 + 380368 X^6 - 104016 X^7 + 14048 X^8 - 784 X^9 \right).
\eea
\noindent
These results for $n=m=1$ do not yet show all the features of the ``generalized instanton'' expansions. As in the Painlev\'e I case, we find that whenever $n=m$, there are no ``open string'' odd $g$ contributions. Also, in these cases, there are no logarithmic contributions yet. Finally, the perturbative series start at $g=0$. All three of these properties disappear when we go to cases where $n \neq m$. For example, when $n=2$ and $m=1$, we find
\bea
P_{0}^{(2|1)[0]} &=& \frac{1}{4} \left( 54 - 45 X - 6 X^2 + 8 X^3 \right), \\
P_{1}^{(2|1)[0]} &=& \frac{9}{32} \left( 324 - 3132 X + 1197 X^2 + 2052 X^3 - 2202 X^4 +940 X^5 - 164 X^6 \right), \\
P_{2}^{(2|1)[0]} &=& \frac{9}{512} \left( -52488 + 317844 X + 961794 X^2 + 7811559 X^3 - 22378842 X^4 + \right. \nonumber \\
&&
\left. + 23547888 X^5 - 13285728 X^6 + 4500468 X^7 - 914760 X^8 + 89840 X^9 \right),
\eea
\noindent
for the logarithm--free contributions, and
\bea
P_{-1}^{(2|1)[1]} &=& \frac{1}{12}, \\
P_{0}^{(2|1)[1]} &=& -\frac{3}{32} \left( 6 + 3 X - 6 X^2 + 2 X^3 \right), \\
P_{1}^{(2|1)[1]} &=& -\frac{27}{512} \left( 36 + 36 X + 1665 X^2 - 2844 X^3 + 1800 X^4 - 536 X^5 + 68 X^6 \right),
\eea
\noindent
for the one--logarithm contributions. Notice that the latter polynomials are essentially the same as the $P_g^{(1|0)}$ reported starting in (\ref{eq:Qpol}). In fact, we find in general that
\be
P_{g}^{(2|1)[1]} = \frac{1}{12}\, P_{g+1}^{(1|0)[0]}.
\ee
\noindent
This relation can be easily derived from the recursion relations that follow from the string equation. In fact, our expression above is simply the analogue of (\ref{eq:propPI}), and it is the first in a sequence of equations that are analogous to the relations (\ref{eq:logprop}) that we have found for the Painlev\'e I transseries coefficients. For the general case, one obtains
\be\label{eq:logrelP}
P^{(n|m)[k]}_g = \frac{1}{k!} \left( \frac{\left( n-m \right)}{12} \right)^k P^{(n-k|m-k)[0]}_{g+k}.
\ee
\noindent
Thus, once again, the logarithmic contributions are simply related to the logarithm--free contributions, and do not seem to constitute new physical sectors.

Using the formula (\ref{eq:QMMds}), the reader can check that, in the double--scaling limit, the data we have presented so far exactly reproduces the expansions for
\be
\Phi_{(0|0)}^{[0]}, \quad \Phi_{(1|0)}^{[0]}, \quad \Phi_{(2|0)}^{[0]}, \quad \Phi_{(1|1)}^{[0]}, \quad \Phi_{(2|1)}^{[0]} \quad \text{and} \quad \Phi_{(2|1)}^{[1]},
\ee
\noindent
that were listed in appendix \ref{app:PI}. For completeness, we also tabulate the coefficients of all other polynomials that are needed to reproduce the expansion coefficients we gave in that appendix.

\begin{table}[t]
\centering
\begin{tabular}{c|rrr}
$g$     & 0 & 1 & 2 \\
\hline
$c$     & $-\frac{3}{8}$ & $\frac{3}{32}$ & $-\frac{27}{256}$ \\
\hline
$X^0$ &    36 &        0 &    $-$11664 \\
$X^1$ &    18 &    23328 &      122472 \\
$X^2$ & $-$38 &    27432 &     3170988 \\
$X^3$ &    11 & $-$73476 &     9125514 \\
$X^4$ &       &    49311 &    25985394 \\
$X^5$ &       & $-$14442 &    24283071 \\
$X^6$ &       &     1667 & $-$11842992 \\
$X^7$ &       &          &     3354462 \\
$X^8$ &       &          &   $-$544628 \\
$X^9$ &       &          &       40996
\end{tabular}
\hspace{2em}
\begin{tabular}{c|rrr}
$g$     & $-$1 & 0 & 1 \\
\hline
$c$     & $-\frac{1}{12}$ & $\frac{1}{16}$ & $-\frac{27}{128}$ \\
\hline
$X^0$ & 1 &     18 &      36 \\
$X^1$ &   &    117 &     468 \\
$X^2$ &   & $-$102 &    5577 \\
$X^3$ &   &     22 & $-$8204 \\
$X^4$ &   &        &    4460 \\
$X^5$ &   &        & $-$1128 \\
$X^6$ &   &        &     116
\end{tabular}
\caption{Prefactor $c$ and coefficients of the polynomials $P^{(3|1)[0]}_g$ (left) and $P^{(3|1)[1]}_g$ (right).}
\end{table}

\begin{table}[t]
\centering
\begin{tabular}{c|rrr}
$g$     & 0 & 2 & 4 \\
\hline
$c$     & $-\frac{9}{4}$ & $\frac{81}{16}$ & $-\frac{6561}{64}$ \\
\hline
$X^0$    & $-$72 &     326592 &   $-$687802752 \\
$X^1$    &    78 &     255636 &  $-$2925199980 \\
$X^2$    & $-$31 & $-$1268946 &     9776740014 \\
$X^3$    &     5 &    1263654 & $-$10514590074 \\
$X^4$    &       &     603801 &     4732494984 \\
$X^5$    &       &     154827 &      148363974 \\
$X^6$    &       &   $-$20062 &  $-$1271607633 \\
$X^7$    &       &        950 &      701712243 \\
$X^8$    &       &            &   $-$203346798 \\
$X^9$    &       &            &       34993318 \\
$X^{10}$ &       &            &     $-$3454976 \\
$X^{11}$ &       &            &         156840
\end{tabular}
\hspace{2em}
\caption{Prefactor $c$ and coefficients of the polynomials $Q^{(2)}_g$.}
\end{table}

\begin{table}[t]
\centering
\begin{tabular}{c|rrr}
$g$     & $-$1 & 0 & 1 \\
\hline
$c$     & $-\frac{1}{16}$ & $\frac{1}{128}$ & $-\frac{9}{2048}$ \\
\hline
$X^0$ &    72 &   $-$7776 &      443232 \\
$X^1$ &    36 &    101088 &  $-$4000752 \\
$X^2$ & $-$90 & $-$137700 &    24782112 \\
$X^3$ &    29 &     44280 & $-$22509576 \\
$X^4$ &       &     24687 &  $-$5930982 \\
$X^5$ &       &  $-$20094 &    21534309 \\
$X^6$ &       &      3941 & $-$17087760 \\
$X^7$ &       &           &     7682442 \\
$X^8$ &       &           &  $-$2022868 \\
$X^9$ &       &           &      240208
\end{tabular}
\hspace{2em}
\begin{tabular}{c|rrr}
$g$     & $-$1 & 0 & 1 \\
\hline
$c$     & $\frac{1}{16}$ & $-\frac{1}{128}$ & $\frac{3}{512}$ \\
\hline
$X^0$ &    72 &      7776 &        69984 \\
$X^1$ &    36 &    194400 &       944784 \\
$X^2$ & $-$90 & $-$106272 &     64513584 \\
$X^3$ &    29 & $-$175392 &  $-$20419776 \\
$X^4$ &       &    197802 & $-$135263034 \\
$X^5$ &       &  $-$73836 &    182249163 \\
$X^6$ &       &      9937 & $-$108164682 \\
$X^7$ &       &           &     35518077 \\
$X^8$ &       &           &   $-$6475770 \\
$X^9$ &       &           &       528388
\end{tabular}
\caption{Prefactor $c$ and coefficients of the polynomials $P^{(3|2)[0]}_g$ (left) and $P^{(4|2)[0]}_g$ (right).}
\end{table}

\begin{table}[t]
\centering
\begin{tabular}{c|rrr}
$g$     & $-$1 & 0 & 1 \\
\hline
$c$     & $\frac{1}{48}$ & $-\frac{3}{128}$ & $-\frac{3}{2048}$ \\
\hline
$X^0$ &    54 &  $-$324 &    $-$52488 \\
$X^1$ & $-$45 &    3132 &      317844 \\
$X^2$ &  $-$6 & $-$1197 &      961794 \\
$X^3$ &     8 & $-$2052 &     7811559 \\
$X^4$ &       &    2202 & $-$22378842 \\
$X^5$ &       &  $-$940 &    23547888 \\
$X^6$ &       &     164 & $-$13285728 \\
$X^7$ &       &         &     4500468 \\
$X^8$ &       &         &   $-$914760 \\
$X^9$ &       &         &       89840
\end{tabular}
\hspace{2em}
\begin{tabular}{c|rrr}
$g$     & $-$1 & 0 & 1 \\
\hline
$c$     & $-\frac{1}{16}$ & $\frac{1}{64}$ & $-\frac{9}{512}$ \\
\hline
$X^0$ &    36 &        0 &    $-$11664 \\
$X^1$ &    18 &    23328 &      122472 \\
$X^2$ & $-$38 &    27432 &     3170988 \\
$X^3$ &    11 & $-$73476 &     9125514 \\
$X^4$ &       &    49311 & $-$25985394 \\
$X^5$ &       & $-$14442 &    24283071 \\
$X^6$ &       &     1667 & $-$11842992 \\
$X^7$ &       &          &     3354462 \\
$X^8$ &       &          &   $-$544628 \\
$X^9$ &       &          &       40996
\end{tabular}
\caption{Prefactor $c$ and coefficients of the polynomials $P^{(3|2)[1]}_g$ (left) and $P^{(4|2)[1]}_g$ (right).}
\end{table}

\begin{table}[t]
\centering
\begin{tabular}{c|rrr}
$g$     & $-$2 & $-$1 & 0 \\
\hline
$c$     & $\frac{1}{288}$ & $-\frac{1}{256}$ & $\frac{9}{4096}$ \\
\hline
$X^0$ & 1 &    6 &      36 \\
$X^1$ &   &    3 &      36 \\
$X^2$ &   & $-$6 &    1665 \\
$X^3$ &   &    2 & $-$2844 \\
$X^4$ &   &      &    1800 \\
$X^5$ &   &      &  $-$536 \\
$X^6$ &   &      &      68 
\end{tabular}
\hspace{2em}
\begin{tabular}{c|rrr}
$g$     & $-$2 & $-$1 & 0 \\
\hline
$c$     & $-\frac{1}{144}$ & $\frac{1}{192}$ & $-\frac{9}{512}$ \\
\hline
$X^0$ & 1 &     18 &      36 \\
$X^1$ &   &    117 &     468 \\
$X^2$ &   & $-$102 &    5577 \\
$X^3$ &   &     22 & $-$8204 \\
$X^4$ &   &        &    4460 \\
$X^5$ &   &        & $-$1128 \\
$X^6$ &   &        &     116
\end{tabular}
\caption{Prefactor $c$ and coefficients of the polynomials $P^{(3|2)[2]}_g$ (left) and $P^{(4|2)[2]}_g$ (right).}
\end{table}

\section{The Double--Scaling Limit: Structural Data}\label{app:QMMFE}

In this appendix, and analogously to the previous one, we present some of the polynomials $\CP^{(n|m)}_g (t)$ that determine the free energy (\ref{eq:Fnmstruct}) of the quartic matrix model. Table \ref{table:gmaxP} shows to which index $g$ we have calculated these polynomials. As will be clear when comparing this table to the analogous table in the previous appendix, the amount of available $\CF^{(n|m)}_g (t)$ data is much smaller than the amount of $R^{(n|m)}_g (x)$ data. The reason for this is that the procedure used to calculate $\CF^{(n|m)}_g (t)$ from $R^{(n|m)}_g (x)$, using the Euler--MacLaurin formula, is rather time consuming. We have therefore chosen to do the tests of resurgence for the quartic matrix model directly at the level of $R^{(n|m)}_g (x)$, where one can construct a sufficient amount of data much more easily. The $\CF^{(n|m)}_g (t)$ for which the data are presented in this appendix mainly serve the purpose of checking that the quartic matrix model free energy gives the $(2,3)$ minimal model free energy in the double--scaling limit.

\begin{table}[t]
\centering
\begin{tabular}{c|ccccccccccc}
\begin{picture}(20,20)(0,0)
\put(17,8){$n$}
\put(3.6,18){\line(1,-1){22.3}}
\put(3,0){$m$}
\end{picture}
&  0 &  1  &  2  &  3  & 4 \\
\hline
0 & 25 & 7/2 &  4  &     &   \\
1 &    &  5  & 7/2 & 4   &   \\
2 &    &     &  4  & 7/2 & 4 
\end{tabular}
\caption{Values of the highest $g$ for which we have calculated $\CP^{(n|m)}_g (t)$.}
\label{table:gmaxP}
\end{table}

As usual, we only present results with $n \geq m$. The results for $n<m$ are related to those by
\be
\CF^{(n|m)}_g (t) = (-1)^{\frac{2g-n-m}{2}}\, \CF^{(m|n)}_g (t).
\ee
\noindent
Results for $n=m=0$ were already listed in (\ref{eq:FQMM1}--\ref{eq:FQMM2}) in the main text. We have also listed two exceptional results in there, (\ref{eq:11exc}) for $n=m=1$ and (\ref{eq:22exc}) for $n=m=2$. For all other (regular) results, we give the noninteger prefactors $c$ alongside with the integer coefficients of the polynomials $\CP^{(n|m)}_g (t)$, defined in (\ref{eq:Fnmstruct}), in tables \ref{table:P11P22}--\ref{table:P32P42}. 

In the first column of each table, we list the monomial $t^n$ that the coefficients in that row multiply. The corresponding power of $\alpha$ is easily derived from the fact that the whole polynomial is homogeneous in $t$ and $\alpha^2$, with the highest power a pure power of $t$. Thus, if $n_{\text{max}}$ is the index of the highest coefficient in a certain column, the coefficient in the row labeled $t^n$ of that column actually multiplies $t^n \alpha^{2(n_{\text{max}}-n)}$.

\begin{table}[t]
\centering
\begin{tabular}{c|rrr}
$g$ & 1 & 3 & 5 \\
\hline
$c$ & $-\frac{1}{6 \sqrt{3}}$ & $\frac{1}{180 \sqrt{3}}$ & $-\frac{1}{378 \sqrt{3}}$ \\
\hline
$t^0   $ & $-$2 &     520 &    $-$61908 \\
$t^1   $ & $-$3 &    2835 &   $-$574056 \\
$t^2   $ &    8 &    3642 &  $-$1614616 \\
$t^3   $ &      &$-$16512 &     1807479 \\
$t^4   $ &      & $-$5472 &     8602998 \\
$t^5   $ &      &    1950 &    17467588 \\
$t^6   $ &      &   51840 & $-$66986172 \\
$t^7   $ &      &$-$36000 &    39683718 \\
$t^8   $ &      &$-$19200 &  $-$60738324 \\
$t^9   $ &      &   16640 &   220690302 \\
$t^{10}$ &      &         &$-$232460928 \\
$t^{11}$ &      &         &    52828048 \\
$t^{12}$ &      &         &    14853888 \\
$t^{13}$ &      &         &    35051520 \\
$t^{14}$ &      &         & $-$38348800 \\
$t^{15}$ &      &         &     9805824
\end{tabular}
\hspace{2em}
\begin{tabular}{c|rr}
$g$ & 2 & 4 \\
\hline
$c$ & $\frac{1}{486}$ & $\frac{1}{43740}$ \\
\hline
$t^0   $    &    130 &      396710 \\
$t^1   $    &    516 &     3402120 \\
$t^2   $    &   2229 &    12327720 \\
$t^3   $    &$-$5408 & $-$20516720 \\
$t^4   $    &   1896 &    12385215 \\
$t^5   $    & $-$816 &$-$230785920 \\
$t^6   $    &   1696 &   536735424 \\
$t^7   $    &        &$-$513929952 \\
$t^8   $    &        &   490487040 \\
$t^9   $    &        &$-$569834240 \\
$t^{10}$    &        &   320398080 \\
$t^{11}$    &        &  $-$6978048 \\
$t^{12}$    &        & $-$34264576
\end{tabular}
\caption{Prefactor $c$ and coefficients of the polynomials $\CP^{(1|1)}_g (t)$ (left) and $\CP^{(2|2)}_g (t)$ (right).}
\label{table:P11P22}
\end{table}

\begin{table}[t]
\centering
\begin{tabular}{c|rrrr}
$g$ & 1/2 & 3/2 & 5/2 & 7/2 \\
\hline
$c$ & $-\frac{\sqrt{2}}{3^{5/4}}$ & $\frac{1}{6 \cdot 3^{3/4} \sqrt{2}}$ & $-\frac{1}{144 \cdot 3^{1/4} \sqrt{2}}$ & $-\frac{1}{8640 \cdot 3^{3/4}\sqrt{2}}$ \\
\hline
$t^0$ & 1 &$-$10 &    676 &     517000 \\
$t^1$ &   &$-$21 &   2820 &    3246300 \\
$t^2$ &   &$-$12 &   2697 &    5408118 \\
$t^3$ &   &   40 &$-$9224 &$-$10506063 \\
$t^4$ &   &      &$-$2208 &$-$15792588 \\
$t^5$ &   &      &   3648 & $-$4743720 \\
$t^6$ &   &      &   1600 &   44745600 \\
$t^7$ &   &      &        &$-$12288960 \\
$t^8$ &   &      &        &$-$17130240 \\
$t^9$ &   &      &        &    6540800
\end{tabular}
\hspace{2em}
\begin{tabular}{c|rrrr}
$g$ & 1 & 2 & 3 & 4 \\
\hline
$c$ & $-\frac{1}{9\sqrt{3}}$ & $\frac{1}{162}$ & $-\frac{1}{648 \sqrt{3}}$ & $\frac{1}{58320}$ \\
\hline
$t^0$    &$-$1 & $-$22 &   $-$316 &     $-$22520 \\
$t^1$    &   4 &   253 &    10564 &      3903200 \\
$t^2$    &     &$-$564 & $-$41715 &  $-$18769266 \\
$t^3$    &     &  1096 &   168044 &    125672865 \\
$t^4$    &     &$-$736 &$-$341936 & $-$421619748 \\
$t^5$    &     &       &   393408 &    941275296 \\
$t^6$    &     &       &$-$281920 &$-$1561721280 \\
$t^7$    &     &       &    93952 &   1764081600 \\
$t^8$    &     &       &          &$-$1258640640 \\
$t^9$    &     &       &          &    530946560 \\
$t^{10}$ &     &       &          & $-$105113600
\end{tabular}
\caption{Prefactor $c$ and coefficients of the polynomials $\CP^{(1|0)}_g (t)$ (top) and $\CP^{(2|0)}_g (t)$ (bottom).}
\label{table:P10P20}
\end{table}

\begin{table}[t]
\centering
\begin{tabular}{c|rrr}
$g$ & 3/2 & 5/2 & 7/2 \\
\hline
$c$ & $-\frac{\sqrt{2}}{27 \cdot 3^{3/4}}$ & $\frac{1}{54 \cdot 3^{1/4} \sqrt{2}}$ & $-\frac{1}{11664 \cdot 3^{3/4} \sqrt{2}}$ \\
\hline
$t^0$ &$-$14 &    184 &   $-$638120 \\
$t^1$ &$-$21 &    888 &  $-$4532580 \\
$t^2$ &$-$96 &   2665 & $-$12820266 \\
$t^3$ &  104 &$-$3972 &    13158375 \\
$t^4$ &      & $-$144 &    10689480 \\
$t^5$ &      &$-$2848 &    70972776 \\
$t^6$ &      &   3200 &$-$114864000 \\
$t^7$ &      &        &    24333120 \\
$t^8$ &      &        &    11455488 \\
$t^9$ &      &        &     2252288
\end{tabular}
\hspace{2em}
\begin{tabular}{c|rrr}
$g$ & 2 & 3 & 4 \\
\hline
$c$ & $\frac{2}{81}$ & $\frac{2}{729 \sqrt{3}}$ & $\frac{1}{2916}$ \\
\hline
$t^0$    &     4 &      341 &        5032 \\
$t^1$    & $-$49 &  $-$6408 &   $-$271696 \\
$t^2$    &   144 &    25197 &     1575868 \\
$t^3$    &$-$310 &$-$113287 & $-$10610537 \\
$t^4$    &   184 &   173664 &    36078160 \\
$t^5$    &       &$-$187692 & $-$86311034 \\
$t^6$    &       &    95792 &   150067240 \\
$t^7$    &       &          &$-$173970320 \\
$t^8$    &       &          &   130130752 \\
$t^9$    &       &          & $-$59481984 \\
$t^{10}$ &       &          &    12789248
\end{tabular}
\caption{Prefactor $c$ and coefficients of the polynomials $\CP^{(2|1)}_g (t)$ (left) and $\CP^{(3|1)}_g (t)$ (right). The case $\CP^{(3|1)}_3 (t)$ is exceptional, in the sense that it factorizes: the polynomial displayed here should be multiplied by $(t-\alpha^2)$ to obtain $\CP^{(3|1)}_3 (t)$.}
\label{table:P21P31}
\end{table}

\begin{table}[t]
\centering
\begin{tabular}{c|rrr}
$g$ & 3/2 & 5/2 & 7/2 \\
\hline
$c$ & $\frac{2\sqrt{2}}{27 \cdot 3^{3/4}}$ & $\frac{1}{729 \cdot 3^{1/4} \sqrt{2}}$ & $\frac{1}{1944 \cdot 3^{3/4} \sqrt{2}}$ \\
\hline
$t^0$ & $-$2 &     896 &   $-$39752 \\
$t^1$ &    6 &    2706 &  $-$292168 \\
$t^2$ &$-$42 &   24537 & $-$1595714 \\
$t^3$ &   29 &$-$30592 &    1325412 \\
$t^4$ &      &   38919 & $-$3580714 \\
$t^5$ &      &$-$79788 &   17588671 \\
$t^6$ &      &   42836 &$-$19468680 \\
$t^7$ &      &         &    8762744 \\
$t^8$ &      &         & $-$5577248 \\
$t^9$ &      &         &    2872832
\end{tabular}
\hspace{2em}
\begin{tabular}{c|rrr}
$g$ & 2 & 3 & 4 \\
\hline
$c$ & $-\frac{4}{243}$ & $-\frac{2}{2187\sqrt{3}}$ & $\frac{1}{39366}$ \\
\hline
$t^0$    & $-$2 &   $-$488 &     $-$75448 \\
$t^1$    &    6 &     9686 &      2192188 \\
$t^2$    &$-$42 & $-$54240 &  $-$15529806 \\
$t^3$    &   29 &   242728 &    105787830 \\
$t^4$    &      &$-$526987 & $-$368392458 \\
$t^5$    &      &   765960 &    949167207 \\
$t^6$    &      &$-$650210 &$-$1732001196 \\
$t^7$    &      &   214280 &   2088163092 \\
$t^8$    &      &          &$-$1656852624 \\
$t^9$    &      &          &    806826880 \\
$t^{10}$ &      &          & $-$179279104
\end{tabular}
\caption{Prefactor $c$ and coefficients of the polynomials $\CP^{(3|2)}_g (t)$ (left) and $\CP^{(4|2)}_g (t)$ (right). The case $\CP^{(4|2)}_2 (t)$ is exceptional, in the sense that it factorizes: the polynomial displayed here should be multiplied by $(4t-\alpha^2)$ to obtain $\CP^{(4|2)}_2 (t)$. It is curious to see that the remaining factor is proportional to $\CP^{(3|2)}_{3/2} (t)$.}
\label{table:P32P42}
\end{table}

\section{Stokes Automorphism of Two--Parameters Instanton Series}\label{app:2-param-stokes-autom}

An expression for the general ordered product of $k$ alien derivatives, of the form $\prod_{i=1}^{k} \Delta_{-\ell_{k+1-i}A} = \Delta_{-\ell_{k}A} \cdots \Delta_{\ell_{1}A}$, acting on $\Phi_{(n|0)}$, was presented in section \ref{sub:Bridge-Eqns-revisited}, namely expression (\ref{k-alien-der-2-param-youngT}). In this appendix we shall outline an inductive proof of this result. First recall what this expression was,
\begin{eqnarray}
\prod_{i=1}^{k} \Delta_{-\ell_{k+1-i}A} \Phi_{(n|0)} &=& \sum_{m=0}^{k}\, \sum_{\delta_{s} \in \Gamma(k,k-m+1)}\, \prod_{s=1}^{k} \bigg\{ \bigg[ \left( s+1-\delta_{s} \right) \widetilde{S}_{-\ell_{s}}^{(\mathbf{d}\delta_{s})} + \label{k-alien-der-2-param-youngT-app} \\
&&
\left. \left. + \left( n-\sum_{i=1}^{s}\ell_{i}+s+1-\delta_{s} \right) S_{-\ell_{s}}^{(\ell_{s}+\mathbf{d}\delta_{s})} \right] \Theta \left( s+1-\delta_{s} \right) \right\} \Phi_{\left. \left( n+m-\sum_{i=1}^{k}\ell_{i} \right| m \right)}. \nonumber
\end{eqnarray}
\noindent
Further recall that in section \ref{sub:Bridge-Eqns-revisited} we have explicitly shown that for the case of $k=2$ (and analogously for the case of $k=1$) this closed form expression correctly reproduced the result we had earlier computed in (\ref{2alien-der-2-param-inst-series}).

Assuming that the above result (\ref{k-alien-der-2-param-youngT-app}) holds true for a particular value of $k>2$, let us apply one more alien derivative $\Delta_{-\ell_{k+1}A}$, with $\ell_{k+1}>0$, to this expression. Notice that this alien derivative, $\Delta_{-\ell_{k+1}A}$, will only act on $\Phi_{(n+m-\sum_{i=1}^{k}\ell_{i}|m)}$, and this action was already computed in (\ref{extra-der-in-general-2par-term}). We thus find
\bea
\Delta_{-\ell_{k+1}A} \prod_{i=1}^{k} \Delta_{-\ell_{k+1-i}A} \Phi_{(n|0)} &=& \sum_{m=0}^{k+1} \Theta \left( n+m-\sum_{i=1}^{k}\ell_{i} \right) \sum_{q=0}^{m} 
\left( \left( n+m-\sum_{i=1}^{k+1}\ell_{i}-q \right) S_{-\ell_{k+1}}^{(\ell_{k+1}+q)} + \right. \nonumber \\
&&
\hspace{-80pt}
+ \left( m-q \right) \widetilde{S}_{-\ell_{k+1}}^{(q)} \bigg) \sum_{\delta_{s}\in\Gamma(k,k-m+2)}\, \prod_{s=1}^{k} \bigg\{ \bigg[ \left( s+1-\delta_{s} \right) \widetilde{S}_{-\ell_{s}}^{(\mathbf{d}\delta_{s})} + \\
&&
\hspace{-80pt}
\left. \left. + \left( n-\sum_{i=1}^{s}\ell_{i}+s+1-\delta_{s} \right) S_{-\ell_{s}}^{(\ell_{s}+\mathbf{d}\delta_{s})} \right] \Theta \left( s+1-\delta_{s} \right) \right\} \Phi_{\left. \left( n+m-\sum_{i=1}^{k+1}\ell_{i}-q \right| m-q \right)}. \nonumber
\eea
\noindent
In order to obtain the expression above, we have changed the variable in the first sum of (\ref{k-alien-der-2-param-youngT-app}) from $\sum_{m=0}^{k} \rightarrow \sum_{m'=1}^{k+1}$, after which one realizes that one may always add the term $m'=0$ as it is zero. The next steps include the change of variables $\sum_{q=0}^{m} = \sum_{q'\equiv m-q=0}^{m}$ and noticing that one can further change the order of the sums as
\begin{equation}
\sum_{m=0}^{k+1}\sum_{q'=0}^{m}=\sum_{q'=0}^{k+1}\sum_{m=q'}^{k+1}.
\end{equation}
\noindent
In this process we thus obtain
\bea
\Delta_{-\ell_{k+1}A} \prod_{i=1}^{k} \Delta_{-\ell_{k+1-i}A} \Phi_{(n|0)} &=& \sum_{q'=0}^{k+1} \sum_{m=q'}^{k+1} \left( \left( n+q'-\sum_{i=1}^{k+1}\ell_{i} \right) S_{-\ell_{k+1}}^{(\ell_{k+1}+m-q')} + q' \cdot \widetilde{S}_{-\ell_{k+1}}^{(m-q')} \right) \times \nonumber \\
&&
\hspace{-90pt}
\times\, \sum_{\delta_{s}\in\Gamma(k,k-m+2)}\, \prod_{s=1}^{k} \left\{ \left[ \left( s+1-\delta_{s} \right) \widetilde{S}_{-\ell_{s}}^{(\mathbf{d}\delta_{s})} + \left( n-\sum_{i=1}^{s}\ell_{i}+s+1-\delta_{s} \right) S_{-\ell_{s}}^{(\ell_{s}+\mathbf{d}\delta_{s})} \right] \times \right. \nonumber \\
&&
\hspace{-90pt}
\times\, \Theta \left( s+1-\delta_{s} \right) \Bigg\}\, \Phi_{\left. \left( n-\sum_{i=1}^{k+1}\ell_{i}+q' \right| q' \right)}.
\eea
\noindent
The final step is to change variables yet one more time, as $\sum_{m=q'}^{k+1} = \sum_{m'=k+2-m=1}^{k+2-q'}$, and introduce a new variable, $\gamma_{k+1}=k+2-q'$. Then
\bea
\Delta_{-\ell_{k+1}A} \prod_{i=1}^{k} \Delta_{-\ell_{k+1-i}A} \Phi_{(n|0)} &=& \sum_{q'=0}^{k+1} \delta_{\gamma_{k+1},k+2-q'} \sum_{m'=1}^{\gamma_{k+1}} \left( \left( n+k+2-\gamma_{k+1}-\sum_{i=1}^{k+1}\ell_{i} \right) \times \right. \nonumber \\
&&
\hspace{-105pt}
\times\, S_{-\ell_{k+1}}^{(\ell_{k+1}+\gamma_{k+1}-m')} + \left( k+2-\gamma_{k+1} \right) \widetilde{S}_{-\ell_{k+1}}^{(\gamma_{k+1}-m')} \Bigg)\, \sum_{\delta_{s}\in\Gamma(k,m')}\, \prod_{s=1}^{k} \Bigg\{ \Bigg[ \widetilde{S}_{-\ell_{s}}^{(\mathbf{d}\delta_{s})} \left( s+1-\delta_{s} \right) + \nonumber \\
&&
\hspace{-105pt}
\left. \left. + S_{-\ell_{s}}^{(\ell_{s}+\mathbf{d}\delta_{s})} \left( n-\sum_{i=1}^{s}\ell_{i}+s+1-\delta_{s} \right) \right] \Theta \left( s+1-\delta_{s} \right) \right\} \Phi_{\left. \left( n-\sum_{i=1}^{k+1}\ell_{i}+q' \right| m-q' \right)}.
\eea
\noindent
Finally recalling that $\delta_{s} \in \Gamma(k,m')$ means that we are summing over all $\delta_{s} : \, 0 < \delta_{1} \le \cdots \le \delta_{n} = m'$, and that now $m' = 1, \cdots, \gamma_{k+1} = k+2-q' \le k+2$, one can naturally rewrite the above expression as a sum over Young diagrams, of length $k+1$, obtaining
\bea
\Delta_{-\ell_{k+1}A} \prod_{i=1}^{k} \Delta_{-\ell_{k+1-i}A} \Phi_{(n|0)} &=& \sum_{q'=0}^{k+1}\, \sum_{\delta_{s}\in\Gamma(k+1,k+2-q')}\, \prod_{s=1}^{k+1} \Bigg\{ \Bigg[ \left( s+1-\delta_{s} \right) \widetilde{S}_{-\ell_{s}}^{(\mathbf{d}\delta_{s})} + \\
&&
\hspace{-60pt}
\left. \left. + \left( n-\sum_{i=1}^{s}\ell_{i}+s+1-\delta_{s} \right) S_{-\ell_{s}}^{(\ell_{s}+\mathbf{d}\delta_{s})} \right] \Theta \left( s+1-\delta_{s} \right) \right\} \Phi_{\left. \left( n-\sum_{i=1}^{k+1}\ell_{i}+q' \right| m-q' \right)}. \nonumber
\eea
\noindent
This is the expected result for the ordered product of $k+1$ alien derivatives, acting on the instanton series $\Phi_{(n|0)}$, as shown in (\ref{k-alien-der-2-param-youngT-app}). It thereby concludes our proof.

\newpage


\bibliographystyle{plain}

\end{document}